\documentclass[a4paper,12pt,Div=12,notitlepage]{scrartcl}
\usepackage[left=2.25cm,right=2.25cm,top=2.25cm,bottom=2.25cm]{geometry}
\usepackage[manualmark]{scrpage2}
\usepackage{fix-cm}
\usepackage[latin1]{inputenc}
\usepackage[ngerman]{babel}
\usepackage[T1]{fontenc}
\usepackage{ae}
\usepackage{amsmath}
\usepackage{amssymb}
\usepackage{graphicx}
\usepackage{pdfpages}
\usepackage[arrow, matrix, curve]{xy}
\usepackage{multirow}
\usepackage{bigdelim}
\usepackage{mathptmx}

\begin{document}

\title{\LARGE Die einheitliche Beschreibung der fundamentalen Objekte und
Wechselwirkungen der Natur in der Quantentheorie der Ur-Alternativen}

\author{Martin Immanuel Kober\footnote{E-mail-Adresse: Martin.Immanuel.Kober@t-online.de}\\
{\normalsize Kettenhofweg 121, 60325 Frankfurt am Main, Deutschland}}

\date{}

\maketitle

\begin{abstract}
\footnotesize{Die Quantentheorie der Ur"=Alternativen des Carl Friedrich
von Weizsäcker begründet die Existenz aller fundamentalen Objekte,
ihrer Wechselwirkungen und des Ortsraumes basierend auf logischen
Alternativen in der Zeit als fundamentalster möglicher Objektivation
der Natur im menschlichen Geist. Wenn man eine Quantenlogik zugrunde
legt und die Alternativen in binäre Alternativen unterteilt, erhält
man Ur"=Alternativen. Die Ur"=Alternativen, dies ist begrifflich
das Entscheidende, befinden sich nicht in einem vorgegebenen Ortsraum,
beziehen sich nicht auf eine von ihnen unterschiedene Substanz und
benötigen keinen physikalischen Träger. Ein solches Bild der
natürlichen Realität, das klassische feldtheoretische Begriffe
basierend auf in einem Raum mit lokaler Kausalitätsstruktur sich
bewegenden voneinander trennbaren Objekten endgültig überwindet,
zeigt sich in empirisch unabweisbarer Weise im EPR"=Paradoxon.
Es muss mit Hilfe einer durch die Kopenhagener Deutung der
Quantentheorie beeinflussten Interpretation der Kantischen
Philosophie tiefer verstanden werden. Das bedeutet, dass sich
in der Quantentheorie der Ur"=Alternativen wie in der Hegelschen
Philosophie die Natur letztendlich aus reiner Logik konstituiert.
Desweiteren muss lediglich die Zeit als Parameter vorausgesetzt werden,
was zu einem Automorphismus in einem komplexen Hilbertraum führt.
Die rein logisch definierten Quantenobjekte stellen sich allerdings
umgekehrt in einem Ortsraum indirekt dar. Diese seitens von Weizsäcker
vollzogene "`Kopernikanische Wende"' wird in dieser Arbeit auch
mathematisch konsistent gefasst. Denn die gegenüber Permutation
symmetrischen Zustände im Tensorraum vieler Ur"=Alternativen können
über die Definition quantenlogischer Operatoren aus den Erzeugungs-
und Vernichtungsoperatoren im Tensorraum, welche der Heisenbergschen
Algebra genügen und daher als Orts- und Impulsoperatoren interpretiert
werden, in einen reellen dreidimensionalen Raum abgebildet werden,
was eine (3+1)"=dimensionale Raum"=Zeit impliziert. Über die
Unterscheidung zwischen symmetrischen und antisymmetrischen
Zuständen vieler Ur"=Alternativen wird die Existenz des
Spin und der inneren Eichsymmetrien der Elementarteilchen
begründet. Denn die Basiszustände im antisymmetrischen Tensorraum
weisen eine Symmetriegruppe $SO(16)$ beziehungsweise $U(8)$ auf.
Dessen Teilräume mit Symmetriegruppen $U(4)$ (Quarksektor),
$U(3)$ (Leptonensektor) und $U(1)$ (skalarer Sektor) können ihrerseits
gemäß $U(4)=SU(3)\otimes SU(2)\otimes SU(2)\otimes U(1) \otimes U(1)$
und $U(3)=SU(2)\otimes SU(2)\otimes U(1)\otimes U(1)\otimes U(1)$
aufgespalten werden. Jeweils eine $U(1)$ kann mit der Hyperladung,
eine $SU(2)$ mit dem Spin, eine $SU(2)$ mit dem schwachen Isospin
und die $SU(3)$ mit dem Farbfreiheitsgrad identifiziert werden.
Zustände im symmetrischen Tensorraum sind zudem symmetrisch unter
beliebigen quantenlogischen Translationen. Über eine Forderung
einer von den quantenlogischen Ortsoperatoren abhängigen abstrakten
Translations"=Symmetrie im symmetrischen Tensorraum und entsprechenden
$SU(N)$"=Symmetrien im antisymmetrischen Tensorraum kann eine
quantenlogische Fassung der Eichtheorien der allgemeinen Relativitätstheorie
und der Elementarteilchenphysik hergeleitet werden. Die darin enthaltenen
Tensorprodukte konstituieren Verschränkungsbeziehungen als abstraktes
Analogon zur Wechselwirkung. Damit ist nicht nur die Gravitation in einem
rein quantentheoretischen Rahmen gefasst, sondern es sind auch alle fundamentalen
Wechselwirkungen der Natur vereinheitlicht. Eine fundamentalere Physik
kann es nicht geben, denn eine sich aus reiner Logik ergebende binäre
Alternative ist die einfachste und abstrakteste Struktur rationalen Denkens.
Schließlich ergibt sich eine erweiterte $E_8$"=Symmetrie, welche sich
auf bestimmte Transformationenen im symmetrischen und antisymmertrischen
Teilraum zugleich gründet.}
\end{abstract}

\fontsize{12.8pt}{14.8pt}\selectfont

\pagestyle{scrheadings}

\chead[\pagemark]{\pagemark}
\ohead[]{}
\cfoot[]{}
\ofoot[]{}

\newpage

\fontsize{15}{17}\selectfont

\null\vfill

\begin{center}
\large
\textbf{\quad\quad\quad gewidmet in ewiger Dankbarkeit und Liebe\newline
\\
\vspace{1cm}
\Large
\quad\ \ Albert Einstein, Niels Bohr, Werner Heisenberg\newline
\\
\quad\quad\ \ und Carl Friedrich von Weizsäcker,\newline
\\
\vspace{1cm}
\large
in deren Tradition und Dienst ich als theoretischer Physiker stehe}
\vspace{1cm}
\end{center}

\null\vfill

\newpage

\fontsize{12.8pt}{14.8pt}\selectfont

\tableofcontents

\newpage

\begin{quote}
\textbf{\textit{"`'Du möchtest also', fügte ich ein, 'die Elementarteilchen, und damit
schließlich die Welt, in der gleichen Weise aus Alternativen aufbauen, wie Plato seine
regulären Körper und damit auch die Welt aus Dreiecken aufbauen wollte. Die Alternativen
sind ebensowenig Materie wie die Dreiecke in Platos >Timaios<. Aber wenn man die Logik
der Quantentheorie zugrunde legt, so ist die Alternative eine Grundform, aus der
kompliziertere Grundformen durch Wiederholung entstehen. Der Weg soll also,
wenn ich dich richtig verstanden habe, von der Alternative zu einer Symmetriegruppe,
das heißt zu einer Eigenschaft führen; die Darstellenden einer oder mehrerer
Eigenschaften sind die mathematischen Formen, die die Elementarteilchen abbilden;
sie sind sozusagen die Ideen der Elementarteilchen, denen dann schließlich das
Objekt Elementarteilchen entspricht. Diese allgemeine Konstruktion ist mir
durchaus verständlich. Auch ist die Alternative sicher eine sehr viel fundamentalere
Struktur unseres Denkens als das Dreieck. Aber die exakte Durchführung deines Programms
stelle ich mir doch außerordentlich schwierig vor. Denn sie wird ein Denken von so hoher
Abstraktheit erfordern, wie sie bisher, wenigstens in der Physik, nie vorgekommen ist.
Mir wäre das sicher zu schwer. Aber die jüngere Generation hat es ja leichter,
abstrakt zu denken. Also solltest du das mit deinen Mitarbeitern unbedingt versuchen.'"'\\
\\
\footnotesize{Werner Heisenberg zu Carl Friedrich von Weizsäcker - "`Der Teil und das Ganze"'
(Gespräche im Umkreis der Atomphysik), Piper Verlag München 1969, Seite 286 \cite{Heisenberg:1969}.}}}
\end{quote}

\section{Erkenntnistheoretisches Prolegomena und Grundidee}

Das Wesen des Verstehens in der Naturwissenschaft besteht darin, die Vielfalt der
Phänomene in der empirisch untersuchbaren Wirklichkeit der Natur unter möglichst
einheitliche Begriffe zu bringen, indem man auf eine den Sinnen nicht unmittelbar
zugängliche dahinter stehende Realitätsebene blickt. Dies entspricht dem Versuch,
die begrifflichen Strukturen in unserem Denken mit jener höheren geistigen Ordnung
in Übereinstimmung zu bringen, die sich in der Natur manifestiert und für welche
Werner Heisenberg den Begriff "`zentrale Ordnung"' \cite{Heisenberg:1969} gebraucht,
in dem sich wiederum seine durch und durch Platonische Sicht auf die
Wirklichkeit ausdrückt. In diesem Sinne meint Heisenberg in einem Vortrag
über das "`Verstehen in der modernen Physik"' \cite{Heisenberg:1967Vortrag},
dass man einen Erfahrungsbereich dann verstanden habe, wenn man die richtigen
Begriffe zur Beschreibung gefunden habe. Eine fundamentale physikalische Theorie
besteht demnach aus einem System grundlegender Begriffe zur Beschreibung eines
Erfahrungsbereiches, in dessen Rahmen bestimmte grundlegende Postulate und mit
Hilfe bestimmter mathematischer Strukturen damit in Zusammenhang stehende
Naturgesetze formuliert werden. Wenn sich ein solches fundamentales begriffliches
System, Heisenberg spricht von abgeschlossenen Theorien \cite{Heisenberg:1948},
basal ändert und durch ein neues ersetzt wird, so entspricht dem eine
wissenschaftliche Revolution, die von dem Prozess der Behandlung einzelner
empirischer und mathematischer Probleme im vorgegebenen Rahmen einer bereits
bestehenden fundamentalen Theorie zu unterscheiden ist, wie es für die normale
Wissenschaft charakteristisch ist \cite{Kuhn:1962}. Ganz entscheidend ist
dabei im Unterschied zur Lösung einzelner spezieller Probleme basierend auf
vorgegebenen Begriffen und Grundannahmen, die im Rahmen der normalen Wissenschaft
nicht in Frage gestellt werden, dass nicht immer wieder neue spezielle
Modelle ausprobiert und angewandt werden, sondern im Rahmen des richtigen
Begriffssystems die fundamentalen Naturgesetze und Gleichungen die
einfachste sowie abstrakteste überhaupt denkbare und zugleich in sich
konsistente Gestalt annehmen. Auf diese Weise hat die theoretische Physik
im Laufe der letzten Jahrhunderte immer mehr Phänomene innerhalb immer
umfassenderer Erfahrungsbereiche der Natur im Rahmen gewisser ziemlich
grundlegender Theorien beschreiben können. Dabei wurden immer wieder
mehrere spezielle Theorien im Rahmen allgemeinerer Theorien vereinheitlicht.

Der Prozess des Verstehens ist von seiner Natur her so angelegt, dass mehr konkrete,
komplexe und spezielle mit unserer gewöhnlichen Sinnenwelt in Übereinstimmung stehende
Begriffe unter mehr abstrakten, einfachen und weit von unserer gewöhnlichen Sinnenwelt
entfernt liegenden Begriffen subsummiert werden. Einfachheit, Abstraktheit und Allgemeinheit
hängen direkt miteinander zusammen und eine Theorie wird umso fundamentaler sein,
je einfacher ihre grundlegenden Begriffe und Strukturen sind. Einfachheit bedeutet dabei
aber keineswegs Einfachheit im Verständnis. Es ist vielmehr so, dass unsere gewöhnliche
Anschauungswelt sehr kompliziert ist. Um also die Beziehung der tief darunter
liegenden Gegebenheiten, die nur durch ganz einfache aber hochabstrakte Begriffe
beschrieben werden können, zu unserer an der Oberfläche liegenden Sinnenwelt
zu verstehen, ist daher eine sehr tiefe und exakte Analyse der Beziehung unseres
gewöhnlichen menschlichen Anschauens und Denkens zur sehr viel abstrakteren und
einfacheren dahinter stehenden unserem gewöhnlichen Anschauen und Denken nicht
zugänglichen fundamentalen Wirklichkeit der Natur notwendig. Die besondere
geistesgeschichtliche Bedeutung der Relativitätstheorie und der Quantentheorie
besteht darin, dass sie mit unseren gewöhnlichen Anschauungsformen und Begriffen
radikal gebrochen haben, ja brechen mussten, da sie in viel tiefer gelegene
Wirklichkeiten der Natur vorgedrungen sind als jede vorherige Theorie.

Die Philosophie des Immanuel Kant aber zeigt genau jene grundlegenden Strukturen
des Anschauens und Denkens auf, die unserem Geist a priori gegeben sind, und ordnet
sie in ein erkenntnistheoretisches System ein \cite{Kant:1781}, \cite{Kant:1783}.
Hinter diesen Strukturen steht für Kant das "`Ding an sich"', also die Natur, wie sie
wirklich ist. Die moderne Physik dringt allerdings weiter in jenen Bereich des "`Dinges an sich"' vor,
der jenseits der Kantischen Anschauungsformen und Kategorien steht. Gerade deshalb aber ist
ein Studium der Kantischen Philosophie unentbehrlich, um diesen Unterschied zwischen der
Erscheinung der Natur im Rahmen unserer Anschauungsformen und Kategorien einerseits,
also vor allem Raum, Zeit und Kausalität, und der dahinter stehenden Realität der Natur
an sich andererseits, zunächst überhaupt zu verstehen, um dann anschließend daran geistig in
den Bereich des "`Dinges an sich"' vorzudringen. Kant hätte dies zwar für unmöglich gehalten,
weil er glaubte, dass die grundlegenden Begriffe in unserem Denken gar nichts
mit der Wirklichkeit zu tun hätten, sondern ausschließlich ein Darstellungsmedium
dieser Wirklichkeit seien, die in ihrem an"=sich"=Sein überhaupt nicht erkennbar sei.
Die moderne Physik hat allerdings mit der Relativitätstheorie und der Quantentheorie
beziehungsweise den diesen Theorien entsprechenden empirischen Phänomenen gezeigt,
dass auch die hinter den spezifischen Gegebenheiten unserer Sinnenwelt liegenden
tieferen Realitäten sozusagen indirekt durch diese Sinnenwelt hindurchschimmern,
sodass man versuchen kann, sie durch eine radikale Veränderung der Art des
Denkens mit Hilfe bestimmter Arten der gedanklichen Abstraktion von unseren
gewöhnlichen Denk- und Wahrnehmungsstrukturen doch wenigstens indirekt zu beschreiben.
Damit zeigt sich aber rückwirkend, dass die a priori gegebenen Begriffe in unserem
Denken auf der oberflächlichen Betrachtungsebene unserer Alltagserfahrung zwar
doch etwas mit der Wirklichkeit zu tun haben, denn sonst würden wir ja schon
auf der Ebene der klassischen Physik bezüglich der Denk- und Wahrnehmungsstrukturen
in unserem Geist mit der Wirklichkeit kollidieren. Sie müssen jedoch durch Hinwendung
zu einer tieferen Betrachtungsebene radikal verändert werden, um auch in Bezug auf
fundamentalere Bereiche der Natur wieder eine Übereinstimmung zwischen der Wirklichkeit
und unserem Denken in diesem verallgemeinerten abstrakteren Sinne herzustellen.
Dies geschieht durch die Bildung entsprechender abstrakter Begriffssysteme,
sodass die Abstraktion von unserer unmittelbaren anschaulichen Denkweise
konkret ermöglicht wird und wir jene tieferen jenseits der Oberfläche der
klassischen Physik gelegenen Wirklichkeitsbereiche der Natur doch zumindest
indirekt gedanklich erfassen können. Die Kantische Erkenntnistheorie wird also
durch die moderne Physik sowohl bestätigt alsauch modifiziert. Sie wird insofern
bestätigt, als wir die empirische Wirklichkeit notwendigerweise in bestimmten
angeborenen Anschauungsformen und Kategorien beschreiben müssen. Dadurch erklärt
sie gerade die Paradoxien der Quantentheorie. Denn diese entstehen daraus, dass der
menschliche Geist auch Realitäten in diesen angeborenen Denk- und Wahrnehmungsstrukturen
darstellt, auf die sie eigentlich gar nicht passen. Die Kantische Erkenntnistheorie wird
aber dadurch gleichzeitig insofern in doppelter Hinsicht erweitert, als diese Kollision
mit dem "`Ding an sich"' einerseits zeigt, dass sich in unserer gewöhnlichen Anschauung
und in unserem gewöhnlichem Denken eben doch Eigenschaften der Wirklichkeit selbst
widerspiegeln, aber diese Spiegelung andererseits nur die Oberfläche jener Wirklichkeit
betrifft, daher nur Näherungscharakter aufweist und bei tieferer Betrachtung ersetzt
werden muss. Die Quantentheorie bestätigt zudem Platon darin, dass das "`Ding an sich"'
einen rein geistigen Charakter hat, das daher durch bestimmte sehr abstrakte Begriffe
und die sich auf diese Begriffe beziehenden mathematischen Eigenschaften charakterisiert
werden kann, und umgekehrt überhaupt erst durch eine solche Platonische Sicht auf die
Wirklichkeit vollständig zu verstehen ist \cite{Platon:Timaios}, \cite{Platon:Parmenides}.

Aus dem Einstein"=Podolsky"=Rosen"=Paradoxon \cite{Einstein:1935}, \cite{Bell:1964},
das in vielen, vielen Experimenten empirisch konkret nachgewiesen wurde und immer
wieder nachgewiesen wird, folgt als Konsequenz mit unausweichlicher Evidenz,
dass sich die Natur nicht basierend auf feldtheoretischen Prinzipien,
also mit Hilfe separierter Objekte in einem vorgegebenen Raum mit lokaler
Kausalitätsstruktur beschreiben lässt, die kontinuierlich sich im Raum
ausbreitende Wirkungen induziert. Zudem kann die Quantentheorie in ihrer ganz
abstrakten Dirac"=von Neumannschen Fassung bereits als ein denkbar abstraktes rein
logisches Schema interpretiert werden. Sie ist demnach eigentlich die Darstellung
einer rein quantenlogischen Struktur in der Zeit ohne irgendwelche konkreten
a priori Annahmen über die Existenz eines physikalischen Ortsraumes, spezieller
Objekte wie Teilchen oder Wellen oder durch punktweise Produkte beschriebener
Wechselwirkungen. Diese abstrakte logische Struktur manifestiert sich in ihrer
ganz einfachen und abstrakten Form explizit in Gestalt der mathematischen
Beschreibung der Quantenzahlen wie Spin, Isospin oder Farbladung.

Carl Friedrich von Weizsäcker versuchte seit den fünfziger Jahren des zwanzigsten
Jahrhunderts, diese abstrakte Fassung der Quantentheorie aus rein erkenntnistheoretischen
Postulaten zu begründen und aus der abstrakten Quantentheorie wiederum die konkrete
Physik herzuleiten \cite{Weizsaecker:1955}, \cite{Weizsaecker:1958},
\cite{WeizsaeckerScheibeSuessmann:1958}. Denn wenn sich die allgemeinen Naturgesetze aus
den Bedingungen der Möglichkeit der Erfahrung herleiten lassen, dann ist damit bewiesen,
dass sie gelten müssen, weil ohne sie Erfahrung und Naturwissenschaft gar nicht
möglich wäre. Im Grunde bedeutet dies, dass von Weizsäcker nicht von einer
konkreten physikalischen Theorie ausgeht, die er dann quantentheoretisch formuliert,
sondern in seinem Ansatz ergeben sich aus dem abstrakten Schema der reinen
Quantentheorie selbst dann auch die konkreten Theorien der Physik. Dies leitete
von Weizsäcker geistig zur Quantentheorie der Ur"=Alternativen, die er damit als
die fundamentalste mögliche objektivierende Beschreibungsweise der Wirklichkeit
statuiert. Das Programm führt dann faktisch dazu, dass die konkrete theoretische
Physik aus einer reinen Quantenlogik in der Zeit interpretiert als basalstes Schema
zur Objektivation der Natur hergeleitet werden muss \cite{Weizsaecker:1971},
\cite{Weizsaecker:1985}, \cite{Weizsaecker:1992}, \cite{Weizsaecker:1999}. Diese
logische Struktur kann sich am einfachsten und abstraktesten in schlichten N"=fachen
Alternativen realisieren, welchen Wahrscheinlichkeiten zugeordnet werden, wobei sich
die Wahrscheinlichkeitswerte dieser Alternativen kontinuierlich mit der Zeit ändern.
Aufgrund schlichter logischer Evidenz lassen sich alle logischen Alternativen über die
Bildung Cartesischer Produkte in binäre Alternativen zerlegen. Diese Zerlegung in binäre
Alternativen führt zur fundamentalsten möglichen Darstellung der abstrakten Quantenlogik
in der Zeit. Hier besteht eine gewisse Verwandtschaft zur Philosophie Wittgensteins, der die Welt
nicht als aus Dingen, sondern als aus reinen Tatsachen bestehend ansah \cite{Wittgenstein:1921}.  
Denn wichtig bei diesem ganzen Programm ist, dass die basalen logischen Einheiten
beziehungsweise Informationseinheiten sich nicht in einer bereits vorgegebenen
physikalischen Realität befinden. Sie befinden sich nicht in einem vorgegebenen
physikalischen Raum, besitzen keinen physikalischen Träger und beziehen sich nicht
auf eine andere von ihnen verschiedene physikalische Realität. Sie sind selbst bereits
die gesamte physikalische Realität. Wie bei Hegel konstituiert sich die Natur demnach
also aus reiner Logik, die sich nicht auf einen von ihr verschiedenen Inhalt
bezieht, sondern aus der die Inhalte selbst hervorgehen. Die Logik schwebt sozusagen
über dem Nichts und damit wird die Natur ganz gemäß Hegel als eine Entäußerung reinen
absoluten Geistes angesehen \cite{Hegel:1812}, der sich als Logik darstellt.

Carl Friedrich von Weizsäcker hat dieses Grundprogramm zunächst bezüglich seiner
philosophischen und begrifflichen Basis exakt formuliert. Desweiteren hat er
grundlegende Teile der Rekonstruktion vollzogen. Dies geschah von den fünfziger
Jahren des zwanzigsten Jahrhunderts ausgehend primär in den sechziger, siebziger
und achtziger Jahren bis in die neunziger Jahre hinein, unterstützt vor allem
durch seine Mitarbeiter Michael Drieschner, Lutz Castell und Thomas Görnitz
\cite{Weizsaecker:1955}, \cite{Weizsaecker:1958}, \cite{WeizsaeckerScheibeSuessmann:1958},
\cite{Weizsaecker:1974}, \cite{Weizsaecker:1978}, \cite{Weizsaecker:1984},
\cite{Drieschner:1967}, \cite{Drieschner:1979}, \cite{Castell:1974}, \cite{Kuenemund:1984},
\cite{Goernitz:1986}, \cite{Goernitz:1987}, \cite{Drieschner:1988}, \cite{Goernitz:1992}.
Umfassende Darstellungen des damit erreichten physikalischen Weltbildes
sind zu finden in \cite{Weizsaecker:1971}, \cite{Weizsaecker:1985},
\cite{Weizsaecker:1992}, \cite{Weizsaecker:1999}, \cite{Goernitz:2002},
\cite{Drieschner:2002}, \cite{Castell:2003}. Später beschäftigte sich
Holger Lyre mit dieser Theorie \cite{Lyre:1994}, \cite{Lyre:1995},
\cite{Lyre:1996}, \cite{Lyre:1998}. Thomas Görnitz liefert bis in die Gegenwart
hinein wichtige Beiträge zur Theorie \cite{Goernitz:2010}, \cite{Goernitz:2012},
\cite{Goernitz:2014}, \cite{Goernitz:2016} und versucht sie desweiteren im
Bewusstsein der Menschen wach zu halten und allgemeinverständlich darzustellen.

In dieser Arbeit glaube ich nun basierend auf dem Vermächtnis vor allem
Carl Friedrich von Weizsäckers selbst, aber zum Teil auch seiner Mitarbeiter
und eigener früherer Beiträge, insbesondere \cite{Kober:2009B}, \cite{Kober:2010},
\cite{Kober:2012}, \cite{Kober:2017}, argumentativ stringent zeigen zu können,
dass sich aus der reinen Quantenlogik in der Zeit und der Forderung möglichst
hoher Symmetrie im abstrakten Raum logischer Alternativen die Existenz der
realen physikalischen Objekte und ihrer Wechselwirkungen ergibt. Es kann
exakt begründet und hergeleitet werden, warum sich die physikalischen Objekte
in einem dreidimensionalen Raum befinden, warum es die internen Eichsymmetrien
der Elementarteilchenphysik und warum es den Spin gibt, der beides verbindet.
Dies erklärt, warum sich die physikalischen Objekte zwar als in einem
mit der Zeit zu einer (3+1)"=dimensionalen Raum"=Zeit verbundenen Raum
befindend darstellen, aber sich ganz offensichtlich so verhalten, als seien sie an
dessen lokale Kausalitätsstrukturen überhaupt nicht gebunden (siehe EPR-Paradoxon).
Durch eine Forderung möglichst hoher Symmetrie des abstrakten logischen Raumes
vieler Ur"=Alternativen ergeben sich in natürlicher Weise nicht nur die mathematische
Grundstruktur der Wechselwirkungen des Standardmodells der Elementarteilchenphysik,
also der starken, der schwachen und der elektromagnetischen Wechselwirkung,
sondern auch eine rein quantentheoretische Fassung der Gravitation in Gestalt
der allgemeinen Relativitätstheorie. Diese wird als eine Eichtheorie einer
Darstellung der lokalen Translationsgruppe im reinen Quanteninformationsraum
formuliert. Zudem löst sich das Hierarchieproblem wahrscheinlich dadurch auf,
dass ein Quantenobjekt im symmetrischen Tensorraum, auf den sich
die Gravitationseichsymmetrie bezieht, in der Regel etwa $10^{40}$
Ur"=Alternativen enthält, während der antisymmetrische Tensorraum
nur maximal vier Ur"=Alternativen enthält. Die Abbildung aus dem
abstrakten Tensorraum der Ur"=Alternativen in den physikalischen
Ortsraum geschieht mit Hilfe der Mathematik des quantentheoretischen
harmonischen Oszillators, was auch die Normiertheit der Wellenfunktionen
gewährleistet. Sehr kondensiert und vereinfacht könnte man sagen,
dass basierend auf der von Weizsäckerschen Theorie gezeigt ist, dass jene
ganz einfachen und hochabstrakten logischen Strukturen, die sich in den
diskreten Quantenzahlen der Elementarteilchenphysik explizit zeigen, letztendlich
auch hinter den raum"=zeitlichen scheinbar kontinuierlichen Freiheitsgraden
von Teilchen und Feldern stehen. Dies geschieht aber eben in keiner Weise
durch eine Diskretisierung der Raum"=Zeit, sondern durch eine radikale
Abkehr von topologischen Verhältnissen und räumlich"=kausalen Strukturen überhaupt
oder auch von abstrakten Netzwerken oder dergleichen zu Gunsten einer Hinwendung
zur einfachst möglichen Darstellung rein quantenlogischer Strukturen ohne
zusätzlichen physikalischen Träger. Damit erhält die in der Quantentheorie
der Ur"=Alternativen seitens Carl Friedrich von Weizsäcker statuierte
"`Kopernikanische Wende"', dergemäß sich nicht Objekte in einem vorgegebenen
Raum bewegen, sondern abstrakte rein logische Objekte umgekehrt die Existenz
des Raumes begründen, nun auch eine meiner Ansicht nach adäquate mathematische
Fassung. Diese ganze Rekonstruktion der Physik kann in folgendem Diagramm
veranschaulicht werden:

\fontsize{12}{12}\selectfont
\begin{center}
\begin{xy}
\xymatrix{\fbox{\textbf{Zeit}}\ar[rd]& &\fbox{\textbf{Logik}}\ar[ld]\\
&\fbox{\parbox[c]{45mm}{\textbf{Alternativen in der Zeit}}}\ar[d]&\\
\quad & \fbox{\parbox[c]{30mm}{\textbf{Tensorraum der Ur-Alternativen}}}
\ar[ld] \ar[rd] \ar[dd] & \quad\\
\fbox{\parbox[c]{40mm}{\textbf{Teilraum der symmetrischen Zustände}}}\ar[rd]\ar[d]\ar@/_2.5cm/[dd] &   &
\fbox{\parbox[c]{40mm}{\textbf{Teilraum der antisymmetrischen Zustände}}}\ar[ld]\ar[d]\ar@/^2.5cm/[dd]\\
\fbox{\parbox[c]{35mm}{\textbf{dreidimensionaler\\ Ortsraum}}}
\ar[r] & \fbox{\textbf{Elementarteilchen}}\ar[d]\ar[ld]\ar[rd]
& \fbox{\parbox[c]{40mm}{\textbf{innere Quantenzahlen\\ und Spinfreiheitsgrad}}}\ar[l]\\
\fbox{\parbox[c]{45mm}{\textbf{Translations-Symmetrie in Abhängigkeit
der abstrakten Ortsoperatoren}}}\ar[d]
& \fbox{\textbf{Vielteilchenzustände}}\ar[dd]\ar[ld]\ar[rd]\ar[r]\ar[l] &
\fbox{\parbox[c]{45mm}{\textbf{SU(N)-Symmetrien in Abhängigkeit
der abstrakten Ortsoperatoren}}}\ar[d]\\
\fbox{\textbf{Gravitation}}\ar[rd] &   & \fbox{\textbf{SU(N)-Wechselwirkungen}}\ar[ld]\\
&\fbox{\parbox[c]{45mm}{\textbf{Beziehungsstruktur \mbox{quantenlogischer
Versch-} ränkungen in der Zeit}}}&}
\end{xy}
\end{center}

\fontsize{12.8pt}{14.8pt}\selectfont

Die theoretische Physik besteht im Wesentlichen aus drei Bereichen, nämlich
A) der Phänomenologie, also der Beziehung der Theorie zu den konkreten
empirischen Ergebnissen; B) der Mathematik, also der Fassung der Grundbegriffe und
Gesetze einer Theorie in konkreten Strukturen; und C) der begrifflichen Basis,
die auch mit Hilfe der Einbeziehung philosophischer und speziell erkenntnistheoretischer
Methoden herausgearbeitet und verstanden werden muss. In jener Auffindung der
richtigen Grundbegriffe, welche die tiefere Wirklichkeit der Natur im
menschlichen Geist überhaupt erst darzustellen gestattet, besteht gemäß dem
zu Beginn erwähnten Zitat Heisenbergs aber das Herzstück zumindest der fundamentalen
theoretischen Physik. Die Quantentheorie der Ur"=Alternativen wird jetzt endgültig in
überwältigender Weise durch alle drei Bereiche gestützt. Die Experimente Zeilingers zeigen
den nichtlokalen Charakter der Quantentheorie ganz explizit auf \cite{Zeilinger:1999},
\cite{Zeilinger:2008}, welcher die radikale Überwindung klassisch"=feldtheoretischer
Begriffe empirisch fordert; die grundlegenden mathematischen Strukturen der Quantentheorie,
der allgemeinen Relativitätstheorie und der Wechselwirkungen des Standardmodells
der Elementarteilchenphysik ergeben sich in einer rein quantentheoretischen Fassung
aus den logischen Alternativen in der Zeit in natürlicher Weise beinahe zwangsläufig;
und es wird mit Hilfe der Kantischen Philosophie \cite{Kant:1781}, \cite{Kant:1783}
und der Kopenhagener Deutung der Quantentheorie deutlich, warum diese Beschreibung
der physikalischen Wirklichkeit basierend auf solch abstrakten durch reine
Logik charakterisierten Begriffen unumgänglich ist. Viel mehr kann eine
naturwissenschaftliche Theorie eigentlich gar nicht mehr leisten.
Es sei darauf hingewiesen, dass damit nicht nur eine rein quantentheoretische
Fassung der allgemeinen Relativitätstheorie gefunden, sondern zudem die
fundamentalen Wechselwirkungen der Natur in einer reinen Quantentheorie
vereinheitlicht beschrieben sind, und zwar basierend und begründet aus
reiner Logik in der Zeit. Dies bedeutet, dass sich die bekannten Theorien
der theoretischen Physik aus den im begrifflichen Rahmen einer Quantenlogik
einfachsten und fundamentalsten Strukturen in natürlicher Weise ergeben.
Es sei desweiteren erwähnt, dass damit die fundamentale theoretische Physik
zumindest bezüglich ihrer Grundlagen an ihr definitives Ende gekommen ist.
Denn im Rahmen einer überhaupt noch in rationalen Strukturen beschreibbaren
naturwissenschaftlichen Theorie ist etwas Abstrakteres, Einfacheres
und Fundamentaleres als eine Auflösung der Wirklichkeit der Natur
in Ja"=Nein"=Entscheidungen überhaupt gar nicht mehr möglich.

Dies bedeutet freilich nicht, dass nicht noch viele Einzelprobleme wie beispielsweise
die Entstehung der Massen und der vielen Parameter sowie konkrete weitere Berechungen
bezüglich der sich ergebenden Verschränkungen gelöst werden müssten. Aber es
bedeutet, dass mir im Grundsatz gezeigt zu sein scheint, dass die Quantentheorie
der Ur"=Alternativen Carl Friedrich von Weizsäckers die begriffliche Basis
liefert, in der die Quantentheorie, die Relativitätstheorie und die
Elementarteilchenphysik zu einer Einheit verschmelzen. Die Relativitätsheorie,
die aus der speziellen Relativitätstheorie \cite{Einstein:1905A} und der
allgemeinen Relativitätstheorie \cite{Einstein:1914A}, \cite{Einstein:1914B},
\cite{Einstein:1915}, \cite{Einstein:1916} besteht, stellt im Wesentlichen
das alleinige Werk Albert Einsteins dar. Die Quantentheorie wurde seitens
Max Planck begründet \cite{Planck:1900}, \cite{Planck:1901A}, \cite{Planck:1901B},
um dann zunächst durch Einstein auf die Beschreibung des Lichtes \cite{Einstein:1905B}
und später durch Niels Bohr auf die Beschreibung der Atome angewandt \cite{Bohr:1913}
sowie begrifflich am tiefsten analysiert zu werden \cite{Bohr:1935}. Der entscheidende
Durchbruch zur grundlegenden Idee der vollendeten Gestalt der Quantentheorie gelang
schließlich Heisenberg \cite{Heisenberg:1925}, der diese Grundidee dann gemeinsam mit
Max Born und Pascual Jordan zu einer vollständigen mathematischen Theorie ausarbeitete
\cite{Heisenberg:1926}, \cite{BornJordan:1925}, \cite{BornHeisenbergJordan:1926}.
Gemeinsam erkannten Niels Bohr und Werner Heisenberg wenig später den physikalischen
und philosophischen Sinn der Quantentheorie, den sie in der Kopenhagener
Deutung der Quantentheorie zum Ausdruck brachten, welche die neue
begriffliche Struktur der Quantentheorie einem adäquaten Verständnis
zugänglich macht. Hierbei spielt die Heisenbergsche Unbestimmtheitsrelation
\cite{Heisenberg:1927}, welche die Grenzen der Anwendbarkeit klassischer
Begriffe aus der Quantentheorie mathematisch exakt herleitet, eine ebenso
bedeutsame Rolle wie der Bohrsche Begriff der Komplementarität \cite{Bohr:1928},
welcher klassische Begriffe wie Teilchen und Welle in eine einander ausschließende
und zugleich ergänzende Beziehung zueinander setzt, um die dahinter liegende
quantentheoretische Wirklichkeit gleichnishaft verstehen zu können. Zudem
lieferten Wolfgang Pauli \cite{Pauli:1926}, \cite{Pauli:1927}, Erwin Schrödinger
\cite{Schroedinger:1926}, Paul Adrien Maurice Dirac \cite{Dirac:1926},
\cite{Dirac:1928} und Louis de Broglie \cite{DeBroglie:1925} entscheidende
Beiträge zur Formulierung der Quantentheorie. Paul Adrien Maurice Dirac
und Johann von Neumann brachten die Quantentheorie in die abstrakteste mathematische
Formulierung \cite{Dirac:1958}, \cite{Neumann:1932}. Zur Elementarteilchenphysik,
die auf der Quantentheorie verbunden mit der speziellen Relativitätstheorie zur
Quantenfeldtheorie basiert \cite{Bjorken:1967}, \cite{Weinberg:1995}, deren
grundlegende Formulierung Heisenberg und Pauli ausarbeiteten \cite{Heisenberg:1929},
lieferten Richard Feynman \cite{Feynman:1942}, \cite{Feynman:1949}, Steven Weinberg
\cite{Weinberg:1967} und Murray Gell"=Mann \cite{GellMann:1964}, \cite{Fritzsch:1973}
die vielleicht wichtigsten Beiträge. Heisenberg allerdings machte mit seiner
einheitlichen Spinorfeldtheorie den bisher grundlegendsten Versuch, zu einer wirklichen
einheitlichen Basis für diese auf der relativistischen Quantenfeldtheorie basierende
Elementarteilchenphysik zu gelangen \cite{Heisenberg:1957}, \cite{Heisenberg:1963}, \cite{Heisenberg:1967},
\cite{Durr:1979}, \cite{Durr:1982}, \cite{Durr:1983}, \cite{Kober:2008}. In diesem Sinne,
dass sich all dies bezüglich der wesentlichen Ideen und mathematischen Strukturen
innerhalb der Quantentheorie der Ur"=Alternativen in den begrifflichen Rahmen einer
reinen Quantenlogik in der Zeit einbetten und daraus begründen lässt, erweist sich
aber Carl Friedrich von Weizsäcker als der begrifflich und philosophisch tiefste
und größte Denker in der theoretischen Physik aller Zeiten. Es erfüllt mich
mit tiefer Dankbarkeit und Freude, dass ich einen entscheidenden Beitrag dazu
leisten konnte, dies nun auch in der mathematischen Formulierung konkret darzulegen.

In dieser Arbeit werden zunächst die grundlegenden Postulate der Quantentheorie
der Ur"=Alternativen dargelegt, bevor anschließend die Rekonstruktion systematisch
durchgeführt wird; von den Alternativen; über die Wahrscheinlichkeiten; den dadurch
entstehenden Hilbertraum; die Einbindung der Zeitentwicklung über einen einparametrigen
Automorphismus, was zur Darstellung in komplexen Hilberträumen führt; über die
Aufspaltung der Darstellung in Ur"=Alternativen, was zu einem Tensorraum vieler
Ur"=Alternativen führt; über die aus der Ununterscheidbarkeit der Ur"=Alternativen
sich ergebende Unterscheidung zwischen unter Permutation symmetrischen und
antisymmetrischen Zuständen; über die Abbildung der symmetrischen Zustände mit
beliebig vielen Ur"=Alternativen in die Raum"=Zeit; das Herausarbeiten der inneren
Symmetrien der Elementarteilchen aus dem antisymmetrischen Anteil des Tensorraumes
und der damit in Zusammenhang stehenden Begründung, warum die inneren Symmetrien
nicht auf Raumtransformationen reagieren, es aber den Spin gibt, der darauf reagiert;
bis hin zur Forderung möglichst hoher Symmetrie in Bezug auf den symmetrischen und
antisymmetrischen Teil des Tensorraumes und im gesamten Tensorraum und der damit
in Zusammenhang stehenden Herleitung einer rein auf quantentheoretischen Begriffen
basierenden Fassung der Gravitation als abstrakte quantenlogische Translationseichtheorie
analog zur allgemeinen Relativitätstheorie und der Wechselwirkungen des Standardmodells der
Elementarteilchenphysik als abstrakte quantenlogische Eichtheorie der inneren Symmetrien.
Der Begriff der Wechselwirkung muss aber durch den der quantentheoretischen Verschränkung
der abstrakten Zustände mehrerer Objekte ersetzt werden, die damit nicht mehr als voneinander
trennbare Objekte angesehen werden können, wobei das Phänomen der Wechselwirkung
nur einer unvollkommenen Betrachtung entspricht, bei der die Welt in voneinander
separierbare Einzelobjekte aufgelöst wird. Damit ist ein wesentlicher Teil der
fundamentalen theoretischen Physik aus der Quantentheorie der Ur"=Alternativen
und damit aus reiner Logik in der Zeit mit einigen Zusatzpostulaten begründet
und systematisch hergeleitet. Schließlich wird durch eine Kombination von
Transformationen im gegenüber Permutation der Ur"=Alternativen symmetrischen
und Transformationen im gegenüber Permutationen antisymmetrischen Teilraum
des Tensorraumes der Ur"=Alternativen, bei welcher die Transformationen
im symmetrischen Teilraum von den Zuständen im antisymmetrischen Teilraum
abhängen aber nicht umgekehrt, eine erweiterte $E_8$"=Symmetriegruppe statuiert.

\section{Die begriffliche Grundbasis der Quantentheorie der Ur-Alternativen}

\subsection{Kopenhagener Deutung und Relativierung klassischer Begriffe}

Um den ganzen inneren Sinn der Quantentheorie in Bezug auf die Beschreibung der
äußeren in einem objektiven geistigen Sinne existierenden Wirklichkeit der Natur
zu verstehen, um die es ja in der theoretischen Physik nur gehen kann, wenn diese
überhaupt einen Sinn haben soll, ist die Kopenhagener Deutung der Quantentheorie
in ihrem inneren Bezug zur Kantischen Philosophie absolut unentbehrlich. Die Mathematik
der Quantentheorie beziehungsweise die Quantisierungsregeln, im speziellen Falle
der Quantenmechanik sind diese in besonderer Weise mit der Heisenbergschen Algebra
$[x,p]=i$ verbunden, die direkt mit der Heisenbergschen Unbestimmtheitsrelation in
Zusammenhang steht, könnten zunächst in der Weise aufgefasst werden, dass man es weiterhin
mit einer physikalischen Realität zu tun hat, in der sich voneinander separierbare räumliche
Objekte bewegen, die miteinander wechselwirken, und sich nur die physikalischen Gesetze,
nach denen dies geschieht, mit der Quantentheorie geändert hätten. Diese Auffassung macht
ein wirkliches Verständnis der Quantentheorie allerdings vollkommen unmöglich und verstellt
damit zugleich die Möglichkeit der Formulierung einer einheitlichen Naturtheorie, in der
alle Objekte und Wechselwirkungen in einem einheitlichen begrifflichen und mathematischen
Rahmen gefasst sind. Denn der wirkliche innere Sinn der Unbestimmtheitsrelation besteht
darin, dass die klassischen Begriffe wie Ort und Impuls im Inneren der Natur, auf der
fundamentalen Ebene, welche die Quantentheorie beschreibt, ihre Gültigkeit vollkommen verlieren,
weil man es mit einer vollkommen anderen Art der Wirklichkeit zu tun hat, auf welche klassische
Begriffe einfach nicht mehr passen. Und die Grenzen, innerhalb derer sie näherungsweise
anwendbar sind, werden mathematisch exakt durch die Unbestimmtheitsrelation definiert.
Demnach beschreibt die Unbestimmtheitsrelation weder veränderte Gesetze für Objekte
im Raum noch ändert sie die konkrete Gestalt dieser Objekte im Raum, sondern sie
drückt aus, dass der Begriff des Objektes im Raum ganz grundsätzlich kein adäquater
Begriff ist, um die fundamentale Wirklichkeit der Natur überhaupt erfassen zu können.
Ort und Impuls können nur in jener Näherung sinnvoll angewandt werden, welche eben
durch die Unbestimmtheitsrelation mathematisch exakt definiert wird. Die Kopenhagener
Deutung der Quantentheorie Niels Bohrs und Werner Heisenbergs besagt ganz in diesem Sinne,
dass die Paradoxien der Quantentheorie daher rühren, dass unser menschlicher Geist,
mit dem wir die Welt erleben und beschreiben, die Welt notwendigerweise klassisch darstellt,
weshalb wir klassische Begriffe bei der Beschreibung von Experimenten verwenden müssen,
aber diese Begriffe auf der fundamentalen Ebene der Realität in Wirklichkeit ihre Gültigkeit
verlieren \cite{Bohr:1930-1961}, \cite{Heisenberg:1969}, \cite{Heisenberg:1958}, \cite{Heisenberg:1979},
\cite{Weizsaecker:1943}, \cite{Mittelstaedt:1963}, \cite{Kober:2009A}. Und aus dieser Diskrepanz
unserer angeborenen Denk- und Wahrnehmungsstrukturen einerseits zur Wirklichkeit der Natur an
sich andererseits erwachsen die Paradoxien. Mit den Paradoxien ist hier etwa die Tatsache gemeint,
dass Quantenobjekte sowohl Teilcheneigenschaften alsauch Welleneigenschaften aufweisen,
oder ein Quantenobjekt beim Doppelspaltexperiment durch beide Spalten gleichzeitig läuft,
während es gleichzeitig einen sehr eng lokalisierten Schwärzungspunkt auf der
Photoplatte erzeugt. Darüber hinaus werden die damit direkt in Zusammenhang stehende
nicht"=Räumlichkeit des physikalischen Geschehens, die nicht"=Trennbarkeit der Objekte
im schlechthinnigen Sinne und damit die Grenzen einer Beschreibungsweise basierend
auf feldtheoretischen Prinzipien, also basierend auf einer lokalen Kausalitätsstruktur
der Welt, explizit im berühmten Einstein"=Podolsky"=Rosen"=Paradoxon
vor Augen geführt \cite{Einstein:1935}.

In der Kantischen Philosophie ist der Raum eine grundlegende Anschauungsform in unserem
menschlichen Geist, in der sich das in seinem an"=sich"=selbst"=Sein nicht erkennbare
"`Ding an sich"' darstellt, dem die Anschauungsform des Raumes für sich selbst betrachtet
nicht zukommt. Die Einsteinsche Relativitätstheorie zeigt zwar, dass die Raum"=Zeit doch
auch in der äußeren Realität existieren muss, denn sonst könnte eine empirische Wissenschaft
nicht herausfinden, dass ihre Strukturen genaugenommen andere sind als diejenigen in unserem
menschlichen Geist. Aber wenn man noch tiefer in die Realität der Natur eindringt, dann
verlieren räumliche Kausalstrukturen schließlich doch ihre Gültigkeit. Dies wurde bereits
im ersten Kapitel ausgeführt. Die evolutionäre Erkenntnistheorie schließlich erklärt,
warum sich in unserem Geist eine Anschauungsform des Raumes im Laufe der Evolution
herausgebildet hat. Denn diese näherungsweise auf einer Oberflächenebene mit der
Natur übereinstimmende Darstellung der Natur war gerade ausreichend,
um das Überleben der Vorfahren des Menschen zu gewährleisten
\cite{Lorenz:1973}, \cite{Vollmer:1975}, \cite{Ditfurth:1976}.

\subsection{Beziehung der Quantentheorie zur allgemeinen Relativitätstheorie}

Basierend auf einer erkenntnistheoretischen Grundlage, welche die Kantische Philosophie
und die Kopenhagener Deutung der Quantentheorie zugrunde legt, muss nun die begriffliche
Grundbasis der Quantentheorie der Ur"=Alternativen dargestellt werden, wie sie seitens
Carl Friedrich von Weizsäcker erdacht und begründet wurde. Dazu muss zunächst die
Beziehung der Quantentheorie zur allgemeinen Relativitätstheorie thematisiert werden.

Es gibt in der bisherigen fundamentalen theoretischen Physik zwei basale Theorien,
nämlich die Quantentheorie und die Relativitätstheorie. Die Verbindung der Quantentheorie
mit der speziellen Relativitätstheorie, also die relativistische Quantenfeldtheorie,
liefert den grundlegenden begrifflichen Rahmen für die Elementarteilchenphysik und ihre
Wechselwirkungen, also die elektromagnetische, die schwache und die starke Wechselwirkung,
wobei hier eine Dualität aus den Wellenfunktionen in der Raum"=Zeit einerseits und den mit
diesen Wellenfunktionen im Sinne von Tensorprodukten verbundenen Quantenzahlen andererseits
auftritt. Die verschiedenen Wechselwirkungen werden über die Forderung lokaler Eichsymmetrien
in Bezug auf die inneren Quantenzahlen als punktweise Produkte von Feldern eingeführt.
Eine Verbindung der Quantentheorie mit der allgemeinen Relativitätstheorie ist
im Standardmodell der Elementarteilchenphysik nicht gelungen, sodass dort auch
die Gravitation nicht in den Rahmen der Quantentheorie integriert ist.

Die wichtigste begriffliche Eigenschaft der allgemeinen Relativitätstheorie besteht
in der sogenannten Hintergrundunabhängigkeit, die mit der mathematischen Eigenschaft
der Diffeomorphismeninvarianz in Zusammenhang steht. Das aber bedeutet, dass in der
allgemeinen Relativitätstheorie den Raum"=Zeit"=Koordinaten überhaupt keine
physikalische Bedeutung mehr zukommt, sondern nur noch Beziehungen zwischen den
Feldern auf der Raum"=Zeit, zu denen auch das metrische Feld gehört, welches das
Gravitationsfeld darstellt. Dies hat damit zu zu tun, dass es in der allgemeinen
Relativitätstheorie überhaupt keine absoluten Größen mehr gibt, da auch die Metrik
im Gegensatz zur speziellen Relativitätstheorie in der allgemeinen Relativitätstheorie
zu einer dymamischen Größe wird. Das aber bedeutet, dass die Raum"=Zeit ohne die in
ihr enthaltenen dynamischen Objekte für sich genommen überhaupt gar nicht existiert.
Allerdings existieren gemäß der allgemeinen Relativitätstheorie in einem relationalen
Sinne räumlich"=kausale Strukturen, da in ihr als den zentralen dynamischen Größen
Felder auf einer Raum"=Zeit beschrieben werden.

In der Quantentheorie hingegen existieren auf der fundamentalen Ebene noch
überhaupt keine räumlichen Kausalstrukturen, was sich am Deutlichsten im berühmten
Einstein"=Podolsky"=Rosen"=Paradoxon zeigt. Es gibt in der Quantentheorie zwar
durchaus Freiheitsgrade, die sich wenigstens indirekt räumlich darstellen, weshalb
Quantenobjekte als kontinuierliche Wellen im Raum erscheinen. Aber die inneren
Freiheitsgrade, nämlich die abstrakten Quantenzahlen wie Spin, Isospin und Farbe,
haben überhaupt gar nichts mehr mit der Raum"=Zeit zu tun. In der abstrakten
Hilbertraum"=Formulierung setzt die Quantentheorie überhaupt keine konkreten
physikalischen Begriffe wie Raum, Körper, Feld, Teilchen, Bewegung, Wechselwirkung,
Masse oder Ladung voraus. Lediglich die Zeit als fundamentalster physikalischer
Begriff, dessen Bedeutung weit über die Physik hinausgeht, muss verbindlich
beibehalten werden. Deshalb kommt dem Raum in der Quantentheorie ein noch
viel geringerer ontologischer Status zu als in der allgemeinen Relativitätstheorie.
Beide Theorien stimmen also darin überein, dass der Raum an sich keine für sich
selbst existierende Realität ist, aber die Quantentheorie ist diesbezüglich noch
deutlich radikaler, da sie noch nicht einmal räumliche Kausalstrukturen voraussetzt
und Freiheitsgrade beschreibt, die nicht feldtheoretisch"=geometrisch,
sondern stattdessen rein abstrakt"=logisch charakterisiert sind. Die abstrakte
Hilbertraum"=Formulierung der Quantentheorie ist so allgemein, dass sie viel
abstraktere Objekte und Beziehungen beschreiben kann, als die der gewöhnlichen
physikalischen Wirklichkeit.

Eine Quantenzahl beziehungsweise ein abstrakter Zustand in einem Hilbertraum
hat keinerlei konkrete Entsprechung in unserer sinnlichen Erfahrung und ist daher
sehr viel abstrakter als jedes Objekt unserer konkreten Erfahrung und daher als
jede raum"=zeitliche Realität überhaupt. Die Quantentheorie ist damit deutlich
abstrakter als die spezielle und die allgemeine Relativitätstheorie.
Genaugenommen stellt die Quantentheorie in ihrer abstrakten Form letztendlich
beinahe nur eine Art eines quantenlogischen Schemas im Sinne einer abstrakten
Beziehungsstruktur dar. Eine Theorie, in der Quantentheorie und allgemeine
Relativitätstheorie vereinheitlicht sind, sollte also auf den Begriffen der
Quantentheorie und nicht auf denen der allgemeinen Relativitätstheorie basieren.
Denn die Begründung der theoretischen Physik muss sich immer vom Abstrakten zum
Konkreten bewegen. Da die Quantentheorie in ihrem abstrakten Kern so etwas
wie ein logisches Schema darstellt, handelt sie letztendlich von reiner Logik.
Die Relativitätstheorie hingegen stellt in ihrem Kern so etwas wie ein
geometrisches Schema dar, und handelt damit letztendlich von Geometrie.
Logik aber ist viel fundamentaler als Geometrie. Denn Logik bezieht sich auf
die Relation von Aussagen, unabhängig davon, welcher spezielle Inhalt den
Aussagen zukommt. Geometrie hingegen bezieht sich auf mathematische Räume
mit ganz bestimmten Eigenschaften. Logische Gesetze gelten auch für Aussagen,
die sich auf Geometrie beziehen. Aber umgekehrt sind logische Gesetze von
geometrischen Gesetzen vollkommen unabhängig. Dies bedeutet, dass die
Quantentheorie fundamentaler ist als die allgemeine Relativitätstheorie 
und nur die Herleitung der Relativitätstheorie aus der Quantentheorie
gelingen kann aber nicht umgekehrt. Einer Begründung der speziellen und der
allgemeinen Relativitätstheorie aus der Quantentheorie entspricht demnach eine
Begründung der Geometrie aus reiner Logik. Bei einer Vereinheitlichung von
Theorien werden immer verschiedene Wirklichkeitsbereiche in eine höhere
Ordnung integriert. Eine der zentralen Thesen der Quantentheorie der
Ur"=Alternativen besteht also darin, dass einer Vereinheitlichung
der Physik die Integration der Geometrie in die Logik entspricht,
und zwar in eine Logik in der Zeit.

In dieser Arbeit wird im Rahmen der Quantentheorie der Ur"=Alternativen
ganz konkret gezeigt werden, dass sich sowohl die abstrakten Quantenzahlen
der Elementarteilchenphysik alsauch der sich raum"=zeitlich darstellende
Anteil der Quantenobjekte und damit zugleich die raum"=zeitliche Wirklichkeit
der allgemeinen Relativitätstheorie tatsächlich mathematisch aus der abstrakten
Quantentheorie aufgefasst als ein rein logisches Schema in der Zeit ergibt.
Dies schließt dann auch eine Einbettung der Wechselwirkungen der
Elementarteilchenphysik einerseits und der Gravitation andererseits
in diesen denkbar einfachen, abstrakten und fundamentalen begrifflichen
Rahmen ein, wobei an die Stelle der Wechselwirkung die Verschränkung
quantentheoretischer Objekte tritt.

\subsection{Das grundlegende geistige Unternehmen von Weizsäckers}

Carl Friedrich von Weizsäcker versucht im Rahmen seines Unternehmens der
Rekonstruktion der theoretischen Physik in der Quantentheorie der Ur"=Alternativen
die Naturbeschreibung ausschließlich aus der Quantentheorie zu begründen. Der
Unterschied zur gewöhnlichen Herangehensweise besteht bei von Weizsäcker darin,
dass nicht von einer speziellen klassischen Theorie ausgegangen wird, welche
bestimmte Objekte und Wechselwirkungen voraussetzt, wobei diese klassische Theorie
dann anschließend quantisiert wird. Vielmehr werden überhaupt gar keine spezielle
Theorie und auch gar keine speziellen physikalischen Begriffe mehr vorausgesetzt.
Denn es wird lediglich die abstrakte Quantentheorie als eine Darstellung einer
zeitlichen Logik und zugleich als formales Medium der Objektivation der Wirklichkeit
der Natur interpretiert, woraus sich dann umgekehrt als Konsequenz in einer indirekten
Weise alle konkreten physikalischen Begriffe nachträglich im Sinne einer Näherung ergeben.
Von Weizsäcker abstrahiert also in der Quantentheorie der Ur"=Alternativen von allen
speziellen auf die konkrete Wirklichkeit bezogenen Begriffen und Vorstellungen unseres
Denkens und auch den damit verbundenen speziellen Annahmen über die Natur und
kondensiert stattdessen die grundlegende Basis unseres theoretischen Bezuges
zur Wirklichkeit im Sinne einer objektivierenden Beschreibungsweise auf den
abstraktesten möglichen Grundgehalt, der genau einer solchen fundamentalen
logischen Struktur entspricht, die sich in Alternativen manifestiert. Diese abstrakte
Weise, die Wirklichkeit zu denken, geht von Kant aus, besitzt aber in ihrer letztendlich
erreichten Abstraktheit der Rückführung auch der Inhalte des Denkens auf Logik
erst in der Hegelschen Philosophie eine wirkliche Entsprechung. Diese logische Basis ist
vollkommen unabhängig von speziellen konkreten Vorurteilen über die Wirklichkeit
und ist damit so abstrakt, dass sie in der Lage ist, auch die ganz fundamentale Ebene
der Wirklichkeit jener geistigen Ordnung zu erfassen, die sich uns in der Natur offenbart.
Dies entspricht der Heisenbergschen Äußerung, die ebenfalls aus dem bereits eingangs
erwähnten Vortrag stammt \cite{Heisenberg:1967Vortrag}, dass das Schwierigste bei
der Formulierung einer neuen fundamentalen Theorie nicht die Auffindung der neuen
Begriffe sei, sondern die Loslösung von den alten Begriffen. Carl Friedrich von
Weizsäcker führt nämlich gar keine wirklich neuen Begriffe und Strukturen ein,
die über die abstrakte Quantentheorie hinausgingen, sondern er versteht sie nur
tiefer und lässt daher die klassischen Begriffe und Strukturen auf der fundamentalen
Ebene der grundlegenden Postulate vollkommen weg, die bei ihm wie erwähnt allenfalls
im Rahmen einer gewissen Näherung aus der anschließenden mathematischen Konstruktion
hervorgehen. Denn die Phänomene der Quantentheorie haben gezeigt, dass die klassischen
Begriffe unseres Denkens nur ausreichen, um die Oberfläche der Wirklichkeit zu
beschreiben, aber nicht, um ins Innere der Natur geistig vorzudringen.

Die Kopenhagener Deutung der Quantentheorie und die abstrakte mathematische
Fassung der Quantentheorie im Sinne der Diracschen beziehungsweise von
Neumannschen Fassung des Hilbertraumes \cite{Dirac:1958}, \cite{Neumann:1932}
wird demnach zugrunde gelegt und auf eine noch tiefere erkenntnistheoretische und
logische Basis gestellt. Darauf aufbauend wird dann im Rahmen jener Auffassung
der Quantentheorie als einer reinen Theorie zeitlicher Logik durch die einfachst
mögliche Darstellung jener Logik durch binäre quantentheoretische Alternativen
die konkrete Physik systematisch konstruiert. Jene binären Alternativen sind die
fundamentalsten begrifflich überhaupt nur denkbaren Objekte in einer beliebigen
Quantentheorie und in einer überhaupt noch in logisch"=rationalen Kategorien
beschreibbaren Theorie. Die ganze Idee der Rekonstruktion der theoretischen Physik
besteht also im Wesentlichen darin, dass sich die abstrakte Quantentheorie und damit
in der Konsequenz die gesamte Beschreibung der Physik aus den Bedingungen
der Möglichkeit von Erfahrung ergibt. Denn unter dieser Voraussetzung müssen die
Naturgesetze notwendigerweise Weise in der Natur gelten, weil sie eben gerade
Bedingungen für jene Erfahrung über die Natur darstellen, es also menschliche
Erfahrung ohne sie gar nicht geben könnte. Dieser Gedanke entstammt der Kantischen
Philosophie, ist aber noch konsequenter, weil Kant nur den allgemeinen Grundrahmen
der Anschauungsformen und Kategorien als a priori bestimmt sah, während die speziellen
Naturgesetze bei Kant durch konkrete Erfahrung gefunden werden müssen. Bei Carl Friedrich
von Weizsäcker folgt nicht nur der Grundrahmen, sondern es folgen auch die speziellen
Naturgesetze aus jenen erkenntnistheoretischen Bedingungen, die im Wesentlichen durch
eine abstrakte zeitliche Logik charakterisiert sind, weshalb dies in der Art und Weise,
die Wirklichkeit zu denken, wie erwähnt bis in bestimmte Aspekte des Hegelschen Idealismus
hineinführt. Er hat die ganze begriffliche Basis der Quantentheorie der Ur"=Alternativen
bereits gelegt \cite{Weizsaecker:1971},\cite{Weizsaecker:1985},\cite{Weizsaecker:1992}.
Zudem hat er auch schon bestimmte Teile der Basis der mathematischen
Herleitung der konkreten Strukturen der Physik formuliert. Beides geschah
wie bereits angesprochen auch mit Hilfe seiner Mitarbeiter, vor allem Michael
Drieschner, Lutz Castell und Thomas Görnitz, der bis heute weiter an der
Theorie arbeitet. Aber von Weizsäcker und seinen Mitarbeitern gelang es
noch nicht, die Zustände in dem rein logischen Raum vieler Ur"=Alternativen,
welchen er als den Tensorraum der Ur"=Alternativen bezeichnet, in die Raum"=Zeit
abzubilden, und die Existenz der Quantenzahlen der Elementarteilchenphysik systematisch
herzuleiten und zu erklären. Desweiteren wurde vorher auch keine Formulierung
einer rein quantentheoretischen Fassung der Gravitation erreicht, die in einem
durch reine Quantenlogik charakterisierten begrifflichen Rahmen mit den Wechselwirkungen
der Elementarteilchenphysik durch abstrakte Symmetrieforderungen vereinheitlicht ist.
Dies ist mir aber nun in dieser Arbeit im entscheidenden Kern gelungen, sowenig
ich leugnen möchte, dass sich hier noch viele Einzelfragen stellen und das Ganze
noch weiter ausgearbeitet werden muss. Es werden in dieser Arbeit also zunächst
in sehr kondensierter und höchstens leicht modifizierter Weise die begrifflichen
Grundlagen der Quantentheorie der Ur"=Alternativen behandelt, wie sie im Wesentlichen
in \cite{Weizsaecker:1985} erarbeitet und dargestellt werden. Anschließend werden meine
eigenen darauf basierenden mathematischen Beiträge zur Quantentheorie der Ur"=Alternativen
dargestellt, die überall auf die zentralen Grundgedanken von Weizsäckers bezogen sind.
Diese meine eigenen Beiträge sind als ein Dienst an Carl Friedrich von Weizsäcker
und seiner visionären Theorie gedacht. Denn es ist und bleibt seine Theorie und
ich helfe ihm in dieser Arbeit als sein treuer Diener lediglich bei der Ausarbeitung.
Wenn er diesen Dienst an ihm und seiner Theorie seiner überlegenen geistigen
und menschlichen Größe als würdig erachtet hätte, dann wäre dies die beglückendste
und erhebendste Frucht meiner nunmehr etwa eineinhalb Dekaden währenden Beschäftigung
mit dieser wohl tiefsten, abstraktesten und fundamentalsten physikalischen Theorie,
die jemals erdacht wurde.

\subsection{Beziehung zu anderen Ansätzen}

Die gegenwärtig existierenden Ansätze zu einer einheitlichen Naturtheorie beziehungsweise
Quantentheorie der Gravitation gehen entweder von einer gewöhnlichen Hintergrund"=Raum"=Zeit
aus oder sie modifizieren die Struktur der Raum"=Zeit. In der Schleifenquantengravitation
etwa, die immerhin die Eigenschaft der Hintergrundunabhängigkeit aufweist, also keine
starre metrische Struktur voraussetzt, tritt an die Stelle der gewöhnlichen Raum"=Zeit
ein sogenanntes Spinnetzwerk, also ein System von Knotenpunkten, die durch Holonomien
verbunden sind, welche quantisiert werden und über denen damit ein Hilbertraum konstruiert wird,
sodass sich etwas ergibt, das man als Quantenschaum bezeichnen kann \cite{Rovelli:1989},
\cite{Rovelli:1994}, \cite{Rovelli:1995}, \cite{Rovelli:2004}, \cite{Kiefer:2004}.
Man hat von Knoten und Holonomien eingeschlossene Zellen, deren Volumen bestimmte
quantisierte Zustände annehmen kann. Dies ist schon ein deutlich abstrakteres und
der Quantentheorie angepassteres Bild des physikalischen Raumes. Aber der Raum
bleibt hier weiterhin, wenn auch in vollkommen veränderter quantentheoretischer Weise,
als eine in sich existierende Realität bestehen. In vielen Betrachtungen wird das Konzept
der Quanteninformation einfach im Rahmen bestehender physikalischer Theorien verwendet,
etwa in Zusammenhang mit der ADS/CFT"=Korrespondenz. Aber dies führt bezüglich der
prinzipiellen Frageebene überhaupt nicht weiter, weil man hier einfach die alte
begriffliche Basis der theoretischen Physik beibehält und das Konzept der Quanteninformation
im Rahmen eines bestehenden Raumes und bezogen auf unabhängige physikalische Träger
verwendet wird. In solchen Zusammenhängen ist dann auch zuweilen von dem holographischen
Prinzip die Rede, das sich eben auf Informationsströme im Raum bezieht, also von einem
solchen Informationsbegriff ausgeht, der an eine räumliche physikalische Realität
gebunden ist. Im letzten Kapitel wurde aber deutlich gemacht, dass der entscheidende
philosophische Sinn der Quantentheorie der Ur"=Alternativen gerade darin besteht,
die Wirklichkeit durch eine rein logische Beziehungsstruktur zu verstehen und
alle weiteren physikalischen Entitäten außer der Zeit als der fundamentalsten
jener Entitäten aus dieser reinen Quantenlogik herzuleiten. Näher an diese
Idee heran kommen die Ansätze von Fotini Markopoulou und einigen ihrer Kollegen,
welche die Welt als ein abstraktes nicht"=räumliches Netzwerk ansehen, in dem Information
ausgetauscht wird \cite{Konopka:2006}, \cite{Konopka:2008}, \cite{Hamma:2009}, \cite{Caravelli:2011},
\cite{Caravelli:2012}, \cite{CaravelliMarkopoulou:2012}, \cite{Markopoulou:2012}.
Aber hier bleibt eine solche von der Information, die ausgetauscht wird, vollkommen
unabhängige Struktur als Medium noch bestehen, die außerdem Annahmen wie eine
bestimmte Geschwindigkeit des Informationsaustausches enthält. Der Raum wird in
keiner Weise durch einen abstrakten Objektbegriff konstituiert. Der Ansatz der fermionischen
Systeme \cite{Finster:2015}, \cite{Finster:2018} betrachtet zwar zunächst abstrakte
mathematische Systeme ohne vorgegebene Raum"=Zeit. Um aber einen Bezug zur Wirklichkeit
der Natur herzustellen, muss dann doch eine vorgegebene Raum"=Zeit mit vorausgesetzt
werden. Zudem ist im Rahmen dieser Ansätze bisher in keiner Weise eine Begründung der
konkreten physikalischen Realitäten der Natur in der Weise geschehen, dass aus diesem
Schema die konkrete Herleitung der Existenz eines dreidimensionalen reellen
Raumes als eines Darstellungsmediums vollzogen worden wäre, oder man das
quantentheoretische Analogon zu den Wechselwirkungen der Natur in einem verallgemeinerten
einheitlichen quantentheoretischen Rahmen konkret ausformuliert und systematisch begründet
hätte. Desweiteren ist in diesen Ansätzen überhaupt keine philosophisch durchdachte begriffliche
Grundbasis gegeben. In \cite{Damour:2007} wird die Idee eines Verschwindens einer gewöhnlichen
räumlichen Beschreibungsweise der physikalischen Wirklichkeit auf fundamentaler Ebene lediglich als
Möglichkeit in Erwägung gezogen. Insofern erweist sich die Quantentheorie der Ur"=Alternativen
nicht nur als am begrifflich konsequentesten in Bezug auf die Realisierung einer physikalischen
Ontologie, welche einen rein quantentheoretischen Realitätsbegriff ohne vorgegebene räumliche
oder sonstige Hintergrundstruktur wirklich ernst nimmt. Sondern sie liefert darüber hinaus als
einzige bisher existierende Theorie in einem solch abstrakten Begriffsrahmen eine Begründung,
warum es physikalische Objekte gibt, die nachträglich in einem dreidimensionalen reellen
Raum dargestellt werden können und sich deshalb für uns innerhalb des Rahmens einer
indirekten Betrachtung als in einem solchen Raum existierend erweisen. Zudem kann nur in
ihrem Rahmen auch der Wechselwirkungsbegriff in einer solchen Weise verallgemeinert werden,
dass er sich als Näherung einer holistischen Betrachtungsweise ergibt, in der die höhere
Ganzheit durch eine Verschränkung nur näherungsweise voneinander separierbarer Alternativen
dargestellt wird, wobei die Gestalt dieser Verschränkung sich aus einem quantenlogisch
verallgemeinerten Symmetrieprinzip ergibt.

\subsection{Postulate der Quantentheorie der Ur-Alternativen}

Die Begründung der abstrakten Quantentheorie aufgefasst als die Darstellung einer
zeitlichen Logik, auf welche sich die fundamentalste mögliche Objektivation der
Natur gründet, basiert auf grundlegenden Postulaten, die unten systematisch
charakterisiert werden. Diese sind natürlich bezüglich ihres grundlegenden
Gehaltes seitens Carl Friedrich von Weizsäcker aufgestellt worden. Sie werden
jedoch hier meinerseits in einer solchen Weise organisiert, formuliert und
kondensiert, wie ich dies aus meiner Perspektive im Rahmen dieser Arbeit
als sinnvoll erachte. Das Wesentliche, was für das grundlegende Verständnis
der Rekonstruktion der theoretischen Physik aus Ur"=Alternativen notwendig ist,
ist damit aus meiner Sicht erfasst. Aber wer die originale Formulierung seitens
Weizsäcker einschließlich seiner philosophisch viel ausführlicheren und
exakteren Begründung kennenlernen will, der sei auf \cite{Weizsaecker:1985} verwiesen.
Dies hier noch weiter auszuführen, würde den Rahmen dieser Arbeit sprengen, in der
es ja im Wesentlichen um die weitere Ausarbeitung der konkreten mathematischen
Strukturen geht, sofern die begriffliche Basis einmal etabliert und begründet ist.
Die Postulate, die hier in leicht modifizierter Form wiedergegeben werden, stellen die
gesamte Basis der theoretischen Physik und damit der Naturwissenschaft überhaupt dar.
Alle weiteren Herleitungen der konkreten Gegebenheiten der Physik, die in dieser
Arbeit vollzogen werden, basieren auf diesen hochabstrakten Postulaten Weizsäckers.
Diese konkreten daraus erst hergeleiteten Gegebenheiten bestehen wie bereits
angesprochen in der Existenz eines dreidimensionalen reellen Ortsraumes als
Darstellung der Zustände im Tensorraum der Ur"=Alternativen; der Aufspaltung in
jene Freiheitsgrade, die räumlich dargestellt werden können und eine Poincare"=Symmetrie
aufweisen und die internen Quantenzahlen der Elementarteilchenphysik mit niedrigdimensionalen
$SU(N)$"=Symmetrien; und schließlich der grundlegenden Struktur der dynamischen Gleichungen
mit Tensorprodukten abstrakter Zustände, die aus der Verallgemeinerung dieser Symmetrien
durch Einbeziehung der abstrakten Ortsoperatoren im Tensorraum der Ur"=Alternativen
folgen, und rein quantentheoretische Strukturen repräsentieren, welche zur Gravitation
beziehungsweise den Wechselwirkungen der Elementarteilchenphysik näherungsweise
isomorph sind und zu Verschränkungen von Zuständen im Tensorraum der Ur"=Alternativen
als rein quantentheoretischem Analogon klassischer Wechselwirkungen führen.
Die Postulate der Quantentheorie der Ur"=Alternativen basieren wiederum auf folgender
Definition einer $N$"=fachen empirisch entscheidbaren logischen Alternative:

\noindent
\fbox{\parbox{163mm}{\textbf{Eine N"=fache empirisch entscheidbare logische Alternative
$A_N$ ist ein Aussagenverband, innerhalb dessen jede der $N$ Aussagen ein mögliches
Ereignis in der Wirklichkeit als wahr charakterisiert. Für diese Alternative gilt,
dass wenn eines Ihrer Elemente sich bei einer empirischen Untersuchung als wahr
herausstellt, alle anderen Elemente notwendigerweise falsch sein müssen.}}}\\
\\

\noindent
Die fundamentalen Postulate, aus denen die abstrakte Quantentheorie und damit
die gesamte theoretische Physik folgen, basieren auf diesem Begriff einer N"=fachen
Alternative, und sie können in der folgenden Weise charakterisiert werden:\\
\\
\noindent
\textbf{A) Postulat der Objektivation durch Alternativen:}

\noindent
Die einfachste, abstrakteste und allgemeinste denkbare formale Schematisierung
und Objektivation der Wirklichkeit besteht in dem logischen Schema einer $N$"=fachen
Alternative möglicher Aussagen bezüglich einer bestimmten empirischen Situation, wie
sie in der obigen Definition charakterisiert wurde. Denn jede beliebige Objektivation
beziehungsweise Beobachtung der Natur kann formal immer irgendwie entweder als eine
$N$"=fache Alternative dargestellt oder in eine Hintereinanderreihung solcher
$N$"=facher Alternativen zerlegt werden, die je für sich entschieden werden
können. Der Beginn der Argumentation liegt daher bei einer schlichten
$N$"=fachen empirischen Alternative $A_N$ mit den Elementen $a_1$,...,$a_N$:

\begin{equation}
A_N=\left(a_1,...,a_N\right).
\end{equation}
Dabei spielt es keinerlei Rolle, um welche Art von Realität es sich handelt, auf die
sich die Alternative bezieht. Das einzige Kriterium besteht darin, dass allgemeine
logische Gesetze für diese Wirklichkeit gelten müssen und sie damit in rational
untersuchbaren Kategorien beschreibbar ist. Die Alternativen beschreiben also eine
wirklich unabhängig von unserem Geist existierende Realität der Natur, stellen diese
aber in unserem Geist in einer bestimmten Weise dar. Damit definieren sie gemäß von
Weizsäcker unseren theoretischen Bezug zur Wirklichkeit und besitzen daher einer
epistemologische und eine ontologische Dimension zugleich. Die logischen Alternativen
sind also zugleich die physikalischen Objekte selbst.\\
\\
\noindent
\textbf{B) Postulat der Quantenlogik:}

\noindent
Für die Alternativen gilt eine Quantenlogik. Dies bedeutet, dass solange keine
konkrete empirische Entscheidung vorgenommen wird, nicht zwangsläufig objektiv
bestimmt ist, dass ein Element der Alternative wahr sein muss, sondern
die Elemente der Alternative beliebige Wahrheitswerte im kontinuierlichen
Zahlenspektrum zwischen $0$ und $1$ annehmen können. Dies entspricht der
Zuordnung von reellen Werten $\varphi\left(a_n\right)\equiv \varphi_n$ zu den
einzelnen Elementen der Alternative $A_N$, deren Quadrate die jeweilige
Wahrscheinlichkeit darstellen, dass sich die entsprechende Aussage bei
einer empirischen Prüfung als wahr erweist, wobei gelten muss:

\begin{equation}
\sum_{n=1}^N |\varphi\left(a_n\right)|^2=\sum_{n=1}^N |\varphi_n|^2=1,
\end{equation}
da die Gesamtwahrscheinlichkeit eins betragen muss. Dies führt
zu einem $N$"=dimensionalen Vektorraum $V_N$ als Menge der möglichen
Wahrscheinlichkeitsverteilungen $\varphi_N$ über der $N$"=fachen
Alternative $A_N$. Wenn man ein inneres Produkt in diesem
Vektorraum auszeichnet:

\begin{equation}
\langle \left(\varphi_B\right)_N|\left(\varphi_A\right)_N \rangle
=\sum_{n=1}^N \left(\varphi_B\right)_n \left(\varphi_A\right)_n,
\end{equation}
so wird der Vektorraum $V_N$ zu einem Hilbertraum $\mathcal{H}_N$.
Das Quadrat dieses inneren Produktes beschreibt die Wahrscheinlichkeit $\mathcal{W}\left[\left(\varphi_B\right)_N,\left(\varphi_A\right)_N\right]$,
den Zustand $\left(\varphi_B\right)_N$ bei einer Beobachtung vorzufinden,
wenn vorher der Zustand $\left(\varphi_A\right)_N$ vorlag:

\begin{equation}
\mathcal{W}\left[\left(\varphi_B\right)_N,\left(\varphi_A\right)_N\right]=
|\langle\left(\varphi_B\right)_N|\left(\varphi_A\right)_N \rangle|^2.
\end{equation}
Bei einer solchen Quantenlogik ist das in der klassischen Logik gültige "`tertium non datur"'
verletzt, der Satz vom ausgeschlossenen Dritten, demgemäß bezüglich einer Aussage gilt,
dass die Aussage entweder definitiv wahr oder definitiv falsch ist, auch wenn man im
Einzelfall nicht weiß, was der Fall ist. Carl Friedrich von Weizsäcker begründet die
Gültigkeit der Quantenlogik aus der Struktur der Zeit mit Vergangenheit, Gegenwart und
Zukunft, denn bei einer Aussage, die sich auf die Zukunft bezieht, muss im Allgemeinen
nicht objektiv bestimmt sein, ob sie wahr oder falsch ist. Dieser Argumentationsweg wird hier
deshalb nicht beschritten, weil zusätzlich in jedem Falle noch ein kontinuierlicher Zeitparameter
und mit ihm eine Dynamik eingeführt werden muss, welche die Entwicklung der sich auf die
Alternativen beziehenden Wahrscheinlichkeitszustände determiniert. Der Messprozess kann auch
als ein dynamischer Vorgang gedeutet werden, sodass dem Indeterminismus in der Quantentheorie
im Gegensatz zur Quantenlogik vielleicht nur epistemologischer, aber kein ontologischer
Charakter zukommt. Der wirklich ontologische Gehalt der Quantentheorie, insbesondere der
Unbestimmtheitsrelation, besteht demnach nicht im Indeterminismus, sondern ausschließlich
in der nicht"=Lokalität, also der Tatsache, dass Logik beziehungsweise Quantenlogik
fundamentaler ist als Geometrie. Dass die Quantenlogik hier als vollkommen unabhängig von der
Zeit bestehend vorausgesetzt wird, ist bezüglich der Grundbasis einer der wenigen Unterschiede
zum ursprünglichen Argumentationsweg von Weizsäckers und diese Frage ist vielleicht mit die
Schwierigste der gesamten Rekonstruktion, hat mit dem berühmten Messproblem zu tun und
wird hier als noch nicht vollständig geklärt angesehen. Dass es eine Offenheit der Zukunft
jenseits einer physikalischen Beschreibungsweise der Wirklichkeit geben mag, das ist eine
vollkommen andere Frage.\\
\\
\noindent
\textbf{C) Postulat des Finitismus:}

\noindent
Die Anzahl $N$ der Elemente einer Alternative $A_N$ muss immer endlich bleiben.
Sie kann beliebig groß sein, aber sie muss in jedem Falle einen endlichen Wert aufweisen.
Dies bedeutet, dass die in der Welt vorhandene Information endlich ist.\\
\\
\noindent
\textbf{D) Postulat der näherungsweisen Trennbarkeit:}

\noindent
In Wirklichkeit sind alle Alternativen des gesamten Kosmos im Sinne
des Holismus in eine komplexe Beziehungsstruktur eingebunden.
Aber als eine Näherung ist zunächst die Trennbarkeit einer
Alternative von allen anderen Alternativen notwendig.
Dies bedeutet, dass die Entscheidung dieser Alternative
als unabhängig von allen anderen Alternativen angesehen
werden muss, um eine empirische Beobachtungssituation
abstrakt zu definieren. Die Korrektur dieser Näherung führt dann
allerdings zum quantentheoretischen Analogon der Wechselwirkung.
Denn gemäß dem Verständnis der natürlichen Wirklichkeit Platons,
welches durch die Quantentheorie bestätigt wird, lässt sich
die Natur im Inneren nur als eine höhere Einheit verstehen
\cite{Platon:Parmenides}, \cite{Weizsaecker:1981}.
Im Parmenides"=Dialog wird diesbezüglich der Begriff des Einen in
Abgrenzung zu einem Ganzen erörtert, wobei das Ganze im Gegensatz
zum Einen aus Teilen besteht. Die Aufspaltung der Natur in voneinander
trennbare Teile entspricht einer Näherung. In der Quantentheorie
manifestiert sich dies konkret darin, dass im Gesamt"=Hilbertraum
$\mathcal{H}_G$ mehrerer Objekte, der durch die Bildung des
Tensorproduktes der Hilberträume der $M$ Einzelobjekte entsteht:
$\mathcal{H}_G=\mathcal{H}_1 \otimes ... \otimes \mathcal{H}_M$,
die Zustände dieses Gesamt"=Hilbertraumes, in denen die Zustände
der Einzelobjekte je für sich selbst ohne Beziehung zu den Zuständen
der anderen Objekte, eindeutig definiert sind, einer Menge des Maßes
null entspricht. In allen anderen Zuständen liegt eine Verschränkung
der Zustände der Einzelobjekte vor, in welcher der Gesamtzustand nicht
einfach als ein Produkt der Einzelzustände dargestellt werden kann,
sondern nur als eine Linearkombination von Produkten.\\
\\
\noindent
\textbf{E) Postulat der Symmetrie:}

\noindent
Mit dem Postulat der Trennbarkeit der Alternativen unmittelbar verbunden
ist die Eigenschaft der Symmetrie bezüglich der verschiedenen Elemente
$a_n, n=1,...,N$ einer Alternative $A_N$. Dies bedeutet nichts anderes,
als dass im Prinzip a priori kein Element vor dem anderen ausgezeichnet ist,
was einer Permutationssymmetrie $\mathcal{P}\left(N\right)$ der Elemente
$a_n, n=1,...,N$ der Alternative $A_N$ untereinander entspricht.
Wenn man den Hilbertraum $\mathcal{H}_N$ der gesamten Menge aller
Wahrscheinlichkeitszustände betrachtet, so entspricht dem
eine $SO(N)$"=Symmetrie.\\
\\
\noindent
\textbf{F) Postulat der kontinuierlichen Zeitentwicklung:}

\noindent
Es gibt eine kontinuierliche Zeit, die als eine von den Alternativen
unabhängig bestehende Entität sowohl in unserem Geist als auch in der
Wirklichkeit der Natur existiert, in welche die Alternativen aber
eingebettet sind. Die Zeit kann mathematisch in einer objektivierenden
Beschreibung der Wirklichkeit der Natur als ein eindimensionaler reeller
Parameter $t$ beschrieben und dargestellt werden. Eine näherungsweise
trennbare Alternative behält in der Zeit ihre grundlegende Identität
und Struktur. Das heißt, die Zahl $N$ der Elemente und der sich aus
den möglichen Wahrscheinlichkeitsverteilungen über diesen Elementen
aufspannende Hilbertraum $\mathcal{H}_N$ bleiben in der Zeit konstant.
Aber der Zustand der Alternative beziehungsweise Vektor im Hilbertraum
$\varphi_N$ entwickelt sich kontinuierlich mit der Zeit,
$\varphi_N=\varphi_N\left(t\right)$. Die Zeitentwicklung entspricht
damit einem einparametrigen Automorphismus des Hilbertraumes $\mathcal{H}_N$
über der Alternative $A_N$ auf sich selbst. Es wird weiter unten
gezeigt werden, wie dies zur Zweckmäßigkeit der Darstellung in einem
komplexen Hilbertraum führt. Es sei darauf hingewiesen, dass dies solange
keine Messung durchführt wird, eine deterministische Beschreibung der
Wirklichkeit impliziert. Da aber auch Messungen als dynamische Prozesse
angesehen werden müssen, impliziert dies generell einen Determinismus,
zumindest innerhalb der Physik. Das eigentliche fundamentale quantentheoretische
Element bei der Beschreibung der Wirklichkeit besteht also wie im Zusammenhang
mit dem Postulat der Quantenlogik bereits erwähnt nicht im Indeterminismus,
sondern in der nicht"=Lokalität, also der Tatsache, dass Logik beziehungsweise
Quantenlogik fundamentaler ist als Geometrie.\\
\\
\noindent
\textbf{G) Postulat der Aufspaltung in Ur-Alternativen:}

\noindent
Jede Alternative $A_N$ kann aus rein logischen Gründen in eine
Kombination binärer Alternativen aufgespalten und entsprechend
dargestellt werden. Dies geschieht über die Bildung des Cartesischen
Produktes beziehungsweise Tensorproduktes einzelner binärer
Alternativen:

\begin{equation}
A_N=\bigotimes_{n=1}^{M} u_n,\quad u_n=\left(u_{n1},u_{n2}\right),\quad N \leq 2^M.
\label{Aufspaltung_Alternativen}
\end{equation}
Eine solche Darstellung ist zweckmäßig, weil eine binäre Alternative
die einfachste und damit fundamentalste logische Einheit und in diesem
Zusammenhang zugleich einfachste und fundamentalste potentielle
Möglichkeit einer empirischen Entscheidung darstellt.

\subsection{Gründe für die innere Überzeugungskraft der Theorie}

Die Quantentheorie der Ur"=Alternativen ist nicht irgendein beliebiger Ansatz zur
Vereinheitlichung der Physik. Vielmehr gibt es eine Fülle prinzipieller Gründe,
welche in beinahe unausweichlicher argumentativer Evidenz die innere Notwendigkeit der
Quantentheorie der Ur"=Alternativen für ein wirkliches Verständnis der Einheit der
Wirklichkeit der Natur aufzeigen. Diese Gründe sollen im Folgenden systematisch
aufgezählt und erläutert werden:\\
\\
\textbf{A) Abstraktheit und Einfachheit der reinen Logik:}

\noindent
Verstehen bedeutet in der Naturwissenschaft die Rückführung der konkreten, speziellen,
komplizierten und daher an der Oberfläche liegenden Wirklichkeit auf eine abstrakte, allgemeine,
einfache und daher fundamentale Wirklichkeit. Dem entspricht eine Subsummierung vieler
konkreter, spezieller und komplizierter Begriffe unter wenige abstrakte, allgemeine und
einfache Begriffe. Die fundamentalste und abstrakteste Basis unseres rationalen Denkens
sind die Strukturen der reinen Logik. Die beiden fundamentalsten Begriffe unseres Denkens
in seinem Bezug zur Wirklichkeit überhaupt sind Sein und Zeit, weshalb Heidegger sein
philosophisches Hauptwerk danach benannt hat \cite{Heidegger:1927}. Sofern die geistige
Ordnung der Natur im Rahmen einer objektivierenden Beschreibung in unserem menschlichen
Geist durch Strukturen erfasst werden soll, sind die fundamentalsten Strukturen des
sich in dieser Weise entbergenden natürlichen Seins die einfachsten logischen Strukturen
in der Zeit. Ur"=Alternativen aber wiederum sind die einfachsten Strukturen einer reinen
Quantenlogik in der Zeit. Eine fundamentalere noch in rationalen Strukturen fassbare
quantentheoretische Beschreibung der natürlichen Wirklichkeit ist also überhaupt gar
nicht mehr denkbar. Demnach ist die Quantentheorie der Ur"=Alternativen auch in ihren
entscheidenden Grundannahmen denkbar sparsam, denn sie basiert im Wesentlichen nur auf
Zeit und Logik mit einigen beinahe evidenten Zusatzpostulaten, welche die Darstellung
dieser Grundrealitäten nur näher charakterisieren. Dies führt dann auf direktem Wege
zur abstrakten Quantentheorie.\\
\\
\textbf{B) Ur-Alternativen sind nicht an räumliche Kausalstrukturen gebunden:}

\noindent
Sowohl die abstrakte mathematische Struktur der Quantentheorie alsauch die konkreten Phänomene
der Quantentheorie wie das Doppelspaltexperiment und das Einstein"=Podolsky"=Rosen"=Paradoxon
zeigen explizit, dass es gemäß der Quantentheorie auf der fundamentalen Ebene keine voneinander
trennbaren Objekte gibt, die sich im physikalischen Ortsraum befinden und an dessen topologisch
definierte Kausalstrukturen gebunden sind. Vielmehr ist der physikalische Ortsraum nur eine Art
eines Darstellungsmediums einer abstrakten dahinter in Wirklichkeit stehenden quantentheoretischen
Beziehungsstruktur, die im Rahmen der Quantentheorie der Ur"=Alternativen basierend auf der
abstrakten Hilbertraum"=Formulierung der Quantentheorie im Sinne einer reinen quantenlogischen
Struktur in der Zeit gedeutet wird. Die Ur"=Alternativen setzen als rein logische Objekte einen
solchen Ortsraum aber gerade nicht voraus. Vielmehr kann man aus ihnen mathematisch stringent
die Existenz eines solchen dreidimensionalen physikalischen Ortsraumes als eines
Darstellungsmediums begründen. Damit ist sehr überzeugend erklärt, warum die physikalischen
Objekte sich einerseits als in einem dreidimensionalen reellen Raum befindend in Form von
Wellenfunktionen darstellen, der sich mit der Zeit, ebenfalls dargestellt als ein reeller
Parameter, zu einer (3+1)"=dimensionalen Raum"=Zeit verbindet, sich aber andererseits
so verhalten, als seien sie an die lokalen Kausalitätsstrukturen dieses Raumes überhaupt
nicht gebunden. Dies steht eben mit der Tatsache in Zusammenhang, dass die Geometrie sich
in diesem begrifflichen Rahmen als Ausfluss einer ontologisch gedeuteten Quantenlogik ergibt.\\
\\
\textbf{C) Schlechthinnige Unteilbarkeit:}

\noindent
Gemäß der zweiten Kantischen Antinomie aus der "`Kritik der reinen Vernunft"' \cite{Kant:1781}
kann es aus prinzipiellen begrifflichen Gründen keine fundamentalen und zugleich räumlich ausgedehnten
Objekte geben. Denn das Volumen jedes noch so kleinen räumlich ausgedehnten Objektes kann zumindest
im Prinzip immer noch weiter in Teilvolumina zerlegt werden. Ur"=Alternativen sind aber
gar keine räumlichen Objekte, sondern als einfache auf reiner Quantenlogik basierende
Ja"=Nein"=Entscheidungen die logisch einfachsten und fundamentalsten Elemente, die überhaupt
denkbar sind. Deshalb sind sie aus rein logischen Gründen unteilbar. Diese prinzipielle
begriffliche Unteilbarkeit kann aufgrund ihres absoluten Charakters als schlechthinnige
Unteilbarkeit bezeichnet werden.\\
\\
\textbf{D) Es ergeben sich die Strukturen der konkreten Physik aus den Ur"=Alternativen:}

\noindent
Basierend auf der Quantentheorie der Ur"=Alternativen kann nicht nur die Existenz eines allen
Objekten gemeinsamen reellen dreidimensionalen Ortsraumes mathematisch exakt hergeleitet werden,
der mit der Zeit als einem reellen Parameter zu einer (3+1)"=dimensionalen Raum"=Zeit
verbunden ist, sondern es können, wie in dieser Arbeit dargelegt werden wird,
auch die inneren Quantenzahlen des Standardmodells der Elementarteilchenphysik
einschließlich des Spin und basierend auf der Forderung möglichst hoher Symmetrie im
rein quantenlogischen Raum der Ur"=Alternativen auch ein rein quantentheoretisches
Analogon zu den Wechselwirkungen der Elementarteilchenphysik sowie eine rein
quantentheoretische Fassung der allgemeinen Relativitätstheorie begründet werden.
Hierbei ist entscheidend, dass an die Stelle der gewöhnlichen punktweisen Produkte
von Feldern auf einer vorgegebenen Raum"=Zeit abstrakte Tensorprodukte rein
quantenlogischer Zustände treten und darauf basierend eine Verschränkung solcher
Zustände verschiedener Objekte als quantentheoretisches Analogon zur klassischen
Wechselwirkung definiert wird.\\
\\
\textbf{E) Diskretheit:}

\noindent
Der Ursprung der Quantentheorie war die Einführung von Wirkungsquanten seitens Max Planck.
Durch diese Diskretisierung wurde die Ultraviolettkatastrophe verhindert, also die
Entstehung unendlicher Energiewerte. Im Rahmen relativistischer Quantenfeldtheorien
kommt aber vor allem durch die punktweisen Wechselwirkungen wieder ein Kontinuum in die
Beschreibung hinein. Es ist zu erwarten, dass durch die Ur"=Alternativen, welche a priori
eine Diskretheit der Beschreibung der Natur voraussetzen, die zudem noch auf einer
grundlegenderen als der raum"=zeitlichen Ebene liegt, das Problem der Entstehung der
Unendlichkeiten von vorneherein gelöst und damit die Notwendigkeit zur Renormierung
umgangen wird.

\subsection{Abstrakte Dynamik und komplexer Hilbertraum}

Die Beschreibung der Zeit wird wie erwähnt in der Quantentheorie schlicht durch einen reellen
Parameter $t$ eingeführt. Sie ist keine gewöhnliche Messgröße, muss als fundamentalste
physikalische Realität neben der reinen Quantenlogik schlicht vorausgesetzt werden und kann sich
demnach auch nicht aus Alternativen ergeben. Die Dynamik muss gemäß dem Postulat der Zeitentwicklung
naturgemäß einem Automorphismus des Hilbertraumes der Alternativen $\mathcal{H}_N$ auf
sich selbst entsprechen. Die Symmetriegruppe muss demnach eine einparametrige Untergruppe
der fundamentalen Symmetriegruppe $SO(N)$ des Hilbertraumes $\mathcal{H}_N$ darstellen.
Dies führt zu einer $N/2$"=dimensionalen Darstellung der Symmetriegruppe $SO(2)$, welche in
zweidimensionalen Unterräumen des $\mathcal{H}_N$ wirkt. Die Zeitentwicklung des $j$"=ten
Elementes einer $N$"=fachen Alternative wird demnach durch eine Gleichung der folgenden
allgemeinen Form beschrieben:

\begin{equation}
\partial_t \varphi_j\left(t\right)=\sum_{k=1}^N H_{jk} \varphi_k\left(t\right),\quad k=1,...,N,
\label{SchroedingerGleichungAlternative}
\end{equation}
wobei der Zeitentwicklungsoperator $H$ die folgende Gestalt aufweist:

\begin{equation}
H_{kl}=\sum_{j=1}^{N/2}\omega_j \theta^{j}_{kl},
\end{equation}
wobei hier wiederum vorausgesetzt wurde, dass $N$ eine gerade Zahl
ist und der Tensor dritter Stufe $\theta$ wie folgt aussieht:

\begin{equation}
\theta_{kl}^j=\left\{\begin{matrix}&+1&\ \text{für}\ k=2j-1,\ l=2j \\&-1&\ \text{für}\ k=2j,\ l=2j-1\\&0&\ \text{ansonsten}\end{matrix}\right. .
\end{equation}
Dies bedeutet, dass der Zeitentwicklungsoperator $H$ als Matrix ausgeschrieben die
folgende Gestalt aufweist:

\begin{equation}
H=\left(\begin{matrix}0 & \omega_1 & 0 & 0 &  ... & 0 & 0\\
                      -\omega_1 & 0 & 0 & 0 & ... & 0 & 0\\
                      0 & 0 & 0 & \omega_2 & ... & ... & ...\\
                      0 & 0 & -\omega_2 & 0 & ... & ... & ...\\
                      ... & ... & ... & .... & ... & ... & ...\\
                      0 & 0 & ... & ... & ... & 0 & \omega_j\\
                      0 & 0 & ... & ... & ... & -\omega_j & 0\\ \end{matrix}\right),\quad j=1,...,N/2.
\end{equation}
Damit besteht die Darstellung des Zeitentwicklungsoperators $H$ als Matrix aus
zweidimensionalen Teilmatrizen, welche die Symmetriegruppe $SO(2)$ in den
entsprechenden zweidimensionalen Unterräumen darstellen. Da sich die durch
den Operator $H$ induzierten Zeittransformationen also in zweidimensionalen
Unterräumen des Gesamt"=Hilbertraumes vollziehen, ist es bezüglich einer
kompakteren Darstellung zweckmäßig, die reellen Alternativen in komplexe
Alternativen von halbierter Dimensionszahl zu überführen. Dies kann in der
folgenden Weise geschehen:

\begin{equation}
\tilde{\varphi}_j=\varphi_{2j-1}+i\varphi_{2j},\quad j=1,...,N/2.
\label{Alternativekomplex}
\end{equation}
Die Zeitentwicklung vollzieht sich in dieser Darstellung in den jeweiligen eindimensionalen
komplexen Unterräumen. Dem entspricht die Isomorphie der $SO(2)$"=Gruppe zur $U(1)$"=Gruppe,
die damit als die fundamentale Zeitentwicklungsgruppe der Quantentheorie statuiert ist.
Die Zeitentwicklung der $j$"=ten Komponente der komplexen Alternative wird demnach
durch die folgende Gleichung beschrieben:

\begin{equation}
i\partial_t \tilde{\varphi}_j\left(t\right)=\omega_{j} \tilde{\varphi}_j\left(t\right),\quad j=1,...,N/2,
\end{equation}
wobei hier über den Index $j$ auf der rechten Seite natürlich nicht summiert wird.
Wenn man nun eine entsprechende auf die komplexe Alternative bezogene Matrix
in der folgenden Weise definiert:

\begin{equation}
\tilde{H}_{kl}=\omega_k \delta_{kl},\quad k,l=1,...,N/2,
\end{equation}
wobei hier über den Index $k$ auf der rechten Seite auch nicht summiert wird,
so kann man die Dynamik der Alternative durch folgende Gleichung ausdrücken:

\begin{equation}
i\partial_t \tilde{\varphi}_j\left(t\right)
=\sum_{k=1}^{N/2}\tilde{H}_{jk} \tilde{\varphi}_k\left(t\right)\quad\Leftrightarrow\quad
i\partial_t \tilde{\varphi}\left(t\right)=\tilde{H}\tilde{\varphi}\left(t\right).
\label{Schroedinger_Gleichung_Alternative}
\end{equation}
Diese Gleichung hat die Gestalt der allgemeinen Formulierung der Schrödingergleichung und stellt
damit zugleich die abstrakteste Gestalt der Schrödingergleichung dar, die in ihrer speziellen
Formulierung im Rahmen der Wellenmechanik in \cite{Schroedinger:1926} zuerst formuliert wurde.
Die entsprechende Zeitentwicklung der Alternative, die sich aus der Schrödingergleichung ergibt,
lautet dementsprechend wie folgt:

\begin{equation}
\tilde{\varphi}\left(t\right)=\exp\left[-i\tilde{H}\left(t-t_0\right)\right]\tilde{\varphi}\left(t_0\right).
\label{Zeitentwicklung_Alternative}
\end{equation}
Bei einer $N$"=dimensionalen reellen Alternative, die einem $N$"=dimensionalen reellen Hilbertraum
entspricht, weist die Zeitentwicklung also $N/2$ reelle Parameter auf, die jeweils eine $SO(2)$"=Transformation
generieren, was sich zweckmäßig in einem $N/2$"=dimensionalen komplexen Hilbertraum über eine zur
$SO(2)$"=Gruppe isomorphe $U(1)$"=Gruppe darstellen lässt. Man hat es also mit einer $N/2$"=fachen
Darstellung der $U(1)$"=Gruppe in einem $N/2$"=dimensionalen komplexen Hilbertraum zu tun.
Diese Beschreibung der Zeitentwicklung liefert die Erklärung dafür, warum in der Quantentheorie
komplexe Hilberträume verwendet werden müssen. Die Symmetriegruppe eines $N$"=dimensionalen
reellen Hilbertraumes ist die $SO(N)$, welche $N\left(N-1\right)/2$ Generatoren aufweist und damit
die gleiche Zahl an reellen Transformationsparametern. Die Symmetriegruppe des entsprechenden komplexen
Hilbertraumes mit $N/2$ Dimensionen ist die $U(N/2)$, die $N^2/4$ Generatoren aufweist. Allerdings enthält
jeder der Generatoren aufgrund der komplexen Zahlen die doppelte Zahl an reellen Transformationsparametern,
also $N^2/2$. Diese Zahl enthält auch die $N/2$ reellen Parameter der Zeitentwicklung in den
eindimensionalen komplexen Unterräumen. Die Zahl der reellen Transformationsparameter
der $U(N/2)$"=Gruppe, also $D_R\left[U(N/2)\right]$, entspricht also der Zahl der reellen
Transformationsparameter der $SO(N)$"=Gruppe als fundamentaler Transformationsgruppe
eines reellen Hilbertraumes mit doppelter Dimensionszahl, also $D_R\left[SO(N)\right]$,
hinzuaddiert zu der Zahl der reellen Transformationsparameter der $N/2$"=dimensionalen
Zeitentwicklungsgruppe $T(N/2)$ als $N/2$"=dimensionaler Darstellung der $U(1)$"=Gruppe,
also $D_R\left[T(N/2)\right]$, denn es gilt das Folgende:

\begin{equation}
D_R\left[U(N/2)\right]=2\left(\frac{N}{2}\right)^2=2\frac{N^2}{4}=\frac{N^2}{2},
\label{Dimensionszahl_U(N/2)}
\end{equation}
und es gilt desweiteren:

\begin{equation}
D_R\left[SO(N)\right]+D_R\left[T(N/2)\right]=\frac{N(N-1)}{2}+\frac{N}{2}=\frac{N^2}{2}-\frac{N}{2}+\frac{N}{2}=\frac{N^2}{2}.
\label{Dimensionszahl_SO(N)+T(N/2)}
\end{equation}

\subsection{Das Bohrsche Korrespondenzprinzip in der Quantentheorie der Ur-Alternativen}

Die allermeisten fundamentalen Strukturen der theoretischen Physik, die in dieser Arbeit ausgehend
von der abstrakten Quantentheorie betrachtet werden, können aus der in diesem Kapitel zusammengefassten
begrifflichen Basis der von Weizsäckerschen Theorie begründet werden. Dies gelingt aber bisher
nicht bei ausnahmslos allen Strukturen, die in der vorliegenden Arbeit behandelt werden. Sofern etwas
nicht in Strenge hergeleitet werden kann, muss das Bohrsche Korrespondenzprinzip herangezogen werden.
Dieses besagt bekanntlich, dass die Naturgesetze im quantentheoretischen Falle in der Weise
postuliert werden müssen, dass sich im klassischen Grenzfall die Struktur der bekannten Naturgesetze
an der Oberfläche ergibt. In diesem Sinne werden in dieser Arbeit bestimmte Konstruktionen aus
Ur"=Alternativen in der Weise vorgenommen, dass sich daraus in einer klassischen Näherung die
Strukturen der bekannten Physik ergeben. Dies bedeutet, dass die Konstruktion aus Ur"=Alternativen
dann als das im Rahmen einer reinen Quantentheorie formuliertes Analogon zum gewöhnlichen Fall
angesehen werden kann. Insofern ist dies dann nichts anderes als eine bestimmte Art einer allerdings
deutlich radikaleren Quantisierung in einem sehr viel prinzipieller gemeinten Sinne. In einer
gewöhnlichen Feldquantisierung werden gewöhnliche physikalische Größen Vertauschungsrelationen
unterworfen, wodurch sie zu Operatoren in Hilberträumen werden. In der Quantentheorie der
Ur"=Alternativen wird überhaupt keine klassische physikalische Wirklichkeit vorausgesetzt,
also eine spezielle Theorie, die dann quantentheoretisch formuliert wird. Sondern aus den
aus der reinen Quantentheorie für sich selbst sich ergebenden Strukturen wird entweder alles
in Strenge hergeleitet, was bei den meisten Strukturen tatsächlich der Fall ist, oder es werden
aus den elementaren in ihr vorkommenden Bausteinen solche Größen und Strukturen konstruiert,
sodass sich in einer bestimmten Darstellung und Näherung die realen Strukturen der empirischen
Wirklichkeit der Natur ergeben. In diesem spezifischen letzteren Sinne ist dann das Bohrsche
Korrespondenzprinzip der Quantentheorie im Rahmen der Quantentheorie der Ur"=Alternativen
anzuwenden und zu interpretieren.

\section{Darstellung durch Ur-Alternativen}

\subsection{Einzelne Ur-Alternativen}

Man kann jede beliebige $N$"=fache Alternative wie im Postulat der Aufspaltung formuliert
natürlich aus rein logischen Gründen in eine Kombination $M$ binärer Alternativen
$u_n,\ n=1,...,M$ überführen, was gemäß ($\ref{Aufspaltung_Alternativen}$) über die
Bildung des Cartesischen Produktes beziehungsweise Tensorproduktes einzelner Alternativen
geschieht. Dem entspricht die Aufspaltung in die elementarsten Bestandteile nicht nur der
Quantentheorie, sondern jedes noch in rationalen Kategorien mit Hilfe von mathematischen
Strukturen beschreibbaren Systems. Denn diese binären Alternativen $u_n,\ n=1,...,M$
sind die einfachsten logischen Objekte, die überhaupt denkbar und möglich sind, und sie
werden aufgrund ihres im Rahmen der von Weizsäckerschen Rekonstruktion der theoretischen
Physik aus der abstrakten Quantentheorie fundamentalen Charakters als Ur"=Alternativen
bezeichnet. Sie besitzen atomistischen Charakter im Sinne schlechthinniger Unteilbarkeit,
denn ihre Unteilbarkeit hat logische Gründe, weil eine binäre Alternative die einfachste
Alternative ist, die logisch noch eine Entscheidungsmöglichkeit in sich enthält.
Eine einfache binäre Alternative wird durch Zuordnung komplexer Wahrheitswerte,
was einer Quantisierung entspricht, zu einem gewöhnlichen normierten zweidimensionalen
komplexen Spinor, der ihren Zustand repräsentiert:

\begin{equation}
u=\left(\begin{matrix} u_{1}\\ u_{2}\end{matrix}\right)\quad\longrightarrow\quad |\varphi\rangle
=\left(\begin{matrix}\varphi\left(u_1\right)\\ \varphi\left(u_2\right)\end{matrix}\right)
=\left(\begin{matrix}\varphi_a+i\varphi_b\\ \varphi_c+i\varphi_d\end{matrix}\right),\quad
\sqrt{\varphi_a^2+\varphi_b^2+\varphi_c^2+\varphi_d^2}=1.
\label{Ur-Alternative}
\end{equation}
Das innere Produkt $\langle \ \cdot\ |\ \cdot\ \rangle$ der zu zwei Ur"=Alternativen
$u$ und $v$ gehörigen Spinorzustände $|\varphi_u\rangle$ und $|\varphi_v\rangle$
ist gegeben durch:

\begin{equation}
\langle \varphi_u|\varphi_v \rangle=\sum_{r=1}^2 \varphi\left(u_r\right) \varphi\left(v_r\right),
\end{equation}
wobei die Basiszustände der einzelnen Ur"=Alternativen orthogonal zueinander sind:

\begin{equation}
\langle u_r|u_s \rangle=\delta_{rs},\quad r,s=1,2.
\end{equation}
Der Aufspaltung der Alternativen $A_N$ in eine Kombination von Ur"=Alternativen $u_n,\ n=1,...,M$
entspricht in der Quantentheorie der Ur"=Alternativen eine Überführung des Hilbertraumes
$\mathcal{H}_{A_N}$ einer $N$"=fachen Alternative in ein $M$"=faches Tensorprodukt
zweidimensionaler komplexer Hilberträume $\mathcal{H}_{u_n}$, deren Elemente
$\varphi_{u_n}$ zweidimensionale Spinoren sind:

\begin{equation}
\mathcal{H}_{A_N}=\bigotimes_{n=1}^M \mathcal{H}_{u_n},\quad N \leq 2^M.
\end{equation}

\subsection{Der Tensorraum vieler Ur-Alternativen}

Man kann also die vielen Teil"=Hilberträume $\mathcal{H}_{u_n}$, durch die sich der Hilbertraum
einer beliebigen Alternative $\mathcal{H}_{A_N}$ darstellen lässt, in einer bestimmten Weise
zu einem Tensorraum vieler Ur"=Alternativen $\mathcal{H}_T$ zusammenfassen. Da hier Alternativen
endlicher aber beliebig hoher Dimensionszahl betrachtet werden können, kann auch der Tensorraum
der Ur"=Alternativen $\mathcal{H}_T$ im Prinzip beliebig viele Ur"=Alternativen enthalten.
Dessen Elemente sind Zustände über den Basiszuständen aller einzelnen Ur"=Alternativen:

\begin{equation}
|\Psi\left(u_{11},u_{12};u_{21},u_{22};...;u_{M1},u_{M2}\right)\rangle \in \mathcal{H}_T,\quad M < \infty.
\end{equation}
Von entscheidender Bedeutung ist in diesem Zusammenhang die Eigenschaft der Ununterscheidbarkeit
der einzelnen Ur"=Alternativen. Natürlich kann eine elementare logische Einheit nicht von
einer anderen unterschieden werden. Denn sobald man solche elementaren Informationseinheiten
einzeln kennzeichnen könnte, was die Voraussetzung einer Unterscheidung wäre, entspräche
dem schon wieder neue Information. Daher entsprechen die wirklichen Zustände im Tensorraum
der Ur"=Alternativen der Summe der Zustände in Bezug auf alle Permutationen der Ur"=Alternativen
untereinander. Aber dies kann auf zwei unterschiedliche Weisen geschehen, was bedeutet,
dass im einen Falle über die Zustände in allen Permutationen der einzelnen Ur"=Alternativen
unter der Voraussetzung summiert wird, dass das Vorzeichen immer gleich bleibt, und im
anderen Falle über die Zustände in allen Permutationen der einzelnen Ur"=Alternativen
unter der Bedingung summiert wird, dass die Anzahl an Vertauschungen der Ur"=Alternativen,
die notwendig ist, um von der Ursprungsanordnung der Ur"=Alternativen in die entsprechende
Anordnung des jeweiligen Termes zu gelangen, über das Vorzeichen des Termes entscheidet.
Bei gerader Anzahl an notwendigen Vertauschungen ist das Vorzeichen positiv und bei
ungerader Anzahl negativ. Der Operator $\mathcal{P}$ bezeichne die Vertauschung zweier
Ur"=Alternativen in einem Zustand im Tensorraum vieler Ur"=Alternativen $\mathcal{H}_T$.
Ein gegenüber beliebigen Permutationen zweier Ur"=Alternativen symmetrischer
Zustand $|\Psi\rangle_S$, für den gilt:

\begin{align}
&\mathcal{P}_{ij}|\Psi\left(...;u_{i1},u_{i2};...;u_{j1},u_{j2};...\right)\rangle_S
=|\Psi\left(...;u_{j1},u_{j2};...;u_{i1},u_{i2};...\right)\rangle_S,\nonumber\\
&\quad i,j=1,...,M\quad i \neq j,
\label{Vertauschung_Ur-Alternativen_symmetrisch}
\end{align}
wird also durch Summierung über alle symmetrischen Permutationen bezüglich aller
Ur"=Alternativen erreicht:

\begin{equation}
|\Psi\left(u_{11},u_{12};u_{21},u_{22};...;u_{n1},u_{n2}\right)\rangle_S
=\bigoplus_{\mathcal{P}_S}|\Psi\left(u_{11},u_{12};u_{21},u_{22};...;u_{n1},u_{n2}\right)\rangle \in \mathcal{H}_{TS},
\label{Symmetrisierung_Ur-Alternativen}
\end{equation}
wobei das Symbol $\mathcal{P}_S$ an dieser Stelle andeuten soll, dass über alle symmetrischen
Permutationen summiert wird. Und ein gegenüber beliebigen Permutationen zweier Ur"=Alternativen
antisymmetrischer Zustand $|\Psi\rangle_{AS}$, für den gilt:

\begin{align}
&\mathcal{P}_{ij}|\Psi\left(...;u_{i1},u_{i2};...;u_{j1},u_{j2};...\right)\rangle_{AS}
=-|\Psi\left(...;u_{j1},u_{j2};...;u_{i1},u_{i2};...\right)\rangle_{AS},\nonumber\\
&\quad i,j=1,...,M\quad i \neq j,
\label{Vertauschung_Ur-Alternativen_antisymmetrisch}
\end{align}
wird also durch Summierung über alle antisymmetrischen Permutationen bezüglich aller
Ur"=Alternativen erreicht:

\begin{equation}
|\Psi\left(u_{11},u_{12};u_{21},u_{22};...;u_{n1},u_{n2}\right)\rangle_{AS}=\bigoplus_{\mathcal{P}_{AS}}
|\psi\left(u_{11},u_{12};u_{21},u_{22};...;u_{n1},u_{n2}\right)\rangle \in \mathcal{H}_{TAS},
\label{Antisymmetrisierung_Ur-Alternativen}
\end{equation}
wobei das $\mathcal{P}_{AS}$ an dieser Stelle andeuten soll, dass über alle antisymmetrischen
Permutationen summiert wird. Aufgrund der Ununterscheidbarkeit der Ur"=Alternativen sind in
beiden Fällen die Basiszustände des Tensorraumes vieler Ur"=Alternativen durch die Anzahl
an Ur"=Alternativen in den verschiedenen Basiszuständen einer einzelnen Ur"=Alternative
gekennzeichnet. Zu dem Tensorraum vieler Ur"=Alternativen gelangt man demnach,
indem man eine iterierte Quantisierung durchführt, bei welcher der zu einer
einzelnen quantentheoretischen Ur"=Alternative $u$ gehörige Spinorzustand
$|\varphi\rangle$, beziehungsweise dessen einzelne Komponenten $\varphi_a$, $\varphi_b$,
$\varphi_c$ und $\varphi_d$, die ja bereits aus einer Quantisierung dieser Alternative
durch Zuordnung komplexer Wahrheitswerte gemäß ($\ref{Ur-Alternative}$) hervorgegangen sind,
in Operatoren überführt werden, welche einzelne Ur"=Alternativen in den jeweiligen
Basiszuständen einer einzelnen Ur"=Alternative erzeugen beziehungsweise vernichten.
Wenn eine Ur"=Alternative weiter quantisiert wird, sie also in einen Operator
umgewandelt wird, der im Tensorraum vieler Ur"=Alternativen wirkt:

\begin{eqnarray}
|\varphi\rangle \longrightarrow \hat \varphi,
\end{eqnarray}
was dementsprechend für die Komponenten das Folgende bedeutet:

\begin{equation}
\varphi_a \longrightarrow \hat \varphi_a\equiv A,
\quad \varphi_b \longrightarrow \hat \varphi_b\equiv B,
\quad \varphi_c \longrightarrow \hat \varphi_c\equiv C,
\quad \varphi_d \longrightarrow \hat \varphi_d\equiv D,
\end{equation}
und damit gilt:

\begin{equation}
\hat \varphi=\left(\begin{matrix} A+Bi\\C+Di \end{matrix}\right),
\label{Operator_Ur-Alternative}
\end{equation}
dann kann dies über Vertauschungsrelationen oder über Antivertauschungsrelationen
geschehen. Es gilt also entweder die folgende Vertauschungsrelation:

\begin{equation}
\left[\hat \varphi_r,\hat \varphi^{\dagger}_s\right]=\hat \varphi_r \hat \varphi^{\dagger}_s
-\hat \varphi^{\dagger}_s \hat \varphi_r=\delta_{rs},
\quad r,s=a,b,c,d,
\label{Vertauschungsrelationen_UrAlternativen}
\end{equation}
oder es gilt die folgende Antivertauschungsrelation:

\begin{equation}
\left\{\hat \varphi_r, \hat \varphi^{\dagger}_s\right\}=\hat \varphi_r \hat \varphi^{\dagger}_s
+\hat \varphi^{\dagger}_s \hat \varphi_r=\delta_{rs},\quad r,s=a,b,c,d.
\label{Antivertauschungsrelationen_UrAlternativen}
\end{equation}
Vertauschungsrelationen entsprechen dem Tensorraum an Zuständen, die gegenüber Permutation
der Ur"=Alternativen symmetrisch sind $\mathcal{H}_{TS}$, und Antivertauschungsrelationen
entsprechen einem Tensorraum an Zuständen $\mathcal{H}_{TAS}$, die gegenüber Permutation
der Ur"=Alternativen antisymmetrisch sind.

\subsection{Operatoren und Zustände im Tensorraum der Ur"=Alternativen}

Die Ununterscheidbarkeit der Ur"=Alternativen als elementarer Quantenobjekte mit der
daraus resultierenden Permutationssymmetrie beziehungsweise Permutationsantisymmetrie
führt also zu einer Aufspaltung des Tensorraumes der Ur"=Alternativen $\mathcal{H}_T$ in einen
symmetrischen Anteil $\mathcal{H}_{TS}$ einerseits und in einen antisymmetrischen Anteil
$\mathcal{H}_{TAS}$ andererseits, wobei die Erzeugungs- und Vernichtungsoperatoren, die im
symmetrischen Anteil wirken, den Vertauschungsrelationen ($\ref{Vertauschungsrelationen_UrAlternativen}$)
gehorchen, während die Erzeugungs- und Vernichtungsoperatoren, die im antisymmetrischen Anteil
wirken, den Antivertauschungsrelationen ($\ref{Antivertauschungsrelationen_UrAlternativen}$)
gehorchen. Dies bedeutet, dass sich folgendes Tensorprodukt ergibt:

\begin{equation}
\mathcal{H}_T=\mathcal{H}_{TS}\otimes \mathcal{H}_{TAS}.
\label{Aufspaltung_symmetrisch_antisymmetrisch}
\end{equation}
Für den symmetrischen Anteil $\mathcal{H}_{TS}$ gilt demnach Bose"=Statistik \cite{Bose:1924}, während für
den antisymmetrischen Anteil $\mathcal{H}_{TAS}$ Fermi"=Statistik \cite{Fermi:1926} gilt. Das bedeutet,
dass in Bezug auf den symmetrischen Anteil eine beliebige Anzahl an Ur"=Alternativen in einem Zustand
existieren kann, und in Bezug auf den antisymmetrischen Anteil aufgrund des Paulischen
Ausschließungsprinzips nur eine Ur"=Alternative in jedem Basiszustand. Die Erzeugungs-
und Vernichtungsoperatoren in Bezug auf die Basiszustände im symmetrischen Anteil des
Tensorraumes der Ur"=Alternativen $\mathcal{H}_{TS}$ seien bezeichnet als:

\begin{equation}
A_S,\quad A_S^{\dagger},\quad B_S,\quad B_S^{\dagger},\quad C_S,\quad C_S^{\dagger},\quad D_S,\quad D_S^{\dagger}.
\label{Operatoren_symmetrisch}
\end{equation}
Sie erfüllen die folgenden Vertauschungsrelationen:

\begin{equation}
\left[A_S,A_S^{\dagger}\right]=1,\quad \left[B_S,B_S^{\dagger}\right]=1,\quad
\left[C_S,C_S^{\dagger}\right]=1,\quad \left[D_S,D_S^{\dagger}\right]=1,
\label{Operatoren_symmetrisch_Kommutatoren}
\end{equation}
wobei alle anderen Vertauschungsrelationen gleich null sind. Und die Erzeugungs- und
Vernichtungsoperatoren in Bezug auf die Basiszustände im antisymmetrischen Anteil
des Tensorraumes der Ur"=Alternativen $\mathcal{H}_{TAS}$ seien bezeichnet als:

\begin{equation}
A_{AS},\quad A_{AS}^{\dagger},\quad B_{AS},\quad B_{AS}^{\dagger},\quad
C_{AS},\quad C_{AS}^{\dagger},\quad D_{AS},\quad D_{AS}^{\dagger}.
\label{Operatoren_antisymmetrisch}
\end{equation}
Sie erfüllen die Anti"=Vertauschungsrelationen:

\begin{equation}
\left\{A_{AS},A_{AS}^{\dagger}\right\}=1,\quad \left\{B_{AS},B_{AS}^{\dagger}\right\}=1,\quad
\left\{C_{AS},C_{AS}^{\dagger}\right\}=1,\quad \left\{D_{AS},D_{AS}^{\dagger}\right\}=1,
\label{Operatoren_antisymmetrisch_Antikommutatoren}
\end{equation}
wobei alle anderen Anti"=Vertauschungsrelationen gleich null sind. Die Basiszustände
des Tensorraumes der Ur"=Alternativen sind durch die Besetzungszahlen an Ur"=Alternativen
bezüglich der Basiszustände einzelner Ur"=Alternativen gegeben. Die Besetzungszahlen
der symmetrischen Basiszustände seien bezeichnet mit $N_A$,$N_B$,$N_C$,$N_D$
und die Besetzungszahlen der antisymmetrischen Zustände seien bezeichnet mit
$\bar N_A$, $\bar N_B$, $\bar N_C$, $\bar N_D$. Diese Besetzungszahlen
kann man zusammenfassen zu:

\begin{equation}
N_{ABCD}=\left(N_A,N_B,N_C,N_D\right),\quad \bar N_{ABCD}=\left(\bar N_A,\bar N_B,\bar N_C,\bar N_D\right).
\end{equation}
Die Basiszustände des symmetrischen Teilraumes des Tensorraumes der Ur"=Alternativen
$\mathcal{H}_{TS}$ sind demnach gegeben durch:

\begin{equation}
|N_{ABCD}\rangle_S=|N_A,N_B,N_C,N_D\rangle_S=|N_A\rangle_S \otimes |N_B\rangle_S \otimes |N_C\rangle_S \otimes |N_D\rangle_S,
\label{Basiszustaende_symmetrisch}
\end{equation}
und die Wirkung der Erzeugungs- und Vernichtungsoperatoren ($\ref{Operatoren_symmetrisch}$) auf
die Basiszustände ($\ref{Basiszustaende_symmetrisch}$) ist demnach gegeben durch die
folgenden Gleichungen:

\begin{eqnarray}
A_S|N_A,N_B,N_C,N_D\rangle_S&=&\sqrt{N_A}|N_A-1,N_B,N_C,N_D\rangle_S,\quad N_A \geq 1 \nonumber\\
A_S^{\dagger}|N_A,N_B,N_C,N_D\rangle_S&=&\sqrt{N_A+1}|N_A+1,N_B,N_C,N_D\rangle_S,\quad N_A \geq 0 \nonumber\\
B_S|N_A,N_B,N_C,N_D\rangle_S&=&\sqrt{N_B}|N_A,N_B-1,N_C,N_D\rangle_S,\quad N_B \geq 1 \nonumber\\
B_S^{\dagger}|N_A,N_B,N_C,N_D\rangle_S&=&\sqrt{N_B+1}|N_A,N_B+1,N_C,N_D\rangle_S,\quad N_B \geq 0 \nonumber\\
C_S|N_A,N_B,N_C,N_D\rangle_S&=&\sqrt{N_C}|N_A,N_B,N_C-1,N_D\rangle_S,\quad N_C \geq 1 \nonumber\\
C_S^{\dagger}|N_A,N_B,N_C,N_D\rangle_S&=&\sqrt{N_C+1}|N_A,N_B,N_C+1,N_D\rangle_S,\quad N_C \geq 0\nonumber\\
D_S|N_A,N_B,N_C,N_D\rangle_S&=&\sqrt{N_D}|N_A,N_B,N_C,N_D-1\rangle_S,\quad N_D \geq 1\nonumber\\
D_S^{\dagger}|N_A,N_B,N_C,N_D\rangle_S&=&\sqrt{N_D+1}|N_A,N_B,N_C,N_D+1\rangle_S,\quad N_D \geq 0.
\end{eqnarray}
Die Anwendung eines der Vernichtungsoperatoren auf einen Zustand
mit jeweiliger verschwindender Besetzungszahl ergibt null:

\begin{eqnarray}
A_S|0,N_B,N_C,N_D\rangle_S=0,\quad B_S|N_A,0,N_C,N_D\rangle_S=0,\nonumber\\
C_S|N_A,N_B,0,N_D\rangle_S=0,\quad D_S|N_A,N_B,N_C,0\rangle_S=0,
\end{eqnarray}
und die Anwendung der Vernichtungsoperatoren auf den vollkommen unbesetzten Zustand
$|0,0,0,0\rangle_S$, also auf das quantenlogische Vakuum, in dem natürlich auch
überhaupt keine Information mehr existiert, ergibt demnach auch null:

\begin{equation}
A_S|0,0,0,0\rangle_S=0,\quad B_S|0,0,0,0\rangle_S=0,\quad C_S|0,0,0,0\rangle_S=0,\quad D_S|0,0,0,0\rangle_S=0.  
\end{equation}
Demnach wirken die Operatoren $A_S^{\dagger} A_S$, $B_S^{\dagger} B_S$, $C_S^{\dagger} C_S$,
$D_S^{\dagger} D_S$ als Besetzungszahloperatoren und es gelten für die Basiszustände des
symmetrischen Teilraumes des Tensorraumes der Ur"=Alternativen $\mathcal{H}_{TS}$ die
folgenden Gleichungen:

\begin{eqnarray}
A_S^{\dagger} A_S|N_A,N_B,N_C,N_D\rangle_S&=&N_A|N_A,N_B,N_C,N_D\rangle_S,\nonumber\\
B_S^{\dagger} B_S|N_A,N_B,N_C,N_D\rangle_S&=&N_B|N_A,N_B,N_C,N_D\rangle_S,\nonumber\\
C_S^{\dagger} C_S|N_A,N_B,N_C,N_D\rangle_S&=&N_C|N_A,N_B,N_C,N_D\rangle_S,\nonumber\\
D_S^{\dagger} D_S|N_A,N_B,N_C,N_D\rangle_S&=&N_D|N_A,N_B,N_C,N_D\rangle_S.
\end{eqnarray}
Analog sind die Basiszustände des antisymmetrischen Teilraumes des Tensorraumes
der Ur"=Alternativen $\mathcal{H}_{TAS}$ gegeben durch:

\begin{equation}
|\bar N_{ABCD}\rangle_{AS}=|\bar N_A,\bar N_B,\bar N_C,\bar N_D\rangle_{AS}
=|\bar N_A\rangle_{AS} \otimes |\bar N_B\rangle_{AS} \otimes |\bar N_C\rangle_{AS} \otimes |\bar N_D\rangle_{AS},
\label{Basiszustaende_antisymmetrisch}
\end{equation}
und die Wirkung der Erzeugungs- und Vernichtungsoperatoren ($\ref{Operatoren_antisymmetrisch}$) auf
die Basiszustände ($\ref{Basiszustaende_antisymmetrisch}$) ist demnach gegeben durch die
folgenden Gleichungen:

\begin{align}
&A_{AS}|1,\bar N_B,\bar N_C,\bar N_D\rangle_{AS}=|0,\bar N_B,\bar N_C,\bar N_D\rangle_{AS},\quad
A_{AS}^{\dagger}|0,\bar N_B,\bar N_C,\bar N_D\rangle_{AS}=|1,\bar N_B,\bar N_C,\bar N_D\rangle_{AS},\nonumber\\
&B_{AS}|\bar N_A,1,\bar N_C,\bar N_D\rangle_{AS}=|\bar N_A,0,\bar N_C,\bar N_D\rangle_{AS},\quad
B_{AS}^{\dagger}|\bar N_A,0,\bar N_C,\bar N_D\rangle_{AS}=|\bar N_A,1,\bar N_C,\bar N_D\rangle_{AS},\nonumber\\
&C_{AS}|\bar N_A,\bar N_B,1,\bar N_D\rangle_{AS}=|\bar N_A,\bar N_B,0,\bar N_D\rangle_{AS},\quad
C_{AS}^{\dagger}|\bar N_A,\bar N_B,0,\bar N_D\rangle_{AS}=|\bar N_A,\bar N_B,1,\bar N_D\rangle_{AS},\nonumber\\
&D_{AS}|\bar N_A,\bar N_B,\bar N_C,1\rangle_{AS}=|\bar N_A,\bar N_B,\bar N_C,0\rangle_{AS},\quad
D_{AS}^{\dagger}|\bar N_A,\bar N_B,\bar N_C,0\rangle_{AS}=|\bar N_A,\bar N_B,\bar N_C,1\rangle_{AS},
\end{align}
beziehungsweise die folgenden Gleichungen:

\begin{eqnarray}
A_{AS}|0,\bar N_B,\bar N_C,\bar N_D\rangle_{AS}&=&0,\quad
A_{AS}^{\dagger}|1,\bar N_B,\bar N_C,\bar N_D\rangle_{AS}=0,\nonumber\\
B_{AS}|\bar N_A,0,\bar N_C,\bar N_D\rangle_{AS}&=&0,\quad
B_{AS}^{\dagger}|\bar N_A,1,\bar N_C,\bar N_D\rangle_{AS}=0,\nonumber\\
C_{AS}|\bar N_A,\bar N_B,0,\bar N_D\rangle_{AS}&=&0,\quad
C_{AS}^{\dagger}|\bar N_A,\bar N_B,1,\bar N_D\rangle_{AS}=0,\nonumber\\
D_{AS}|\bar N_A,\bar N_B,\bar N_C,0\rangle_{AS}&=&0,\quad
D_{AS}^{\dagger}|\bar N_A,\bar N_B,\bar N_C,1\rangle_{AS}=0.
\end{eqnarray}
Die Anwendung des jeweiligen Erzeugungsoperators auf Zustände in Bezug auf den
antisymmetrischen Teilraum des Tensorraumes der Ur"=Alternativen mit einer jeweiligen
Besetzungszahl eins ergibt null. Denn aufgrund des Paulischen Ausschließungsprinzips
\cite{Pauli:1940} kann bei antisymmetrischen Zuständen jeder Basiszustand maximal mit
einer einzelnen Ur"=Alternative besetzt sein, da sich die einzelnen Terme der Summe
bei gleichen Zuständen zweier Ur"=Alternativen ansonsten gegenseitig aufheben.
Die Anwendung der Vernichtungsoperatoren auf den komplett unbesetzten Zustand
$|0,0,0,0\rangle_{AS}$, also das quantenlogische Vakuum, in dem auch keine
Information mehr existiert, ergibt demnach auch null:

\begin{equation}
A_{AS}|0,0,0,0\rangle_{AS}=0,\quad B_{AS}|0,0,0,0\rangle_{AS}=0,\quad
C_{AS}|0,0,0,0\rangle_{AS}=0,\quad D_{AS}|0,0,0,0\rangle_{AS}=0.
\end{equation}
Demnach wirken die Operatoren $A_{AS}^{\dagger} A_{AS}$, $B_{AS}^{\dagger} B_{AS}$, $C_{AS}^{\dagger} C_{AS}$,
$D_{AS}^{\dagger} D_{AS}$ als Besetzungszahloperatoren und es gelten für die Basiszustände des antisymmetrischen
Teilraumes des Tensorraumes der Ur"=Alternativen $\mathcal{H}_{TAS}$ die folgenden Gleichungen:

\begin{eqnarray}
A_{AS}^{\dagger} A_{AS}|\bar N_A,\bar N_B,\bar N_C,\bar N_D\rangle_{AS}
=\bar N_A|\bar N_A,\bar N_B,\bar N_C,\bar N_D\rangle_{AS},\nonumber\\
B_{AS}^{\dagger} B_{AS}|\bar N_A,\bar N_B,\bar N_C,\bar N_D\rangle_{AS}
=\bar N_B|\bar N_A,\bar N_B,\bar N_C,\bar N_D\rangle_{AS},\nonumber\\
C_{AS}^{\dagger} C_{AS}|\bar N_A,\bar N_B,\bar N_C,\bar N_D\rangle_{AS}
=\bar N_C|\bar N_A,\bar N_B,\bar N_C,\bar N_D\rangle_{AS},\nonumber\\
D_{AS}^{\dagger} D_{AS}|\bar N_A,\bar N_B,\bar N_C,\bar N_D\rangle_{AS}
=\bar N_D|\bar N_A,\bar N_B,\bar N_C,\bar N_D\rangle_{AS}.
\end{eqnarray}
Ein allgemeiner Zustand im gesamten Tensorraum der Ur"=Alternativen
ist demnach als das Tensorprodukt eines allgemeinen Zustandes im
symmetrischen und im antisymmetrischen Teilraum des Tensorraum der
Ur"=Alternativen in der folgenden Weise definiert:

\begin{equation}
|\Psi \rangle=|\Psi \rangle_{S}\otimes |\Psi \rangle_{AS}
=\left[\sum_{N_{ABCD}=0}^{N}\psi_S\left(N_{ABCD}\right)|N_{ABCD}\rangle_S\right] \otimes
\left[\sum_{\bar N_{ABCD}=0}^{1}\psi_{AS}\left(\bar N_{ABCD}\right)|\bar N_{ABCD}\rangle_{AS}\right],
\label{Zustand_gesamt}
\end{equation}
wobei für die Besetzungszahlen $\bar N_{A}$, $\bar N_{B}$, $\bar N_{C}$ und $\bar N_{D}$
aufgrund des Paulischen Ausschließungsprinzips natürlich gilt, dass sie jeweils nur
von $0$ bis $1$ laufen, weil ein gegenüber Permutation antisymmetrischer Zustand bei
Besetzungszahlen bezüglich eines Basiszustandes größer als eins verschwindet. Aufgrund
der Tatsache, dass die Gesamtwahrscheinlichkeit gleich eins sein muss, muss zudem die
folgende Normierungsbedingung gelten:

\begin{equation}
\sum_{N_{ABCD}=0}^{N}|\psi_S\left(N_{ABCD}\right)|^2=1,\quad
\sum_{\bar N_{ABCD}=0}^{1}|\psi_{AS}\left(\bar N_{ABCD}\right)|^2=1.
\end{equation}
Zudem gilt natürlich für die Gesamtbesetzungszahlen $N$ und $\bar N$:

\begin{equation}
N=N_A+N_B+N_C+N_D,\quad \bar N=\bar N_A+\bar N_B+\bar N_C+\bar N_D.
\label{Gesamtzahl_Ur-Alternativen}
\end{equation}
Das innere Produkt $\langle\ \cdot\ |\ \cdot\ \rangle$ zwischen zwei Zuständen $|\Psi_A\rangle$
und $|\Psi_B\rangle$ im Tensorraum der Ur"=Alternativen ist gegeben durch:

\begin{align}
&\langle \Psi_B|\Psi_A \rangle=\left(\langle \Psi_B|_S\otimes\langle \Psi_B|_{AS}\right)
\cdot\left(|\Psi_A \rangle_S\otimes|\Psi_A \rangle_{AS}\right)
=\langle \Psi_B|\Psi_A \rangle_S \otimes \langle \Psi_B|\Psi_A \rangle_{AS}\\
&=\left[\sum_{N_{ABCD}=0}^N \left(\psi_B\right)_S\left(N_{ABCD}\right)\left(\psi_A\right)_S\left(N_{ABCD}\right)\right]
\otimes \left[\sum_{\bar N_{ABCD}=0}^1 \left(\psi_B\right)_{AS}\left(\bar N_{ABCD}\right)
\left(\psi_A\right)_{AS}\left(\bar N_{ABCD}\right)\right]\nonumber,
\end{align}
wobei die Basiszustände des symmetrischen und des antiisymmerischen Anteiles des Tensorraumes
der Ur"=Alternativen, die jeweils durch eine fest definierte Besetzungszahl an Ur"=Alternativen
in den vier Basiszuständen einer einzelnen Ur"=Alternative definiert sind, natürlich jeweils
orthogonal zueinander sind:

\begin{eqnarray}
\langle N_{ABCD}|N_{ABCD}^{'}\rangle_S
&=&\delta_{N_A {N_A}^{'}}\delta_{N_B {N_B}^{'}}\delta_{N_C {N_C}^{'}}\delta_{N_D {N_D}^{'}},
\quad N_A, N_B, N_C, N_D=0,...,N,\nonumber\\
\langle \bar N_{ABCD}|\bar N_{ABCD}^{'}\rangle_{AS}
&=&\delta_{\bar N_A {\bar N_A}^{'}}\delta_{\bar N_B {\bar N_B}^{'}}
\delta_{\bar N_C {\bar N_C}^{'}}\delta_{\bar N_D {\bar N_D}^{'}},
\quad \bar N_A, \bar N_B, \bar N_C, \bar N_D=0,1.
\end{eqnarray}
Es wird sich in den folgenden beiden Kapiteln zeigen, dass der symmetrische Anteil des Tensorraumes
der Ur"=Alternativen $\mathcal{H}_{TS}$ dem räumlichen Anteil des Zustandes eines Quantenobjektes
entspricht, der demnach als Wellenfunktion in einem dreidimensionalen Ortsraum dargestellt werden
kann, und der antisymmetrische Anteil $\mathcal{H}_{TAS}$ dem Spin sowie den inneren Quantenzahlen
eines Quantenobjektes gemäß der Elementarteilchenphysik entspricht. Ein allgemeiner Zustand im
Tensorraum der Ur"=Alternativen ($\ref{Zustand_gesamt}$) muss natürlich als zeitabhängiger Zustand
auch der allgemeinen Schrödingergleichung als dynamischer Grundgleichung der Quantentheorie in ihrer
abstraktesten Form ($\ref{Schroedinger_Gleichung_Alternative}$) genügen und dies bedeutet:

\begin{equation}
i\partial_t|\Psi(t)\rangle=H|\Psi(t)\rangle\quad \Leftrightarrow\quad
i\partial_t\left[|\Psi(t)\rangle_S \otimes |\Psi(t)\rangle_{AS}\right]
=H\left[|\Psi(t)\rangle_S \otimes |\Psi(t)\rangle_{AS}\right].
\label{Schroedingergleichung_Tensorraum}
\end{equation}
wobei an dieser Stelle der Hamiltonoperator $H$ noch nicht genau spezifiziert wird.
Er muss sich allerdings grundsätzlich wie alle Operatoren in der Quantentheorie der
Ur"=Alternativen, die sich auf den Zustandsraum eines einzelnen Quantenobjektes
beziehen, aus den Erzeugungs- und Vernichtungsoperatoren ($\ref{Operatoren_symmetrisch}$),
($\ref{Operatoren_antisymmetrisch}$) im Tensorraum der Ur"=Alternativen zusammensetzen.
Ein beliebiger Operator $\mathcal{A}$ im Tensorraum der Ur"=Alternativen weist folgende
allgemeine Gestalt auf:

\begin{align}
&\mathcal{A}=\sum_{n}\left[a\left(n_{AS},\bar n_{AS},n_{BS},\bar n_{BS},n_{CS},\bar n_{CS},n_{DS},\bar n_{DS},
n_{AAS},\bar n_{AAS},n_{BAS},\bar n_{BAS},n_{CAS},\bar n_{CAS},n_{DAS},\bar n_{DAS}\right)
\nonumber\right.\\&\left.\quad\quad\times
\left(A_S\right)^{n_{AS}} \left(A_S^{\dagger}\right)^{\bar n_{AS}}
\left(B_S\right)^{n_{BS}} \left(B_S^{\dagger}\right)^{\bar n_{BS}}
\left(C_S\right)^{n_{CS}} \left(C_S^{\dagger}\right)^{\bar n_{CS}}
\left(D_S\right)^{n_{DS}} \left(D_S^{\dagger}\right)^{\bar n_{DS}}
\right.\\&\left.\quad\quad\times
\left(A_{AS}\right)^{n_{AAS}} \left(A_{AS}^{\dagger}\right)^{\bar n_{AAS}}
\left(B_{AS}\right)^{n_{BAS}} \left(B_{AS}^{\dagger}\right)^{\bar n_{BAS}}
\left(C_{AS}\right)^{n_{CAS}} \left(C_{AS}^{\dagger}\right)^{\bar n_{CAS}}
\left(D_{AS}\right)^{n_{DAS}} \left(D_{AS}^{\dagger}\right)^{\bar n_{DAS}}\right],\nonumber
\end{align}
wobei die Summe $\sum_n$ hier in kompakter Schreibweise die Summierung über
alle beliebigen Kombinationen ganzer Zahlen als Werte der Exponenten andeuten soll,
also über $n_{AS}$, $\bar n_{AS}$, $n_{BS}$, $\bar n_{BS}$, $n_{CS}$, $\bar n_{CS}$,
$n_{DS}$, $\bar n_{DS}$, $n_{AAS}$, $\bar n_{AAS}$, $n_{BAS}$, $\bar n_{BAS}$,
$n_{CAS}$, $\bar n_{CAS}$, $n_{DAS}$, $\bar n_{DAS}$.

\section{Darstellung der abstrakten Zustände im physikalischen Ortsraum}

\subsection{Der symmetrische Teilraum des Tensorraumes der Ur-Alternativen}

In diesem Kapitel soll gezeigt werden, dass die Zustände des symmetrischen Teilraumes des
Tensorraumes der Ur"=Alternativen $\mathcal{H}_{TS}$ als Wellenfunktionen in einem
dreidimensionalen reellen Ortsraum dargestellt werden können. Da die Zustände des Tensorraumes
gemäß dem Postulat der Zeitentwicklung auch von der Zeit als einem reellen Parameter abhängen,
kann auf diese Weise die Existenz einer (3+1)"=dimensionalen Raum"=Zeit als Darstellungsmedium
aller physikalischen Objekte aus der abstrakten Quantentheorie begründet werden. Dies habe ich
bereits in \cite{Kober:2017} durchgeführt. Zu diesem Behuf werden zunächst die folgenden
Operatoren als Linearkombinationen der Erzeugungs- und Vernichtungsoperatoren im symmetrischen
Anteil des Tensorraumes der Ur"=Alternativen ($\ref{Operatoren_symmetrisch}$) definiert:

\begin{eqnarray}
A_x&=&\frac{1}{2}\left(A_S+B_S-C_S-D_S\right),\quad A_x^{\dagger}=\frac{1}{2}\left(A^{\dagger}_S+B^{\dagger}_S-C^{\dagger}_S-D^{\dagger}_S\right),\nonumber\\
A_y&=&\frac{1}{2}\left(A_S-B_S+C_S-D_S\right),\quad A_y^{\dagger}=\frac{1}{2}\left(A^{\dagger}_S-B^{\dagger}_S+C^{\dagger}_S-D^{\dagger}_S\right),\nonumber\\
A_z&=&\frac{1}{2}\left(A_S-B_S-C_S+D_S\right),\quad A_z^{\dagger}=\frac{1}{2}\left(A^{\dagger}_S-B^{\dagger}_S-C^{\dagger}_S+D^{\dagger}_S\right),\nonumber\\
A_n&=&\frac{1}{2}\left(A_S+B_S+C_S+D_S\right),\quad
A_n^{\dagger}=\frac{1}{2}\left(A_S^{\dagger}+B_S^{\dagger}+C_S^{\dagger}+D_S^{\dagger}\right),
\label{ErzeugungsVernichtungsOperatorenXYZ}
\end{eqnarray}
welche zu ($\ref{Operatoren_symmetrisch_Kommutatoren}$) analoge Vertauschungsrelationen erfüllen:

\begin{equation}
\left[A_x,A_x^{\dagger}\right]=1,\quad \left[A_y,A_y^{\dagger}\right]=1,\quad
\left[A_z,A_z^{\dagger}\right]=1,\quad \left[A_n,A_n^{\dagger}\right]=1,
\end{equation}
wobei alle anderen Vertauschungsrelationen zwischen den Erzeugungs- und Vernichtungsoperatoren
($\ref{ErzeugungsVernichtungsOperatorenXYZ}$) gleich null sind. Hieraus geht eine alternative
Darstellung der Basiszustände des symmetrischen Anteiles des Tensorraumes der
Ur"=Alternativen hervor:

\begin{equation}
|N_{xyzn}\rangle_S=|N_x,N_y,N_z,N_n\rangle_S=|N_x\rangle_S \otimes |N_y\rangle_S \otimes |N_z\rangle_S \otimes |N_n\rangle_S,
\end{equation}
was dem folgenden Übergang entspricht:

\begin{equation}
|N_A,N_B,N_C,N_D\rangle_S \quad\longleftrightarrow\quad |N_x,N_y,N_z,N_n\rangle_S.
\end{equation}
Die Wirkung der Erzeugungs- und Vernichtungsoperatoren ($\ref{ErzeugungsVernichtungsOperatorenXYZ}$)
auf die Basiszustände lautet:

\begin{eqnarray}
A_x|N_x,N_y,N_z,N_n\rangle_S&=&\sqrt{N_x}|N_x-1,N_y,N_z,N_n\rangle_S,\quad N_x \geq 1\nonumber\\
A_x^{\dagger}|N_x,N_y,N_z,N_n\rangle_S&=&\sqrt{N_x+1}|N_x+1,N_y,N_z,N_n\rangle_S,\quad N_x \geq 0\nonumber\\
A_y|N_x,N_y,N_z,N_n\rangle_S&=&\sqrt{N_y}|N_x,N_y-1,N_z,N_n\rangle_S,\quad N_y \geq 1\nonumber\\
A_y^{\dagger}|N_x,N_y,N_z,N_n\rangle_S&=&\sqrt{N_y+1}|N_x,N_y+1,N_z,N_n\rangle_S,\quad N_y \geq 0\nonumber\\
A_z|N_x,N_y,N_z,N_n\rangle_S&=&\sqrt{N_z}|N_x,N_y,N_z-1,N_n\rangle_S,\quad N_z \geq 1\nonumber\\
A_z^{\dagger}|N_x,N_y,N_z,N_n\rangle_S&=&\sqrt{N_z+1}|N_x,N_y,N_z+1,N_n\rangle_S,\quad N_z \geq 0\nonumber\\
A_n|N_x,N_y,N_u,N_n\rangle_S&=&\sqrt{N_n}|N_x,N_y,N_z,N_n-1\rangle_S,\quad N_n \geq 1\nonumber\\
A_n^{\dagger}|N_x,N_y,N_z,N_n\rangle_S&=&\sqrt{N_n+1}|N_x,N_y,N_z,N_n+1\rangle_S,\quad N_n \geq 0.
\label{WirkungOperatorenxyzn}
\end{eqnarray}
Die Anwendung eines der Vernichtungsoperatoren auf einen Zustand mit jeweiliger
verschwindender Besetzungszahl ergibt null:

\begin{eqnarray}
A_x|0,N_y,N_z,N_n\rangle_S&=&0,\quad A_y|N_x,0,N_z,N_n\rangle_S=0,\nonumber\\
A_z|N_x,N_y,0,N_D\rangle_S&=&0,\quad A_n|N_A,N_B,N_C,0\rangle_S=0,
\end{eqnarray}
und die Anwendung der Vernichtungsoperatoren auf den vollkommen unbesetzten Zustand
$|0,0,0,0\rangle_S$, also das quantenlogische Vakuum, in dem auch keine
Information mehr existiert, ergibt demnach auch null:

\begin{equation}
A_x|0,0,0,0\rangle_S=0,\quad A_y|0,0,0,0\rangle_S=0,\quad A_z|0,0,0,0\rangle_S=0,\quad A_n|0,0,0,0\rangle_S=0.  
\end{equation}
Die Operatoren $A_x^{\dagger}A_x$, $A_y^{\dagger}A_y$, $A_z^{\dagger}A_z$, $A_n^{\dagger}A_n$
wirken dann in folgender Weise als Besetzungszahloperatoren auf die Basiszustände
im Tensorraum der Ur"=Alternativen:

\begin{eqnarray}
A_x^{\dagger}A_x|N_x,N_y,N_z,N_n\rangle_S&=&N_x|N_x,N_y,N_z,N_n\rangle_S,\nonumber\\
A_y^{\dagger}A_y|N_x,N_y,N_z,N_n\rangle_S&=&N_y|N_x,N_y,N_z,N_n\rangle_S,\nonumber\\
A_z^{\dagger}A_z|N_x,N_y,N_z,N_n\rangle_S&=&N_z|N_x,N_y,N_z,N_n\rangle_S,\nonumber\\
A_n^{\dagger}A_n|N_x,N_y,N_u,N_n\rangle_S&=&N_n|N_x,N_y,N_z,N_n\rangle_S.
\label{Eigenwertgleichungenxyzn}
\end{eqnarray}
Wenn man die Besetzungszahloperatoren $A_x^{\dagger}A_x$, $A_y^{\dagger}A_y$,
$A_z^{\dagger}A_z$ und $A_n^{\dagger}A_n$ durch die Operatoren $A_S$, $A_S^{\dagger}$,
$B_S$, $B_S^{\dagger}$, $C_S$, $C_S^{\dagger}$, $D_S$, $D_S^{\dagger}$
($\ref{Operatoren_symmetrisch}$) ausdrückt, dann ergeben sich die folgenden Ausdrücke:

\begin{eqnarray}
A_{x}^{\dagger}A_x&=&\frac{1}{4}\left(A_S^{\dagger}A_S+A_S^{\dagger}B_S-A_S^{\dagger}C_S
-A_S^{\dagger}D_S+B_S^{\dagger}A_S+B_S^{\dagger}B_S-B_S^{\dagger}C_S-B_S^{\dagger}D_S\right.\nonumber\\
&&\left.-C_S^{\dagger}A_S-C_S^{\dagger}B_S+C_S^{\dagger}C_S+C_S^{\dagger}D_S
-D_S^{\dagger}A_S-D_S^{\dagger}B_S+D_S^{\dagger}C+D_S^{\dagger}D_S\right),\nonumber\\
A_{y}^{\dagger}A_y&=&\frac{1}{4}\left(A_S^{\dagger}A_S-A_S^{\dagger}B_S+A_S^{\dagger}C_S-A_S^{\dagger}D_S
-B_S^{\dagger}A_S+B_S^{\dagger}B_S-B_S^{\dagger}C_S+B_S^{\dagger}D_S\right.\nonumber\\
&&\left.+C_S^{\dagger}A_S-C_S^{\dagger}B_S+C_S^{\dagger}C_S-C_S^{\dagger}D_S
-D_S^{\dagger}A_S+D_S^{\dagger}B_S-D_S^{\dagger}C_S+D_S^{\dagger}D_S\right),\nonumber\\
A_{z}^{\dagger}A_z&=&\frac{1}{4}\left(A_S^{\dagger}A_S-A_S^{\dagger}B_S-A_S^{\dagger}C_S
+A_S^{\dagger}D_S-B_S^{\dagger}A_S+B_S^{\dagger}B_S+B_S^{\dagger}C_S-B_S^{\dagger}D_S\right.\nonumber\\
&&\left.-C_S^{\dagger}A_S+C_S^{\dagger}B_S+C_S^{\dagger}C_S-C_S^{\dagger}D_S
+D_S^{\dagger}A_S-D_S^{\dagger}B_S-D_S^{\dagger}C_S+D_S^{\dagger}D_S\right),\nonumber\\
A_{n}^{\dagger}A_n&=&\frac{1}{4}\left(A_S^{\dagger}A_S+A_S^{\dagger}B_S+A_S^{\dagger}C_S
+A_S^{\dagger}D_S+B_S^{\dagger}A_S+B_S^{\dagger}B_S+B_S^{\dagger}C_S+B_S^{\dagger}D_S\right.\nonumber\\
&&\left.+C_S^{\dagger}A_S+C_S^{\dagger}B_S+C_S^{\dagger}C_S+C_S^{\dagger}D_S
+D_S^{\dagger}A_S+D_S^{\dagger}B_S+D_S^{\dagger}C_S+D_S^{\dagger}D_S\right).
\label{Besetzungszahloperatorenxyzn}
\end{eqnarray}
Ein allgemeiner Zustand im symmetrischen Anteil des Tensorraumes der Ur"=Alternativen
lautet gemäß dieser neuen Darstellung wie folgt:

\begin{equation}
|\Psi \rangle_S=\sum_{N_{ABCD}=0}^{N}\psi_S\left(N_{ABCD}\right)|N_{ABCD}\rangle_S
=\sum_{N_{xyzn}=0}^{N}\psi_S\left(N_{xyzn}\right)|N_{xyzn}\rangle.
\label{Basiszustand_Tensorraum_xyzn}
\end{equation}
Die neuen Basiszustände sind natürlich wieder orthogonal zueinander, was bedeutet:

\begin{equation}
\langle N_{xyzn}|N_{xyzn}^{'}\rangle_S=\delta_{{N_x}{N_x}^{'}}\delta_{{N_y} {N_y}^{'}}
\delta_{{N_z} {N_z}^{'}}\delta_{{N_n} {N_n}^{'}}.
\end{equation}

\subsection{Abstrakte Orts- und Impulsoperatoren im Tensorraum der Ur-Alternativen}

Um die Existenz eines dreidimensionalen reellen Ortsraumes verbunden mit der Zeit zu
einer (3+1)"=dimensionalen Raum"=Zeit aus dem Tensorraum der Ur"=Alternativen zu begründen,
müssen dessen Zustände in diesen Ortsraum abgebildet werden. Zu diesem Behuf müssen
allerdings zunächst in Bezug auf die Darstellung des symmetrischen Anteiles des
Tensorraumes der Ur"=Alternativen gemäß ($\ref{ErzeugungsVernichtungsOperatorenXYZ}$),
($\ref{WirkungOperatorenxyzn}$) und ($\ref{Basiszustand_Tensorraum_xyzn}$) die folgenden
Operatoren als Linearkombinationen der entsprechenden Erzeugungs- und Vernichtungsoperatoren
im Tensorraum der Ur"=Alternativen ($\ref{ErzeugungsVernichtungsOperatorenXYZ}$)
definiert werden:

\begin{eqnarray}
X&=&\frac{1}{\sqrt{2}}\left(A_x+A_x^{\dagger}\right),\quad P_x=-\frac{i}{\sqrt{2}}\left(A_x-A_x^{\dagger}\right),\nonumber\\
Y&=&\frac{1}{\sqrt{2}}\left(A_y+A_y^{\dagger}\right),\quad P_y=-\frac{i}{\sqrt{2}}\left(A_y-A_y^{\dagger}\right),\nonumber\\
Z&=&\frac{1}{\sqrt{2}}\left(A_z+A_z^{\dagger}\right),\quad P_z=-\frac{i}{\sqrt{2}}\left(A_z-A_z^{\dagger}\right).
\label{Ort_Impuls_Operatoren}
\end{eqnarray}
Diese in ($\ref{Ort_Impuls_Operatoren}$) definierten Operatoren, die natürlich ebenfalls
im Tensorraum der Ur"=Alternativen wirken, sind hermitesch:

\begin{equation}
X=X^{\dagger},\quad P_x=P_x^{\dagger},\quad
Y=Y^{\dagger},\quad P_y=P_y^{\dagger},\quad
Z=Z^{\dagger},\quad P_z=P_z^{\dagger},
\end{equation}
erfüllen die Heisenbergschen Vertauschungsrelationen von Orts- und Impulsoperatoren und sollen
aufgrund dessen auch mit diesen identifiziert werden:

\begin{equation}
\left[X,P_x\right]=i,\quad \left[Y,P_y\right]=i,\quad \left[Z,P_z\right]=i.
\label{Heisenbergsche_Algebra}
\end{equation}
Dies bedeutet, dass abstrakte Orts- und Impulsoperatoren in sehr einfacher Weise in einem rein quantenlogischen
Raum dargestellt werden beziehungsweise sich in einer beinahe natürlichen Weise aus diesem ergeben. 
Ausgedrückt durch die ursprünglichen Erzeugungs- und Vernichtungsoperatoren im symmetrischen Anteil
des Tensorraumes der Ur"=Alternativen ($\ref{Operatoren_symmetrisch}$) lauten sie wie folgt:

\begin{eqnarray}
X&=&\frac{1}{2\sqrt{2}}\left(A_S+B_S-C_S-D_S+A^{\dagger}_S+B^{\dagger}_S-C^{\dagger}_S-D^{\dagger}_S\right),\nonumber\\
Y&=&\frac{1}{2\sqrt{2}}\left(A_S-B_S+C_S-D_S+A^{\dagger}_S-B^{\dagger}_S+C^{\dagger}_S-D^{\dagger}_S\right),\nonumber\\
Z&=&\frac{1}{2\sqrt{2}}\left(A_S-B_S-C_S+D_S+A^{\dagger}_S-B^{\dagger}_S-C^{\dagger}_S+D^{\dagger}_S\right),\nonumber\\
P_x&=&-\frac{i}{2\sqrt{2}}\left(A_S+B_S-C_S-D_S-A^{\dagger}_S-B^{\dagger}_S+C^{\dagger}_S+D^{\dagger}_S\right),\nonumber\\
P_y&=&-\frac{i}{2\sqrt{2}}\left(A_S-B_S+C_S-D_S-A^{\dagger}_S+B^{\dagger}_S-C^{\dagger}_S+D^{\dagger}_S\right),\nonumber\\
P_z&=&-\frac{i}{2\sqrt{2}}\left(A_S-B_S-C_S+D_S-A^{\dagger}_S+B^{\dagger}_S+C^{\dagger}_S-D^{\dagger}_S\right).
\label{Ort_Impuls_Operatoren_ABCD}
\end{eqnarray}

\subsection{Der Freiheitsgrad der Gesamtinformationsmenge}

Der Grund der Wahl der Definition ($\ref{ErzeugungsVernichtungsOperatorenXYZ}$)
liegt darin, dass in der daraus hervorgehenden Darstellung der Freiheitsgrade
des Tensorraumes der Ur"=Alternativen aufgrund der mit den Operatoren
$A_n$ und $A_n^{\dagger}$ verbundenen Besetzungszahl der Freiheitsgrad der
Gesamtinformationsmenge $N$ ($\ref{Gesamtzahl_Ur-Alternativen}$) implizit
in einem Teilraum des Tensorraumes enthalten ist. Denn der Tensorraum enthält
vier unabhängige Teilräume, also vier unabhängige Dimensionen. Die Teilräume
im Tensorraum, auf welche die Erzeugungs- und Vernichtungsoperatoren
$A_x$, $A_x^{\dagger}$, $A_y$, $A_y^{\dagger}$, $A_z$, $A_z^{\dagger}$ wirken,
die gemäß ($\ref{ErzeugungsVernichtungsOperatorenXYZ}$) definiert sind,
können aufgrund der Möglichkeit der Definition der Orts- und Impulsoperatoren
($\ref{Ort_Impuls_Operatoren}$) mit den drei Richtungen im physikalischen
Ortsraum identifiziert werden. Der Teilraum, auf den sich die Operatoren
$A_n$, $A_n^{\dagger}$ beziehen, spielt eine Sonderrolle. Da die Gesamtzahl
$N$ der Ur"=Alternativen in einem Zustand gemäß ($\ref{Gesamtzahl_Ur-Alternativen}$)
definiert ist, was bedeutet:

\begin{align}
&\hat N=A^{\dagger}_S A_S+B^{\dagger}_S B_S+C^{\dagger}_S C_S+D^{\dagger}_S D_S
=A_{x}^{\dagger}A_x+A_{y}^{\dagger}A_y+A_{z}^{\dagger}A_z+A_{n}^{\dagger}A_n,\nonumber\\
&\Leftrightarrow\quad A_n^{\dagger}A_n=\hat N-A_x^{\dagger}A_x-A_y^{\dagger}A_y-A_z^{\dagger}A_z.
\label{Besetzungszahl_A_n}
\end{align}
kann man den Besetzungszahloperator $A_n^{\dagger} A_n$ gemäß ($\ref{Besetzungszahl_A_n}$)
in Abhängigkeit des Gesamtbesetzungszahloperators und der Besetzungszahloperatoren der
anderen Teilräume ausdrücken. Für die entsprechenden Besetzungszahlen in den
Basiszuständen bedeutet dies:

\begin{equation}
N_n=N-N_x-N_y-N_z.
\label{Besetzungszahlen_Informationsmenge}
\end{equation}
Wenn also die Besetzungszahlen in den anderen Teilräumen definiert sind, dann ist die
Besetzungszahl $N_n$ direkt mit der Gesamtbesetzungszahl $N$ korreliert. Und dies wiederum
bedeutet, dass der Teilraum, in dem $A_n$ und $A_n^{\dagger}$ wirken, nicht mit einer
weiteren Raumdimension identifiziert werden darf, sondern dann, wenn die anderen Besetzungszahlen
gegeben sind, implizit die Information über die Gesamtbesetzungszahl $N$ und damit die
Gesamtinformationsmenge in einem Tensorraumzustand enthält. Dass der in der Besetzungszahl
$N_n$ enthaltene Freiheitsgrad keiner Raumrichtung entsprechen kann, dies kann man
schon alleine daraus ersehen, dass $A_n$ beziehungsweise $A_n^{\dagger}$ gemäß
($\ref{ErzeugungsVernichtungsOperatorenXYZ}$) einer Addition der Erzeugungs-
beziehungsweise Vernichtungsoperatoren ($\ref{Operatoren_symmetrisch}$) im symmetrischen
Teilraum des Tensorraumes der Ur"=Alternativen $\mathcal{H}_{TS}$ mit gleichem
Vorzeichen entsprechen, welche ihrerseits den Komponenten einer quantisierten Ur"=Alternative 
($\ref{Operator_Ur-Alternative}$) entsprechen. Diese Operatoren ($\ref{Operatoren_symmetrisch}$)
treten in $A_n$ und $A_n^{\dagger}$ demnach in zueinander symmetrischer Weise auf.
Damit hat also eine $SU(2)$"=Transformation aller Ur"=Alternativen des
symmetrischen Teilraumes des Tensorraumes, die einer $SO(3)$"=Transformation
isomorph ist, also einer Drehung im dreidimensionalen reellen Raum, keine Auswirkung auf
den in der Besetzungszahl $N_n$ enthaltenen Freiheitsgrad. Sie hat aber sehr wohl eine
Auswirkung auf die in den Besetzungszahlen $N_x$, $N_y$ und $N_z$ enthaltenen Freiheitsgrade,
da die Operatoren $A_x$, $A_x^{\dagger}$, $A_y$, $A_y^{\dagger}$, $A_z$, $A_z^{\dagger}$
gemäß ($\ref{ErzeugungsVernichtungsOperatorenXYZ}$) einer Addition der Komponenten der Erzeugungs-
beziehungsweise Vernichtungsoperatoren symmetrischen Teilraum des Tensorraumes der Ur"=Alternativen
($\ref{Operatoren_symmetrisch}$) mit unterschiedlichem Vorzeichen entsprechen. Denn damit
transformiert eine $SU(2)$"=Transformation aller Ur"=Alternativen des symmetrischen
Teilraumes des Tensorraumes der Ur"=Alternativen $\mathcal{H}_{TS}$ diese Operatoren
als Komponenen einer quantisierten Ur"=Alternativen ($\ref{Operator_Ur-Alternative}$)
ineinander, was einer Drehung in einem dreidimensionalen reellen Raum entspricht.
Dieser dreidimensionale Raum wird hier mit dem physikalischen Ortsraum identifiziert.
Der grundlegende Charakter der Möglichkeit einer Darstellung der physikalischen Wirklichkeit
in einem dreidimensionalen Raum, wie sie ja in der Natur und in unserem erkennenden Bezug
zur Natur wirklich vorkommt, steht demnach also mit dieser fundamentalen Symmetrieigenschaft
der fundamentalsten quantenlogischen Objekte unmittelbar in Zusammenhang. Denn es ist
tatsächlich äußerst erstaunlich, dass das einfachste quantentheoretisch überhaupt denkbare
Objekt eine zum realen physikalischen Ortsraum isomorphe Symmetrieeigenschaft aufweist.

Die Zeit $t$ kann nicht einer eigenständigen Dimension entsprechen, weil die Zeitentwicklung
einem Automorphismus des Tensorraumes der Ur"=Alternativen auf sich selbst entspricht. Alleine
die Auszeichnung einer Dynamik über die Schrödingergleichung ($\ref{Schroedingergleichung_Tensorraum}$)
bedeutet schon, dass wenn der Zustand eines Quantenobjektes zu einem Zeitpunkt bekannt ist,
er zu jedem Zeitpunkt bekannt ist, solange keine Messung stattfindet. Die Zeitentwicklung
enthält also keine eigenständige Information und kann demnach nicht einer eigenen
Dimension im Tensorraum der Ur"=Alternativen entsprechen. Dies wäre lediglich unter
der Voraussetzung möglich, dass man eine direkte Korrelation zwischen der Informationsmenge
und der Zeit herstellt, also die Dynamik in direkter Weise mit einem Wachstum der Information
in der Zeit verbindet. Carl Friedrich von Weizsäcker glaubte, dass es ein Wachstum der
Informationsmenge mit der Zeit gebe und er begründete diese Anschauung damit, dass bei
Messungen neue Fakten entstünden. Andererseits sagte er an anderer Stelle auch, dass man
zwischen Ur"=Alternativen unterscheiden müsse, an denen man Messungen vornehme und einer
einmal getroffenen Entscheidung. Aber das bedeutet eigentlich, dass sich bei Messungen
nur Zustände von Ur"=Alternativen verändern, aber dabei im Prinzip keine neuen
Ur"=Alternativen entstehen. Insofern muss man zwischen der Information unterscheiden,
die in den Ur"=Alternativen, die sich als eine Objektivation der Natur auf eine
wirkliche Realität beziehen, an objektiver Information enthalten ist, und der Menge
an Information, die dem menschlichen Geist konkret zugänglich ist. Die elementare
objektive Information ändern sich demnach bei einem Messprozess nicht, aber es
ändert sich die Information, welche für den menschlichen Geist zugänglich ist.
Da die Ur"=Alternativen aber die Gesamtheit der Information in der realen Wirklichkeit
der Natur schematisieren und objektivieren sollen, halte ich es für sinnvoll, davon auszugehen,
dass die Zahl der Ur"=Alternativen nicht wächst, jedenfalls nicht systematisch mit der Zeit.
Auch die Expansion des Kosmos ist kein Indiz für das Wachstum der Information, denn dieses
bezieht sich auf metrische Beziehungen und nicht auf die im Kosmos enthaltene Informationsmenge,
wobei zu sagen ist, dass die Existenz des Raumes einschließlich seiner metrischen Struktur
sich in der Quantentheorie der Ur"=Alternativen ohnehin erst nachträglich als indirekte
Konsequenz aus den Ur"=Alternativen ergibt, was ja in diesem Kapitel gerade
gezeigt werden soll.

\subsection{Abbildung in einen reellen dreidimensionalen Ortsraum}

Um nun die Zustände im Tensorraum der Ur"=Alternativen in den physikalischen Ortsraum abzubilden,
muss von der Möglichkeit Gebrauch gemacht werden, die Operatoren ($\ref{Ort_Impuls_Operatoren}$),
welche eine Heisenbergsche Algebra bilden ($\ref{Heisenbergsche_Algebra}$) und daher als
Orts- und Impulsoperatoren interpretiert werden sollen, in der folgenden Weise darzustellen:

\begin{eqnarray}
X&=&\frac{1}{\sqrt{2}}\left(A_x+A_x^{\dagger}\right)=x,\quad P_x=-\frac{i}{\sqrt{2}}\left(A_x-A_x^{\dagger}\right)=-i\partial_x,\nonumber\\
Y&=&\frac{1}{\sqrt{2}}\left(A_y+A_y^{\dagger}\right)=y,\quad P_y=-\frac{i}{\sqrt{2}}\left(A_y-A_y^{\dagger}\right)=-i\partial_y,\nonumber\\
Z&=&\frac{1}{\sqrt{2}}\left(A_z+A_z^{\dagger}\right)=z,\quad
P_z=-\frac{i}{\sqrt{2}}\left(A_z-A_z^{\dagger}\right)=-i\partial_z,
\label{Ort_Impuls_Darstellung}
\end{eqnarray}
wobei $x$, $y$ und $z$ hier einfach reelle Koordinaten sind und $\partial_x$, $\partial_y$
und $\partial_z$ die dazugehörigen Ableitungen. Im Rahmen dieser Darstellung weisen die
Besetzungszahlzustände eine zur quantenmechanischen Beschreibung des harmonischen
Oszillators isomorphe Gestalt auf:

\begin{eqnarray}
|N_x \rangle \quad\longleftrightarrow\quad
w_{N_x}\left(x\right)&=&\frac{1}{N_x!2^{N_x}\pi^2}\left(x-\partial_x\right)^{N_x}\exp\left(-\frac{x^2}{2}\right),\nonumber\\
|N_y \rangle \quad\longleftrightarrow\quad
w_{N_y}\left(y\right)&=&\frac{1}{N_y!2^{N_y}\pi^2}\left(y-\partial_y\right)^{N_y}\exp\left(-\frac{y^2}{2}\right),\nonumber\\
|N_z \rangle \quad\longleftrightarrow\quad
w_{N_z}\left(z\right)&=&\frac{1}{N_z!2^{N_z}\pi^2}\left(z-\partial_z\right)^{N_z}\exp\left(-\frac{z^2}{2}\right).
\label{Darstellung_Besetzungszahlzustände_Ortsraum}
\end{eqnarray}
Wenn man weiter miteinbezieht, dass bei gegebenen Besetzungszahlen $N_x$, $N_y$, $N_z$,
die Besetzungszahl $N_n$ über die Gesamtzahl an Ur"=Alternativen $N$ in einem
Tensorraumbasiszustand definiert ist ($\ref{Besetzungszahlen_Informationsmenge}$),
dann kann man einen Tensorraumbasiszustand $|N_{xyzn}\rangle$ wie folgt darstellen:

\begin{equation}
|N_{xyzn}\rangle \quad\longleftrightarrow\quad \left[w_{N_x}(x)w_{N_y}(y)w_{N_z}(z)\right]_N
\equiv f_N\left(N_x,N_y,N_z,x,y,z\right)\equiv f_{N_{xyzn}}\left(\mathbf{x}\right),
\label{Basiszustand_Tensorraum_Ortsdarstellung}
\end{equation}
wobei der Index $N$ die Gesamtinformationsmenge in dem entsprechenden Basiszustand darstellt.
Die Wellenfunktionen $f_{N_{xyzn}}\left(\mathbf{x}\right)$ sind als räumliche
Darstellung der Basiszustände des Tensorraumes normierbar. Dies ist ein weiterer
interessanter Vorzug der Quantentheorie der Ur"=Alternativen gegenüber Quantenfeldtheorien,
bei denen die Impulseigenzustände der einzelnen Teilchen nicht normierbar sind und daher
nicht zum Hilbertraum der quadratintegrablen Funktionen gehören. Ein allgemeiner
symmetrischer Zustand im Tensorraum der Ur"=Alternativen kann mit Hilfe von
($\ref{Basiszustand_Tensorraum_Ortsdarstellung}$) in der folgenden Weise dargestellt werden:

\begin{eqnarray}
\label{Zustand_Darstellung_Ortsraum}
&&|\Psi\rangle_S=\sum_{N_{ABCD}}\psi_S\left(N_{ABCD}\right)|N_{ABCD}\rangle
=\sum_{N_{xyzn}}\psi_S\left(N_{xyzn}\right)|N_{xyzn}\rangle
\quad\longleftrightarrow\\
&&\sum_{N_x,N_y,N_z,N_n}\psi_S\left(N_x,N_y,N_z,N_n\right)f_N\left(N_x,N_y,N_z,x,y,z\right)
=\sum_{N_{xyzn}}\psi_S\left(N_{xyzn}\right)f_{N_{xyzn}}(\mathbf{x})=\Psi_N\left(\mathbf{x}\right).\nonumber
\end{eqnarray}
An dieser Stelle wird der Übergang der rein quantenlogischen Wirklichkeit zur Darstellung innerhalb
des physikalischen Ortsraumes vollzogen. Wenn man alle Ur"=Alternativen des Tensorraumes mit einer
$SU(2)$"=Transformation transformiert, so entspricht dem eine Drehung im dreidimensionalen Ortsraum,
weshalb hier implizit auch die Isomorphie zwischen der $SU(2)$ und der $SO(3)$ enthalten ist.
Denn es kann ja eigentlich kein Zufall sein, dass die fundamentale Symmetriegruppe des
fundamentalsten quantentheoretischen Objektes, nämlich die $U(2)=SU(2)\otimes U(1)$ direkt
zur Drehgruppe des physikalischen Raumes $SO(3)$ multipliziert mit der Zeitentwicklungsgruppe $U(1)$
isomorph ist. Dies bedeutet, dass sich Zustände vieler Ur"=Alternativen, die quantentheoretischen
Objekten entsprechen, als Objekte in einem dreidimensionalen reellen Raum darstellen. Ganz entscheidend
hierbei aber ist, dass die Wellenfunktionen im Raum, also $\Psi_N\left(\mathbf{x}\right)$,
keine klassischen Felder sind. Denn klassische Felder weisen eine innere räumliche
Kausalstruktur auf, bei der sich Wirkungen im Raum kontinuierlich ausbreiten.
Diese Wellenfunktionen, die hier betrachtet werden, sind lediglich die Darstellung
einer dahinter stehenden rein quantenlogischen Beziehungsstruktur, welche sich in
den Zuständen des Tensorraumes der Ur"=Alternativen manifestiert. Dem entspricht der
Bohrsche Begriff der Individualität, demgemäß ein Quantenobjekt oder eine Wellenfunktion,
die ein solches darstellt, sich als eine schlechthin unteilbare Einheit erweist,
die gerade keine innere räumliche Kausalstruktur aufweist und sich nur im
gesamten Raum gleichzeitig ändern kann. Eben deshalb liefert die Quantentheorie
der Ur"=Alternativen eine Erklärung für das EPR"=Paradoxon, nämlich dass sich
scheinbar weit voneinander entfernte Objekte so verhalten, als seien sie
gar nicht getrennt voneinander. Denn die räumliche Darstellung ist nur eine
oberflächliche Betrachtung der dahinter in Wahrheit stehenden quantenlogischen
Beziehungsstruktur, die an raum"=zeitliche Kausalbeziehungen überhaupt gar nicht gebunden
ist und auch gar nicht gebunden sein kann, weil Logik unabhängig von Geometrie und
räumlicher Kausalität ist. Die räumliche Wirklichkeit kann sich allenfalls umgekehrt
indirekt und näherungsweise als ein Ausfluss der reinen Quantenlogik ergeben.
Und in diesem Sinne ist die räumliche Darstellung der Zustände des Tensorraumes
der Ur"=Alternativen gemäß ($\ref{Zustand_Darstellung_Ortsraum}$) gemeint.
An dieser Stelle könnte man zwar einwenden, dass man nicht sicher wissen könne,
dass der dreidimensionale reelle Raum, der sich als Darstellungsraum der Zustände im
Tensorraum der Ur"=Alternativen ergibt, auch tatsächlich mit dem realen Ortsraum
identifiziert werden könne. Aber diesbezüglich muss geantwortet werden, dass man in
der mathematischen Physik ganz grundsätzlich an irgendeiner Stelle eine Beziehung
zwischen den mathematischen Strukturen und der physikalischen Realität vornehmen muss.
Dies geschieht hier bereits auf der Ebene der Ur"=Alternativen, die in diesem Rahmen
als elementare Einheiten der Objektivation der wirklich existierenden äußeren Natur
im menschlichen Geist angesehen werden. Und wenn sich dann herausstellt, dass man
die Zustände vieler solcher Ur"=Alternativen in einen dreidimensionalen reellen
Raum abbilden kann, dann liegt es doch auf der Hand, diesen Raum auch mit dem
physikalischen Raum als einer in der Natur auf einer Oberflächenebene real
existierenden Struktur zu identifizieren.

\subsection{Die Kopernikanische Wende bezüglich Objekt und Raum}

Es sei darauf hingewiesen, dass an dieser Stelle eine fundamentale sozusagen "`Kopernikanische Wende"'
im Weltbild vollzogen wird. Nicht der Raum ist ein für sich bestehendes Medium, in dem die Objekte
definiert sind, sondern die Existenz der Objekte ist unabhängig vom Raum und dieser konstituiert
sich über die Existenz der Objekte:

\begin{align}
&\textbf{Raum}\quad \xrightarrow{konstituiert}\quad \textbf{Objekte}
\quad\textrm{(gewöhnliche Anschauung)}\nonumber\\
&\downarrow \textrm{Kopernikanische\ Wende}\downarrow\nonumber\\
&\textbf{Objekte}\quad \xrightarrow{konstituieren}\quad \textbf{Raum}\quad\textrm{(Quantentheorie der Ur-Alternativen)}.
\end{align}
Diese war zwar in der Quantentheorie und in gewisser Weise auch in der allgemeinen Relativitätstheorie
implizit schon immer enthalten, was seitens Albert Einstein in Bezug auf die allgemeine
Relativitätstheorie und seitens Werner Heisenberg in Bezug auf die Quantentheorie auch deutlich formuliert wurde.
Aber sowohl Einstein alsauch Heisenberg lösten sich nicht wirklich vom feldtheoretischen Denken,
indem sie doch immer eine Raum"=Zeit"=Mannigfaltigkeit als gegeben voraussetzten. In der allgemeinen
Relativitätstheorie ergibt sich dann zwar der relationale Charakter der Raum"=Zeit als mit der
Diffeomorphismeninvarianz verknüpfte Eigenschaft und Einstein weist explizit darauf hin, dass
in der allgemeinen Relativitätstheorie nur raum"=zeitliche Koinzidenzen real sind und auch die
Raum"=Zeit verschwände, wenn man die Körper aus ihr herausnähme. Auch im Rahmen der Quantentheorie
redet Heisenberg in Bezug auf die Wellenfunktionen von Vorgängen außerhalb von Raum und Zeit,
aber im Rahmen der konkreten Realisierung innerhalb des theoretischen Rahmens relativistischer
Quantenfeldtheorien, wie sie erstmals seitens Heisenberg und Pauli \cite{Heisenberg:1929} formuliert wurde,
und dann auch der nichtlinearen Spinorfeldtheorie Heisenbergs und dem Standardmodell der Elementarteilchenphysik
zugrunde liegt, werden die Wechselwirkungen durch punktweise Produkte von Feldern definiert und
überhaupt bereits eine Raum"=Zeit vorausgesetzt, auf der die Felder definiert sind, die anschließend
quantisiert werden. Erst Carl Friedrich von Weizsäcker geht von reinen logischen Alternativen aus,
die keinerlei physikalischen Raum voraussetzen, seine Existenz umgekehrt aber zu begründen gestatten. 
Er hat dies mathematisch ursprünglich in der Weise zu vollziehen versucht, dass er die
Topologie der $SU(2)$ als Symmetriegruppe einer einzelnen Ur"=Alternative, die einer
$\mathbb{S}^3$ entspricht, also einer dreidimensionalen Sphäre, die man sich als in einen
vierdimensionalen Raum eingebettet vorstellen kann, als die Topologie des Kosmos interpretierte.
Zudem definierte er die abstrakten Impulse im Gegensatz zu der vorliegenden Arbeit nicht
als lineare, sondern als bilineare Ausdrücke bezüglich der Erzeugungs- und
Vernichtungsoperatoren im Tensorraum der Ur"=Alternativen. Ich glaube aber,
dass in Wirklichkeit erst meine Abbildung der Zustände im Tensorraum der Ur"=Alternativen
in einen dreidimensionalen Ortsraum, wie sie erstmals in \cite{Kober:2012} versucht, in \cite{Kober:2017}
konsequent formuliert und in diesem Kapitel dargestellt wird, die richtige mathematische
Fassung dieses entscheidenden und zentralen begrifflichen Gedankens von Weizsäckers
liefert, der natürlich das eigentlich Entscheidende ist, nämlich dass es nicht
einen unabhängig für sich existierenden physikalischen Raum gibt, in dem sich
räumlich charakterisierte physikalische Objekte bewegen, sondern auf der fundamentalen Ebene
rein logisch definierte Objekte existieren, die umgekehrt die Existenz des physikalischen
Raumes erst bedingen. Carl Friedrich von Weizsäcker alleine gebührt also die Ehre für
diese fundamentale sowie zentrale Erkenntnis und "`Kopernikanische Wende"' im physikalischen
Weltbild. Ich begnüge mich damit, dankbar dafür zu sein, dass ich als sein treuer Diener
dazu beitragen konnte, diesen großen Gedanken mathematisch adäquat zu fassen.
Die Bezeichnung "`Kopernikanische Wende"', das sei an dieser Stelle in aller Deutlichkeit
erwähnt, ist nicht nur deshalb angemessen, weil hier wie bei Kopernikus beziehungsweise Aristarch
eine grundlegende Veränderung im naturwissenschaftlichen Weltbild vollzogen wird, sondern vor
allem auch deshalb, weil auch Immanuel Kant seine Verschiebung des menschlichen Erkenntnisprozesses
als Beziehung zwischen objektiver Wirklichkeit und menschlichem Geist zu Gunsten des
menschlichen Geistes in eine Analogie zur Kopernikanischen Wende setzte. Und die Einbeziehung
dieser Kantischen Perspektive ist ja die Voraussetzung dafür, dass auch die seitens Carl Friedrich
von Weizsäcker vollzogene Wende überhaupt möglich wird und verstanden werden kann. Denn sie basiert
auf der Erkenntnis, dass die Räumlichkeit und die darauf basierende lokale Kausalstruktur der Welt
Teil unserer Wahrnehmung und unseres Denkens ist. Die Räumlichkeit hat zwar auf der Oberfläche auch
in der Natur eine reale Entsprechung, aber auf der fundamentalen Ebene kann sie keineswegs absolute
Gültigkeit beanspruchen, obwohl sie auch dort von uns zunächst als essentieller Teil unseres
konkreten menschlichen Bezuges zur Wirklichkeit vorausgesetzt wird und daher in einem Denkprozess
nachträglich mühsam aus diesem Bezug zur Wirklichkeit herausgenommen werden muss, um der
Wirklichkeit der Natur an sich selbst näher zu kommen. Diese fundamentale "`Kopernikanische Wende"'
im physikalischen Weltbild, die mit der Quantentheorie der Ur"=Alternativen vollzogen wird,
soll in folgender Formel zusammengefasst werden:

\noindent
\fbox{\parbox{163mm}{\textbf{In der Wirklichkeit der Natur befinden sich nicht
geometrisch charakterisierte Objekte in einem a priori gegebenen physikalischen
Raum, sondern rein logisch charakterisierte Objekte begründen umgekehrt die
Existenz des physikalischen Raumes als einer Möglichkeit der Darstellung der
durch jene Objekte sich konstituierenden abstrakten Beziehungsstruktur.}}}

\section{Dynamik und Raum-Zeit}

\subsection{Die Sonderrolle der Zeit}

Um die Art und Weise zu deuten, in der Raum und Zeit sich in der Quantentheorie
der Ur"=Alternativen zur Raum"=Zeit in einem zur Relativitätstheorie analogen wenn auch
modifizierten Sinne verbinden, muss zunächst das besondere Wesen der Zeit näher geklärt werden.
Es wurde bereits erwähnt, dass der Zeit gegenüber dem Raum eine Sonderrolle in der
Beschreibung der Natur zukommen muss. Daran hat auch die Relativitätstheorie nicht
viel geändert, denn sie verbindet Raum und Zeit zwar in neuer Weise, aber der
Wesensunterschied des Raumes zur Zeit ist damit nicht angetastet. Dies wird schon alleine
daraus deutlich, dass der Zeit auch in der Philosophie eine existentielle Dimension zukommt,
die der Raum nicht hat. Bei Kant liegt sie nicht nur aller äußeren sondern auch aller inneren Anschauung
zugrunde. Denn auch rein seelische Begebenheiten sind an die Zeit nicht jedoch an den Raum gebunden.
Martin Heidegger nennt sein Hauptwerk sogar "`Sein und Zeit"' \cite{Heidegger:1927} und verdeutlicht
damit den fundamentalen Charakter der Zeit. Aber auch in der theoretischen Physik manifestiert sich
die Sonderrolle der Zeit sehr konkret. Die Raum- beziehungsweise Zeitartigkeit eines Abstandes zweier
Ereignisse im Minkowski"=Raum der speziellen Relativitätstheorie ist eine Lorentz"=invariante Größe.
Zudem ist wie bereits angesprochen durch die Dynamik die Zeitentwicklung determiniert.
In der Quantentheorie gilt dies über die Schrödingergleichung wenigstens solange keine
Messung durchgeführt wird. Dies aber bedeutet, dass der Zeit kein unabhängiger physikalischer
Freiheitsgrad entspricht und die Zeitdimension daher keine zusätzliche Information enthält,
die nicht schon zu einem bestimmten Zeitpunkt existierte. Es ist zwar durchaus denkbar,
dass in der zeitlichen Entwicklung neue Information entsteht. Diese neue Information
bezieht sich aber dann auf etwas anderes und nicht auf den Vollzug der Zeitentwicklung, weshalb
die Zeitdimension dann dennoch keiner Darstellung eines Informationsfreiheitsgrades entspricht,
während die Raumdimensionen einer Darstellung dreier der vier Teilräume im Informationsraum
entsprechen, dem Tensorraum der Ur"=Alternativen. Die vierte Dimension beschreibt demnach nicht
die Zeit, sondern die Informationsmenge in den Zuständen im Tensorraum der Ur"=Alternativen
($\ref{Besetzungszahlen_Informationsmenge}$). Aus diesem Grunde kann der Zeit also weder ein
Freiheitsgrad im abstrakten Informationsraum noch ein Operator entsprechen. Insofern muss
die Zeit gemäß dem Postulat der Zeitentwicklung als ein Automorphismus des Tensorraumes der
Ur"=Alternativen aufgefasst werden. Dieser Automorphismus ist durch die Dynamik determiniert
($\ref{Schroedingergleichung_Tensorraum}$). Und wenn die Dynamik in einer solchen Weise
konstituiert ist, dass der Energieoperator mit der relativistischen Beziehung zwischen
Ort und Impuls übereinstimmt ($\ref{Energie-Impuls-Relation}$), dann ergibt sich,
sofern man die Zeit als zusätzliche Dimension auffasst, faktisch diejenige Struktur,
die wir als Raum"=Zeit"=Kontinuum bezeichnen, also als Minkowski"=Raum oder
Riemannsche Raum"=Zeit"=Mannigfaltigkeit.

\subsection{Die Zeitentwicklung im Tensorraum der Ur-Alternativen}

Die Zeitentwicklung im Tensorraum der Ur"=Alternativen muss gemäß der Formulierung der
allgemeinen Schrödingergleichung im Tensorraum der Ur"=Alternativen ($\ref{Schroedingergleichung_Tensorraum}$)
natürlich durch einen reellen Zeitparameter $t$ mit einem dementsprechenden Hamiltonoperator
als Generator dieser Zeitentwicklung beschrieben werden. Das bedeutet, dass ein Hamiltonoperator
im Tensorraum der Ur"=Alternativen mit Hilfe der Erzeugungs- und Vernichtungsoperatoren ($\ref{ErzeugungsVernichtungsOperatorenXYZ}$) definiert werden muss. Entsprechend dem
Korrespondenzprinzip wird der Energieoperator $E$ eines freien Objektes:

\begin{equation}
E=-i\partial_t,
\label{Energieoperator_Darstellung}
\end{equation}
der gemäß ($\ref{Energieoperator_Darstellung}$) über den Zeitparameter $t$ definiert
ist und gemäß der allgemeinen Schrödingergleichung im Tensorraum der Ur"=Alternativen
($\ref{Schroedingergleichung_Tensorraum}$) gleich dem Hamilton"=Operator $H$
gesetzt werden muss:

\begin{equation}
E=H,
\end{equation}
in Isomorphie zur Beziehung zwischen Energie und Impuls in der speziellen Relativitätstheorie
in dieser Arbeit in der folgenden Weise definiert:

\begin{eqnarray}
E^2=P_x^2+P_y^2+P_z^2 \quad\Leftrightarrow\quad E=\pm \sqrt{P_x^2+P_y^2+P_z^2}.
\label{Energie-Impuls-Relation}
\end{eqnarray}
Zur Berechnung der konkreten Gestalt des Energieoperators $E$ ($\ref{Energie-Impuls-Relation}$) in
Abhängigkeit von den Operatoren ($\ref{ErzeugungsVernichtungsOperatorenXYZ}$) müssen zunächst einmal
die Quadrate der Impulsoperatoren $P_x$, $P_y$ und $P_z$ gebildet und berechnet werden:

\begin{eqnarray}
P_x^2&=&\frac{1}{2}\left(-A_x A_x+A_x A_x^{\dagger}+A_x^{\dagger}A_x-A_x^{\dagger}A_x^{\dagger}\right)
=\frac{1}{2}\left(-A_x A_x+2 A_x^{\dagger}A_x-A_x^{\dagger}A_x^{\dagger}+1\right)\nonumber\\
&=&\frac{1}{8}\left[-A_S A_S-B_S B_S-C_S C_S-D_S D_S-2 A_S B_S+2 A_S C_S+2 A_S D_S+2 B_S C_S+2 B_S D_S-2 C_S D_S
\right.\nonumber\\ &&\left.
-A^{\dagger}_S A^{\dagger}_S-B^{\dagger}_S B^{\dagger}_S-C^{\dagger}_S C^{\dagger}_S-D^{\dagger}_S D^{\dagger}_S
-2A^{\dagger}_S B^{\dagger}_S+2A^{\dagger}_S C^{\dagger}_S+2A^{\dagger}_S D^{\dagger}_S
+2B^{\dagger}_S C^{\dagger}_S+2B^{\dagger}_S D^{\dagger}_S-2C^{\dagger}_S D^{\dagger}_S
\right.\nonumber\\ &&\left.
+2\left(A^{\dagger}_S A_S+B^{\dagger}_S B_S+C^{\dagger}_S C_S+D^{\dagger}_S D_S
+A^{\dagger}_S B_S-A^{\dagger}_S C_S-A^{\dagger}_S D_S+B^{\dagger}_S A_S-B^{\dagger}_S C_S-B^{\dagger}_S D_S
\right.\right.\nonumber\\ &&\left.\left.
-C^{\dagger}_S A_S-C^{\dagger}_S B_S+C^{\dagger}_S D_S
-D^{\dagger}_S A_S-D^{\dagger}_S B_S+D^{\dagger}_S C_S\right)+1\right],
\end{eqnarray}

\begin{eqnarray}
P_y^2&=&\frac{1}{2}\left(-A_y A_y+A_y A_y^{\dagger}+A_y^{\dagger}A_y-A_y^{\dagger}A_y^{\dagger}\right)
=\frac{1}{2}\left(-A_y A_y+2 A_y^{\dagger}A_y-A_y^{\dagger}A_y^{\dagger}+1\right)\nonumber\\
&=&\frac{1}{8}\left[-A_S A_S-B_S B_S-C_S C_S-D_S D_S+2 A_S B_S-2 A_S C_S+2 A_S D_S+2 B_S C_S-2 B_S D_S+2 C_S D_S
\right.\nonumber\\ &&\left.
-A^{\dagger}_S A^{\dagger}_S-B^{\dagger}_S B^{\dagger}_S-C^{\dagger}_S C^{\dagger}_S-D^{\dagger}_S D^{\dagger}_S
+2A^{\dagger}_S B^{\dagger}_S-2A^{\dagger}_S C^{\dagger}_S+2A^{\dagger}_S D^{\dagger}_S
+2B^{\dagger}_S C^{\dagger}_S-2B^{\dagger}_S D^{\dagger}_S+2C^{\dagger}_S D^{\dagger}_S
\right.\nonumber\\ &&\left.
+2\left(A^{\dagger}_S A_S+B^{\dagger}_S B_S+C^{\dagger}_S C_S+D^{\dagger}_S D_S
-A^{\dagger}_S B_S+A^{\dagger}_S C_S-A^{\dagger}_S D_S-B^{\dagger}_S A_S-B^{\dagger}_S C_S+B^{\dagger}_S D_S
\right.\right.\nonumber\\ &&\left.\left.
+C^{\dagger}_S A_S-C^{\dagger}_S B_S-C^{\dagger}_S D_S-D^{\dagger}_S A_S +D^{\dagger}_S B_S-D^{\dagger}_S C_S\right)+1\right],
\end{eqnarray}

\begin{eqnarray}
P_z^2&=&\frac{1}{2}\left(-A_z A_z+A_z A_z^{\dagger}+A_z^{\dagger}A_z-A_z^{\dagger}A_z^{\dagger}\right)
=\frac{1}{2}\left(-A_z A_z+2 A_z^{\dagger}A_z-A_z^{\dagger}A_z^{\dagger}+1\right)\nonumber\\
&=&\frac{1}{8}\left[-A_S A_S-B_S B_S-C_S C_S-D_S D_S+2 A_S B_S+2 A_S C_S-2 A_S D_S-2 B_S C_S+2 B_S D_S+2 C_S D_S
\right.\nonumber\\ &&\left.
-A^{\dagger}_S A^{\dagger}_S-B^{\dagger}_S B^{\dagger}_S-C^{\dagger}_S C^{\dagger}_S-D^{\dagger}_S D^{\dagger}_S
+2A^{\dagger}_S B^{\dagger}_S+2A^{\dagger}_S C^{\dagger}_S-2A^{\dagger}_S D^{\dagger}_S
-2B^{\dagger}_S C^{\dagger}_S+2B^{\dagger}_S D^{\dagger}_S+2C^{\dagger}_S D^{\dagger}_S
\right.\nonumber\\ &&\left.
+2\left(A^{\dagger}_S A_S+B^{\dagger}_S B_S+C^{\dagger}_S C_S+D^{\dagger}_S D_S
-A^{\dagger}_S B_S-A^{\dagger}_S C_S+A^{\dagger}_S D_S-B^{\dagger}_S A_S+B^{\dagger}_S C_S-B^{\dagger}_S D_S
\right.\right.\nonumber\\ &&\left.\left.
-C^{\dagger}_S A_S+C^{\dagger}_S B_S-C^{\dagger}_S D_S+D^{\dagger}_S A_S-D^{\dagger}_S B_S-D^{\dagger}_S C_S\right)+1\right].
\end{eqnarray}
Das über ($\ref{Energie-Impuls-Relation}$) definierte Quadrat des Energieoperators erhält
damit die folgende Gestalt:

\begin{eqnarray}
E^2&=&P_x^2+P_y^2+P_z^2=\frac{1}{2}\left(-A_x A_x+A_x A_x^{\dagger}+A_x^{\dagger}A_x-A_x^{\dagger}A_x^{\dagger}
-A_y A_y+A_y A_y^{\dagger}+A_y^{\dagger}A_y-A_y^{\dagger}A_y^{\dagger}\right.\nonumber\\
&&\left.-A_z A_z+A_z A_z^{\dagger}+A_z^{\dagger}A_z-A_z^{\dagger}A_z^{\dagger}\right)\nonumber\\
&=&\frac{1}{2}\left(-A_x A_x+2 A_x^{\dagger}A_x-A_x^{\dagger}A_x^{\dagger}-A_y A_y+2 A_y^{\dagger}A_y-A_y^{\dagger}A_y^{\dagger}
-A_z A_z+2 A_z^{\dagger}A_z-A_z^{\dagger}A_z^{\dagger}+3\right)\nonumber\\
&=&\frac{1}{8}\left[-3A_S A_S-3B_S B_S-3C_S C_S-3D_S D_S-3A^{\dagger}_S A^{\dagger}_S
-3B^{\dagger}_S B^{\dagger}_S-3C^{\dagger}_S C^{\dagger}_S-3D^{\dagger}_S D^{\dagger}_S
\right.\nonumber\\ &&\left.
+2A_S B_S +2A_S C_S+2A_S D_S+2B_S C_S+2B_S D_S+2C_S D_S
\right.\nonumber\\ &&\left.
+2A^{\dagger}_S B^{\dagger}_S+2A^{\dagger}_S C^{\dagger}_S +2A^{\dagger}_S D^{\dagger}_S
+2B^{\dagger}_S C^{\dagger}_S+2B^{\dagger}_S D^{\dagger}_S +2C^{\dagger}_S D^{\dagger}_S
\right.\nonumber\\ &&\left.
+2\left(3A^{\dagger}_S A_S+3B^{\dagger}_S B_S+3C^{\dagger}_S C_S+3D^{\dagger}_S D_S
\right.\right.\nonumber\\ &&\left.\left.
-A^{\dagger}_S B_S-A^{\dagger}_S C_S-A^{\dagger}_S D_S-B^{\dagger}_S A_S-B^{\dagger}_S C_S-B^{\dagger}_S D_S
\right.\right.\nonumber\\ &&\left.\left.
-C^{\dagger}_S A_S-C^{\dagger}_S B_S-C^{\dagger}_S D_S-D^{\dagger}_S A_S-D^{\dagger}_S B_S-D^{\dagger}_S C_S\right)+3\right],
\end{eqnarray}
und die positive Komponente des Energieoperators lautet damit:

\begin{eqnarray}
E&=&\sqrt{P_x^2+P_y^2+P_z^2}=\frac{1}{\sqrt{2}}\left(-A_x A_x+A_x A_x^{\dagger}+A_x^{\dagger}A_x-A_x^{\dagger}A_x^{\dagger}
-A_y A_y+A_y A_y^{\dagger}+A_y^{\dagger}A_y-A_y^{\dagger}A_y^{\dagger}\right.\nonumber\\
&&\left.-A_z A_z+A_z A_z^{\dagger}+A_z^{\dagger}A_z-A_z^{\dagger}A_z^{\dagger}+3\right)^{\frac{1}{2}}\nonumber\\
&=&\frac{1}{\sqrt{2}}\left(-A_x A_x+2 A_x^{\dagger}A_x-A_x^{\dagger}A_x^{\dagger}-A_y A_y+2 A_y^{\dagger}A_y-A_y^{\dagger}A_y^{\dagger}
-A_z A_z+2 A_z^{\dagger}A_z-A_z^{\dagger}A_z^{\dagger}+3\right)^{\frac{1}{2}}\nonumber\\
&=&\frac{1}{2\sqrt{2}}\left[-3A_S A_S-3B_S B_S-3C_S C_S-3D_S D_S
-3A^{\dagger}_S A^{\dagger}_S-3B^{\dagger}_S B^{\dagger}_S-3C^{\dagger}_S C^{\dagger}_S-3D^{\dagger}_S D^{\dagger}_S
\right.\nonumber\\ &&\left.
+2A_S B_S +2A_S C_S+2A_S D_S+2B_S C_S+2B_S D_S+2C_S D_S
\right.\nonumber\\ &&\left.
+2A^{\dagger}_S B^{\dagger}_S+2A^{\dagger}_S C^{\dagger}_S+2A^{\dagger}_S D^{\dagger}_S
+2B^{\dagger}_S C^{\dagger}_S+2B^{\dagger}_S D^{\dagger}_S+2C^{\dagger}_S D^{\dagger}_S
\right.\nonumber\\ &&\left.
+2\left(3A^{\dagger}_S A_S+3B^{\dagger}_S B_S+3C^{\dagger}_S C_S+3D^{\dagger}_S D_S
\right.\right.\nonumber\\ &&\left.\left.
-A^{\dagger}_S B_S-A^{\dagger}_S C_S-A^{\dagger}_S D_S-B^{\dagger}_S A_S-B^{\dagger}_S C_S-B^{\dagger}_S D_S
\right.\right.\nonumber\\ &&\left.\left.
-C^{\dagger}_S A_S-C^{\dagger}_S B_S-C^{\dagger}_S D_S-D^{\dagger}_S A_S
-D^{\dagger}_S B_S-D^{\dagger}_S C_S\right)+3\right]^{\frac{1}{2}}.
\label{Energie_Operator}
\end{eqnarray}
Um die Zeitentwicklung explizit zu bestimmen, muss zunächst der Energieoperator auf
einen Tensorraumzustand angewandt werden. Es ergibt sich diesbezüglich das Folgende:

\begin{eqnarray}
E|\Psi\rangle&=&\sum_{N_{ABCD}}\psi(N_{ABCD})E|N_{ABCD}\rangle\nonumber\\
&=&\sum_{N_{ABCD}}\psi(N_{ABCD})\sqrt{P_x^2+P_y^2+P_z^2}|N_{ABCD}\rangle\nonumber\\
&\equiv&\sum_{N_{ABCD}}\mathcal{E}_\psi\left(N_{ABCD}\right)|N_{ABCD}\rangle.\nonumber\\
&\equiv&|\Psi\rangle_E,
\label{Anwendung_Energie_Zustand}
\end{eqnarray}
wenn man die Größe $\mathcal{E}_\psi\left(N_{ABCD}\right)$ in der folgenden
Weise definiert:

\begin{align}
&\mathcal{E}_\psi\left(N_{ABCD}\right)
=\frac{1}{2\sqrt{2}}\left[-3\psi(N_A+2,N_B,N_C,N_D)\sqrt{N_A+2}\sqrt{N_A+1}
\right.\nonumber\\&\left.
-3\psi(N_A,N_B+2,N_C,N_D)\sqrt{N_B+2}\sqrt{N_B+1}\right.\nonumber\\
&\left.-3\psi(N_A,N_B,N_C+2,N_D)\sqrt{N_C+2}\sqrt{N_C+1}-3\psi(N_A,N_B,N_C,N_D+2)\sqrt{N_D+2}\sqrt{N_D+1}\right.\nonumber\\
&\left.-3\psi(N_A-2,N_B,N_C,N_D)\sqrt{N_A}\sqrt{N_A-1}-3\psi(N_A,N_B-2,N_C,N_D)\sqrt{N_B}\sqrt{N_B-1}\right.\nonumber\\
&\left.-3\psi(N_A,N_B,N_C-1,N_D)\sqrt{N_C}\sqrt{N_C-1}-3\psi(N_A+2,N_B,N_C,N_D-2)\sqrt{N_D}\sqrt{N_D-1}\right.\nonumber\\
&\left.+2\psi(N_A+1,N_B+1,N_C,N_D)\sqrt{N_A+1}\sqrt{N_B+1}+2\psi(N_A+1,N_B,N_C+1,N_D)\sqrt{N_A+1}\sqrt{N_C+1}\right.\nonumber\\
&\left.+2\psi(N_A+1,N_B,N_C,N_D+1)\sqrt{N_A+1}\sqrt{N_D+1}+2\psi(N_A,N_B+1,N_C+1,N_D)\sqrt{N_B+1}\sqrt{N_C+1}\right.\nonumber\\
&\left.+2\psi(N_A,N_B+1,N_C,N_D+1)\sqrt{N_B+1}\sqrt{N_D+1}+2\psi(N_A,N_B,N_C+1,N_D+1)\sqrt{N_C+1}\sqrt{N_D+1}\right.\nonumber\\
&\left.+2\psi(N_A-1,N_B-1,N_C,N_D)\sqrt{N_A-1}\sqrt{N_B-1}+2\psi(N_A-1,N_B,N_C-1,N_D)\sqrt{N_A-1}\sqrt{N_C-1}\right.\nonumber\\
&\left.+2\psi(N_A-1,N_B,N_C,N_D-1)\sqrt{N_A-1}\sqrt{N_D-1}+2\psi(N_A,N_B-1,N_C-1,N_D)\sqrt{N_B-1}\sqrt{N_C-1}\right.\nonumber\\
&\left.+2\psi(N_A,N_B-1,N_C,N_D-1)\sqrt{N_B-1}\sqrt{N_D-1}+2\psi(N_A,N_B,N_C-1,N_D-1)\sqrt{N_C-1}\sqrt{N_D-1}\right.\nonumber\\
&\left.+2\left(3\psi(N_A,N_B,N_C,N_D)N_A+3\psi(N_A,N_B,N_C,N_D)N_B
\right.\right.\nonumber\\
&\left.\left.+3\psi(N_A,N_B,N_C,N_D)N_C+3\psi(N_A,N_B,N_C,N_D)N_D
\right.\right.\nonumber\\
&\left.\left.-\psi(N_A+1,N_B-1,N_C,N_D)\sqrt{N_A+1}\sqrt{N_B}-\psi(N_A+1,N_B,N_C-1,N_D)\sqrt{N_A+1}\sqrt{N_C}
\right.\right.\nonumber\\
&\left.\left.-\psi(N_A+1,N_B,N_C,N_D-1)\sqrt{N_A+1}\sqrt{N_D}-\psi(N_A-1,N_B+1,N_C,N_D)\sqrt{N_A}\sqrt{N_B+1}
\right.\right.\nonumber\\
&\left.\left.-\psi(N_A,N_B+1,N_C-1,N_D)\sqrt{N_B+1}\sqrt{N_C}-\psi(N_A,N_B+1,N_C,N_D-1)\sqrt{N_B+1}\sqrt{N_D}
\right.\right.\nonumber\\
&\left.\left.-\psi(N_A-1,N_B,N_C+1,N_D)\sqrt{N_A}\sqrt{N_C+1}-\psi(N_A,N_B-1,N_C+1,N_D)\sqrt{N_B}\sqrt{N_C+1}
\right.\right.\nonumber\\
&\left.\left.-\psi(N_A,N_B,N_C+1,N_D-1)\sqrt{N_C+1}\sqrt{N_D}-\psi(N_A-1,N_B,N_C,N_D+1)\sqrt{N_A}\sqrt{N_D+1}
\right.\right.\nonumber\\
&\left.\left.-\psi(N_A,N_B-1,N_C,N_D+1)\sqrt{N_B}\sqrt{N_D+1}-\psi(N_A,N_B,N_C-1,N_D+1)\sqrt{N_C}\sqrt{N_D+1}\right)\right.
\nonumber\\&\left.
+3\psi(N_A,N_B,N_C,N_D)\right]\times\left[\psi(N_{ABCD})\right]^{\frac{1}{2}}.
\end{align}
Es sei noch einmal in aller Deutlichkeit darauf hingewiesen, dass im Einklang mit den obigen
Erläuterungen die Operatoren $X$, $Y$, $Z$, $P_x$, $P_y$, $P_z$ und $E$ alle im abstrakten
Tensorraum der Ur"=Alternativen wirken und keineswegs auf einen vorgegebenen Orts-
beziehungsweise Impulsraum bezogen sind. Denn sie sind gemäß ($\ref{Ort_Impuls_Operatoren}$)
in Abhängigkeit der Erzeugungs- und Vernichtungsoperatoren ($\ref{Operatoren_symmetrisch}$)
beziehungsweise gemäß ($\ref{Ort_Impuls_Operatoren_ABCD}$) in Abhängigkeit der Erzeugungs-
und Vernichtungsoperatoren ($\ref{ErzeugungsVernichtungsOperatorenXYZ}$) definiert.
Vielmehr ergibt sich aufgrund der algebraischen Eigenschaften die Struktur des
physikalischen Ortsraumes als Darstellungsraum erst nachträglich. Man kann nun
die Komponenten des Impulsoperators aus ($\ref{Ort_Impuls_Operatoren}$) gemeinsam
mit dem Energieoperator ($\ref{Energieoperator_Darstellung}$) zu einem
Energie"=Impuls"=Operator zusammenfassen:

\begin{equation}
P_{ABCD}=\left(E,P_x,P_y,P_z\right).
\label{Viererimpuls_Tensorraum}
\end{equation}
In Bezug auf spätere Betrachtungen dieser Arbeit ist es sinnvoll, in Analogie zum
abstrakten Energie"=Impuls"=Operator im Tensorraum der Ur"=Alternativen $P_{ABCD}$
definiert in ($\ref{Viererimpuls_Tensorraum}$) einen abstrakten Viererortsoperator
$X_{ABCD}$ in Bezug auf den Tensorraum der Ur"=Alternativen in der folgenden
Weise zu definieren:

\begin{equation}
X_{ABCD}=\left(t,X,Y,Z\right),
\label{Ortsoperator_ABCD}
\end{equation}
wobei $t$ den Zeitparameter darstellt, während die Ortsoperatoren $X$, $Y$ und $Z$ gemäß
der Definition ($\ref{Ort_Impuls_Operatoren}$) im Tensorraum der Ur"=Alternativen wirken.
Mit Hilfe von ($\ref{Viererimpuls_Tensorraum}$) kann die Gleichung ($\ref{Energie-Impuls-Relation}$)
in der folgenden Weise geschrieben werden:

\begin{equation}
\left(P_{ABCD}\right)^\mu \left(P_{ABCD}\right)_\mu=0,\quad \mu=0,...,3,\quad
0\widehat{=}t, 1\widehat{=}x, 2\widehat{=}y, 3\widehat{=}z.
\label{Viererimpuls_Tensorraum_Gleichung}
\end{equation}
In Gleichung ($\ref{Viererimpuls_Tensorraum_Gleichung}$) und von nun an auch weiter in
dieser Arbeit wird die Einsteinsche Summenkonvention vorausgesetzt, also über doppelt
auftretende Indizes summiert.

\subsection{Die abstrakte Klein-Gordon-Gleichung im Tensorraum der Ur-Alternativen}

Um nun entsprechend ($\ref{Zustand_Darstellung_Ortsraum}$) auch die Zeitentwicklung
explizit in die Beschreibung und Darstellung der Zustände im Tensorraum der
Ur"=Alternativen hineinzubekommen und damit zur gesamten Raum"=Zeit zu gelangen,
muss man zunächst von der Schrödingergleichung in ihrer ganz abstrakten allgemeinen
Gestalt ausgehen, wie sie in ($\ref{Schroedinger_Gleichung_Alternative}$) formuliert
und in ($\ref{Schroedingergleichung_Tensorraum}$) in Bezug auf die Darstellung
einer beliebigen Alternative im Tensorraum der Ur"=Alternativen formuliert wurde.
Unter Beschränkung auf den symmetrischen Teilraum lautet sie wie folgt:

\begin{equation}
i\partial_t|\Psi\left(t\right)\rangle_S=H|\Psi\left(t\right)\rangle_S,
\label{Schroedingergleichung_Tensorraum_symmetrisch}
\end{equation}
wobei $|\Psi\rangle_S$ gemäß ($\ref{Basiszustand_Tensorraum_xyzn}$) definiert ist.
Einen allgemeinen zeitabhängigen gegenüber Permutation der Ur"=Alternativen symmetrischen
Zustand im Tensorraum der Ur"=Alternativen kann man demnach mit ($\ref{Zustand_Darstellung_Ortsraum}$)
in die Raum"=Zeit abbilden:

\begin{eqnarray}
\label{Zustand_Darstellung_Raum-Zeit}
&&|\Psi(t)\rangle_S=\sum_{N_{ABCD}}\psi\left(N_{ABCD},t\right)|N_{ABCD}\rangle
=\sum_{N_{xyzn}}\psi\left(N_{xyzn},t\right)|N_{xyzn}\rangle
\quad\longleftrightarrow\\
&&\sum_{N_x,N_y,N_z,N_n}\psi\left(N_x,N_y,N_z,N_n,t\right)f_N\left(N_x,N_y,N_z,x,y,z\right)
=\sum_{N_{xyzn}}\psi\left(N_{xyzn},t\right)f_{N_{xyzn}}(\mathbf{x})=\Psi_N\left(\mathbf{x},t\right).\nonumber
\end{eqnarray}
Wenn man den Operator auf der linken Seite der Gleichung ($\ref{Viererimpuls_Tensorraum_Gleichung}$)
auf einen beliebigen Zustand im symmetrischen Anteil des Tensorraumes der Ur"=Alternativen
anwendet, der in ($\ref{Zustand_gesamt}$) enthalten ist, dann erhält man das quantenlogische
Analogon zur gewöhnlichen Klein"=Gordon"=Gleichung im abstrakten Tensorraum der Ur"=Alternativen:

\begin{equation}
\left(P_{ABCD}\right)^\mu \left(P_{ABCD}\right)_\mu |\Psi\left(t\right)\rangle_S=0.
\label{Klein-Gordon-Gleichung}
\end{equation}
Und wenn man diese Gleichung nun desweiteren gemäß ($\ref{Energieoperator_Darstellung}$),
($\ref{Ort_Impuls_Darstellung}$) und ($\ref{Zustand_Darstellung_Raum-Zeit}$) in den
Ortsraum beziehungsweise die Raum"=Zeit abbildet, dann erhält man die gewöhnliche
Gestalt der Klein"=Gordon"=Gleichung, die aber hier lediglich einer Darstellung
der dahinter in Wirklichkeit stehenden abstrakten quantenlogischen Beziehungen
aus ($\ref{Klein-Gordon-Gleichung}$) entspricht:

\begin{equation}
\partial^\mu \partial_\mu \Psi_N\left(\mathbf{x},t\right)=0.
\label{Klein-Gordon-Gleichung-Raum-Zeit}
\end{equation}
Damit ist dann durch die Kombination von ($\ref{Schroedingergleichung_Tensorraum}$),
($\ref{Energieoperator_Darstellung}$), ($\ref{Energie-Impuls-Relation}$),
($\ref{Ort_Impuls_Darstellung}$) und ($\ref{Zustand_Darstellung_Ortsraum}$)
im Rahmen der Dynamik eine raum"=zeitliche Darstellung der abstrakten Operatoren
und Zustände des symmetrischen Anteiles des Tensorraumes der Ur"=Alternativen erreicht.

\subsection{Die abstrakte Dirac-Gleichung im Tensorraum der Ur-Alternativen}

Um die Klein"=Gordon"=Gleichung ($\ref{Klein-Gordon-Gleichung}$) als dynamische Grundgleichung
im Tensorraum der Ur"=Alternativen in die Gestalt der allgemeinen Schrödingergleichung 
($\ref{Schroedingergleichung_Tensorraum}$) zu bringen, ist wie im gewöhnlichen Falle 
\cite{Dirac:1928}, \cite{Bjorken:1966} eine Linearisierung notwendig,
die hier in einem quantenlogischen Rahmen in Analogie zum gewöhnlichen feldtheoretischen
Rahmen vollzogen wird. Hierzu muss von der folgenden Relation Gebrauch gemacht werden:

\begin{equation}
\gamma^\mu \gamma^\nu+\gamma^\nu \gamma^\mu=2\eta^{\mu\nu}\mathbf{1},
\label{Dirac-Matrizen-Relation}
\end{equation}
wobei $\mathbf{1}$ die Einheitsmatrix in vier Dimensionen, $\eta^{\mu\nu}$
die Minkowski"=Metrik und die $\gamma^\mu$ die Dirac"=Matrizen beschreiben,
welche folgende Gestalt aufweisen:

\begin{equation}
\gamma^0=\left(\begin{matrix} 0 & \sigma^0\\ \sigma^0 & 0\end{matrix}\right),\quad
\gamma^1=\left(\begin{matrix} 0 & -\sigma^1\\ \sigma^1 & 0\end{matrix}\right),\quad
\gamma^2=\left(\begin{matrix} 0 & -\sigma^2\\ \sigma^2 & 0\end{matrix}\right),\quad
\gamma^3=\left(\begin{matrix} 0 & -\sigma^3\\ \sigma^3 & 0\end{matrix}\right),
\label{Dirac-Matrizen}
\end{equation}
wobei $\sigma^0$ die Einheitsmatrix in zwei Dimensionen beschreibt und
$\sigma^1$, $\sigma^2$ und $\sigma^3$ die Pauli"=Matrizen sind:

\begin{equation}
\sigma^0=\left(\begin{matrix}1 & 0 \\ 0 & 1 \end{matrix}\right),\quad
\sigma^1=\left(\begin{matrix}0 & 1 \\ 1 & 0 \end{matrix}\right),\quad
\sigma^2=\left(\begin{matrix}0 & -i \\ i & 0 \end{matrix}\right),\quad
\sigma^3=\left(\begin{matrix}1 & 0 \\ 0 & -1 \end{matrix}\right).
\label{Pauli-Matrizen}
\end{equation}
Mit Hilfe der Relation ($\ref{Dirac-Matrizen-Relation}$) kann man den Operator
$\left(P_{ABCD}\right)^\mu \left(P_{ABCD}\right)_\mu$ in der folgenden
Weise ausdrücken:

\begin{equation}
\left(P_{ABCD}\right)^\mu \left(P_{ABCD}\right)_\mu \mathbf{1}=
\gamma^\mu \gamma^\nu \left(P_{ABCD}\right)_\mu \left(P_{ABCD}\right)_\nu.
\label{Dirac_Impuls_Relation}
\end{equation}
Um die abstrakte Klein"=Gordon"=Gleichung im Tensorraum der Ur"=Alternativen
($\ref{Klein-Gordon-Gleichung}$) umzuformulieren, muss der Zustand $|\Psi \rangle_S$
zunächst in der Weise erweitert werden, dass er als aus vier Komponenten bestehend
betrachtet wird. Zunächst gibt es eine Komponente zu positiver und eine zu negativer
Energie, $|\Psi_{E}\rangle_S$ und $|\Psi_{-E}\rangle_S$. Desweiteren muss mit
diesen beiden Komponenten noch jeweils mindestens eine einzelne Ur"=Alternative
$|\varphi_{E}\rangle$ beziehungsweise $|\varphi_{-E}\rangle$ über das Tensorprodukt
verbunden werden, wobei dieser Ur"=Alternative gemäß ($\ref{Ur-Alternative}$) ein
zweidimensionaler Spinor entspricht, welcher den Spin beschreibt, sodass sich unter
Einbeziehung der Zeitabhängigkeit ein erweiterter vierdimensionaler Spinor"=Zustand
$|\Psi_{\Gamma}\left(t\right)\rangle$ ergibt, wobei das $\Gamma$ die darin
enthaltenen zusätzlichen inneren Freiheitsgrade andeuten soll:

\begin{equation}
|\Psi_{\Gamma}\left(t\right)\rangle=
\left(\begin{matrix} |\Psi_{E}\left(t\right)\rangle_S
\otimes |\varphi_{E}\left(t\right)\rangle\\
|\Psi_{-E}\left(t\right)\rangle_S
\otimes |\varphi_{-E}\left(t\right)\rangle\end{matrix}\right).
\label{Dirac-Zustand}
\end{equation}
An dieser Stelle wird der zusätzliche Spin"=Freiheitsgrad zunächst künstlich eingeführt.
Es wird sich im nächsten Kapitel zeigen, dass er in Wirklichkeit zusammen mit weiteren
inneren Freiheitsgraden durch den antisymmetrischen Anteil im Tensorraum der
Ur"=Alternativen $|\Psi\rangle_{AS}$ geliefert wird. Genaugenommen handelt es sich
bei den zusätzlichen Freiheitsgraden in ($\ref{Dirac-Zustand}$) noch nicht wirklich
um innere Freiheitsgrade, denn die Aufspaltung in Komponenten zu positiver und negativer
Energie sowie der Spin stehen noch in einer Beziehung zur Raum"=Zeit. Es wird aber weiter
unten gezeigt werden, dass der antisymmetrische Teilraum des Tensorraumes der Ur"=Alternativen
nicht nur in natürlicher Weise den Spin"=Freiheitsgrad, sondern auch die wirklichen
inneren Symmetrien ohne jeden Bezug zum physikalischen Ortsraum liefert. Wenn man
nun ($\ref{Dirac_Impuls_Relation}$) in ($\ref{Klein-Gordon-Gleichung}$) verwendet
und auf den Zustand ($\ref{Dirac-Zustand}$) bezieht, dann ergibt sich die
abstrakte Dirac"=Gleichung im Tensorraum der Ur"=Alternativen:

\begin{equation}
\gamma^\mu \left(P_{ABCD}\right)_\mu |\Psi_{\Gamma}\left(t\right)\rangle=0.
\label{Dirac-Gleichung}
\end{equation}
In der Form der allgemeinen Schrödingergleichung ($\ref{Schroedingergleichung_Tensorraum}$)
geschrieben lautet sie:

\begin{equation}
i\partial_t |\Psi_{\Gamma}\left(t\right)\rangle=i\gamma^0 \gamma^j \left(P_{ABCD}\right)_j
|\Psi_{\Gamma}\left(t\right)\rangle=H_D |\Psi_{\Gamma}\left(t\right)\rangle,\quad j=1,...,3.
\label{Dirac-Schroedinger-Gleichung_Tensorraum}
\end{equation}
Dies bedeutet, dass der Dirac"=Hamilton"=Operator im Tensorraum der Ur"=Alternativen
die folgende Form aufweist:

\begin{equation}
H_D=i\gamma^0 \gamma^j \left(P_{ABCD}\right)_j,\quad j=1,...,3.
\label{Dirac-Hamilton-Operator}
\end{equation}
Und die zur in ($\ref{Dirac-Schroedinger-Gleichung_Tensorraum}$) gegebenen
Dirac"=Schrödinger"=Gleichung beziehungsweise zum entsprechenden in
($\ref{Dirac-Hamilton-Operator}$) gegebenen Dirac"=Hamilton"=Operator
im Tensorraum der Ur"=Alternativen gehörige Zeitentwicklung eines
Dirac"=Zustandes im Tensorraum ($\ref{Dirac-Zustand}$) wird
dementsprechend in der folgenden Weise beschrieben:

\begin{equation}
|\Psi_{\Gamma}\left(t\right)\rangle=\exp\left[-iH_D t\right]|\Psi_{\Gamma}\left(0\right)\rangle.
\end{equation}
Wenn man nun die abstrakte Dirac"=Gleichung im Tensorraum wie die abstrakte
Klein"=Gordon"=Gleichung mit Hilfe von ($\ref{Energieoperator_Darstellung}$),
($\ref{Ort_Impuls_Darstellung}$) und ($\ref{Zustand_Darstellung_Raum-Zeit}$) in die
Raum"=Zeit abbildet, dann ergibt sich die gewöhnliche Form der Dirac"=Gleichung,
wobei hinter dieser raum"=zeitlichen Darstellung die abstrakten quantenlogischen
Beziehungen aus ($\ref{Dirac-Gleichung}$) stehen:

\begin{equation}
i\gamma^\mu \partial_\mu \left(\Psi_\Gamma\right)_N\left(\mathbf{x},t\right)=0.
\end{equation}
In der Form der allgemeinen Schrödingergleichung ($\ref{Schroedingergleichung_Tensorraum}$)
geschrieben lautet sie:

\begin{equation}
i\partial_t\left(\Psi_\Gamma\right)_N\left(\mathbf{x},t\right)=\gamma^0 \gamma^j \partial_j
\left(\Psi_\Gamma\right)_N\left(\mathbf{x},t\right),\quad j=1,...,3.
\end{equation}

\subsection{Die Poincare-Symmetrie im Tensorraum der Ur-Alternativen}

Im Tensorraum der Ur"=Alternativen muss basierend auf den im symmetrischen Teilraum dieses
Raumes wirkenden Erzeugungs- und Vernichtungsoperatoren ($\ref{Operatoren_symmetrisch}$)
die Poincare"=Gruppe der gewöhnlichen Raum"=Zeit dargestellt werden können. Denn die Zustände
im Tensorraum der Ur"=Alternativen können ja gemäß ($\ref{Zustand_Darstellung_Raum-Zeit}$)
in die gewöhnliche Minkowski"=Raum"=Zeit abgebildet werden. Der einzige allerdings entscheidende
Unterschied besteht jedoch darin, dass sich die entsprechenden Transformationen auf die
diskreten Basiszustände der rein quantenlogischen Struktur beziehen, welche durch die
Ur"=Alternativen dargestellt wird. Die Poincare"=Gruppe besteht bekanntlich aus den
Raumtranslationen, der Zeittranslation, den räumlichen Drehungen und den eigentlichen
Lorentz"=Transformationen. Das rein quantenlogische Analogon zu den Raumtranslationen wird
einfach durch die aus Erzeugungs- und Vernichtungsoperatoren im Tensorraum der Ur"=Alternativen,
($\ref{Operatoren_symmetrisch}$) beziehungsweise ($\ref{ErzeugungsVernichtungsOperatorenXYZ}$),
gebildeten Impulsoperatoren beschrieben, wie sie in ($\ref{Ort_Impuls_Operatoren}$)
beziehungsweise ($\ref{Ort_Impuls_Operatoren_ABCD}$) definiert sind,
$P_x\left(A_S, B_S, C_S, D_S\right)$, $P_y\left(A_S, B_S, C_S, D_S\right)$
und $P_z\left(A_S, B_S, C_S, D_S\right)$, wobei hier die Abhängigkeit
von den Operatoren im Tensorraum ($\ref{Operatoren_symmetrisch}$) explizit
ausgedrückt wird. Die Zeittranslation wird durch den entsprechenden
Hamiltonoperator ($\ref{Dirac-Hamilton-Operator}$) beschrieben,
$\left(P_{ABCD}\right)_0=E_{ABCD}=H_{ABCD}=i\gamma^0 \gamma^j \left(P_{ABCD}\right)_j$,
wobei $\left(P_{ABCD}\right)_1=P_x\left(A_S,B_S,C_S,D_S\right)$,
$\left(P_{ABCD}\right)_2=P_y\left(A_S,B_S,C_S,D_S\right)$ und
$\left(P_{ABCD}\right)_3=P_z\left(A_S,B_S,C_S,D_S\right)$.
Diese Operatoren kommutieren alle miteinander:

\begin{equation}
\left[\left(P_{ABCD}\right)_i, \left(P_{ABCD}\right)_j\right]=0,\quad
\left[\left(P_{ABCD}\right)_i, H_{ABCD}\right]=0,\quad i,j=1,...,3.
\label{Kommutatoren_Energie-Impuls}
\end{equation}
Desweiteren müssen die Generatoren der quantenlogischen räumlichen Drehungen sowie
der quantenlogischen eigentlichen Lorentz"=Transformationen aus den Erzeugungs- und
Vernichtungsoperatoren im Tensorraum der Ur"=Alternativen ($\ref{Operatoren_symmetrisch}$)
konstruiert werden. Die quantenlogischen räumlichen Drehungen werden durch abstrakte
Drehimpulsoperatoren erzeugt, welche wiederum aus den Operatoren in ($\ref{Ort_Impuls_Operatoren}$)
und ($\ref{Ort_Impuls_Operatoren_ABCD}$) konstruiert werden können. Dazu gehören neben den
Komponenten des abstrakten Energie"=Impuls"=Operators $P_{ABCD}$ ($\ref{Viererimpuls_Tensorraum}$)
auch die des abstrakten Viererortsoperators $X_{ABCD}$ ($\ref{Ortsoperator_ABCD}$),
$\left(X_{ABCD}\right)_1=X\left(A_S,B_S,C_S,D_S\right)$, $\left(X_{ABCD}\right)_2=Y\left(A_S,B_S,C_S,D_S\right)$
und $\left(X_{ABCD}\right)_3=Z\left(A_S,B_S,C_S,D_S\right)$. Die Komponenten des quantenlogischen
Drehimpulsoperators lauten demnach:

\begin{eqnarray}
\left(L_{ABCD}\right)_1&=&\left(X_{ABCD}\right)_2 \left(P_{ABCD}\right)_3
-\left(X_{ABCD}\right)_3 \left(P_{ABCD}\right)_2,\nonumber\\
\left(L_{ABCD}\right)_2&=&\left(X_{ABCD}\right)_3 \left(P_{ABCD}\right)_1
-\left(X_{ABCD}\right)_1 \left(P_{ABCD}\right)_3,\nonumber\\
\left(L_{ABCD}\right)_3&=&\left(X_{ABCD}\right)_1 \left(P_{ABCD}\right)_2
-\left(X_{ABCD}\right)_2 \left(P_{ABCD}\right)_1.
\end{eqnarray}
Diese erfüllen die Vertauschungsrelationen:

\begin{equation}
\left[\left(L_{ABCD}\right)_i, \left(L_{ABCD}\right)_j\right]=\epsilon_{ijk} \left(L_{ABCD}\right)_k,\quad i,j=1,...,3,
\end{equation}
wobei $\epsilon_{ijk}$ der total antisymmetrische Tensor dritter Stufe ist. Die Generatoren
der eigentlichen Lorentz"=Gruppe spielen hier in gewissem Sinne eine Sonderrolle, weil die
Zeit als ein reeller Parameter auftritt und nicht aus Erzeugungs- und Vernichtungsoperatoren
konstruiert wird. Diesbezüglich ist entscheidend, dass der Energieoperator $E$ gemäß
der Schrödingergleichung ($\ref{Schroedingergleichung_Tensorraum}$) beziehungsweise
($\ref{Dirac-Schroedinger-Gleichung_Tensorraum}$) aufgrund seiner Gleichheit mit dem
Hamiltonoperator $H_{ABCD}$, $E=H_{ABCD}$, nicht nur Transformationen im Tensorraum
der Ur"=Alternativen generiert, sondern damit verknüpft auch Transformationen in
Bezug auf den Zeitparameter $t$ generiert, also Zeittranslationen, und deshalb gemäß
($\ref{Energieoperator_Darstellung}$) $E=-i\partial_t$ gilt. Die Generatoren der
eigentlichen Lorentz"=Transformationen können wie folgt ausgedrückt werden:

\begin{eqnarray}
\left(K_{ABCD}\right)_1&=&t \left(P_{ABCD}\right)_1+\left(X_{ABCD}\right)_1 E,\nonumber\\
\left(K_{ABCD}\right)_2&=&t \left(P_{ABCD}\right)_2+\left(X_{ABCD}\right)_2 E,\nonumber\\
\left(K_{ABCD}\right)_3&=&t \left(P_{ABCD}\right)_3+\left(X_{ABCD}\right)_3 E.
\end{eqnarray}
Sie erfüllen untereinander die folgenden Vertauschungsrelationen:

\begin{equation}
\left[\left(K_{ABCD}\right)_i, \left(K_{ABCD}\right)_j\right]=-i\epsilon_{ijk} \left(L_{ABCD}\right)_k,\quad i,j=1,...,3.
\end{equation}
Die Vertauschungsrelationen zwischen den Generatoren der Drehungen und den Generatoren
der eigentlichen Lorentz"=Transformationen lauten:

\begin{equation}
\left[\left(L_{ABCD}\right)_i,\left(K_{ABCD}\right)_j\right]=i\epsilon_{ijk} \left(K_{ABCD}\right)_k,\quad i,j=1,...,3.
\end{equation}
Schließlich müssen noch die Vertauschungsrelationen zwischen den Generatoren der Drehungen
beziehungsweise den Generatoren der quantenlogischen eigentlichen Lorentz"=Transformationen
und den Translationsoperatoren formuliert werden:

\begin{eqnarray}
\left[\left(L_{ABCD}\right)_i,\left(P_{ABCD}\right)_j\right]&=&i\epsilon_{ijk} \left(P_{ABCD}\right)_k,\quad
\left[\left(K_{ABCD}\right)_i,\left(P_{ABCD}\right)_j\right]=i\delta_{ij}E,\nonumber\\
\left[\left(L_{ABCD}\right)_i,E\right]&=&0,\quad
\left[\left(K_{ABCD}\right)_i,E\right]=\left(P_{ABCD}\right)_i,\quad i,j=1,...,3.
\end{eqnarray}
Wenn man nun die folgende Definition macht:

\begin{equation}
\left(M_{ABCD}\right)^{ij}=-\epsilon^{ijk} \left(L_{ABCD}\right)^k,\quad
\left(M_{ABCD}\right)^{0i}=\left(K_{ABCD}\right)^i,\quad i,j=1,...,3,
\label{Lorentz+Drehgruppe_Operatoren}
\end{equation}
dann kann man die Vertauschungsrelationen der Generatoren der Poincare"=Gruppe
im Tensorraum der Ur"=Alternativen wie folgt ausdrücken:

\begin{eqnarray}
\left[\left(M_{ABCD}\right)^{\mu\nu},\left(M_{ABCD}\right)^{\rho\sigma}\right]
&=&\eta^{\mu\rho}\left(M_{ABCD}\right)^{\nu\sigma}-\eta^{\mu\sigma}\left(M_{ABCD}\right)^{\nu\rho}\nonumber\\
&&-\eta^{\nu\rho}\left(M_{ABCD}\right)^{\mu\sigma}+\eta^{\nu\sigma}\left(M_{ABCD}\right)^{\mu\rho},\nonumber\\
\left[\left(M_{ABCD}\right)^{\mu\nu},\left(P_{ABCD}\right)^{\rho}\right]&=&
\eta^{\mu\rho}\left(P_{ABCD}\right)^{\nu}-\eta^{\nu\rho}\left(P_{ABCD}\right)^{\mu},\nonumber\\
\left[\left(P_{ABCD}\right)^{\mu},\left(P_{ABCD}\right)^{\nu}\right]&=&0,\quad \mu,\nu,\rho,\sigma=0,...,3.
\label{Poincare_Algebra}
\end{eqnarray}
Da man also in dieser Weise ($\ref{Poincare_Algebra}$) die Poincare"=Gruppe im Tensorraum
der Ur"=Alternativen darstellen kann und die Zustände der Quantenobjekte im Tensorraum
der Ur"=Alternativen ($\ref{Zustand_gesamt}$) sowie die dynamischen Grundgleichungen
dieser Objekte dieser Symmetriegruppe gehorchen, ist damit die Existenz von Elementarteilchen
im Sinne irreduzibler Darstellungen der Poincare"=Gruppe, wie sie Eugene Wigner definiert
\cite{Wigner:1939}, aus Ur"=Alternativen in der Zeit als basalster Darstellung der Natur
im menschlichen Geist hergeleitet.

\section{Die Eichsymmetrien der Elementarteilchenphysik}

\subsection{Der antisymmetrische Teilraum des Tensorraumes der Ur-Alternativen}

Im letzten Kapitel wurden in Zusammenhang mit der Formulierung der abstrakten
Dirac"=Gleichung im Tensorraum der Ur"=Alternativen ($\ref{Dirac-Gleichung}$),
($\ref{Dirac-Schroedinger-Gleichung_Tensorraum}$) durch die jeweilige Bildung
des Tensorproduktes der beiden Komponenten des Zustandes im symmetrischen Anteil
des Tensorraumes der Ur"=Alternativen zu positiver und zu negativer Energie mit
dem Zustand einer zusätzlichen Ur"=Alternative weitere Freiheitsgrade eingeführt,
welche durch den Index $\Gamma$ des entsprechenden Zustandes angedeutet werden
($\ref{Dirac-Zustand}$). In diesem Kapitel soll gezeigt werden, dass die zusätzlichen
Freiheitsgrade in Wirklichkeit durch den antisymmetrischen Anteil des Tensorraumes
der Ur"=Alternativen geliefert werden. Denn man kann neben den gegenüber Permutation
einzelner Ur"=Alternativen symmetrischen Zuständen mit Index $S$ nun auch die gegenüber
Permutation einzelner Ur"=Alternativen antisymmetrischen Zustände mit Index $AS$
näher betrachten und untersuchen. Die Operatoren, welche auf diese Zustände wirken,
sind die Erzeugungs- und Vernichtungsoperatoren aus ($\ref{Operatoren_antisymmetrisch}$),
welche durch die Antivertauschungsrelationen ($\ref{Operatoren_antisymmetrisch_Antikommutatoren}$)
definiert sind. Die antisymmetrischen Zustände sind natürlich implizit in ($\ref{Zustand_gesamt}$)
enthalten. Ein allgemeiner Zustand im antisymmetrischen Anteil des Tensorraumes der
Ur"=Alternativen, dessen Basiszustände gemäß ($\ref{Basiszustaende_antisymmetrisch}$)
mit $|\bar N_{ABCD}\rangle_{AS}=|\bar N_A,\bar N_B,\bar N_C,\bar N_D\rangle_{AS}$
bezeichnet seien, weist damit die folgende Gestalt auf:

\begin{equation}
|\Psi\rangle_{AS}=\sum_{\bar N_{ABCD}=0}^1 \psi_{AS}\left(\bar N_{ABCD}\right)|\bar N_{ABCD}\rangle_{AS},
\end{equation}
wobei die Summierung wie erwähnt natürlich deshalb jeweils nur von $0$ bis $1$ läuft, da sich aufgrund
des Paulischen Ausschließungsprinzips nur maximal eine einzige Ur"=Alternative in jedem der
Basiszustände einer einzelnen Ur"=Alternative befinden darf. \pagebreak Demnach gibt es gemäß der
Kombinatorik genau $16$ Basiszustände im antisymmetrischen Anteil des Tensorraumes der
Ur"=Alternativen, welche ausgeschrieben wie folgt lauten:

\begin{eqnarray}
|0,0,0,0\rangle_{AS},\quad |1,0,0,0\rangle_{AS},\quad |0,1,0,0\rangle_{AS},\quad |0,0,1,0\rangle_{AS},\nonumber\\
|0,0,0,1\rangle_{AS},\quad |1,1,0,0\rangle_{AS},\quad |1,0,1,0\rangle_{AS},\quad |1,0,0,1\rangle_{AS},\nonumber\\
|0,1,1,0\rangle_{AS},\quad |0,1,0,1\rangle_{AS},\quad |0,0,1,1\rangle_{AS},\quad |1,1,1,0\rangle_{AS},\nonumber\\
|1,1,0,1\rangle_{AS},\quad |1,0,1,1\rangle_{AS},\quad |0,1,1,1\rangle_{AS},\quad |1,1,1,1\rangle_{AS}.
\label{Basiszustaende_antisymmetrisch_A}
\end{eqnarray}
Die Menge der Basiszustände im antisymmetrischen Anteil des Tensorraumes der Ur"=Alternativen
entspricht also einer $16$"=fachen Alternative. Einem allgemeinen Zustand $|\Psi \rangle_{AS}$
entspricht demnach ein $16$"=dimensionaler reeller Vektor oder ein $8$"=dimensionaler komplexer
Spinor, da jede komplexe Dimension an sich schon einer binären Alternative entspricht.
Allerdings können die $16$ Basiszustände nach der Anzahl der in ihnen enthaltenen Ur"=Alternativen
klassifiziert werden. Sie enthalten entweder $0$, $1$, $2$, $3$ oder $4$ Ur"=Alternativen:

\begin{align}
&\textrm{0 Ur-Alternativen:}\ |0,0,0,0\rangle_{AS},\nonumber\\
&\textrm{1 Ur-Alternative:}\ |1,0,0,0 \rangle_{AS}, |0,1,0,0 \rangle_{AS},
|0,0,1,0 \rangle_{AS}, |0,0,0,1 \rangle_{AS},\nonumber\\
&\textrm{2 Ur-Alternativen:}\ |1,1,0,0\rangle_{AS}, |1,0,1,0\rangle_{AS}, |1,0,0,1\rangle_{AS},
|0,1,1,0\rangle_{AS}, |0,1,0,1\rangle_{AS}, |0,0,1,1\rangle_{AS},\nonumber\\
&\textrm{3 Ur-Alternativen:}\ |1,1,1,0\rangle_{AS}, |1,1,0,1\rangle_{AS}, |1,0,1,1\rangle_{AS},
|0,1,1,1\rangle_{AS},\nonumber\\
&\textrm{4 Ur-Alternativen:}\ |1,1,1,1\rangle_{AS}.
\label{Basiszustaende_antisymmetrisch_B}
\end{align}
Zunächst kann man damit einen beliebigen Zustand im antisymmetrischen Anteil des Tensorraumes
der Ur"=Alternativen in der folgenden Weise als reellen Vektor ausdrücken:

\begin{align}
|\Psi\rangle_{AS}=&\alpha_1 |0,0,0,0\rangle_{AS}+\alpha_2 |1,1,1,1\rangle_{AS}
+\alpha_3 |1,0,0,0 \rangle_{AS}+\alpha_4|0,1,0,0 \rangle_{AS}\nonumber\\
&+\alpha_5 |0,0,1,0 \rangle_{AS}+\alpha_6|0,0,0,1 \rangle_{AS}
+\alpha_7 |1,1,1,0\rangle_{AS}+\alpha_8|1,1,0,1\rangle_{AS}\nonumber\\
&+\alpha_9 |1,0,1,1\rangle_{AS}+\alpha_{10}|0,1,1,1\rangle_{AS}
+\alpha_{11} |1,1,0,0\rangle_{AS}+\alpha_{12} |1,0,1,0\rangle_{AS}\nonumber\\
&+\alpha_{13} |1,0,0,1\rangle_{AS}+\alpha_{14} |0,1,1,0\rangle_{AS}
+\alpha_{15} |0,1,0,1\rangle_{AS}+\alpha_{16} |0,0,1,1\rangle_{AS},
\label{Linearkombination_antisymmetrische_Basiszustaende}
\end{align}
wobei hier bewusst eine ganz spezifische Nummerierung und Anordnung gewählt wurde.
In ($\ref{Linearkombination_antisymmetrische_Basiszustaende}$) sind
zu den $16$ Basiszuständen gehörige reelle Koeffizienten enthalten,
die mit $\alpha_n$, $n=1,...,16$, bezeichnet sind, womit ein $16$"=dimensionaler
reeller Hilbertraum statuiert ist, dessen fundamentale Symmetriegruppe die
$SO(16)$ ist, zu welcher $16(16-1)/2=120$ Generatoren und damit $120$ reelle
Transformationsparameter gehören. Die $SO(16)$"=Gruppe ist isomorph zur
$SU(11)$"=Gruppe, welche $11^2-1=120$ Generatoren aufweist, also die gleiche
Anzahl, aber aufgrund des Körpers der komplexen Zahlen, über denen sie aufgebaut ist,
in Wirklichkeit die doppelte Anzahl an reellen Transformationsparametern enthält,
nämlich $240$. Man kann die $16$ Koeffizienten aber auch zu einem $8$"=dimensionalen
komplexen Spinor $|\chi_8\rangle$ zusammenfassen. Dies begründet wie gewöhnlich
eine $U(8)$"=Symmetriegruppe. Diese enthält $8^2=64$ Generatoren, weist aber
aufgrund des Körpers der komplexen Zahlen, über denen sie aufgebaut ist,
die doppelte Anzahl an reellen Transformationsparametern auf, nämlich $128$.
Die $16/2=8$ zusätzlichen Transformationsparameter beziehen sich gemäß
($\ref{Dimensionszahl_U(N/2)}$) und ($\ref{Dimensionszahl_SO(N)+T(N/2)}$) auf
die Transformationsparameter der Zeitentwicklungsgruppe als $8$"=dimensionaler
Darstellung der $U(1)$"=Gruppe, welche durch die Dynamik determiniert sind,
aber durch die Darstellung der $16$ Freiheitsgrade innerhalb eines
$8$"=dimensionalen komplexen Spinors zusätzlich mit in die
Beschreibung kommen. Das bedeutet, dass gilt:

\begin{equation}
|\Psi\rangle_{AS}\quad \widehat{=}\quad |\chi_8\rangle=\left(\begin{matrix}\alpha_1+\alpha_2 i\\ \alpha_3+\alpha_4 i\\
\alpha_5+\alpha_6 i\\ \alpha_7+\alpha_8 i\\
\alpha_9+\alpha_{10} i\\ \alpha_{11}+\alpha_{12} i\\
\alpha_{13}+\alpha_{14} i\\ \alpha_{15}+\alpha_{16} i\end{matrix}\right).
\label{chi8}
\end{equation}
Demnach kann ein Zustand im gesamten Tensorraum der Ur"=Alternativen ($\ref{Zustand_gesamt}$)
in der folgenden Weise dargestellt werden:

\begin{equation}
|\Psi\rangle=\left[\sum_{N_{ABCD}=0}^{N}\psi_S\left(N_{ABCD}\right)|N_{ABCD}\rangle_S\right] \otimes |\chi_8\rangle.
\label{Zustand_gesamt_chi8}
\end{equation}

\subsection{Die abstrakte Dirac-Gleichung mit inneren Quantenzahlen}

Wenn man nun von ($\ref{Zustand_gesamt_chi8}$) ausgehend einen zeitabhängigen
Zustand mit positiver und negativer Energiekomponente als Lösung der
Dirac"=Gleichung ($\ref{Dirac-Zustand}$) betrachtet, so erhält man:

\begin{equation}
|\Psi_{\Gamma}\left(t\right)\rangle=\left(\begin{matrix}|\Psi_{E}\left(t\right)\rangle_S
\otimes |\left(\chi_{8}\right)_{E}\left(t\right)\rangle\\
|\Psi_{-E}\left(t\right)\rangle_S \otimes |\left(\chi_{8}\right)_{-E}\left(t\right)\rangle
\end{matrix}\right).
\label{Dirac-Zustand_chi8}
\end{equation}
Der zusätzliche Dirac"=Spinor"=Freiheitsgrad, welcher den Spin beschreibt, wird durch den antisymmetrischen
Zustand geliefert und zusätzlich noch weitere interne Quantenzahlen, die als die Quantenzahlen mit den
damit verbundenen Eichsymmetrien des Standardmodells der Elementarteilchenphysik angesehen werden können,
wie unten gezeigt werden wird. Der Index $\Gamma$ in $|\Psi_{\Gamma}\left(t\right)\rangle$ deutet jetzt
also nicht nur die Aufspaltung in die beiden Komponenten zu positiver und negativer Energie und den Spin an,
sondern zugleich auch die in $|\chi_8\left(t\right)\rangle$ enthaltenen wirklichen inneren Freiheitsgrade,
die zu entsprechenden inneren Symmetrien führen. Dies wird weiter unten noch ausführlich erläutert werden.
Die abstrakte Dirac"=Gleichung im Tensorraum der Ur"=Alternativen ($\ref{Dirac-Gleichung}$) lautet damit:

\begin{align}
&\gamma^\mu_{8x8} \gamma^\nu_{8x8} \left(P_{ABCD}\right)_\mu \left(P_{ABCD}\right)_\nu |\Psi\left(t\right)\rangle=0
\quad \Leftrightarrow\quad
\gamma^\mu_{8x8} \left(P_{ABCD}\right)_\mu |\Psi\left(t\right)\rangle=0
\quad \Leftrightarrow\quad\nonumber\\
&\gamma^\mu_{8x8} \left(P_{ABCD}\right)_\mu \left[|\Psi\left(t\right)\rangle_S \otimes |\Psi\left(t\right) \rangle_{AS}\right]=0
\quad \Leftrightarrow\quad \gamma^\mu_{8x8} \left(P_{ABCD}\right)_\mu \left(\begin{matrix}
|\Psi_E\left(t\right)\rangle_S \otimes |\left(\chi_8\right)_E\left(t\right)\rangle\nonumber\\
|\Psi_{-E}\left(t\right)\rangle_S \otimes |\left(\chi_8\right)_{-E}\left(t\right)\rangle\end{matrix}\right)=0\\
&\Leftrightarrow\quad\gamma^\mu_{8x8} \left(P_{ABCD}\right)_\mu |\Psi_\Gamma (t)\rangle=0,
\label{Dirac-Gleichung_chi8}
\end{align}
wobei die speziellen Dirac"=Matrizen $\gamma_{8x8}$ definiert sind als:

\begin{equation}
\gamma^0_{8x8}=\left(\begin{matrix} 0 & \sigma^0_{8x8}\\ \sigma^0_{8x8} & 0\end{matrix}\right),\
\gamma^1_{8x8}=\left(\begin{matrix} 0 & -\sigma^1_{8x8}\\ \sigma^1_{8x8} & 0\end{matrix}\right),\
\gamma^2_{8x8}=\left(\begin{matrix} 0 & -\sigma^2_{8x8}\\ \sigma^2_{8x8} & 0\end{matrix}\right),\
\gamma^3_{8x8}=\left(\begin{matrix} 0 & -\sigma^3_{8x8}\\ \sigma^3_{8x8} & 0\end{matrix}\right),
\label{Dirac-Matrizen_8x8}
\end{equation}
und die darin enthaltenen Matrizen $\sigma_{8x8}$ achtdimensionale komplexe Matrizen sind, die auf
die achtdimensionalen komplexen Spinoren ($\ref{chi8}$) als Elemente des antisymmetrischen Teilraumes
des Tensorraumes der Ur"=Alternativen $\mathcal{H}_{TAS}$ wirken und in welche die gewöhnlichen
Pauli"=Matrizen ($\ref{Pauli-Matrizen}$) als Generatoren der $SU(2)$"=Symmetriegruppe in Bezug
auf den Spin einbettet sind, auf dessen Rolle in der Quantentheorie der Ur"=Alternativen weiter
unten noch näher eingegangen wird. Hierbei enthält $|\left(\chi_{8}\right)_E\left(t\right)\rangle$
beziehungsweise $|\left(\chi_{8}\right)_{-E}\left(t\right)\rangle$, um es noch einmal ausführlich
zu betonen, nicht nur den $SU(2)$"=Freiheitsgrad, welcher den Spin des Quantenobjektes
darstellt, und $U(1)$"=Freiheitsgrade, welche die Zeitentwicklung im antisymmetrischen
Teilraum darstellen, sondern es sind aufgrund der acht komplexen Dimensionen und der damit
verbundenen $U(8)$"=Symmetrie im antisymmetrischen Teilraum noch zusätzliche interne Freiheitsgrade
enthalten, die mit den internen Quantenzahlen der Elementarteilchenphysik und den entsprechenden
$U(N)$"=Symmetrien beziehungsweise $SU(N)$"=Symmetrien, wobei $N < 8$, in Verbindung gebracht
werden müssen. Einer Spin"=Transformation entspricht also eine $SU(2)$"=Transformation aller
Ur"=Alternativen des antisymmetrischen Teilraumes gegenüber dem symmetrischen Teilraum und den
internen Quantenzahlen eine relative Transformation einzelner Ur"=Alternativen des antisymmetrischen
Teilraumes in Bezug aufeinander. Jedoch ist eine direkte Kongruenz der $U(8)$"=Symmetrie mit einer
Kombination der Symmetrien geringerer Dimensionszahl der Elementarteilchenphysik zunächst nicht
unmittelbar erkennbar. Allerdings kann man diesen durch die $16$ Basiszustände
($\ref{Basiszustaende_antisymmetrisch_A}$), ($\ref{Basiszustaende_antisymmetrisch_B}$)
ausgezeichneten antisymmetrischen Anteil des Tensorraumes der Ur"=Alternativen, dessen
Zustände man demgemäß als achtdimensionale komplexe Spinoren darstellen kann, seinerseits
in Teilräume aufspalten. Hierbei erscheint eine Aufspaltung in eine Summe dreier Teilräume
als zweckmäßig, die bezeichnet seien als $T_{AS1}$, $T_{AS2}$ und $T_{AS3}$, sodass gilt:

\begin{equation}
T_{AS}=T_{AS1}\oplus T_{AS2}\oplus T_{AS3}.
\end{equation}
Der erste Teilraum $T_{AS1}$ soll aus dem vollkommen unbesetzten Zustand $|0,0,0,0\rangle_{AS}$
und dem vollbesetzten Zustand $|1,1,1,1\rangle_{AS}$ gebildet werden, sodass ein beliebiger
Zustand in diesem Raum wie folgt lautet:

\begin{equation}
|\Psi \rangle_{AS1}=\alpha_1 |0,0,0,0\rangle_{AS}+\alpha_2 |1,1,1,1\rangle_{AS}.
\end{equation}
Die Kombination dieser beiden Basiszustände kann dadurch gerechtfertigt werden,
dass diese Zustände von außen in Bezug auf den symmetrischen Tensorraum gleich
aussehen, sich also nur innerhalb des antisymmetrischen Teilraumes überhaupt
unterscheiden lassen. Eine gleichzeitige Drehung aller Ur-Alternativen des
antisymmetrischen Teilraumes, die einer Spin"=Transformation entspricht, ändert diese
Zustände also nicht. Ein solcher Zustand kann daher mit seinen zwei reellen Freiheitsgraden
als ein komplexer Skalar $\chi$ dargestellt werden, der eine $U(1)$"=Symmetrie aufweist:

\begin{equation}
|\Psi \rangle_{AS1}\quad \widehat{=}\quad |\chi\rangle=\alpha_1+\alpha_2 i.
\label{chi}
\end{equation}
Der zweite Teilraum $T_{AS2}$ soll aus den vier Zuständen gebildet werden, welche eine einzige
Ur"=Alternative enthalten, und aus den vier Zuständen, welche drei Ur"=Alternativen enthalten,
sodass ein beliebiger Zustand in diesem Raum lautet:

\begin{eqnarray}
|\Psi \rangle_{AS2}&=&\alpha_3 |1,0,0,0 \rangle_{AS}+\alpha_4 |0,1,0,0 \rangle_{AS}
+\alpha_5 |0,0,1,0 \rangle_{AS}+\alpha_6 |0,0,0,1 \rangle_{AS}\nonumber\\
&&+\alpha_7 |1,1,1,0\rangle_{AS}+\alpha_8 |1,1,0,1\rangle_{AS}
+\alpha_9 |1,0,1,1\rangle_{AS}+\alpha_{10}|0,1,1,1\rangle_{AS}.
\end{eqnarray}
Die Kombination dieser Basiszustände ist deshalb gerechtfertigt, weil sich durch Drehung aller
Ur"=Alternativen des antisymmetrischen Teilraumes, die einer Spin"=Transformation entspricht,
die Ausrichtung in Bezug auf den symmetrischen Teilraum in der gleichen Weise ändert, die
jeweils durch den einen besetzten beziehungsweise unbesetzten Basiszustand einer einzelnen
Ur"=Alternative definiert ist. Ein solcher Zustand kann mit seinen acht reellen Freiheitsgraden
als ein vierdimensionaler komplexer Spinor $\chi_4$ dargestellt werden, der eine
$U(4)$"=Symmetrie aufweist:

\begin{equation}
|\Psi \rangle_{AS2}\quad \widehat{=}\quad |\chi_4\rangle=\left(\begin{matrix}\alpha_3+\alpha_4 i\\ \alpha_5+\alpha_6 i\\
\alpha_7+\alpha_8 i\\ \alpha_9+\alpha_{10} i\end{matrix}\right).
\label{chi4}
\end{equation}
Der dritte Teilraum $T_{AS3}$ soll aus den sechs Zuständen gebildet werden, welche zwei
Ur"=Alternativen enthalten, sodass ein beliebiger Zustand in diesem Raum lautet:

\begin{eqnarray}
|\Psi \rangle_{AS3}&=&\alpha_{11}|1,1,0,0\rangle_{AS}+\alpha_{12}|1,0,1,0\rangle_{AS}+\alpha_{13}|1,0,0,1\rangle_{AS}\nonumber\\
&&+\alpha_{14}|0,1,1,0\rangle_{AS}+\alpha_{15}|0,1,0,1\rangle_{AS}+\alpha_{16}|0,0,1,1\rangle_{AS}.
\end{eqnarray}
Die Kombination dieser Basiszustände ist deshalb gerechtfertigt, weil die Ausrichtung aller
dieser Zustände relativ zum symmetrischen Tensorraum jeweils durch die Besetzung zweier
Basiszustände einer einzelnen Ur"=Alternative definiert ist. Ein solcher Zustand kann mit
seinen sechs reellen Freiheitsgraden als ein dreidimensionaler komplexer Spinor $\chi_3$
dargestellt werden, der eine $U(3)$"=Symmetrie aufweist:

\begin{equation}
|\Psi \rangle_{AS3}\quad \widehat{=}\quad |\chi_3\rangle=\left(\begin{matrix}\alpha_{11}+\alpha_{12} i\\
\alpha_{13}+\alpha_{14} i\\ \alpha_{15}+\alpha_{16} i\end{matrix}\right).
\label{chi3}
\end{equation}
Damit kann also ein beliebiger Zustand im antisymmetrischen Anteil des Tensorraumes
der Ur"=Alternativen in der folgenden Weise ausgedrückt werden:

\begin{equation}
|\Psi\rangle_{AS}=|\chi\rangle \oplus |\chi_4\rangle \oplus |\chi_3\rangle.
\label{Zustand_antisymmetrisch_chi-chi4-chi3}
\end{equation}
Wenn man nun den Zustand als Lösung der abstrakten Dirac"=Gleichung im Tensorraum der
Ur"=Alternativen ($\ref{Dirac-Zustand_chi8}$) mit Hilfe von ($\ref{Zustand_antisymmetrisch_chi-chi4-chi3}$)
umschreibt, so lautet er wie folgt:

\begin{equation}
|\Psi_{\Gamma}\left(t\right)\rangle=\left(\begin{matrix}|\Psi_{E}\left(t\right)\rangle_S
\otimes \left[|\chi_E\left(t\right)\rangle\oplus
|\left(\chi_{4}\right)_{E}\left(t\right)\rangle\oplus |\left(\chi_{3}\right)_{E}\left(t\right)\rangle\right]\\
|\Psi_{-E}\left(t\right)\rangle_S \otimes \left[|\chi_{-E}\left(t\right)\rangle
\oplus |\left(\chi_{4}\right)_{-E}\left(t\right)\rangle\oplus |\left(\chi_{3}\right)_{-E}\left(t\right)\rangle\right]
\end{matrix}\right).
\label{Dirac-Zustand_chi-chi4-chi3}
\end{equation}
Dies wiederum kann man in der folgenden Weise umformen zu:

\begin{align}
|\Psi_{\Gamma}\left(t\right)\rangle=&\left(\begin{matrix}|\Psi_{E}\left(t\right)\rangle_S
\otimes |\chi_E\left(t\right)\rangle\\
|\Psi_{-E}\left(t\right)\rangle_S \otimes |\chi_{-E}\left(t\right)\rangle
\end{matrix}\right)
\oplus
\left(\begin{matrix}|\Psi_{E}\left(t\right)\rangle_S
\otimes |\left(\chi_{4}\right)_{E}\left(t\right)\rangle\\
|\Psi_{-E}\left(t\right)\rangle_S \otimes |\left(\chi_{4}\right)_{-E}\left(t\right)\rangle
\end{matrix}\right)\nonumber\\
&\oplus \left(\begin{matrix}|\Psi_{E}\left(t\right)\rangle_S
\otimes |\left(\chi_{3}\right)_{E}\left(t\right)\rangle\\
|\Psi_{-E}\left(t\right)\rangle_S \otimes |\left(\chi_{3}\right)_{-E}\left(t\right)\rangle
\end{matrix}\right).
\label{Dirac-Zustand_chi-chi4-chi3_getrennt}
\end{align}
Wenn man nun den symmetrischen Anteil dieses allgemeinen Tensorraumzustandes
mit Hilfe von ($\ref{Zustand_Darstellung_Raum-Zeit}$) in die Raum"=Zeit abbildet,
so erhält man die folgende Darstellung eines allgemeinen Zustandes im
Tensorraum der Ur"=Alternativen:

\begin{align}
|\Psi_{\Gamma}\left(t\right)\rangle \quad\longleftrightarrow\quad
\left(\Psi_{\Gamma}\right)_N\left(\mathbf{x},t\right)
=&\left(\begin{matrix}\left[\Psi_N\left(\mathbf{x},t\right)\right]_E
\otimes |\chi_E\left(t\right)\rangle\\
\left[\Psi_N\left(\mathbf{x},t\right)\right]_{-E} \otimes |\chi_{-E}\left(t\right)\rangle
\end{matrix}\right)
\oplus
\left(\begin{matrix}\left[\Psi_N\left(\mathbf{x},t\right)\right]_E
\otimes |\left(\chi_{4}\right)_{E}\left(t\right)\rangle\\
\left[\Psi_N\left(\mathbf{x},t\right)\right]_{-E} \otimes |\left(\chi_{4}\right)_{-E}\left(t\right)\rangle
\end{matrix}\right)\nonumber\\
&\oplus \left(\begin{matrix}\left[\Psi_N\left(\mathbf{x},t\right)\right]_E
\otimes |\left(\chi_{3}\right)_{E}\left(t\right)\rangle \\
\left[\Psi_N\left(\mathbf{x},t\right)\right]_{-E} \otimes |\left(\chi_{3}\right)_{-E}\left(t\right)\rangle
\end{matrix}\right).
\label{Dirac-Zustand_chi-chi4-chi3_getrennt_Ortsraum}
\end{align}
Damit bestehen also drei Sektoren eines allgemeinen Zustandes im Tensorraum der Ur"=Alternativen.
Da $|\chi\rangle$ einfach ein komplexer Skalar ist, entspricht der erste Sektor einem
skalaren Quantenobjekt. Dies könnte demnach als das Higgs"=Teilchen zu interpretieren
sein oder zumindest etwas damit zu tun haben, auch wenn der Freiheitsgrad des Isospin
beziehungsweise der Hyperladung darin zunächst nicht enthalten zu sein scheint.
Die darin enthaltene $U(1)$"=Gruppe hat vermutlich mit der Zeitentwicklung zu tun.

\subsection{Die inneren Symmetrien im Tensorraum der Ur-Alternativen}

Man kann nun weiter die komplexere mathematische Struktur des Spinors $|\chi_4\rangle$
betrachten. Die Symmetriegruppe dieser vierfachen komplexen quantentheoretischen
Alternative ist die \mbox{$U(4)=SU(4)\otimes U(1)$}"=Gruppe. Diese Symmetriegruppe
kann in vier unabhängige \mbox{$U(2)=SU(2)\otimes U(1)$}"=Gruppen aufgespalten werden:

\begin{equation}
U(4)\longleftrightarrow U(2)\otimes U(2)\otimes U(2)\otimes U(2),
\end{equation}
beziehungsweise in die Relation:

\begin{equation}
U(4)\longleftrightarrow SU(4)\otimes U(1)\longleftrightarrow SU(2)\otimes U(1) \otimes SU(2) \otimes U(1)
\otimes SU(2) \otimes U(1)\otimes SU(2)\otimes U(1).
\label{Relation_U4_A}
\end{equation}
Die $SU(3)$"=Gruppe besitzt acht Generatoren und kann daher durch die Kombination zweier
$U(2)=SU(2)\otimes U(1)$"=Gruppen dargestellt werden. Dies wurde in Zusammenhang mit
der Quantentheorie der Ur"=Alternativen erstmals in \cite{Goernitz:2016} verwendet.
Die $SU(3)$"=Struktur wird damit also durch zwei überlagerte Ur"=Alternativen repräsentiert:

\begin{equation}
SU(3) \longleftrightarrow SU(2)\otimes U(1)\otimes SU(2)\otimes U(1).
\label{Relation_SU3}
\end{equation}
In $\cite{Kober:2017}$ wird die $G_2$"=Gruppe, die Automorphismus"=Gruppe der Oktonionen,
auf dem Raum einer vierfachen quantentheoretischen Alternative dargestellt. Diese lässt
sich aufspalten in:

\begin{equation}
G_2 \longleftrightarrow SU(3)\otimes SU(2)\otimes SU(2),
\end{equation}
was zum Beispiel aus den Betrachtungen in \cite{Guenaydin:1973}, \cite{Guenaydin:1978}, \cite{LeBlanc:1988},
\cite{Bincer:1993}, \cite{Donaldson:2007} folgt. Es fehlen diesbezüglich noch zwei $U(1)$"=Gruppen,
um zur vollen $U(4)$"=Gruppe zu gelangen:

\begin{equation}
U(4) \longleftrightarrow G_2 \otimes U(1) \otimes U(1)
\longleftrightarrow SU(3) \otimes SU(2)\otimes SU(2)\otimes U(1)\otimes U(1).
\label{Relation_U4_B}
\end{equation}
Wenn man nun die Relation ($\ref{Relation_SU3}$) in ($\ref{Relation_U4_B}$) verwendet,
so erhält man genau die Relation ($\ref{Relation_U4_A}$). In $\cite{Kober:2017}$ wird
die $SU(3)$"=Symmetrie als Farbfreiheitsgrad gedeutet, eine der beiden $SU(2)$"=Symmetrien
als schwacher Isospin und die andere der beiden $SU(2)$"=Symmetrien als gewöhnlicher Spin.
Hier muss nun zusätzlich die eine der beiden verbleibenden $U(1)$"=Gruppen als Hyperladung
und die andere der beiden verbleibenden $U(1)$"=Gruppen einfach als eine Untergruppe der
Zeitentwicklungsgruppe in Bezug auf die Alternative interpretiert werden. Einer der
$SU(2)$"=Freiheitsgrade muss die relative Ausrichtung der Ur"=Alternativen, die in der vierfachen
Alternative implizit enthalten sind, zu den Ur"=Alternativen im symmetrischen Teil des Tensorraumes
der Ur"=Alternativen und auch zu den Ur"=Alternativen aller anderen Objekte des Kosmos beschreiben,
damit aber zu allen anderen Ur"=Alternativen des gesamten Kosmos. Die anderen
Freiheitsgrade beschreiben währenddessen eine relative Ausrichtung zu diesem
einen Freiheitsgrad und stellen deshalb innere Symmetrien dar, die keinen direkten Bezug
zu den raum"=zeitlichen Symmetrien aufweisen. Die relative Ausrichtung zu allen anderen
Ur"=Alternativen des Kosmos muss bezüglich der entsprechenden $SU(2)$"=Gruppe daher den
Spin beschreiben. Dies ist der Grund, warum es den Spin als eine Art Mischung aus
innerer und äußerer Symmetrie einerseits gibt, der von räumlichen Drehungen
beeinflusst wird, und die wirklichen inneren Symmetrien andererseits, die nicht
von räumlichen Transformationen beeinflusst werden. Mindestens eine $U(1)$"=Gruppe muss
gemäß der Schrödingergleichung immer mit der Zeitentwicklung zu tun haben, sodass die
verbleibende $SU(4)$ also als fundamentale Symmetriegruppe dieses Sektors angesehen
werden kann, aus der sich der Spin mit $SU(2)$"=Symmetrie ebenso wie die internen
Symmetrien ergeben, auf denen die Eichtheorien der Wechselwirkungen der gewöhnlichen
Elementarteilchenphysik basieren, also $SU(3)\otimes SU(2)\otimes U(1)$.
Die Idee, die hier zugrunde gelegt wird, nämlich dass der Spin die relative Ausrichtung
der Ur"=Alternativen eines an den symmetrischen Hilbertraum vieler Ur"=Alternativen,
der mit Hilfe von ($\ref{Zustand_Darstellung_Ortsraum}$) in den reellen dreidimensionalen
Ortsraum abgebildet wird, multiplizierten zusätzlichen antisymmetrischen Hilbertraumes
beschreibt, während die rein inneren Quantenzahlen relative Ausrichtungen der Ur"=Alternativen
dieses zusätzlichen antisymmetrischen Hilbertraumes in Bezug aufeinander beschreiben,
wurde bereits in \cite{Kober:2009B} in Betracht gezogen und auch in \cite{Kober:2017}
verwendet. In dieser Arbeit wird diese Idee nun mit der Aufspaltung des Tensorraumes
der Ur"=Alternativen in einen gegenüber Permutation der Ur"=Alternativen symmetrischen
und antisymmetrischen Anteil verbunden, wobei der antisymmetrische Anteil genau als
der an den symmetrischen Hilbertraum des Tensorraumes multiplizierte zusätzliche Teilraum
interpretiert wird, welcher die Quantenzahlen der Elementarteilchenphysik repräsentiert.
Dass die räumlichen Freiheitsgrade, denen der symmetrische Anteil des Tensorraumes
der Ur"=Alternativen zugrunde liegt, im Gegensatz zu den Freiheitsgraden der Quantenzahlen,
denen der antisymmetrische Anteil des Tensorraumes der Ur"=Alternativen zugrunde liegt,
kontinuierlich wirken, liegt daran, dass der symmetrische Anteil des Tensorraumes der
Ur"=Alternativen viel mehr Ur"=Alternativen fassen kann, im Prinzip sogar beliebig viele,
sodass die Diskretheit aufgrund der ungeheuren Zahl der Zustände sich irgendwann näherungsweise
wie ein Kontinuum verhält. Zudem können die diskreten Zustände des symmetrischen Teilraumes
des Tensorraumes der Ur"=Alternativen gemäß ($\ref{Zustand_Darstellung_Ortsraum}$) als
kontinuierliche Wellenfunktionen in einem kontinuierlichen reellen dreidimensionalen
Raum dargestellt werden und sind normierbar.

Um explizit zu zeigen, dass die $SU(3)$"=Gruppe tatsächlich durch zwei miteinander
kombinierte $U(2)=SU(2)\otimes U(1)$"=Gruppen dargestellt werden kann, ist es sinnvoll,
die Generatoren der $SU(3)$"=Symmetriegruppe in modifizierter Weise darzustellen.
Sie werden gewöhnlich durch die Gell"=Mann"=Matrizen dargestellt, die folgende
Gestalt aufweisen:

\begin{eqnarray}
\tau_1&=&\frac{1}{2}\left(\begin{matrix}0 & 1 & 0\\ 1 & 0 & 0\\ 0 & 0 & 0 \end{matrix}\right),\
\tau_2=\frac{1}{2}\left(\begin{matrix}0 & -i & 0\\ i & 0 & 0\\ 0 & 0 & 0 \end{matrix}\right),\
\tau_3=\frac{1}{2}\left(\begin{matrix}1 & 0 & 0\\ 0 & 1 & 0\\ 0 & 0 & 0 \end{matrix}\right),\
\tau_4=\frac{1}{2}\left(\begin{matrix}0 & 0 & 1\\ 0 & 0 & 0\\ 1 & 0 & 0 \end{matrix}\right),\\
\tau_5&=&\frac{1}{2}\left(\begin{matrix}0 & 0 & -i\\ 0 & 0 & 0\\ i & 0 & 0 \end{matrix}\right),\
\tau_6=\frac{1}{2}\left(\begin{matrix}0 & 0 & 0\\ 0 & 0 & 1\\ 0 & 1 & 0 \end{matrix}\right),\
\tau_7=\frac{1}{2}\left(\begin{matrix}0 & 0 & 0\\ 0 & 0 & -i\\ 0 & i & 0 \end{matrix}\right),\
\tau_8=\frac{1}{2\sqrt{3}}\left(\begin{matrix}1 & 0 & 0\\ 0 & 1 & 0\\ 0 & 0 & -2 \end{matrix}\right).\nonumber\\
\nonumber
\end{eqnarray}
In der folgenden Darstellung kann man explizit erkennen, dass sich die $SU(3)$"=Symmetriegruppe
in zwei $SU(2) \otimes U(1)$"=Symmetriegruppen aufspalten lässt:

\begin{eqnarray}
\tau_1&=&\frac{1}{2}\left(\begin{matrix}0 & 1 & 0\\ 1 & 0 & 0\\ 0 & 0 & 0 \end{matrix}\right),\
\tau_2=\frac{1}{2}\left(\begin{matrix}0 & -i & 0\\ i & 0 & 0\\ 0 & 0 & 0 \end{matrix}\right),\
\tau_3=\frac{1}{2}\left(\begin{matrix}1 & 0 & 0\\ 0 & -1 & 0\\ 0 & 0 & 0 \end{matrix}\right),\
\tau_4=\frac{1}{2}\left(\begin{matrix}1 & 0 & 0\\ 0 & 1 & 0\\ 0 & 0 & 0 \end{matrix}\right),\nonumber\\
\tau_5&=&\frac{1}{2}\left(\begin{matrix}0 & 0 & 0\\ 0 & 0 & 1\\ 0 & 1 & 0 \end{matrix}\right),\
\tau_6=\frac{1}{2}\left(\begin{matrix}0 & 0 & 0\\ 0 & 0 & -i\\ 0 & i & 0 \end{matrix}\right),\
\tau_7=\frac{1}{2}\left(\begin{matrix}0 & 0 & 0\\ 0 & 1 & 0\\ 0 & 0 & -1 \end{matrix}\right),\
\tau_8=\frac{1}{2}\left(\begin{matrix}0 & 0 & 0\\ 0 & 1 & 0\\ 0 & 0 & 1 \end{matrix}\right).\nonumber\\
\label{Darstellung_SU3_neu}
\end{eqnarray}
Hierbei beschreiben in beiden Reihen jeweils die ersten drei Generatoren eine $SU(2)$"=Symmetriegruppe
und der letzte Generator eine $U(1)$"=Symmetriegruppe. Eine beliebige Symmetrietransformation der
$SU(3)$"=Symmetriegruppe kann demnach wie folgt formuliert werden:

\begin{equation}
U\left[\alpha\right]_{SU(3)}=\exp\left(i\sum_{a=1,2,3}\alpha_a \tau_a\right)\otimes\exp\left(i\alpha_4 \tau_4\right)
\otimes\exp\left(i\sum_{a=5,6,7}\alpha_a \tau_a\right)\otimes\exp\left(i\alpha_8 \tau_8\right).
\end{equation}
Da nur die Quarks nicht aber die Leptonen einen internen $SU(3)$"=Farbfreiheitsgrad aufweisen,
liegt es sehr nahe, diesen mit dem Spinor $|\chi_4\rangle$ verbundenen zweiten Sektor eines
allgemeinen Zustandes im Tensorraum der Ur"=Alternativen mit dem Quarksektor des
Standardmodells der Elementarteilchenphysik zu identifizieren.

Es bleibt schließlich der dritte Sektor zu analysieren, der mit dem Spinor $|\chi_3\rangle$
in Zusammenhang steht. Diese dreifache komplexe quantentheoretische Alternative weist die
Symmetriegruppe \mbox{$U(3)=SU(3)\otimes U(1)$} auf, die sich gemäß ($\ref{Relation_SU3}$)
und ($\ref{Darstellung_SU3_neu}$) aufspalten lässt:

\begin{equation}
U(3) \longleftrightarrow SU(3)\otimes U(1) \longleftrightarrow SU(2)\otimes U(1)\otimes SU(2)\otimes U(1)\otimes U(1).
\label{Relation_U3}
\end{equation}
Analog zum Fall $|\chi_4\rangle$ muss eine $SU(2)$"=Gruppe als die relative Ausrichtung der
Zustände des antisymmetrischen Anteiles des Tensorraumes der Ur"=Alternativen zum symmetrischen
Anteil des Tensorraumes der Ur"=Alternativen interpretiert werden. Und da der symmetrische Anteil
in die Raum"=Zeit abgebildet werden kann, muss diese $SU(2)$"=Symmetrie etwas mit einer Ausrichtung
in Bezug auf den Raum zu tun haben und kann daher als Spin interpretiert werden. Wie im Fall
$|\chi_4\rangle$ kann die weitere $SU(2)$"=Gruppe mit dem schwachen Isospin und eine $U(1)$"=Gruppe
mit der Hyperladung identifiziert werden. Die beiden anderen $U(1)$"=Gruppen müssen erneut jeweils
als Untergruppe der Zeitentwicklungsgruppe identifiziert werden. Der einzige Unterschied zum
Sektor bezüglich $|\chi_4\rangle$ besteht also im nicht"=Bestehen eines darüber hinaus existierenden
$SU(3)$"=Farbfreiheitsgrades und in einem stattdessen bestehenden zusätzlichen $U(1)$"=Freiheitsgrad.
Da ein darüber hinaus existierender $SU(3)$"=Farbfreiheitsgrad im $|\chi_3\rangle$"=Sektor also
nicht besteht, aber sowohl Spin als auch schwacher Isospin und Hyperladung vorhanden sind,
liegt es nahe, diesen Sektor mit dem Leptonensektor der Elementarteilchenphysik zu identifizieren.
Die Tatsache, dass hier neben der $U(1)$"=Hyperladung anstatt einer nun zwei $U(1)$"=Untergruppen der
Zeitentwicklungsgruppe bestehen, kann einstweilen ebensowenig in eindeutiger Weise gedeutet werden
wie die Tatsache, dass insgesamt, also im gesamten antisymmetrischen Teilraum, vier zusätzliche
$U(1)$"=Gruppen auftauchen, die als Untergruppen der Zeitentwicklungsgruppe gedeutet werden.
Die anderen vier $U(1)$"=Freiheitsgrade, die notwendig sind, um auf die notwendige Dimension
von acht unabhängigen Zeitentwicklungsfreiheitsgraden zu gelangen, könnte mit Transformationen
zwischen den Sektoren zu tun haben. Aber es ist einstweilen überhaupt noch nicht klar,
ob diese Aufspaltung in dieser Weise überhaupt richtig ist, zumal das Higgsteilchen
ja auch einen schwachen Isospin aufweisen müsste. Aber, dass der antisymmetrische
Anteil des Tensorraumes der Ur"=Alternativen irgendwie mit den inneren $SU(N)$"=Eichsymmetrien
zu tun haben muss, das erscheint mir nach dieser Betrachtung als nahezu sicher.
Die Flavour"=Symmetrie kann an dieser Stelle natürlich überhaupt noch gar nicht auftauchen,
weil diese sich ja auf eine Unterscheidung unterschiedlicher Massen bezieht. Da man die Masse
aber als Konsequenz einer Wechselwirkung ansieht, im Standardmodell der Elementarteilchenphysik
einer Wechselwirkung mit dem Higgs"=Feld, und in der Heisenbergschen nichtlinearen
Spinorfeldtheorie als Konsequenz der Selbstwechselwirkung des fundamentalen Urfeldes,
können die mit der Flavour"=Symmetrie verbundenen Freiheitsgrade erst durch unterschiedliche
Massenzustände mit in die Beschreibung gelangen. Es sollte also in keiner Weise irritieren,
dass diese Freiheitsgrade ohne die Beschreibung der Wechselwirkung noch nicht auftauchen.

\section{Systeme mit mehreren Quantenobjekten}

\subsection{Zustände vieler freier Quantenobjekte}

Wenn man nun aus mehreren Quantenobjekten zusammengesetzte Systeme beschreiben möchte, so müssen
Tensorprodukte mehrerer Zustände $|\Psi\rangle$ im Tensorraum der Ur"=Alternativen $\mathcal{H}_T$
betrachtet werden. Diese Produktzustände $|\Phi\rangle$ in dem entsprechenden Gesamt"=Hilbertraum
$\mathcal{H}_G$ hängen von den Besetzungszahlen der Einzelzustände ab und sind gegeben durch:

\begin{equation}
|\Phi\left(N_{ABCD}^1,N_{ABCD}^2,...,N_{ABCD}^M\right)\rangle \in \mathcal{H}_G.
\end{equation}
Da die Zustandsräume der aus den Ur"=Alternativen gebildeten Quantenobjekte, die jeweils dem Tensorraum
vieler Ur"=Alternativen mit gegenüber Permutation symmetrischem und antisymmetrischem Anteil entsprechen,
isomorph zueinander sind, müssen auch solche Quantenobjekte als ununterscheidbar voneinander angesehen
werden. Die Basiszustände des entsprechenden Hilbertraumes vieler Quantenobjekte $\mathcal{H}_G$ sind
demnach in Bezug auf gegenüber Permutation der einzelnen Ur"=Alternativen symmetrische Zustände durch
die Anzahl $\mathcal{N}\left(N_{ABCD}\right)$ der Quantenobjekte gekennzeichnet, die sich in den
Basiszuständen $|N_{ABCD}\rangle$ des $\mathcal{H}_{TS}$ befinden, die wiederum durch eine fest
definierte Zahl an Ur"=Alternativen $N_A$, $N_B$, $N_C$, $N_D$ in den Basiszuständen einzelner
Ur"=Alternativen gekennzeichnet sind. Dem entspricht eine weitere Quantisierung, bei welcher
die Koeffizienten der Basiszustände bezüglich des entsprechenden Tensorraumzustandes zu
Erzeugungs- und Vernichtungsoperatoren werden, welche Quantenobjekte mit vielen Ur"=Alternativen
erzeugen und vernichten. Auch hier können bei der Bildung der Tensorprodukte der verschiedenen
Tensorraumzustände bezüglich Permutation symmetrische und antisymmetrische Tensorprodukte
betrachtet werden. Im symmetrischen Falle gilt für einen Zustand,
der aus $M$ Quantenobjekten besteht:

\begin{align}
&\mathcal{P}_{mn}|\Phi\left(...,N_{ABCD}^m,...,N_{ABCD}^n,...\right)\rangle_S
=|\Phi\left(...,N_{ABCD}^n,...,N_{ABCD}^m,...\right)\rangle_S,\nonumber\\
&m,n=1,...,M\quad m \neq n,
\label{Vertauschung_Quantenobjekte_symmetrisch}
\end{align}
wobei der Operator $\mathcal{P}$ in Analogie zu ($\ref{Vertauschung_Ur-Alternativen_symmetrisch}$)
und ($\ref{Vertauschung_Ur-Alternativen_antisymmetrisch}$) die Vertauschung zweier Quantenobjekte
bezeichnet, die ihrerseits aus vielen Ur"=Alternativen gebildet sind. Ein gegenüber beliebiger
Permutation symmetrischer Zustand als Element des $\mathcal{H}_{GS}$ wird in Analogie zu
($\ref{Symmetrisierung_Ur-Alternativen}$) durch Summierung über alle Permutationen der
Quantenobjekte gebildet:

\begin{equation}
|\Phi\left(N_{ABCD}^1,N_{ABCD}^2,...,N_{ABCD}^M\right)\rangle_S
=\bigoplus_{\mathcal{P}_S}|\Phi\left(N_{ABCD}^1,N_{ABCD}^2,...,N_{ABCD}^M\right)\rangle_S \in \mathcal{H}_{GS},
\label{Symmetrisierung_Quantenobjeke}
\end{equation}
wobei $\mathcal{P}_S$ in Analogie zu ($\ref{Symmetrisierung_Ur-Alternativen}$) die symmetrische
Permutation zwischen den $M$ Quantenobjekten andeutet. Im antisymmetrischen Falle gilt für
einen Zustand, der aus $M$ Quantenobjekten besteht:

\begin{align}
&\mathcal{P}_{mn}|\Phi\left(...,N_{ABCD}^m,...,N_{ABCD}^n,...\right)\rangle_{AS}
=-|\Phi\left(...,N_{ABCD}^n,...,N_{ABCD}^m,...\right)\rangle_{AS},\nonumber\\
&m,n=1,...,M\quad m \neq n.
\label{Vertauschung_Quantenobjekte_antisymmetrisch}
\end{align}
Ein gegenüber beliebiger Permutation zweier Quantenobjekte antisymmetrischer Zustand als Element
des Hilbertraumes $\mathcal{H}_{GAS}$ wird in Analogie zu ($\ref{Antisymmetrisierung_Ur-Alternativen}$)
durch Summierung über alle Permutationen der Quantenobjekte unter der Voraussetzung gebildet,
dass der entsprechende Term der Summe bei einer geraden Zahl von Permutationen gegenüber dem
Ausgangszustand ein positives Vorzeichen erhält und bei ungerader Anzahl ein negatives Vorzeichen:

\begin{equation}
|\Phi\left(N_{ABCD}^1,N_{ABCD}^2,...,N_{ABCD}^M\right)\rangle_{AS}
=\bigoplus_{\mathcal{P}_{AS}}|\Phi\left(N_{ABCD}^1,N_{ABCD}^2,...,N_{ABCD}^M\right)\rangle_{AS} \in \mathcal{H}_{GAS},
\label{Antisymmetrisierung_Quantenobjekte}
\end{equation}
wobei $\mathcal{P}_{AS}$ in Analogie zu ($\ref{Antisymmetrisierung_Ur-Alternativen}$)
die antisymmetrische Permutation zwischen den $M$ Quantenobjekten andeutet.
Dementsprechend kann diese weitere Quantisierung entweder durch die Forderung
von Vertauschungsrelationen oder von Antivertauschungsrelationen zwischen
den Koeffizienten der Basiszustände des Tensorraumes der Ur"=Alternativen
realisiert werden. Dies hängt wie gewöhnlich davon ab, welchen Spin das Quantenobjekt
aufweist \cite{Pauli:1940}. Und dies wiederum hängt in der Quantentheorie der
Ur"=Alternativen gemäß ($\ref{Zustand_gesamt}$) beziehungsweise ($\ref{Zustand_gesamt_chi8}$)
vom antisymmetrischen Anteil des Zustandsraumes der Ur"=Alternativen ab. Wenn man
zunächst von Zuständen ausgeht, bei denen der antisymmetrische Anteil des Zustandes
mit Ur"=Alternativen vollkommen besetzt oder unbesetzt ist, dann hat man es mit
Bose"=Statistik zu tun, und ansonsten mit Fermi"=Statistik. Im Falle
symmetrischer Tensorprodukte der Tensorraumzustände der einzelnen Quantenobjekte
($\ref{Vertauschung_Quantenobjekte_symmetrisch}$) weisen die entsprechenden
Vertauschungsrelationen die folgende Gestalt auf:

\begin{equation}
\left[\hat \psi\left(N_{ABCD}\right),\hat \psi^{\dagger}\left(N^{'}_{ABCD}\right)\right]
=\delta_{N_A N^{'}_A}\delta_{N_B N^{'}_B}\delta_{N_ C N^{'}_C}\delta_{N_D N^{'}_D},
\label{Operatoren_Quantenobjekte_Vertauschungsrelationen}
\end{equation}
wobei die Besetzungszahlen $N_{ABCD}$ beziehungsweise $N^{'}_{ABCD}$ sich natürlich
auf den symmetrischen Teil des Tensorraumes der Ur"=Alternativen beziehen.
Im Falle antisymmetrischer Tensorprodukte der Tensorraumzustände der einzelnen
Quantenobjekte ($\ref{Vertauschung_Quantenobjekte_antisymmetrisch}$) weisen
die entsprechenden Antivertauschungsrelationen die folgende Gestalt auf:

\begin{equation}
\left\{\hat \psi\left(N_{ABCD}\right),\hat \psi^{\dagger}\left(N^{'}_{ABCD}\right)\right\}
=\delta_{N_A N^{'}_A}\delta_{N_B N^{'}_B}\delta_{N_ C N^{'}_C}\delta_{N_D N^{'}_D},
\label{Operatoren_Quantenobjekte_Antivertauschungsrelationen}
\end{equation}
wobei die Besetzungszahlen $N_{ABCD}$ beziehungsweise $N^{'}_{ABCD}$ sich natürlich
auch hier auf den symmetrischen Teil des Tensorraumes der Ur"=Alternativen beziehen.
Ein Zustand vieler Quantenobjekte im Tensorraum der Ur"=Alternativen entspricht also
dem Tensorprodukt der Zustände der einzelnen Objekte, wobei im Falle der Bosestatistik
mit den Vertauschungsrelationen ($\ref{Operatoren_Quantenobjekte_Vertauschungsrelationen}$)
einfach die Summe über alle Permutationen der Quantenobjekte gebildet wird
($\ref{Symmetrisierung_Quantenobjeke}$), und im Falle der Fermi"=Statistik mit den
Antivertauschungsrelationen ($\ref{Operatoren_Quantenobjekte_Antivertauschungsrelationen}$)
die Summe unter der Bedingung, dass sich bei den einzelnen Termen bei jeder Permutation
in Bezug auf den Zustand mit der Grundanordnung der Einzelzustände das Vorzeichen umkehrt
($\ref{Antisymmetrisierung_Quantenobjekte}$). Im Falle der Bosestatistik hat man es also mit
einem gegenüber Permutation symmetrischen Produkt gegenüber Permutation symmetrischer
Zustände von Ur"=Alternativen zu tun. Das bedeutet, dass die Zugehörigkeit der einzelnen
Ur"=Alternativen zu den verschiedenen in einem Gesamtzustand vieler Objekte befindlichen
Objektzuständen dadurch gekennzeichnet ist, dass der Gesamtzustand gegenüber Vertauschung
der zum gleichen Objekt gehörigen Ur"=Alternativen symmetrisch ist, aber keinesfalls
gegenüber der Vertauschung von Ur"=Alternativen, die zu unterschiedlichen Objekten gehören.
Denn ein symmetrisches Produkt symmetrischer Zustände ist keineswegs wieder symmetrisch,
was gleich noch weiter geführt werden wird. Ein allgemeiner Zustand im Hilbertraum vieler
Quantenobjekte, die ihrerseits viele Ur"=Alternativen enthalten, ist im Falle der
Bosestatistik demnach gekennzeichnet durch:

\begin{eqnarray}
|\Phi_{\mathcal{N}}\rangle_S&=&\sum_{N_{ABCD}}\sum_{\mathcal{N}\left(N_{ABCD}\right)}
\phi_{\mathcal{N}S}\left[\mathcal{N}\left(N_{ABCD}\right)\right]|\mathcal{N}\left(N_{ABCD}\right)\rangle_S\nonumber\\
&=&\sum_{N_{xyzn}}\sum_{\mathcal{N}\left(N_{xyzn}\right)}\phi_{\mathcal{N}S}
\left[\mathcal{N}\left(N_{xyzn}\right)\right]|\mathcal{N}\left(N_{xyzn}\right)\rangle_S,
\end{eqnarray}
und im Falle der Fermi"=Statistik gekennzeichnet durch:

\begin{eqnarray}
|\Phi_{\mathcal{N}}\rangle_{AS}&=&\sum_{N_{ABCD}}\sum_{\mathcal{N}\left(N_{ABCD}\right)=0}^{1}
\phi_{\mathcal{N}AS}\left[\mathcal{N}\left(N_{ABCD}\right)\right]|\mathcal{N}\left(N_{ABCD}\right)\rangle_{AS}\nonumber\\
&=&\sum_{N_{xyzn}}\sum_{\mathcal{N}\left(N_{xyzn}\right)=0}^{1}\phi_{\mathcal{N}AS}
\left[\mathcal{N}\left(N_{xyzn}\right)\right]|\mathcal{N}\left(N_{xyzn}\right)\rangle_{AS},
\end{eqnarray}
da sich im letzteren Falle aufgrund des Paulischen Ausschließungsprinzips in jedem
Basiszustand maximal ein Quantenobjekt befinden kann. Dem Übergang von einer klassischen
zu einer quantentheoretischen Ur"=Alternative, von einer Ur"=Alternative zu einem Zustand
vieler Ur"=Alternativen, und von einem Zustand vieler Ur"=Alternativen zu einem Zustand
vieler Quantenobjekte mit jeweils vielen in ihnen enthaltenen Ur"=Alternativen entspricht
jeweils ein Vorgang der Quantisierung. Dies ist gleichbedeutend mit einer Iteration des
Vorganges der Quantisierung, der wie folgt schematisiert dargestellt werden soll:

\begin{align}
&\textbf{binäre Alternative} &\xrightarrow{Quantisierung}& \quad\textbf{Zustand Ur-Alternative}\nonumber\\
&\textbf{Zustand Ur-Alternative} &\xrightarrow{Quantisierung}& \quad\textbf{Zustand vieler Ur-Alternativen}
\ \widehat{=}\ \textbf{Quantenobjekt}\nonumber\\
&\textbf{Quantenobjekt} &\xrightarrow{Quantisierung}& \quad\textbf{Zustand vieler Quantenobjekte}
\end{align}
Da wie bereits erwähnt ein symmetrisches beziehungsweise antisymmetrisches Produkt
symmetrischer beziehungsweise antisymmetrischer Zustände nicht wieder ein symmetrischer
beziehungsweise antisymmetrischer Zustand ist, entspricht einem solchen Zustand zwar in
Bezug auf die Quantenobjekte, die jeweils aus vielen Ur"=Alternativen bestehen, nicht
aber in Bezug auf die einzelnen Ur"=Alternativen eine Bose"=Statistik beziehungsweise
Fermi"=Statistik, sondern eine sogenannte Parabosestatistik. Um die Eigenschaften der
Parabose"=Statistik zu kennzeichnen, soll folgende Definition vorgenommen werden:

\begin{eqnarray}
&&a_1=A,\quad a_2=B,\quad a_3=C,\quad a_4=D,\nonumber\\
&&a_1^{\dagger}=A^{\dagger},\quad a_2^{\dagger}=B^{\dagger},\quad a_3^{\dagger}=C^{\dagger},\quad a_4^{\dagger}=D^{\dagger}.
\label{a1234ABCD}
\end{eqnarray}
Basierend auf ($\ref{a1234ABCD}$) kann man die Algebra der Parabose"=Statistik
wie folgt formulieren:

\begin{equation}
\left[\frac{1}{2}\left\{a_r, a_s^{\dagger}\right\}, a_t\right]=-\delta_{st} a_r,\quad
\left[\left\{a_r, a_s\right\},a_t\right]=\left[\left\{a_r^{\dagger}, a_s^{\dagger}\right\}, a_t^{\dagger}\right]=0,
\quad r,s,t=1,...,4.
\label{Parabose_Statistik_Algebra_A}
\end{equation}
Wenn man nun desweiteren die folgenden Relationen zugrunde legt:

\begin{equation}
a_r=\sum_{\alpha=1}^{p} b_r^{\alpha},\quad a_r^{\dagger}=\sum_{\alpha=1}^p b_r^{\alpha\dagger},\quad
r=1,...,4,\quad p=\mathcal{N},
\end{equation}
wobei der Parabose"=Ordnung $p$ die Zahl der Quantenobjekte $\mathcal{N}$ entspricht, dann kann man
die Algebra ($\ref{Parabose_Statistik_Algebra_A}$) in der folgenden Weise ausdrücken:

\begin{eqnarray}
&&\left[b_r^{\alpha}, b_s^{\alpha \dagger}\right]=\delta_{rs},\quad
\left[b_r^{\alpha},b_s^{\alpha}\right]=\left[b_r^{\alpha\dagger},b_s^{\alpha\dagger}\right]=0,\quad r,s=1,...,4,\nonumber\\
&&\left\{b_r^{\alpha}, b_s^{\beta\dagger}\right\}=\left\{b_r^{\alpha}, b_s^{\beta}\right\}
=\left\{b_r^{\alpha\dagger}, b_s^{\beta\dagger}\right\}=0\quad \textrm{für}\quad \alpha \neq \beta.
\label{Parabose_Statistik_Algebra_B}
\end{eqnarray}
Ein Zustand vieler Quantenobjekte stellt genaugenommen ein Gesamtquantenobjekt dar, dass in
sich Teilräume enthält, deren Zustände unter Permutation der einzelnen Ur"=Alternativen jeweils
symmetrisch sind, falls es sich um bosonische Zustände handelt, oder aus einem symmetrischen und
einem antisymmetrischen Anteil bestehen, falls es sich um fermionische Zustände handelt.
Man bildet in diesem Sinne also einen symmetrischen Zustand gegenüber Permutation der
darin enthaltenen Einzelobjekte, die ihrerseits gegenüber Permutation der einzelnen in
ihnen enthaltenen Ur"=Alternativen in sich symmetrische beziehungsweise antisymmetrische
Zustände darstellen.

\subsection{Verschränkung als quantentheoretisches Analogon zur Wechselwirkung}

Zunächst wurde im Rahmen der Rekonstruktion der theoretischen Physik als ein Postulat die
näherungsweise Trennbarkeit der Alternativen gefordert. Da die Basiszustände des zu einem
Quantenobjekt gehörigen Tensorraumes der Ur"=Alternativen eine Alternative darstellen,
die in viele Ur"=Alternativen aufgespalten ist, entspricht dem Postulat der näherungsweisen
Trennbarkeit der Alternativen in der Darstellung durch Ur"=Alternativen die näherungsweise
Trennbarkeit der zu den verschiedenen Quantenobjekten gehörigen Tensorräume vieler Ur"=Alternativen.
Aber es ist hier eben nur von einer Näherung die Rede. Denn wenn man das quantentheoretische
Gesamtsystem aus $M$ Quantenobjekten betrachtet, dessen Hilbertraum $\mathcal{H}_{G}$ dem
Tensorprodukt der Hilberträume der einzelnen Quantenobjekte entspricht, also im Falle der
Ur"=Alternativen der jeweiligen Tensorräume vieler Ur"=Alternativen, die ihrerseits einen
gegenüber Permutation symmetrischen und antisymmetrischen Teilraum aufweisen:

\begin{equation}
\mathcal{H}_{G}=\mathcal{H}_{T1}\otimes ... \otimes\mathcal{H}_{TM},
\label{Tensorprodukt_Hilbertraeume}
\end{equation}
dann entspricht den Zuständen dieses Gesamthilbertraumes, in denen sich die Einzelobjekte in
jeweils für sich selbst wohldefinierten Zuständen befinden, die unabhängig von den Zuständen
der anderen Quantenobjekte sind, eine Menge des Maßes null. Dies bedeutet, dass die
Einzelobjekte innerhalb eines Gesamtzustandes in der Regel in eine Beziehungsstruktur
eingebunden sind, die man im Allgemeinen mit dem Begriff der Verschränkung bezeichnet.
Demnach sind die quantentheoretischen Alternativen, denen ein gegenüber Permutation symmetrischer
Tensorraumzustand entspricht, eben in der Regel nicht wirklich trennbar, weshalb das Postulat
nur ein Näherungspostulat darstellt, um zunächst einzelne Objekte beziehungsweise Alternativen
näherungsweise zu definieren. In Wirklichkeit ist jede einzelne nur näherungsweise separierbare
Alternative Teil einer größeren Alternative beziehungsweise einer Beziehungsstruktur vieler
Alternativen. Und innerhalb dieser Gesamtalternative beziehungsweise Gesamtstruktur sind die
Zustände der Teilalternativen eben gewöhnlich für sich selbst genommen nicht wohldefiniert,
weshalb eine Aufspaltung in diese Teilalternativen unmöglich ist, und man eigentlich nur die
Gesamtalternative als ein quantentheoretisches Objekt sinnvoll betrachten kann. Dennoch sind
die verschiedenen Teilalternativen implizit in der Gesamtalternative enthalten, sodass es weiterhin
durchaus sinnvoll bleibt, von verschiedenen Teilalternativen zu reden, auch wenn diese nicht voneinander
trennbar sind. Denn bei einer Messung zerfallen sie wieder in voneinander unabhängige Teilalternativen.
Genaugenommen muss diese Beziehungsstruktur eigentlich alle Alternativen des gesamten Kosmos
enthalten. Denn jedes Herausseparieren einer Teilbeziehung, die nur bestimmte Alternativen
enthält, stellt schon wieder nur eine Näherung dar. Diesbezüglich ergibt sich die begriffliche
Schwierigkeit, dass der objektivierbare Teil des Beobachters selbst in diese quantenlogische
Beziehungsstruktur der Alternativen eingebunden ist und der Begriff des Kosmos daher aus
dieser Perspektive heraus gesehen problematisch erscheint. Jedenfalls ist diese abstrakte
quantentheoretische Beziehungsstruktur des Eingebundenseins verschiedener voneinander
nicht separierbarer Teilzustände in einen Gesamtzustand, die als Verschränkung von Zuständen
im Tensorraum der Ur"=Alternativen bezeichnet wird, jene Gegebenheit, welche dem Phänomen
letztendlich zu Grunde liegt, das wir auf einer oberflächlichen Betrachtungsebene als
Wechselwirkung zwischen verschiedenen Objekten beschreiben. Die Verschränkung zwischen
den Zuständen $M$ verschiedener Quantenobjekte mit bezüglich Permutation symmetrischen
Zuständen kann man explizit in der folgenden Weise schreiben:

\begin{equation}
|\Phi(t)\rangle_{N^1_{ABCD},...,N^M_{ABCD}}=\bigoplus_{\mathcal{P}_S}
\sum_{N^1_{ABCD},...,N^M_{ABCD}}\phi\left(N^{1}_{ABCD},...,N^{M}_{ABCD},t\right)
\left[|N^{1}_{ABCD}\rangle \otimes ... \otimes |N^{M}_{ABCD}\rangle\right],
\label{Zustand_N_Objekte_A}
\end{equation}
wobei $\mathcal{P}_S$ eine zu ($\ref{Symmetrisierung_Quantenobjeke}$) analoge
Bedeutung hat. Damit in einem solchen Zustand vieler Quantenobjekte eine
Verschränkung vorliegt, muss natürlich gelten:

\begin{equation}
\phi\left(N_{ABCD}^1, ... ,N_{ABCD}^M,t\right)\neq \phi_1\left(N_{ABCD}^1,t\right)
\times ... \times \phi\left(N_{ABCD}^M,t\right).
\end{equation}
Wenn man das M"=Tupel $\mathbf{N}_{ABCD}^M$ definiert als:

\begin{equation}
\mathbf{N}_{ABCD}^M=\left(N_{ABCD}^1, ... ,N_{ABCD}^M\right),
\end{equation}
dann kann man ($\ref{Zustand_N_Objekte_A}$) schreiben als:

\begin{equation}
|\Phi(t)\rangle_{\mathbf{N}_{ABCD}^M}=\bigoplus_{\mathcal{P}_S}\sum_{\mathbf{N}_{ABCD}^M} \phi(\mathbf{N}_{ABCD}^M,t)
\left[|N^{1}_{ABCD}\rangle \otimes ... \otimes |N^{M}_{ABCD}\rangle\right],
\label{Zustand_N_Objekte_B}
\end{equation}
Um bei einem System zweier oder mehrerer Quantenobjekte die Verschränkungsbeziehungen genau
zu bestimmen, muss das rein quantentheoretische Analogon zu den gewöhnlichen Wechselwirkungstheorien
mit punktweisen Produkten im feldtheoretischen Sinne gefunden werden. Dies soll in dieser Arbeit
in Analogie zu der Beschreibung im Rahmen gewöhnlicher Quantenfeldtheorien durch die Verwendung
von Eichsymmetrien geschehen, im Speziellen in Bezug auf die Symmetrien, die oben bereits im
Rahmen des Tensorraumes der Ur"=Alternativen ausgedrückt wurden, nämlich die Translationssymmetrie als
Teil der Poincare"=Symmetrie im symmetrischen Anteil und die $SU(N)$"=Symmetrien im antisymmetrischen Anteil
des Tensorraumes der Ur"=Alternativen. Der entscheidende Unterschied zu gewöhnlichen Eichtheorien besteht
allerdings darin, dass sich diese Eichsymmetrien, auch die Translationssymmetrie, hier auf den abstrakten
Tensorraum der Ur"=Alternativen beziehen und die Transformationen beziehungsweise die entsprechenden
Generatoren mit Hilfe von Erzeugungs- und Vernichtungsoperatoren in diesem rein quantenlogischen
Raum formuliert werden. In dieser abstrakteren Weise der Formulierung dessen, was dem zugrunde liegt,
was wir im Allgemeinen als Wechselwirkung näherungsweise trennbarer Objekte bezeichnen, besteht überhaupt
der wichtigste Unterschied zu gewöhnlichen Quantenfeldtheorien. Denn in den Quantenfeldtheorien wird
die Wechselwirkung über punktweise Produkte von Feldern definiert, wodurch sich das Element der
lokalen Kausalität konkret manifestiert, das in der Quantentheorie für sich selbst genommen
überhaupt nicht enthalten ist. Durch die Bildung eines solchen Hybriden aus Quantentheorie
und klassischer Feldtheorie kommt also sowohl die lokale Kausalität alsauch das Kontinuum
in die Beschreibung der Natur hinein, dessen Überwindung ja gerade den Kern der Planckschen
Quantenhypothese darstellt, wodurch dann auch die unendlichen Werte entstehen, die gewöhnlich
über das ein wenig konstruiert erscheinende Verfahren der Renormierung künstlich herausgerechnet
werden müssen.

\subsection{Dynamik und Verschränkung der Zustände}

Um nun die Dynamik dessen zu beschreiben, was wir auf einer oberflächlichen Betrachtungsebene als
Wechselwirkung näherungsweise voneinander separierbarer Objekte bezeichnen, muss die allgemeine
Schrödingergleichung im Tensorraum der Ur"=Alternativen ($\ref{Schroedingergleichung_Tensorraum}$)
erweitert werden. Hierzu müssen $M$ Hamiltonoperatoren $H_m,\ m=1,...,M$ definiert werden,
die sich jeweils ausschließlich auf den $m$"=ten Teilraum des $m$"=ten Quantenobjektes beziehen
und dessen freie Dynamik beschreiben, und zudem ein Hamiltonoperator $H_W$, der in den
$M$ Teilräumen, welche die Zustände der einzelnen Objekte enthalten, zugleich wirkt.
Der Gesamthamiltonoperator $H_G$ besteht aus der Summe dieser $M+1$ Hamiltonoperatoren:

\begin{equation}
H_G=\sum_{m=1}^{M}H_m+H_W.
\label{Hamiltonoperator_Viele_Objekte}
\end{equation}
Damit besitzt die abstrakte Schrödingergleichung in Bezug auf Zustände im gesamten Hilbertraum
$\mathcal{H}_G$ ($\ref{Zustand_N_Objekte_B}$), der sich aus dem Tensorprodukt der einzelnen
zu den $M$ Quantenobjekten gehörigen Tensorräume vieler Ur"=Alternativen ergibt
($\ref{Tensorprodukt_Hilbertraeume}$), die folgende Gestalt:

\begin{equation}
i\partial_t|\Phi(t)\rangle_{\mathbf{N}_{ABCD}^M}=H_G|\Phi(t)\rangle_{\mathbf{N}_{ABCD}^M}
=\left[\sum_{m=1}^M H_m+H_W\right]|\Phi(t)\rangle_{\mathbf{N}_{ABCD}^M},
\label{Schroedingergleichung_viele_Objekte}
\end{equation}
wobei $|\Phi(t)\rangle_{\mathbf{N}_{ABCD}^M}$ gemäß ($\ref{Zustand_N_Objekte_B}$) definiert ist.
Die entsprechende Zeitentwicklung eines Zustandes wird demnach bezüglich ihrer
allgemeinen Gestalt wie folgt beschrieben:

\begin{align}
&|\Phi(t)\rangle_{\mathbf{N}_{ABCD}^M}=\exp\left[-i H_G t\right]|\Phi(0)\rangle_{\mathbf{N}_{ABCD}^M}
\label{Zeitentwicklung_N_Objekte}\\
&=\exp\left[-i\left(\sum_{m=1}^M H_m+H_W\right)t\right]\bigoplus_{\mathcal{P}_S}
\sum_{\mathbf{N}_{ABCD}^M}\phi\left(\mathbf{N}_{ABCD}^M,0\right)
\left[|N^{1}_{ABCD}\rangle \otimes ... \otimes |N^{M}_{ABCD}\rangle\right]\nonumber\\
&=\bigoplus_{\mathcal{P}_S}\sum_{\mathbf{N}_{ABCD}^M}\phi\left(\mathbf{N}_{ABCD}^M,0\right)
\exp\left[-i H_W t\right]\left[\exp\left[-iH_1 t\right]|N^{1}_{ABCD}\rangle \otimes ... \otimes
\exp\left[-iH_M t\right]|N^{M}_{ABCD}\rangle\right]\nonumber\\
&=\bigoplus_{\mathcal{P}_S}\sum_{\mathbf{N}_{ABCD}^M}\phi\left(\mathbf{N}_{ABCD}^M,0\right)
\exp\left[-i H_W t\right]\left[\omega\left(N^{1}_{ABCD},t\right)|N^{1}_{ABCD}\rangle \otimes ... \otimes
\omega\left(N^{M}_{ABCD},t\right)|N^{M}_{ABCD}\rangle\right]\nonumber\\
&=\bigoplus_{\mathcal{P}_S}\sum_{\mathbf{N}_{ABCD}^M}\phi(\mathbf{N}_{ABCD}^M,0)
f_W\left[\omega\left(N^{1}_{ABCD},t\right),...,\omega\left(N^{M}_{ABCD},t\right),t\right]
\left[|N^{1}_{ABCD}\rangle \otimes ... \otimes |N^{M}_{ABCD}\rangle\right].\nonumber
\end{align}
Nun bleibt natürlich noch die Form der Hamiltonoperatoren $H_m,\ m=1,...,M$ und $H_W$ zu bestimmen,
was das eigentlich Entscheidende ist. Hierzu sollen die zu den $M$ Teilräumen gehörigen Erzeugungs-
und Vernichtungsoperatoren jeweils mit einem Index bezeichnet werden, um sie voneinander zu unterscheiden.
Jeder der Teilräume kann natürlich gemäß ($\ref{Aufspaltung_symmetrisch_antisymmetrisch}$)
seinerseits in einen symmetrischen und in einen antisymmetrischen Teil aufgespalten werden:

\begin{equation}
\mathcal{H}_{Tm}=\mathcal{H}_{TmS}\otimes \mathcal{H}_{TmAS},\quad m=1,...,M.
\end{equation}
Demnach existieren zu jedem Teilraum $\mathcal{H}_{Tm}$ ein Satz an Erzeugungs-
und Vernichtungsoperatoren in Bezug auf den jeweils symmetrischen Anteil:

\begin{equation}
A_S^m,\quad B_S^m,\quad C_S^m,\quad D_S^m,\quad
A^{m\dagger}_S,\quad B^{m\dagger}_S,\quad C^{m\dagger}_S,\quad D^{m\dagger}_S,\quad  m=1,...,M,
\label{Operatoren_Sm}
\end{equation}
welche die folgenden Vertauschungsrelationen erfüllen:

\begin{equation}
\left[A_S^m, A_S^{n\dagger}\right]=\delta^{mn},\
\left[B_S^m, B_S^{n\dagger}\right]=\delta^{mn},\
\left[C_S^m, C_S^{n\dagger}\right]=\delta^{mn},\
\left[D_S^m, D_S^{n\dagger}\right]=\delta^{mn},\ m,n=1,...,M,
\end{equation}
und ein Satz an Erzeugungs- und Vernichtungsoperatoren in Bezug auf
den jeweils antisymmetrischen Anteil:

\begin{equation}
A^m_{AS},\quad B^m_{AS},\quad C^m_{AS},\quad D^m_{AS},\quad
A^{m\dagger}_{AS},\quad B^{m\dagger}_{AS},\quad C^{m\dagger}_{AS},\quad D^{m\dagger}_{AS},\quad  m=1,...,M,
\end{equation}
welche die folgenden Antivertauschungsrelationen erfüllen:

\begin{equation}
\left\{A_{AS}^m, A_{AS}^{n\dagger}\right\}=\delta^{mn},\ 
\left\{B_{AS}^m, B_{AS}^{n\dagger}\right\}=\delta^{mn},\ 
\left\{C_{AS}^m, C_{AS}^{n\dagger}\right\}=\delta^{mn},\
\left\{D_{AS}^m, D_{AS}^{n\dagger}\right\}=\delta^{mn},\ m,n=1,...,M.
\end{equation}
Die Hamiltonoperatoren $H_m$, $m=1,...,M$ die nur in den entsprechenden Teilräumen wirken,
hängen natürlich nur von den entsprechenden Erzeugungs- und Vernichtungsoperatoren ab,
die sich auf den jeweiligen Teilraum $\mathcal{H}_{Tm}$, $m=1,...,M$ beziehen,
was bedeutet:

\begin{align}
&H_m=H_m\left(A_S^m, B_S^m, C_S^m, D_S^m, A^{m\dagger}_S, B^{m\dagger}_S, C^{m\dagger}_S, D^{m\dagger}_S,
A^m_{AS}, B^m_{AS}, C^m_{AS}, D^m_{AS}, A^{m\dagger}_{AS}, B^{m\dagger}_{AS}, C^{m\dagger}_{AS}, D^{m\dagger}_{AS}\right).
\nonumber\\
&m=1,...,M
\end{align}
Im Spezialfall, dass die nur auf die Teilräume der einzelnen Quantenobjekte
bezogenen Hamiltonoperatoren $H_m$ ausschließlich von den auf den symmetrischen
Teilraum bezogenen Erzeugungs- und Vernichtungsoperatoren ($\ref{Operatoren_Sm}$)
abhängen, werden bei einem Zustand mit einem antisymmetrischen Anteil, also mit
inneren Symmetrien, die Hamiltonoperatoren $H_m$ im Allgemeinen die gewöhnliche
Gestalt des Dirac"=Hamiltonoperators ($\ref{Dirac-Hamilton-Operator}$) in dem
jeweiligen $m$"=ten Tensorraum aufweisen. Ein Hamiltonoperator $H_W$, der eine
Wechselbeziehung zwischen den Zuständen verschiedener Objekte induziert,
muss hingegen von Erzeugungs- und Vernichtungsoperatoren abhängen, die sich
auf unterschiedliche Teilräume beziehen:

\begin{align}
&H_W\left(A_S^m, B_S^m, C_S^m, D_S^m,
A^{m\dagger}_S, B^{m\dagger}_S, C^{m\dagger}_S, D^{m\dagger}_S,
A^m_{AS}, B^m_{AS}, C^m_{AS}, D^m_{AS},
A^{m\dagger}_{AS}, B^{m\dagger}_{AS}, C^{m\dagger}_{AS}, D^{m\dagger}_{AS}\right),\nonumber\\
&\neq H_W^1\left(A_S^1, B_S^1, C_S^1, D_S^1,
A^{1\dagger}_S, B^{1\dagger}_S, C^{1\dagger}_S, D^{1\dagger}_S,
A^1_{AS}, B^1_{AS}, C^1_{AS}, D^1_{AS},
A^{1\dagger}_{AS}, B^{1\dagger}_{AS}, C^{1\dagger}_{AS}, D^{1\dagger}_{AS}\right)+...\nonumber\\
&\quad+H_W^M\left(A_S^M, B_S^M, C_S^M, D_S^M,
A^{M\dagger}_S, B^{M\dagger}_S, C^{M\dagger}_S, D^{M\dagger}_S,
A^M_{AS}, B^M_{AS}, C^M_{AS}, D^M_{AS},
A^{M\dagger}_{AS}, B^{M\dagger}_{AS}, C^{M\dagger}_{AS}, D^{M\dagger}_{AS}\right).\nonumber\\
&m=1,...,M
\end{align}

\subsection{Dynamik mit Hamiltonoperatoren als Produkten abstrakter Zustände}

Weiter unten wird gezeigt werden, wie über das Eichprinzip formuliert im Tensorraum der
Ur"=Alternativen Hamiltonoperatoren definiert werden können, welche dynamische Gleichungen mit
Produkten verschiedener Zustände definieren. Solche Produkte von Zuständen in den dynamischen
Gleichungen induzieren ebenfalls Verschränkungen. Die spezifische Gestalt dieser
Produkte in einer rein quantenlogischen Fassung sowohl der $SU(N)$"=Eichtheorien der
Elementarteilchenphysik alsauch der Translationseichtheorie der Gravitation, wird erst in den
nächsten drei Kapiteln bestimmt. In diesem Kapitel soll zunächst nur die allgemeine Gestalt
solcher Produkte von Zuständen einzelner Quantenobjekte im Tensorraum erörtert werden.
Grundsätzlich weist ein entsprechender Gesamt"=Hamiltonoperator auch die Gestalt
($\ref{Hamiltonoperator_Viele_Objekte}$) auf, wobei der Hamiltonoperator $H_W$,
welcher eine Beziehung zwischen den Zuständen der $M$ verschiedenen Objekte herstellt,
im Allgemeinen anders aussieht:

\begin{equation}
\label{Hamiltonoperator_Wechselwirkung_Produkt}
H_W=\sum_{p1,...,pM}\Omega\left(p1,...,pM\right)\left[\left[|\Psi_{N^1}(t)\rangle\right]^{p1} \otimes ... \otimes
\left[|\Psi_{N^M}(t)\rangle\right]^{pM}\right],
\end{equation}
wobei die Exponenten $p1,...,pM$ beliebige ganzzahlige Werte annehmen können. Dies bedeutet,
dass ein solcher Hamiltonoperator im Prinzip beliebige Produkte der Zustände der verschiedenen
Objekte untereinander aufweist. Dies stellt das rein quantentheoretische Analogon zu punktweisen
Produkten im Rahmen relativistischer Quantenfeldtheorien dar. Die abstrakte Schrödingergleichung
vieler Quantenobjekte ($\ref{Schroedingergleichung_viele_Objekte}$) mit dem
Wechselbeziehungs"=Hamiltonoperator der allgemeinen Form
($\ref{Hamiltonoperator_Wechselwirkung_Produkt}$) nimmt
dann folgende Gestalt an:

\begin{align}
\label{Schroedingergleichung_viele_Objekte_Produkt}
&i\partial_t |\Phi(t)\rangle_{\mathbf{N}_{ABCD}^M}\\
&=\left[\sum_{m=1}^M H_m+\sum_{p1,...,pM}\Omega\left(p1,...,pM\right)
\left[\left[|\Psi_{N_{ABCD}^1}(t)\rangle\right]^{p1} \otimes ...
\otimes\left[|\Psi_{N_{ABCD}^M}(t)\rangle\right]^{pM}\right]\right]
|\Phi(t)\rangle_{\mathbf{N}_{ABCD}^M}.\nonumber
\end{align}
Wenn man nun davon ausgeht, dass zum Zeitpunkt $t=0$ keine Verschränkung zwischen
den Zuständen vorliegt und damit gilt:

\begin{equation}
|\Phi(t=0)\rangle_{\mathbf{N}_{ABCD}^M}=\bigotimes_{\mathcal{P}_S}\left[|\Psi_1(t=0)\rangle_{N^1}\otimes ...
\otimes |\Psi_M\left(t=0\right)\rangle_{N^M}\right],
\end{equation}
und man desweiteren folgende Einzelzustände definiert, welche nur die in jedem der Einzelhilberträume
$\mathcal{H}_{Tm}$ bestehende freie Dynamik enthalten:

\begin{equation}
\label{Zeitentwicklung_frei}
|\Psi_{F}(t)\rangle_{N^m}=\mathcal{O}_t\exp\left[-i\int_0^t H_m(t) dt\right]|\Psi(t=0)\rangle_{N^m},
\end{equation}
wobei $\mathcal{O}_t$ den Dysonschen Zeitordnungsoperator darstellt, dann kann man die aus der
Schrödingergleichung der Gestalt ($\ref{Schroedingergleichung_viele_Objekte_Produkt}$)
sich ergebende Zeitentwicklung in einem System von $M$ Gleichungen ausdrücken:

\begin{align}
&|\Psi(t)\rangle_{N^1}=\mathcal{O}_t\exp\left[-i\int_0^t \sum_{p1,...,pM}\Omega\left(p1,...,pM\right)
\left[\left[|\Psi(t)\rangle_{N^1}\right]^{p1} \otimes ... \otimes
\left[|\Psi(t)\rangle_{N^M}\right]^{pM}\right] dt\right]|\Psi_{F}(t)\rangle_{N^1},\nonumber\\
&\quad\quad\quad\quad.....\nonumber\\
&|\Psi(t)\rangle_{N^M}=\mathcal{O}_t\exp\left[-i\int_0^t \sum_{p1,...,pM}\Omega\left(p1,...,pM\right)
\left[\left[|\Psi(t)\rangle_{N^1}\right]^{p1} \otimes ... \otimes
\left[|\Psi(t)\rangle_{N^M}\right]^{pM}\right] dt\right]|\Psi_{F}(t)\rangle_{N^M}.
\label{Gleichungssystem_Wechselwirkung}
\end{align}
Die Zeitentwicklung jedes der $M$ Einzelzustände hängt also im allgemeinsten Falle von den Einzelzuständen
der anderen $M-1$ Einzelzustände und von ihm selbst ab, wobei alle anderen $M-1$ Einzelzustände ja ihrerseits
wieder von den anderen $M-1$ Einzelzuständen und ihnen selbst anhängen. Dies führt zu einer sehr komplizierten
Verschränkung. Wenn man einen Hamiltonoperator mit Produkten quantentheoretischer Zustände im Tensorraum
der Ur"=Alternativen, wie er in ($\ref{Hamiltonoperator_Wechselwirkung_Produkt}$) in der allgemeinen
Gestalt formuliert ist, gemäß ($\ref{Zustand_Darstellung_Ortsraum}$) in den Ortsraum abbildet und damit
für einen zeitabhängigen Zustand gemäß ($\ref{Zustand_Darstellung_Raum-Zeit}$) eine raum"=zeitliche
Darstellung erhält, so ergibt sich für den entsprechenden Hamiltonoperator der Wechselwirkung Folgendes:

\begin{equation}
H_W\left(\mathbf{x},t\right)=\sum_{p1,...,pM}\Omega\left(p1,...,pM\right)
\left[\left[\Psi_{N^1}\left(\mathbf{x},t\right)\right]^{p1}\cdot ...
\cdot\left[\Psi_{N^M}\left(\mathbf{x},t\right)\right]^{pM}\right],
\label{Hamiltonoperator_Wechselwirkung_Produkt_Raum-Zeit}
\end{equation}
und für die entsprechende abstrakte Schrödingergleichung der Gestalt
($\ref{Schroedingergleichung_viele_Objekte_Produkt}$) ergibt sich:

\begin{align}
\label{Schroedingergleichung_viele_Objekte_Produkt_Raum-Zeit}
&i\partial_t \Phi_{\mathbf{N}_{ABCD}^M}\left(\mathbf{x},t\right)\\
&=\left[\sum_{m=1}^M H_m+\sum_{p1,...,pM}\Omega\left(p1,...,pM\right)
\left[\left[\Psi_{N^1}\left(\mathbf{x},t\right)\right]^{p1} \cdot ...
\cdot\left[\Psi_{N^M}\left(\mathbf{x},t\right)\right]^{pM}\right]\right]
\Phi_{\mathbf{N}_{ABCD}^M}\left(\mathbf{x},t\right).\nonumber
\end{align}
Entscheidend sowohl in Bezug auf ($\ref{Hamiltonoperator_Wechselwirkung_Produkt_Raum-Zeit}$)
alsauch in Bezug auf ($\ref{Schroedingergleichung_viele_Objekte_Produkt_Raum-Zeit}$) ist
natürlich wie grundsätzlich in dieser Arbeit, dass die darin auftretenden Produkte keine
punktweisen Produkte im Sinne von Quantenfeldtheorien darstellen. Vielmehr handelt es sich
nur um eine Darstellung der dahinter gemäß ($\ref{Hamiltonoperator_Wechselwirkung_Produkt}$)
und ($\ref{Schroedingergleichung_viele_Objekte_Produkt}$) in Wirklichkeit stehenden Produkte
abstrakter Zustände im Tensorraum der Ur"=Alternativen, die in keiner Weise an räumliche
Lokalitäts- und Kausalitätsbeziehungen gebunden sind. Dies gilt dann auch für die daraus
sich ergebenden Wechselwirkungen beziehungsweise Verschränkungen verschiedener Quantenobjekte.

\subsection{Störungstheorie im Tensorraum der Ur-Alternativen}

Die Störungstheorie stellt selbstverständlich eine Näherung dar, in der man die Verschränkung
von Zuständen wie eine Wechselwirkung behandelt. Über die durch die Vertauschungsrelationen
($\ref{Operatoren_Quantenobjekte_Vertauschungsrelationen}$) beziehungsweise ($\ref{Operatoren_Quantenobjekte_Antivertauschungsrelationen}$) definierten Operatoren,
welche Quantenobjekte mit vielen Ur"=Alternativen erzeugen und vernichten, ist aufgrund
der Möglichkeit der Abbildung in die Raum"=Zeit ($\ref{Zustand_Darstellung_Raum-Zeit}$)
zugleich ein rein quantentheoretisches Analogon zu Feldoperatoren definiert.
Diese repräsentieren Erzeugungs"= und Vernichtungsoperatoren für aus vielen
Ur"=Alternativen zusammengesetzte Quantenobjekte dargestellt in der Raum"=Zeit
durch Wellenfunktionen und weisen folgende Gestalt auf:

\begin{align}
&\hat \Psi\left(t\right)=\sum_{N_{xyzn}}\hat \psi\left(N_{xyzn},t\right)|N_{xyzn}\rangle
\quad\longleftrightarrow\quad
\hat \Psi_N\left(\mathbf{x},t\right)=\sum_{N_{xyzn}}\hat \psi\left(N_{xyzn},t\right)f_{N_{xyzn}}\left(\mathbf{x}\right),
\nonumber\\
&\hat \Psi^{\dagger}\left(t\right)=\sum_{N_{xyzn}}\hat \psi^{\dagger}\left(N_{xyzn},t\right)|N_{xyzn}\rangle
\quad\longleftrightarrow\quad
\hat \Psi_N^{\dagger}\left(\mathbf{x},t\right)=\sum_{N_{xyzn}}\hat \psi^{\dagger}
\left(N_{xyzn},t\right)f_{N_{xyzn}}\left(\mathbf{x}\right),
\label{Feldoperatoren_Tensorraum_Ur-Alternativen}
\end{align}
wobei hier auf ($\ref{Zustand_Darstellung_Raum-Zeit}$) Bezug genommen wurde. Wenn man nun mit
$\left(\hat \Psi_F\right)_{N}\left(t\right)$ die Feldoperatoren bezeichnet, welche nur den
freien Teil der Dynamik enthalten:

\begin{align}
&\left(\hat \Psi_F\right)\left(t\right)=\exp\left[-i H_F \left(t-t_0\right)\right]
\sum_{N_{xyzn}}\hat \psi_F\left(N_{xyzn},t_0\right)|N_{xyzn}\rangle,\quad t > t_0\nonumber\\
&\longleftrightarrow\quad \left(\hat \Psi_F\right)_{N}\left(\mathbf{x},t\right)=\sum_{N_{xyzn}}
\hat \psi_F\left(N_{xyzn},t\right)f_{N_{xyzn}}\left(\mathbf{x}\right),\nonumber\\
&\left(\hat \Psi_F^{\dagger}\right)\left(t\right)=\exp\left[-i H_F \left(t-t_0\right)\right]
\sum_{N_{xyzn}}\hat \psi_F^{\dagger}\left(N_{xyzn},t_0\right)|N_{xyzn}\rangle,\quad t > t_0\nonumber\\
&\longleftrightarrow\quad \left(\hat \Psi_F^{\dagger}\right)_{N}\left(\mathbf{x},t\right)=\sum_{N_{xyzn}}
\hat \psi_F^{\dagger}\left(N_{xyzn},t\right)f_{N_{xyzn}}\left(\mathbf{x}\right),
\label{Feldoperatoren_Tensorraum_Ur-Alternativen_Dynamik}
\end{align}
wobei $H_F$ hier zeitunabhängig ist, dann ist die Übergangsamplitude zwischen den Zuständen
zweier Quantenobjekte mit den Indizes $m$ und $n$ ohne Wechselwirkung gegeben durch:

\begin{equation}
\Delta\left(N_n,t,N_m,t_0\right)=\langle 0|\left(\hat \Psi_F\right)_{N^n}\left(t\right)
\left(\hat \Psi_F^{\dagger}\right)_{N^m}\left(t_0\right)|0\rangle,\quad t > t_0.
\label{Propagator_Tensorraum_Ur-Alternativen_frei}
\end{equation}
Wenn man nun einen zeitunabhängigen Wechselwirkungsterm $H_W$ in die
Zeitentwicklung integriert, so ergibt sich das Folgende:

\begin{equation}
\Delta\left(N_n,t,N_m,t_0\right)=\langle 0|\left(\hat \Psi_F\right)_{N^n}\left(t\right)
\exp\left[iH_W\left(t-t_0\right)\right]
\left(\hat \Psi_F^{\dagger}\right)_{N^m}\left(t_0\right)|0\rangle,\quad t > t_0.
\label{Propagator_Tensorraum_Ur-Alternativen_Wechselwirkung}
\end{equation}

\section{Eichtheorien in der Quantentheorie der Ur-Alternativen}

In Bezug auf die Dynamik im Tensorraum der Ur"=Alternativen existieren zwei fundamentale
Symmetriegruppen. Die eine Symmetriegruppe bezieht sich auf den Teilraum, welcher die
gegenüber Permutation der Ur"=Alternativen symmetrischen Zustände enthält, und die andere
Symmetriegruppe bezieht sich auf den Teilraum, der die gegenüber Permutation
der Ur"=Alternativen antisymmetrischen Zustände enthält. Im Raum der zeitabhängigen
symmetrischen Zustände ist die Poincare"=Gruppe ($\ref{Poincare_Algebra}$) die fundamentale
Symmetriegruppe und bezüglich der antisymmetrischen Zustände die $U(8)$"=Symmetriegruppe,
welche im Sinne von ($\ref{Zustand_antisymmetrisch_chi-chi4-chi3}$) wiederum aufgespalten
werden kann in die Summe von Teilräumen mit einer $U(1)$"=Symmetrie beziehungsweise
$U(4)$"=Symmetrie beziehungsweise $U(3)$"=Symmetrie, wobei diese Symmetrien
ihrerseits aufgespalten und die Teilräume im Sinne der Ausführungen des vorletzten
Kapitels demnach als die verschiedenen Sektoren der Elementarteilchenphysik
interpretiert werden können. Dies statuiert also die Poincare"=Gruppe und die $U(8)$"=Gruppe
als fundamentale Symmetriegruppen dargestellt in den Teilräumen des Tensorraumes der Ur"=Alternativen.
Um die mathematische Struktur der Wechselwirkungen zu erhalten, muss das Eichprinzip, dessen Grundidee
in \cite{Weyl:1918}, \cite{Weyl:1929} entwickelt wurde, in die Quantentheorie der Ur"=Alternativen
integriert werden. Im Rahmen gewöhnlicher Quantenfeldtheorien erhält man die Wechselwirkungsterme
durch die Forderung lokaler Eichinvarianz, also die Ausdehnung der bereits existierenden
globalen Symmetrien auf lokale Symmetrien \cite{Weinberg:1967}, \cite{Fritzsch:1973}, \cite{Yang:1954}.
In der Quantentheorie der Ur"=Alternativen existiert aber ein solcher kontinuierlicher
physikalischer Raum, in Bezug auf den die Lokalität der Eichtransformationen durchgeführt
werden könnte, gerade nicht. Dieser kann nur als ein Darstellungsmedium der rein quantenlogischen
Zustände des symmetrischen Teilraumes indirekt hergeleitet werden. Und dass auf der Basis der
Ur"=Alternativen als den basalen quantenlogischen Einheiten nur eine rein abstrakte
Informationsstruktur existiert, darin liegt ja gewissermaßen gerade der entscheidende
Gehalt der Theorie. Es muss also das Analogon im Tensorraum der Ur"=Alternativen zur
Lokalitätsforderung im Rahmen gewöhnlicher Quantenfeldtheorien formuliert werden.
Dies wird hier dadurch geschehen, dass die Parameter zu den Generatoren der jeweiligen
Gruppen nicht von einem kontinuierlichen Parameter, sondern von den Ortsoperatoren
im Tensorraum der Ur"=Alternativen, ($\ref{Ort_Impuls_Operatoren}$) beziehungsweise
($\ref{Ort_Impuls_Operatoren_ABCD}$), abhängig gemacht werden, die ja eine Linearkombination der 
entsprechenden Erzeugungs- und Vernichtungsoperatoren im Tensorraum der Ur"=Alternativen darstellen,
($\ref{ErzeugungsVernichtungsOperatorenXYZ}$) beziehungsweise ($\ref{Operatoren_symmetrisch}$),
und daher im quantenlogischen Raum operieren, also dem Tensorraum der Ur"=Alternativen.
Dies bedeutet, dass nicht lokale Eichtransformationen in Bezug auf eine vorgegebene
kontinuierliche Raum"=Zeit durchgeführt werden, sondern diese im abstrakten und
diskreten Tensorraum vieler Ur"=Alternativen durchgeführt werden. Man muss nun
zunächst von der entsprechenden  kten Dirac"=Gleichung im Tensorraum der
Ur"=Alternativen mit den inneren Symmetrien ausgehen, die gegeben ist in
($\ref{Dirac-Gleichung}$) beziehungsweise in ($\ref{Dirac-Gleichung_chi8}$).
Der Ausgangspunkt der weiteren Betrachtungen ist demnach die dadurch definierte
freie Dynamik. Die raum"=zeitliche Symmetrie, welche durch die Poincare"=Gruppe
respräsentiert wird, ist darin durch den symmetrischen Teilraum $\mathcal{H}_{TS}$
enthalten. Die inneren Symmetrien stehen wie dargelegt mit dem antisymmetrischen
Teilraum $\mathcal{H}_{TAS}$ in Zusammenhang, ohne dass die Symmetriestruktur gemäß
der Aufspaltung des vorletzten Kapitels zunächst genau spezifiziert werden müsste.
Da diese inneren Symmetrien nicht nur alle durch Lie"=Algebren, sondern grundsätzlich
ganz spezifisch durch $U(N)$"=Symmetrien beziehungsweise $SU(N)$"=Symmetrien
repräsentiert werden, spielt die Frage der mit der Aufspaltung der $U(8)$"=Symmetriegruppe
zusammenhängenden konkreten Symmetrien für die grundsätzliche Betrachtung zunächst keine Rolle.

\subsection{Poincare-Invarianz der abstrakten Dirac-Gleichung}

Die Dirac"=Gleichung mit den inneren Symmetrien im antisymmetrischen Anteil des Tensorraumes
der Ur"=Alternativen ($\ref{Dirac-Gleichung}$) beziehungsweise ($\ref{Dirac-Gleichung_chi8}$)
gehorcht zunächst einer globalen Transformation in Bezug auf die Poincare"=Gruppe
($\ref{Poincare_Algebra}$), welche sich auf den symmetrischen Anteil des Tensorraumes
der Ur"=Alternativen bezieht. Die entsprechenden Generatoren sind gegeben durch
($\ref{Viererimpuls_Tensorraum}$) und ($\ref{Lorentz+Drehgruppe_Operatoren}$)
und erfüllen die Algebra ($\ref{Poincare_Algebra}$). Eine solche Transformation
wird durch einen unitären Transformationsoperator ausgedrückt, der aus einem
Transformationsoperator besteht, der sich auf die Translationsgruppe bezieht,
und einem Transformationsoperator, der sich auf die Lorentzgruppe bezieht.
Dieser weist folgende Gestalt auf:

\begin{align}
&U_P\left[\lambda\right]=\exp\left[i\lambda^m \left(P_{ABCD}\right)_m\right],\quad
U_M\left[\Sigma\right]=\exp\left[i\Sigma^{mn}\left(M_{ABCD}\right)_{mn}\right],
\label{Transformationsoperator_Poincare-Gruppe}\\
&U_{PM}\left[\lambda,\Sigma\right]=U_P\left[\lambda\right]\otimes U_M\left[\Sigma\right]
=\exp\left[i\lambda^m \left(P_{ABCD}\right)_m
+i\Sigma^{mn}\left(M_{ABCD}\right)_{mn}\right],\quad m,n=0,...,3,\nonumber
\end{align}
wobei $\lambda$ als Vektor und $\Sigma$ als Tensor jeweils aus reellen Parametern bestehen.
Eine dementsprechende Anwendung auf Zustände im Tensorraum der Ur"=Alternativen
($\ref{Dirac-Zustand_chi8}$):

\begin{equation}
|\Psi_\Gamma(t)\rangle\quad \longrightarrow\quad U_{PM}\left[\lambda,\Sigma\right]|\Psi_\Gamma(t)\rangle,
\label{Transformation_Poincare-Gruppe}
\end{equation}
führt aufgrund der Tatsache, dass der Operator $U_P\left[\lambda\right]$ mit den
abstrakten Energie"=Impuls"=Operatoren $P_{ABCD}$ kommutiert und $P_{ABCD}$ sich
unter Lorentztransformationen daher gemäß:

\begin{equation}
P_{ABCD}\quad \longrightarrow\quad U_{PM}\left[\lambda,\Sigma\right] P_{ABCD} U^{\dagger}_{PM}\left[\lambda,\Sigma\right]
\quad\longrightarrow\quad U_M \left[\Sigma\right] P_{ABCD} U^{\dagger}_M\left[\Sigma\right]
\end{equation}
transformiert, also zu folgender Transformation der abstrakten Dirac"=Gleichung ($\ref{Dirac-Gleichung_chi8}$):

\begin{align}
\gamma^\mu_{8x8} \left(P_{ABCD}\right)_\mu|\Psi_\Gamma (t)\rangle=0 \quad \longrightarrow
\quad &\gamma^\mu_{8x8} U_M\left[\Sigma\right] \left(P_{ABCD}\right)_\mu U^{\dagger}_M\left[\Sigma\right]
U_{PM}\left[\lambda,\Sigma\right]|\Psi_\Gamma(t)\rangle\nonumber\\
&=\gamma^\mu_{8x8} U_M\left[\Sigma\right] \left(P_{ABCD}\right)_\mu U_P\left[\lambda\right]|\Psi_\Gamma(t)\rangle\nonumber\\
&=U_M\left[\Sigma\right] U_P\left[\lambda\right] \gamma^\mu_{8x8} \left(P_{ABCD}\right)_\mu|\Psi_\Gamma(t)\rangle=0,
\label{Transformation_Poincare-Gruppe_Dirac-Gleichung}
\end{align}
wobei in der zweiten Zeile ausgenutzt wurde, dass bei unitären Operatoren $U^{\dagger}=U^{-1}$
und damit $U^{\dagger}U=\mathbf{1}$ gilt, und in der dritten Zeile, dass $U_P\left[\lambda\right]$
nur $P_{ABCD}$ enthält, dessen Komponenten untereinander kommutieren. Damit ist also gezeigt,
dass die abstrakte Dirac"=Gleichung im Tensorraum invariant unter Transformationen
bezüglich der Poincaregruppe ist.

\subsection{Invarianz unter internen Symmetrien der abstrakten Dirac-Gleichung}

Die $\tau_{8x8}^a,\ a=1,...,N_\tau^2-1,\ N_\tau < 8$ sollen die Generatoren der inneren
Symmetriegruppen darstellen, also der internen Symmetriegruppen $U(N_\tau)$ beziehungsweise
$SU(N_\tau)$, die im antisymmetrischen Teilraum enthalten sind. Demnach werden die die
entsprechenden Generatoren $\tau^a,\ a=1,...,N_\tau^2-1,\ N_\tau < 8$ analog zu den
Dirac"=Matrizen ($\ref{Dirac-Matrizen_8x8}$) in dem achtdimensionalen Spinorraum der
Elemente ($\ref{chi8}$) des antisymmetrischen Teilraumes des Tensorraumes der Ur"=Alternativen
dargestellt, in dem man die diesen Generatoren entsprechenden $N_\tau$"=dimensionalen komplexen
Matrizen in achtdimensionale komplexe Matrizen einbettet. Diese Generatoren erfüllen naturgemäß
die allgemeine Relation einer Lie"=Algebra:

\begin{equation}
\left[\tau_{8x8}^a,\tau_{8x8}^b\right]=if^{abc}\tau_{8x8}^c,\quad a,b,c=1,...,N_\tau^2-1,
\label{Generatoren_antisymmetrischer_Tensorraum}
\end{equation}
wobei die $f^{abc}$ die Strukturkonstanten bezeichnen. Natürlich kann man die Generatoren
$\tau_{8x8}^a$ im antisymmetrischen Teilraum des Tensorraumes der Ur"=Alternativen
($\ref{Generatoren_antisymmetrischer_Tensorraum}$) aus den entsprechenden
Erzeugungs- und Vernichtungsoperatoren im antisymmetrischen Teilraum
($\ref{Operatoren_antisymmetrisch}$) bilden. Eine dementsprechende
Transformation wird durch den folgenden unitären Transformationsoperator ausgedrückt:

\begin{equation}
U_\tau\left[\alpha\right]=\exp\left[i\alpha^a \tau_{8x8}^a\right],\quad a=1,...,N_\tau^2-1,
\label{Transformationsoperator_AS}
\end{equation}
wobei der Vektor $\alpha$ aus komplexen Parametern besteht. Ein Dirac"=Zustand
($\ref{Dirac-Zustand_chi8}$) mit innerer Symmetrie transformiert sich
demnach wie folgt:

\begin{equation}
|\Psi_\Gamma(t)\rangle\quad \longrightarrow\quad U_\tau\left[\alpha\right]|\Psi_\Gamma(t)\rangle.
\end{equation}
Da $U_\tau\left[\alpha\right]$ mit dem abstrakten Energie"=Impuls"=Operator $P_{ABCD}$ kommutiert,
gilt Folgendes:

\begin{align}
\gamma_{8x8}^\mu (P_{ABCD})_\mu|\Psi_\Gamma (t)\rangle=0 \longrightarrow
&\gamma_{8x8}^\mu (P_{ABCD})_\mu U_\tau\left[\alpha\right]|\Psi_\Gamma(t)\rangle
=U_\tau\left[\alpha\right] \gamma_{8x8}^\mu \left(P_{ABCD}\right)_\mu|\Psi_\Gamma (t)\rangle=0.
\label{Symmetrie_AS}
\end{align}
Damit ist gezeigt, dass die abstrakte Dirac"=Gleichung im Tensorraum der Ur"=Alternativen
($\ref{Dirac-Gleichung_chi8}$) auch invariant unter Transformationen bezüglich der
inneren Symmetrien ist.

\subsection{Der Übergang zum abstrakten Analogon der lokalen Symmetrien}

Um die Beschreibung von Wechselwirkungen zu integrieren, muss in gewöhnlichen
Quantenfeldtheorien eine Symmetrie unter lokalen Transformationen gefordert werden.
In der Quantentheorie der Ur"=Alternativen muss an die Stelle der gewöhnlichen
Wechselwirkungen eine Verschränkung der quantentheoretischen Zustände im Tensorraum
der Ur"=Alternativen der verschiedenen Objekte treten ($\ref{Zustand_N_Objekte_B}$),
welche durch eine erweiterte Symmetrieforderung bezüglich der dynamischen
Grundgleichungen der Zustände im Tensorraum definiert wird. Da hier kein
feldtheoretischer Raum"=Zeit"=Hintergrund vorausgesetzt werden kann, muss an
die Stelle der Abhängigkeit der Transformationsparameter von der Ortskoordinate,
welche in der gewöhnlichen Quantenmechanik den Ortsoperator darstellt, eine
Abhängigkeit der Transformationsparameter von jenen Operatoren im Tensorraum
treten ($\ref{Ort_Impuls_Operatoren}$), welche als Ortsoperatoren interpretiert werden,
weil sie gemeinsam mit den Impulsoperatoren ($\ref{Ort_Impuls_Operatoren}$) die Heisenbergsche
Algebra der Quantenmechanik erfüllen ($\ref{Heisenbergsche_Algebra}$). Diesbezüglich erscheint
es bezüglich der Kompaktheit der Darstellung als sinnvoll, sich auf die Definition
$X_{ABCD}$ gemäß ($\ref{Ortsoperator_ABCD}$) zu beziehen, welche zudem auch
den Zeitparameter enthält. Dies bedeutet, dass für einen beliebigen unitären
Transformationsoperator $U_{\Lambda}\left[\alpha\right]$, der von einem $N$"=Tupel
$\alpha$ an Parametern abhängig ist, die zu den $N_\Lambda$ Generatoren $\Lambda^{a},\ a=1,...,N_\Lambda$
einer Lie"=Algebra gehören, $\alpha=\left(\alpha_1,...,\alpha_{N_\Lambda}\right)$, beziehe diese
sich nun auf den Teilraum der symmetrischen oder der antisymmetrischen Zustände,
folgender Übergang vollzogen werden muss:

\begin{equation}
U_\Lambda\left[\alpha\right]=\exp\left[i\alpha^a \Lambda^a\right]
\longrightarrow U_\Lambda\left[\alpha\left(X_{ABCD}\right)\right]
=\exp\left[i\alpha^a\left(X_{ABCD}\right)\Lambda^a\right],\quad a=1,...,N_\Lambda.
\label{Uebergang_Transformation}
\end{equation}
Sofern sich die Generatoren $\Lambda^a$ auf den antisymmetrischen Teilraum des
Tensorraumes der Ur"=Alternativen beziehen, hat der Übergang zu Parametern
$\alpha\left(X_{ABCD}\right)$, die von dem vierdimensionalen Vektor
$X_{ABCD}$ mit den Ortsoperatoren und dem Zeitparameter abhängen:

\begin{equation}
\alpha \quad\longrightarrow\quad \alpha\left(X_{ABCD}\right),
\end{equation}
zur Folge, dass hier die Forderung nach einer übergreifenden Symmetrie statuiert wird,
deren Transformationen sich auf den gegenüber Permutation symmetrischen Teil des
Tensorraumes der Ur"=Alternativen $\mathcal{H}_{TS}$ und den gegenüber Permutation
antisymmetrischen Teil des Tensorraumes der Ur"=Alternativen $\mathcal{H}_{TAS}$
zugleich beziehen.

\section{Die Wechselwirkungen der Elementarteilchenphysik}

\subsection{Lokale Eichtransformationen im Tensorraum der Ur-Alternativen}

Um nun zu den konkreten Wechselwirkungen des Standardmodells der Elementarteilchenphysik
zu gelangen, muss im Sinne des letzten Kapitels die Symmetrie in Bezug auf die internen
Freiheitsgrade, also den antisymmetrischen Teil des Tensorraumes der Ur"=Alternativen,
wie sie sich in ($\ref{Symmetrie_AS}$) manifestiert, auf eine Symmetrie erweitert werden,
bei welcher die Parameter der Transformationen vom Viererortsoperator $X_{ABCD}$ abhängen.
Im Sinne von ($\ref{Uebergang_Transformation}$) kann man die Transformationsoperatoren
in Bezug auf ($\ref{Transformationsoperator_AS}$) nun verallgemeinern zu:

\begin{equation}
U_\tau\left[\alpha\right]=\exp\left[i\alpha^a \tau_{8x8}^a\right]\quad \longrightarrow \quad
U_\tau\left[\alpha\left(X_{ABCD}\right)\right]=\exp\left[i\alpha^a\left(X_{ABCD}\right)\tau_{8x8}^a\right].
\label{Transformationsoperator_SUN_Ortsoperator}
\end{equation}
Ein Dirac"=Zustand mit internen Freiheitsgraden transformiert sich also wie folgt:

\begin{equation}
|\Psi_\Gamma(t)\rangle\quad \longrightarrow \quad U_\tau\left[\alpha\left(X_{ABCD}\right)\right]|\Psi_\Gamma(t)\rangle.
\label{Transformation_SUN_Ortsoperator}
\end{equation}
Durch eine vom Viererortsoperator $X_{ABCD}$ abhängige Eichtransformation
($\ref{Transformation_SUN_Ortsoperator}$) ergibt sich aus der abstrakten
Dirac"=Gleichung im Tensorraum der Ur"=Alternativen ($\ref{Dirac-Gleichung_chi8}$):

\begin{align}
\gamma_{8x8}^\mu (P_{ABCD})_\mu|\Psi_\Gamma (t)\rangle=0 \quad\longrightarrow\quad
&\gamma_{8x8}^\mu \left(P_{ABCD}\right)_\mu U_\tau\left[\alpha\left(X_{ABCD}\right)\right]|\Psi_\Gamma (t)\rangle
\label{Eichtransformation_mit_Ortsoperator_Dirac-Gleichung}\\
&=\gamma_{8x8}^\mu \left[\left(P_{ABCD}\right)_\mu, U_\tau\left[\alpha\left(X_{ABCD}\right)\right]\right]
|\Psi_\Gamma (t)\rangle\nonumber\\
&\quad+U_\tau\left[\alpha\left(X_{ABCD}\right)\right]\gamma_{8x8}^\mu \left(P_{ABCD}\right)_\mu
|\Psi_\Gamma (t)\rangle\nonumber\\
&=\gamma_{8x8}^\mu \left[\left(P_{ABCD}\right)_\mu, U_\tau\left[\alpha\left(X_{ABCD}\right)\right]\right]
|\Psi_\Gamma (t)\rangle \neq 0.\nonumber
\end{align}
Im letzten Schritt wurde die Dirac"=Gleichung im Tensorraum der Ur"=Alternativen in ihrer
usprünglichen Gestalt verwendet ($\ref{Dirac-Gleichung_chi8}$). Da der Energie"=Impuls"=Operator
im Tensorraum der Ur"=Alternativen $P_{ABCD}$ nicht mit dem Viererortsoperator im Tensorraum
der Ur"=Alternativen $X_{ABCD}$ kommutiert, verschwindet ähnlich zu gewöhnlichen
Eichtheorien der obige Kommutator nicht. Denn gewöhnlich muss diesbezüglich ein Eichfeld
beziehungsweise eine Wellenfunktion eines Vektorbosons eingeführt werden, dass sich in der
Weise transformiert, dass der durch die Lokalität der Transformation entstehende zusätzliche
Term exakt kompensiert wird. In der Quantentheorie der Ur"=Alternativen muss an Stelle eines
Feldes ein weiterer quantenlogischer Zustand aus vielen Ur"=Alternativen eingefügt werden,
der einem weiteren Quantenobjekt entspricht. Dieser muss selbstredend auch aus Zuständen
im gegenüber Permutation symmetrischen beziehungsweise antisymmetrischen Teilraum des
Tensorraumes der Ur"=Alternativen aufgebaut werden. Der einzige Unterschied zu den
gewöhnlichen Dirac"=Zuständen im Tensorraum ($\ref{Dirac-Zustand_chi8}$) besteht darin,
dass dies ein wenig anders geschehen muss, um zu einem Zustand zu gelangen,
der einerseits einen Minkowski"=Vektor darstellt und andererseits ein Element bezüglich
der Lie"=Algebra der jeweiligen $SU(N)$"=Symmetriegruppe als Eichgruppe der entsprechenden
Wechselwirkung. Zunächst muss dazu allerdings die Einbettung der Generatoren
in Bezug auf den antisymmetrischen Teilraum näher betrachtet werden.
Die entsprechenden Zustände im antisymmetrischen Teilraum beziehungsweise in
dessen Unterräumen wurden oben bereits formuliert. Sie sind gegeben durch $|\chi_8\rangle$ ($\ref{chi8}$),
das gemäß ($\ref{Zustand_antisymmetrisch_chi-chi4-chi3}$) aufgespalten werden kann in
$|\chi\rangle$ ($\ref{chi}$), $|\chi_4\rangle$ ($\ref{chi4}$) und $|\chi_3\rangle$ ($\ref{chi3}$).
Ein allgemeiner Zustand im Tensorraum der Ur"=Alternativen $|\Psi_{\Gamma}\left(t\right)\rangle$
wurde in entsprechender Weise in ($\ref{Zustand_gesamt_chi8}$) formuliert, wobei $|\chi_8\rangle$
zum Spin und den inneren Symmetrien führt, und kann als Lösung der abstrakten
Dirac"=Gleichung im Tensorraum mit positiver und negativer Energiekomponente
gemäß ($\ref{Dirac-Zustand_chi8}$) geschrieben werden. Wenn man die Zerlegung des
internen Raumes gemäß ($\ref{Zustand_antisymmetrisch_chi-chi4-chi3}$) betrachtet,
dann kann man ($\ref{Dirac-Zustand_chi8}$) auch als ($\ref{Dirac-Zustand_chi-chi4-chi3}$)
beziehungsweise ($\ref{Dirac-Zustand_chi-chi4-chi3_getrennt}$) schreiben. Um das Ganze
möglichst kompakt darstellen zu können, sei hier zunächst gemäß ($\ref{Dirac-Zustand_chi8}$)
nicht nur der gesamte symmetrische, sondern auch der gesamte antisymmetrische Teilraum
betrachtet. Aus einem Zustand dieser Form muss selbstredend auch der Zustand
des Eichvektorbosons aufgebaut werden, das zugleich Element bezüglich der Lie"=Algebra
der internen $SU(N_\tau)$"=Eichgruppe ist. Dies soll in den sich anschließenden
Betrachtungen geschehen. Die Gesamtsymmetriegruppe ist die $U(8)$"=Gruppe.
Der Teilraum des antisymmetrischen Teilraumes, auf den sich die entsprechende
Wechselwirkung bezieht, gehorcht einer $U(N_\mathcal{G})$"=Symmetrie, wobei gilt, dass
$2 \leq N_\mathcal{G} \leq 8$. Diese $U(N_\mathcal{G})$"=Symmetrie enthält immer eine
bestimmte $U(2)=SU(2)\otimes U(1)$"=Symmetrie als Teilsymmetrie, wobei die $SU(2)$"=Symmetrie
die relative Ausrichtung des Zustandes im antisymmetrischen Teilraum in Bezug auf den Zustand
im symmetrischen Teilraum beschreibt, der seinerseits gemäß ($\ref{Zustand_Darstellung_Ortsraum}$)
in einem reellen dreidimensionalen Raum als Wellenfunktion dargestellt werden kann.
Deshalb kann dieser $SU(2)$"=Freiheitsgrad wie weiter oben bereits ausgeführt mit
dem Spin identifiziert werden, der ja eine Ausrichtung im Ortsraum beschreibt,
während die $U(1)$"=Symmetrie mit der Zeitentwicklung zu tun hat. Die entsprechende
$U(N_\mathcal{G})$"=Gruppe enthält im Allgemeinen gemäß ($\ref{Relation_U4_A}$), ($\ref{Relation_U4_B}$)
und ($\ref{Relation_U3}$) noch weitere $SU(N_\tau)$"=Gruppen, wobei $2 \leq N_\tau \leq N_\mathcal{G}$ gilt,
welche inneren Symmetrien entsprechen und welche als die zu den Wechselwirkungen gehörigen Eichsymmetrien
interpretiert werden. Die $N_\tau^2-1$ Generatoren einer $SU(N_\tau)$"=Untergruppe der $U(N_\mathcal{G})$"=Gruppe
als Eichgruppe sind $N_\tau$"=dimensionale Matrizen, welche in $N_\mathcal{G}$"=dimensionale Matrizen
eingebettet sind, die ihrerseits in achtdimensionale Matrizen eingebettet sind. Sie seien bezeichnet
als $(\tau_{8x8}^1, ...,\tau_{8x8}^{N_\tau})$. Da diese Generatoren, gegebenenfalls neben Generatoren
weiterer Untergruppen, alle zur gesamten $U(N_\mathcal{G})$"=Gruppe gehören,
kann man folgende Definition vornehmen:

\begin{align}
&\mathcal{G}_{8x8}^1=\left(\sigma_{N_\mathcal{G}}\right)_{8x8}^1,\quad \mathcal{G}_{8x8}^2=\left(\sigma_{N_\mathcal{G}}\right)_{8x8}^2,\quad
\mathcal{G}_{8x8}^3=\left(\sigma_{N_\mathcal{G}}\right)_{8x8}^3,\quad \mathcal{G}_{8x8}^4=\left(\sigma_{N_\mathcal{G}}\right)_{8x8}^0,\nonumber\\
&\mathcal{G}_{8x8}^5=\left(\tau_{N_\mathcal{G}}\right)_{8x8}^1=\tau_{8x8}^1,\quad ...,\quad \mathcal{G}^{4+N_\tau^2-1}=\left(\tau_{N_\mathcal{G}}\right)_{8x8}^{N_\tau^2-1}=\tau_{8x8}^{N_\tau^2-1},\quad ...\ .
\label{Generatoren_antisymmetrischer_Tensorraum_insgesamt}
\end{align}
Die Pauli"=Matrizen $\left(\sigma_{N_\mathcal{G}}\right)_{8x8}^1$, $\left(\sigma_{N_\mathcal{G}}\right)_{8x8}^2$, $\left(\sigma_{N_\mathcal{G}}\right)_{8x8}^3$, $\left(\sigma_{N_\mathcal{G}}\right)_{8x8}^0$ sind nicht
exakt diejenigen Pauli"=Matrizen in den Dirac"=Matrizen ($\ref{Dirac-Matrizen_8x8}$). Denn die 
höherdimensionalen Pauli"=Matrizen in den Dirac"=Matrizen ($\ref{Dirac-Matrizen_8x8}$) wirken
in zwei Unterräumen mit $U(N_\mathcal{G})$"=Symmetriegruppe zugleich, $2 \leq N_\mathcal{G} \leq 8$
des achtdimensionalen Spinorraumes der Zustände ($\ref{Dirac-Zustand_chi8}$), während die
Pauli"=Matrizen aus ($\ref{Generatoren_antisymmetrischer_Tensorraum_insgesamt}$) nur im
Unterraum der einen $U(N_\mathcal{G})$"=Symmetriegruppe wirken, die jeweils betrachet wird.
Natürlich müssen auch weiterhin die Dirac"=Matrizen aus ($\ref{Dirac-Matrizen_8x8}$) mit den
Pauli"=Matrizen verwendet werden, die in beiden Unterräumen zugleich wirken. Die $\tau_{8x8}$"=Generatoren
der internen $SU(N_\tau)$"=Untergruppe hingegen wirken grundsätzlich nur in dem entsprechenden Teilraum,
weshalb die Kennzeichnung mit $N_\mathcal{G}$ beiseite gelassen werden kann. Die Punkte hinter den
Generatoren in ($\ref{Generatoren_antisymmetrischer_Tensorraum_insgesamt}$) deuen an, dass die 
$U(N_\mathcal{G})$"=Symmetriegruppe im Fall von $N_\mathcal{G}$=4 noch eine weitere
$SU(N_\tau)$"=Unergruppe enthält. Es sei noch einmal in aller Deutlichkeit darauf hingewiesen, dass die
$\mathcal{G}_{8x8}$"=Generatoren aus ($\ref{Generatoren_antisymmetrischer_Tensorraum_insgesamt}$),
die im antisymmetrischen Teilraum des Tensorraumes der Ur"=Alternativen wirken, sich aus den
entsprechenden Erzeugungs- und Vernichtungsoperatoren im antisymmetrischen Teilraum des
Tensorraumes der Ur"=Alternativen ($\ref{Operatoren_antisymmetrisch}$) bilden.

\subsection{Abstrakte Zustände der Eichvektorbosonen im Tensorraum der Ur-Alternativen}

Ein allgemeiner abstrakter Dirac"=Zustand im Tensorraum der Ur"=Alternativen,
also $|\Psi_{\Gamma}\left(t\right)\rangle$, wie er in ($\ref{Dirac-Zustand_chi8}$)
beziehungsweise in ($\ref{Dirac-Zustand_chi-chi4-chi3}$) und
($\ref{Dirac-Zustand_chi-chi4-chi3_getrennt}$) gegeben ist, kann auf einen
allgemeinen abstrakten Zustand eines Eichvektorbosons mit internem Lie"=Algebra"=Freiheitsgrad
abgebildet werden. Dazu müssen die in ($\ref{Generatoren_antisymmetrischer_Tensorraum_insgesamt}$)
und ($\ref{Dirac-Matrizen_8x8}$) gegebenen mathematischen Größen, die als Operatoren
in Bezug auf den antisymmetrischen Teilraum des Tensorraumes der Ur"=Alternativen
beziehungsweise im Falle von ($\ref{Dirac-Matrizen_8x8}$) zugleich in Bezug auf die
Aufspaltung in eine Komponente zu positiver und zu negativer Energie der Zustände
des symmetrischen Teilraumes des Tensorraumes der Ur"=Alternativen wirken,
verwendet werden. Diese Abbildung kann in der folgenden Weise geschehen:

\begin{equation}
|A_\mu^a(t)\rangle=\langle \Psi_{\Gamma}\left(t\right)|\left[\gamma^{8x8}_0 \gamma^{8x8}_\mu \tau_{8x8}^a\right]
|\Psi_{\Gamma}\left(t\right)\rangle,\quad a=1,...,N_\tau^2-1.
\label{Zustand_Eichvektorboson}
\end{equation}
Hierbei beschreibt $\langle \Psi_{\Gamma}\left(t\right)|$ ein Element des Dualraumes zu
$\mathcal{H}_{T}$. Man kann diese Definition ($\ref{Zustand_Eichvektorboson}$) mit Hilfe der
in ($\ref{Generatoren_antisymmetrischer_Tensorraum_insgesamt}$) und ($\ref{Dirac-Matrizen_8x8}$)
gegebenen Definitionen auch etwas ausführlicher schreiben, sodass sich Folgendes ergibt:

\begin{align}
|A_0^a(t)\rangle=&\left(\langle \Psi_{E}\left(t\right)|_S \otimes \langle\left(\chi_{8}\right)_{E}\left(t\right)|,
\langle\Psi_{-E}\left(t\right)|_S \otimes \langle\left(\chi_{8}\right)_{-E}\left(t\right)|\right)\\
&\cdot\left(\begin{matrix} 0 & \sigma_{8x8}^0 \\ \sigma_{8x8}^0 & 0 \end{matrix}\right)
\cdot\left(\begin{matrix} 0 & \sigma_{8x8}^0 \tau_{8x8}^a\\ \sigma_{8x8}^0
\tau_{8x8}^a & 0 \end{matrix}\right)
\cdot\left(\begin{matrix}|\Psi_{E}\left(t\right)\rangle_S \otimes |\left(\chi_{8}\right)_{E}\left(t\right)\rangle\\
|\Psi_{-E}\left(t\right)\rangle_S \otimes |\left(\chi_{8}\right)_{-E}\left(t\right)\end{matrix}\rangle\right),\nonumber\\
|A_1^a(t)\rangle=&\left(\langle \Psi_{E}\left(t\right)|_S \otimes \langle\left(\chi_{8}\right)_{E}\left(t\right)|,
\langle\Psi_{-E}\left(t\right)|_S \otimes \langle\left(\chi_{8}\right)_{-E}\left(t\right)|\right)\nonumber\\
&\cdot\left(\begin{matrix} 0 & \sigma_{8x8}^0 \\ \sigma_{8x8}^0 & 0 \end{matrix}\right)
\cdot\left(\begin{matrix} 0 & -\sigma_{8x8}^1 \tau_{8x8}^a\\ \sigma_{8x8}^1 \tau_{8x8}^a & 0 \end{matrix}\right)
\cdot\left(\begin{matrix}|\Psi_{E}\left(t\right)\rangle_S \otimes |\left(\chi_{8}\right)_{E}\left(t\right)\rangle\\
|\Psi_{-E}\left(t\right)\rangle_S \otimes |\left(\chi_{8}\right)_{-E}\left(t\right)\rangle\end{matrix}\right),\nonumber\\
|A_2^a(t)\rangle=&\left(\langle \Psi_{E}\left(t\right)|_S \otimes \langle\left(\chi_{8}\right)_{E}\left(t\right)|,
\langle\Psi_{-E}\left(t\right)|_S \otimes \langle\left(\chi_{8}\right)_{-E}\left(t\right)|\right)\nonumber\\
&\cdot\left(\begin{matrix} 0 & \sigma_{8x8}^0 \\ \sigma_{8x8}^0 & 0 \end{matrix}\right)
\cdot\left(\begin{matrix} 0 & -\sigma_{8x8}^2 \tau_{8x8}^a\\ \sigma_{8x8}^2
\tau_{8x8}^a & 0 \end{matrix}\right)
\cdot\left(\begin{matrix}|\Psi_{E}\left(t\right)\rangle_S \otimes |\left(\chi_{8}\right)_{E}\left(t\right)\rangle\\
|\Psi_{-E}\left(t\right)\rangle_S \otimes |\left(\chi_{8}\right)_{-E}\left(t\right)\rangle\end{matrix}\right),\nonumber\\
|A_3^a(t)\rangle=&\left(\langle \Psi_{E}\left(t\right)|_S \otimes \langle\left(\chi_{8}\right)_{E}\left(t\right)|,
\langle\Psi_{-E}\left(t\right)|_S \otimes \langle\left(\chi_{8}\right)_{-E}\left(t\right)|\right)\nonumber\\
&\cdot\left(\begin{matrix} 0 & \sigma_{8x8}^0 \\ \sigma_{8x8}^0 & 0 \end{matrix}\right)
\cdot\left(\begin{matrix} 0 & -\sigma_{8x8}^3 \tau_{8x8}^a\\ \sigma_{8x8}^3
\tau_{8x8}^a & 0 \end{matrix}\right)
\cdot\left(\begin{matrix}|\Psi_{E}\left(t\right)\rangle_S \otimes |\left(\chi_{8}\right)_{E}\left(t\right)\rangle\\
|\Psi_{-E}\left(t\right)\rangle_S \otimes |\left(\chi_{8}\right)_{-E}\left(t\right)\rangle\end{matrix}\right).\nonumber
\end{align}
Dies kann man in der folgenden Weise umschreiben:

\begin{align}
\label{Potential_Komponenten}
|A_0^a(t)\rangle=&\left[\langle \Psi_{E}\left(t\right)|_S \otimes
\langle \left(\chi_{8}\right)_{E}\left(t\right)|\right]
\left[\tau_{8x8}^a\right] \left[|\Psi_{E}\left(t\right)\rangle_S \otimes
|\left(\chi_{8}\right)_{E}\left(t\right)\rangle\right]\\
&+\left[\langle\Psi_{-E}\left(t\right)|_S \otimes \langle\left(\chi_{8}\right)_{-E}\left(t\right)|\right]
\left[\tau_{8x8}^a\right] \left[|\Psi_{-E}\left(t\right)\rangle_S
\otimes |\left(\chi_{8}\right)_{-E}\left(t\right)\rangle\right],\nonumber\\
|A_1^a(t)\rangle=&\left[\langle \Psi_{E}\left(t\right)|_S \otimes \langle \left(\chi_{8}\right)_{E}\left(t\right)|\right]
\left[\sigma_{8x8}^1 \tau_{8x8}^a\right] \left[|\Psi_{E}\left(t\right)\rangle_S \otimes
|\left(\chi_{8}\right)_{E}\left(t\right)\rangle\right]\nonumber\\
&-\left[\langle\Psi_{-E}\left(t\right)|_S \otimes \langle\left(\chi_{8}\right)_{-E}\left(t\right)|\right]
\left[\sigma_{8x8}^1 \tau_{8x8}^a\right] \left[|\Psi_{-E}\left(t\right)\rangle_S \otimes
|\left(\chi_{8}\right)_{-E}\left(t\right)\rangle\right],\nonumber\\
|A_2^a(t)\rangle=&\left[\langle \Psi_{E}\left(t\right)|_S \otimes \langle \left(\chi_{8}\right)_{E}\left(t\right)|\right]
\left[\sigma_{8x8}^2 \tau_{8x8}^a\right] \left[|\Psi_{E}\left(t\right)\rangle_S \otimes
|\left(\chi_{8}\right)_{E}\left(t\right)\rangle\right]\nonumber\\
&-\left[\langle\Psi_{-E}\left(t\right)|_S \otimes \langle\left(\chi_{8}\right)_{-E}\left(t\right)|\right]
\left[\sigma_{8x8}^2 \tau_{8x8}^a\right] \left[|\Psi_{-E}\left(t\right)\rangle_S \otimes
|\left(\chi_{8}\right)_{-E}\left(t\right)\rangle\right],\nonumber\\
|A_3^a(t)\rangle=&\left[\langle \Psi_{E}\left(t\right)|_S \otimes \langle \left(\chi_{8}\right)_{E}\left(t\right)|\right]
\left[\sigma_{8x8}^3 \tau_{8x8}^a\right] \left[|\Psi_{E}\left(t\right)\rangle_S \otimes
|\left(\chi_{8}\right)_{E}\left(t\right)\rangle\right]\nonumber\\
&-\left[\langle\Psi_{-E}\left(t\right)|_S \otimes \langle\left(\chi_{8}\right)_{-E}\left(t\right)|\right]
\left[\sigma_{8x8}^3 \tau_{8x8}^a\right] \left[|\Psi_{-E}\left(t\right)\rangle_S \otimes
|\left(\chi_{8}\right)_{-E}\left(t\right)\rangle\right].\nonumber
\end{align}
Man kann desweiteren noch die Summe über die Produkte der einzelnen Komponenten
bezüglich der Lie"=Algebra der Eichgruppe des abstrakten Zustandes eines Eichvektorbosons
($\ref{Zustand_Eichvektorboson}$) mit den jeweiligen Generatoren der $SU(N_T)$"=Lie"=Gruppe
bilden und damit die folgende Größe definieren:

\begin{equation}
|A_\mu^\tau(t)\rangle=|A_\mu^a(t)\rangle \tau_{8x8}^a.
\label{Zustand_Eichvektorboson_Summierung}
\end{equation}
Da gemäß ($\ref{Klein-Gordon-Gleichung}$) für die positive und die negative Energiekomponente 
des symmetrischen Anteiles eines Dirac"=Zustandes ($\ref{Dirac-Zustand_chi8}$) natürlich
folgende Gleichungen gelten:

\begin{eqnarray}
\left(P_{ABCD}\right)^\nu \left(P_{ABCD}\right)_\nu |\Psi_{E}\left(t\right)\rangle_S&=&0,\quad\nonumber\\
\left(P_{ABCD}\right)^\nu \left(P_{ABCD}\right)_\nu |\Psi_{-E}\left(t\right)\rangle_S&=&0,
\end{eqnarray}
gehorcht ein Zustand $|A_\mu^a(t)\rangle$ definiert in ($\ref{Zustand_Eichvektorboson}$) selbstredend
dem quantenlogischen Analogon zur Wellengleichung im Tensorraum der Ur"=Alternativen:

\begin{equation}
\left(P_{ABCD}\right)^\nu \left(P_{ABCD}\right)_\nu |A_\mu^a(t)\rangle=0.
\end{equation}

\subsection{Abstrakte Dirac-Gleichung mit Wechselwirkungsterm}

Um nun die gemäß ($\ref{Eichtransformation_mit_Ortsoperator_Dirac-Gleichung}$) zunächst nicht gegebene
Symmetrie der abstrakten Dirac"=Gleichung im Tensorraum der Ur"=Alternativen ($\ref{Dirac-Gleichung_chi8}$)
unter Transformationen der Form ($\ref{Transformation_SUN_Ortsoperator}$) zu gewährleisten, kann man die
abstrakte Dirac"=Gleichung durch die Einbindung des Zustandes eines Eichvektorbosons im Tensorraum der
Ur"=Alternativen ($\ref{Zustand_Eichvektorboson}$) erweitern, indem man einen Term hinzuaddiert,
welcher das Tensorprodukt des Zustandes des Eichvektorbosons mit dem Dirac"=Zustand
im Tensorraum der Ur"=Alternativen ($\ref{Dirac-Zustand_chi8}$) enthält:

\begin{equation}
\gamma_{8x8}^\mu \left[\left(P_{ABCD}\right)_\mu+\kappa_A|A_\mu^\tau(t)\rangle\right]|\Psi_\Gamma(t)\rangle=0,
\label{Dirac-Gleichung_Gamma_Wechselwirkung}
\end{equation}
wobei $|A_\mu^\tau(t)\rangle |\Psi_\Gamma(t)\rangle=|A_\mu^\tau(t)\rangle \otimes |\Psi_\Gamma(t)\rangle$
gilt und $\kappa_A$ ein reeller Parameter ist, der einer Kopplungskonstante entspricht, welche natürlich die
Stärke der Wechselwirkung bestimmt. Denn das in der Gleichung ($\ref{Dirac-Gleichung_Gamma_Wechselwirkung}$)
zusätzlich auftauchende Produkt $\kappa_A\left[|A_\mu^\tau(t)\rangle \otimes |\Psi_\Gamma(t)\rangle\right]$
definiert eine quantenlogische Verschränkung zwischen zwei Zuständen im Tensorraum der Ur"=Alternativen.
Wenn man für den Zustand $|A_\mu^\tau(t)\rangle$ bei einer Eichtransformation die folgende
Transformationsregel fordert:

\begin{align}
|A_\mu^\tau(t)\rangle \quad\longrightarrow\quad &U_\tau\left[\alpha\left(X_{ABCD}\right)\right]|A_\mu^\tau(t)\rangle
U_\tau^{\dagger}\left[\alpha\left(X_{ABCD}\right)\right]\nonumber\\
&-\frac{1}{\kappa_A}\left[\left(P_{ABCD}\right)_\mu
U_\tau\left[\alpha\left(X_{ABCD}\right)\right]\right] U_\tau^{\dagger}\left[\alpha\left(X_{ABCD}\right)\right],
\end{align}
dann sieht eine Eichtransformation insgesamt wie folgt aus:

\begin{align}
\label{Eichtransformationen}
|\Psi_\Gamma (t)\rangle \quad\longrightarrow\quad&
U_\tau\left[\alpha\left(X_{ABCD}\right)\right]|\Psi_\Gamma (t)\rangle\nonumber\\
|A_\mu^\tau(t)\rangle \quad\longrightarrow\quad& U_\tau\left[\alpha\left(X_{ABCD}\right)\right]|A_\mu^\tau(t)\rangle
U_\tau^{\dagger}\left[\alpha\left(X_{ABCD}\right)\right]\nonumber\\
&-\frac{1}{\kappa_A}\left[\left(P_{ABCD}\right)_\mu U_\tau\left[\alpha\left(X_{ABCD}\right)\right]\right] U_\tau^{\dagger}\left[\alpha\left(X_{ABCD}\right)\right].
\end{align}
Bei einer Transformation der abstrakten Dirac"=Gleichung im Tensorraum der Ur"=Alternativen
mit Wechselwirkungsterm ($\ref{Dirac-Gleichung_Gamma_Wechselwirkung}$) gemäß den
Eichtransformationen ($\ref{Eichtransformationen}$) ergibt sich:

\begin{align}
\label{Eichtransformation_Dirac-Gleichung_Wechselwirkung}
&\gamma_{8x8}^\mu \left[\left(P_{ABCD}\right)_\mu+\kappa_A|A_\mu^\tau(t)\rangle\right]|\Psi_\Gamma(t)\rangle=0\nonumber\\
&\longrightarrow\quad \gamma_{8x8}^\mu \left\{\left(P_{ABCD}\right)_\mu+\kappa_A U_\tau\left[\alpha\left(X_{ABCD}\right)\right]
|A^\tau_\mu(t)\rangle U_\tau^{\dagger}\left[\alpha\left(X_{ABCD}\right)\right]\right.\nonumber\\
&\left.\quad\quad\quad\quad-\left[\left(P_{ABCD}\right)_\mu U_\tau\left[\alpha\left(X_{ABCD}\right)\right]\right]
U_\tau^{\dagger}\left[\alpha\left(X_{ABCD}\right)\right]\right\}
U_\tau\left[\alpha\left(X_{ABCD}\right)\right]|\Psi_\Gamma(t)\rangle\nonumber\\
&\quad\quad\quad=\gamma_{8x8}^\mu \left\{U_\tau\left[\alpha\left(X_{ABCD}\right)\right]\left(P_{ABCD}\right)_\mu
+\left[\left(P_{ABCD}\right)_\mu U_\tau\left[\alpha\left(X_{ABCD}\right)\right]\right]\right.\nonumber\\
&\left.\quad\quad\quad\quad\quad
+\kappa_A U_\tau\left[\alpha\left(X_{ABCD}\right)\right]|A^\tau_\mu(t)\rangle
U_\tau^{\dagger}\left[\alpha\left(X_{ABCD}\right)\right]U_\tau\left[\alpha\left(X_{ABCD}\right)\right]\right.\nonumber\\
&\left.\quad\quad\quad\quad\quad
-\left[\left(P_{ABCD}\right)_\mu U_\tau\left[\alpha\left(X_{ABCD}\right)\right]\right]
U_\tau^{\dagger}\left[\alpha\left(X_{ABCD}\right)\right]
U_\tau\left[\alpha\left(X_{ABCD}\right)\right]\right\}|\Psi_\Gamma(t)\rangle\nonumber\\
&\quad\quad\quad=\gamma_{8x8}^\mu \left\{U_\tau\left[\alpha\left(X_{ABCD}\right)\right] \left(P_{ABCD}\right)_\mu
+\left[\left(P_{ABCD}\right)_\mu U_\tau\left[\alpha\left(X_{ABCD}\right)\right]\right]
\right.\nonumber\\
&\left.\quad\quad\quad\quad\quad
+\kappa_A U_\tau\left[\alpha\left(X_{ABCD}\right)\right]|A^\tau_\mu(t)\rangle
-\left[\left(P_{ABCD}\right)_\mu U_\tau\left[\alpha\left(X_{ABCD}\right)\right]\right]\right\}|\Psi_\Gamma(t)\rangle\nonumber\\
&\quad\quad\quad=U_\tau\left[\alpha\left(X_{ABCD}\right)\right]
\gamma_{8x8}^\mu \left[\left(P_{ABCD}\right)_\mu
+\kappa_A|A_\mu^\tau(t)\rangle\right]|\Psi_\Gamma(t)\rangle=0.
\end{align}
In ($\ref{Eichtransformation_Dirac-Gleichung_Wechselwirkung}$) wurde im
zweiten Umformungsschritt nach der Eichtransformation verwendet, dass
$U_\tau^{\dagger}\left[\alpha\left(X_{ABCD}\right)\right]=U_\tau^{-1}\left[\alpha\left(X_{ABCD}\right)\right]$ und damit
$U_\tau^{\dagger}\left[\alpha\left(X_{ABCD}\right)\right]U_\tau\left[\alpha\left(X_{ABCD}\right)\right]=\mathbf{1}$ gilt.
Durch das Einfügen des Eichvektorbosonenzustandes im Tensorraum der Ur"=Alternativen ($\ref{Zustand_Eichvektorboson}$)
in die quantenlogische Dirac"=Gleichung ($\ref{Dirac-Gleichung_Gamma_Wechselwirkung}$) ist die gemäß
($\ref{Eichtransformation_mit_Ortsoperator_Dirac-Gleichung}$) in ($\ref{Dirac-Gleichung_chi8}$)
zunächst gebrochene Symmetrie unter von den quantenlogischen Viererortsoperatoren $X_{ABCD}$
abhängigen Eichtransformationen ($\ref{Transformation_SUN_Ortsoperator}$) wieder hergestellt.
Damit ist also gezeigt, dass die abstrakte Dirac"=Gleichung im Tensorraum der Ur"=Alternativen
mit Wechselwirkungsterm in der allgemeinen Gestalt ($\ref{Dirac-Gleichung_Gamma_Wechselwirkung}$)
invariant unter Eichtransformationen der Gestalt ($\ref{Eichtransformationen}$) ist,
welche sich auf den antisymmetrischen Teilraum des Tensorraumes beziehen und durch
die Abhängigkeit von den abstrakten quantenlogischen Viererortsoperatoren $X_{ABCD}$
zugleich mit dem symmetrischen Teil des Tensorraumes verbunden sind. Da der Eichvektorbosonenzustand
($\ref{Zustand_Eichvektorboson}$) auch aus gewöhnlichen Tensorraumzuständen konstruiert wird,
handelt es sich hierbei in gewisser Weise um eine Art der Selbstwechselbeziehung und damit der
Verschränkung einer einzigen Art von Zuständen im Tensorraum der Ur"=Alternativen. Dies hat
zur Folge, dass die abstrakte Dirac"=Gleichung im Tensorraum mit Selbstwechselwirkungsterm
($\ref{Dirac-Gleichung_Gamma_Wechselwirkung}$) beim Übergang in die Raum"=Zeit"=Darstellung
gemäß ($\ref{Energieoperator_Darstellung}$), ($\ref{Ort_Impuls_Darstellung}$) und
($\ref{Zustand_Darstellung_Raum-Zeit}$) der fundamentalen Heisenbergschen Feldgleichung
in der nichtlinearen Spinorfeldtheorie bezüglich der grundlegenden Form erstaunlich ähnlich ist.
Wenn man die folgende Definition vornimmt:

\begin{equation}
\left(\mathcal{D}_{ABCD}^\tau\right)_\mu=\left(P_{ABCD}\right)_\mu+\kappa_A|A_\mu^\tau(t)\rangle,
\label{Kovarianter_Impulsoperator}
\end{equation}
dann kann man die abstrakte Dirac"=Gleichung mit Wechselwirkungsterm im Tensorraum
der Ur"=Alternativen ($\ref{Dirac-Gleichung_Gamma_Wechselwirkung}$) auch in der
folgenden Weise formulieren:

\begin{equation}
\gamma_{8x8}^\mu \left(\mathcal{D}_{ABCD}^\tau\right)_\mu |\Psi_\Gamma(t)\rangle=0.
\end{equation}
Der kovariante Energie"=Impuls"=Operator ($\ref{Kovarianter_Impulsoperator}$) transformiert
sich unter den Eichtransformationen der in ($\ref{Eichtransformationen}$) gegebenen Form
in der folgenden Weise:

\begin{equation}
\left(\mathcal{D}_{ABCD}^\tau\right)_\mu \quad \longrightarrow \quad
U_\tau\left[\alpha\left(X_{ABCD}\right)\right] \left(\mathcal{D}_{ABCD}^\tau\right)_\mu
U_\tau^{\dagger}\left[\alpha\left(X_{ABCD}\right)\right].
\label{Kovarianter_Impulsoperator_Transformation}
\end{equation}
Dies wiederum kann in der folgenden Weise gezeigt werden:

\begin{align}
&\left(\mathcal{D}_{ABCD}^\tau\right)_\mu=\left(P_{ABCD}\right)_\mu \mathbf{1}+\kappa_A|A_\mu^\tau(t)\rangle
\nonumber\\&\longrightarrow \left(P_{ABCD}\right)_\mu U_\tau\left[\alpha\left(X_{ABCD}\right)\right]
U_\tau^{\dagger}\left[\alpha\left(X_{ABCD}\right)\right]
+\kappa_A U_\tau\left[\alpha\left(X_{ABCD}\right)\right]|A_\mu^\tau(t)\rangle
U_\tau^{\dagger}\left[\alpha\left(X_{ABCD}\right)\right]
\nonumber\\&\quad\quad
-\left[\left(P_{ABCD}\right)_\mu U_\tau\left[\alpha\left(X_{ABCD}\right)\right]\right]
U_\tau^{\dagger}\left[\alpha\left(X_{ABCD}\right)\right]
\nonumber\\&\quad\quad
=\left[\left(P_{ABCD}\right)_\mu U_\tau\left[\alpha\left(X_{ABCD}\right)\right]\right] U_\tau^{\dagger}\left[\alpha\left(X_{ABCD}\right)\right]
+U_\tau\left[\alpha\left(X_{ABCD}\right)\right] 
\left(P_{ABCD}\right)_\mu U_\tau^{\dagger}\left[\alpha\left(X_{ABCD}\right)\right]
\nonumber\\&\quad\quad\quad
+\kappa_A U_\tau\left[\alpha\left(X_{ABCD}\right)\right]|A_\mu^\tau(t)\rangle
U_\tau^{\dagger}\left[\alpha\left(X_{ABCD}\right)\right]
-\left[\left(P_{ABCD}\right)_\mu U_\tau\left[\alpha\left(X_{ABCD}\right)\right]\right]
U_\tau^{\dagger}\left[\alpha\left(X_{ABCD}\right)\right]
\nonumber\\&\quad\quad
=U_\tau\left[\alpha\left(X_{ABCD}\right)\right] 
\left[\left(P_{ABCD}\right)_\mu U_\tau^{\dagger}\left[\alpha\left(X_{ABCD}\right)\right]\right]
+\kappa_A U_\tau\left[\alpha\left(X_{ABCD}\right)\right]|A_\mu^\tau(t)\rangle
U_\tau^{\dagger}\left[\alpha\left(X_{ABCD}\right)\right]
\nonumber\\&\quad\quad
=U_\tau\left[\alpha\left(X_{ABCD}\right)\right]\left[\left(P_{ABCD}\right)_\mu
+\kappa_A|A_\mu^\tau(t)\rangle\right]U_\tau^{\dagger}\left[\alpha\left(X_{ABCD}\right)\right]
\nonumber\\&\quad\quad
=U_\tau\left[\alpha\left(X_{ABCD}\right)\right]\left(\mathcal{D}_{ABCD}^\tau\right)_\mu
U_\tau^{\dagger}\left[\alpha\left(X_{ABCD}\right)\right],
\label{Beweis_Kovarianter_Impulsoperator_Transformation}
\end{align}
wobei $\mathbf{1}$ hier den Einsoperator beziehungsweise die Einheitsmatrix in Bezug auf den
antisymmetrischen Teilraum beschreibt, dessen Zustände durch $|\chi_8\rangle$ ($\ref{chi8}$) gegeben sind,
und $U_\tau\left[\alpha\left(X_{ABCD}\right)\right]U_\tau^{\dagger}\left[\alpha\left(X_{ABCD}\right)\right]
=U_\tau\left[\alpha\left(X_{ABCD}\right)\right]U_\tau^{-1}\left[\alpha\left(X_{ABCD}\right)\right]
=\mathbf{1}$ benutzt wurde. Wenn man nun mit Hilfe von ($\ref{Energieoperator_Darstellung}$),
($\ref{Ort_Impuls_Darstellung}$) und ($\ref{Zustand_Darstellung_Raum-Zeit}$) den Übergang in
die Raum"=Zeit"=Darstellung vornimmt, so ergibt sich die folgende Gleichung:

\begin{equation}
i\gamma_{8x8}^\mu\left[\partial_\mu+i\kappa_A\left(A_{\mu}^\tau\right)_{N2}\left(\mathbf{x},t\right)\right]
\left(\Psi_{\Gamma}\right)_{N1}\left(\mathbf{x},t\right)=0.
\label{Dirac-Gleichung_Gamma_Wechselwirkung_Raum-Zeit-Darstellung}
\end{equation}

\subsection{Dynamik des Zustandes der Eichvektorbosonen}

Es kann aus dem Zustand des Eichvektorbosons im Tensorraum der Ur"=Alternativen
($\ref{Zustand_Eichvektorboson}$) ein Zustand konstruiert werden, der das
abstrakte rein quantentheoretische Analogon zu einem Feldstärketensor bezüglich
des Eichvektorbosons darstellt. Dessen Komponenten werden über den Kommutator
der Komponenten des kovarianten Energie"=Impuls"=Operators ($\ref{Kovarianter_Impulsoperator}$)
in Bezug auf die Eichtransformation ($\ref{Eichtransformationen}$) im Tensorraum
der Ur"=Alternativen formuliert:

\begin{eqnarray}
\label{Analogon_Feldstaerketensor}
|F_{\mu\nu}^\tau(t)\rangle&=&|F_{\mu\nu}^a(t)\rangle \tau^a
=\frac{1}{\kappa_A}\left[\left(\mathcal{D}_{ABCD}^\tau\right)_\mu,\left(\mathcal{D}_{ABCD}^\tau\right)_\nu\right]\\
&=&\left[\left(P_{ABCD}\right)_\mu |A_\nu^\tau(t)\rangle-\left(P_{ABCD}\right)_\nu |A_\mu^\tau(t)\rangle
+\kappa_A|A_\mu^\tau(t)\rangle |A_\nu^\tau(t)\rangle-\kappa_A|A_\nu^\tau(t)\rangle |A_\mu^\tau(t)\rangle\right]\nonumber\\
&=&\left[\left(P_{ABCD}\right)_\mu |A_\nu^{a}(t)\rangle \tau^a
-\left(P_{ABCD}\right)_\nu |A_\mu^{a}(t)\rangle \tau^a
+\kappa_A|A_\mu^{a}(t)\rangle |A_\nu^{b}(t)\rangle \left[\tau^a,\tau^b\right]\right]\nonumber\\
&=&\left[\left(P_{ABCD}\right)_\mu |A_\nu^{a}(t)\rangle \tau^a
-\left(P_{ABCD}\right)_\nu |A_\mu^{a}(t)\rangle \tau^a
+i\kappa_A f^{abc}|A_\mu^{a}(t)\rangle |A_\nu^{b}(t)\rangle \tau^c \right].\nonumber
\end{eqnarray}
Aufgrund der Jacobi"=Identität gilt damit, als rein quantenlogisches Analogon zur gewöhnlichen
Formulierung im Rahmen relativistischer Quantenfeldtheorien, die folgende dynamische
Grundgleichung für den Eichvektorbosonenzustand ($\ref{Zustand_Eichvektorboson}$):

\begin{equation}
\left[\left(\mathcal{D}_{ABCD}^\tau\right)_\mu, |F_{\nu\rho}^\tau(t)\rangle\right]
+\left[\left(\mathcal{D}_{ABCD}^\tau\right)_\nu, |F_{\rho\mu}^\tau(t)\rangle\right]
+\left[\left(\mathcal{D}_{ABCD}^\tau\right)_\rho, |F_{\mu\nu}^\tau(t)\rangle\right]=0.
\label{Jacobi_Identitaet_Dynamik}
\end{equation}
Ein Zustand im Tensorraum der Ur"=Alternativen der speziellen in ($\ref{Analogon_Feldstaerketensor}$)
konstruierten Form transformiert sich unter den in ($\ref{Eichtransformationen}$) gegebenen
Eichtransformationen in folgender Weise:

\begin{align}
&|F_{\mu\nu}^\tau(t)\rangle=\frac{1}{\kappa_A}
\left[\left(\mathcal{D}_{ABCD}^\tau\right)_\mu,\left(\mathcal{D}_{ABCD}^\tau\right)_\nu\right]
\nonumber\\ &\longrightarrow \frac{1}{\kappa_A}
\left[U_\tau\left[\alpha\left(X_{ABCD}\right)\right]\left(\mathcal{D}_{ABCD}^\tau\right)_\mu
U^{\dagger}_\tau\left[\alpha\left(X_{ABCD}\right)\right], U_\tau\left[\alpha\left(X_{ABCD}\right)\right]
\left(\mathcal{D}_{ABCD}^\tau\right)_\nu U^{\dagger}_\tau\left[\alpha\left(X_{ABCD}\right)\right]\right]\nonumber\\
&\quad\quad=\frac{1}{\kappa_A} U_\tau\left[\alpha\left(X_{ABCD}\right)\right]\left[\left(\mathcal{D}_{ABCD}^\tau\right)_\mu,
\left(\mathcal{D}_{ABCD}^\tau\right)_\nu\right] U_\tau^{\dagger}\left[\alpha\left(X_{ABCD}\right)\right]\nonumber\\
&\quad\quad=U_\tau\left[\alpha\left(X_{ABCD}\right)\right]|F_{\mu\nu}^\tau(t)\rangle
U^{\dagger}_\tau\left[\alpha\left(X_{ABCD}\right)\right].
\end{align}
Aufgrund der erneuten Bedingung der Eichinvarianz unter den in ($\ref{Eichtransformationen}$) gegebenen
Eichtransformationen kann eine entsprechende quantenlogische Wirkung formuliert werden, aus der sich
die dynamischen Grundgleichungen für den Dirac"=Zustand ($\ref{Dirac-Gleichung_Gamma_Wechselwirkung}$)
und den Eichvektorbosonenzustand ergeben. Diese lautet wie folgt:

\begin{equation}
S_{YMQ}=\int dt \left[\langle \Psi_\Gamma(t)|i \gamma_{8x8}^\mu \left(\mathcal{D}_{ABCD}^\tau\right)_\mu |\Psi_\Gamma(t)\rangle
+\frac{1}{4}\left[|F^{\mu\nu a}(t)\rangle |F_{\mu\nu}^a(t)\rangle\right]\right].
\label{Wirkung_Yang-Mills}
\end{equation}
Eine infinitesimale Transformation einer Komponente des Eichvektorbosonenzustandes
($\ref{Zustand_Eichvektorboson}$) der Form ($\ref{Eichtransformationen}$),
also unter der Bedingung, dass $\alpha\left(X_{ABCD}\right)\longrightarrow 0$,
ist gegeben durch:

\begin{equation}
|A_\mu^a(t)\rangle \quad\longrightarrow\quad |A_\mu^a(t)\rangle
-\frac{i}{\kappa_A}\left[\left(P_{ABCD}\right)_\mu \alpha^a\left(X_{ABCD}\right)\right]
+f^{abc}|A_\mu^b(t)\rangle \alpha^c\left(X_{ABCD}\right).
\end{equation}
Dies führt bezüglich einer infinitesimalen Transformation einer Komponente des
quantenlogischen Analogons eines Feldstärketensor ($\ref{Analogon_Feldstaerketensor}$)
der Form ($\ref{Eichtransformationen}$) zu Folgendem:

\begin{equation}
|F_{\mu\nu}^a(t) \rangle \quad\longrightarrow\quad f^{abc}|A_\mu^b(t)\rangle \alpha^c\left(X_{ABCD}\right). 
\end{equation}
In dieser Weise kann man also sehen, dass ($\ref{Wirkung_Yang-Mills}$) invariant unter Eichtransformationen
der Form ($\ref{Eichtransformationen}$) ist. Damit ergibt sich neben ($\ref{Jacobi_Identitaet_Dynamik}$)
die folgende weitere dynamische Grundgleichung für die Komponenten des Eichvektorbosonenzustandes
($\ref{Zustand_Eichvektorboson}$) mit einer Beziehung zu einem Dirac"=Zustand im Tensorraum der
Ur"=Alternativen ($\ref{Dirac-Zustand_chi8}$):

\begin{equation}
\label{Dynamik_Eichvektorbosonenzustand}
i\left(P_{ABCD}\right)^\mu |F_{\mu\nu}^a(t)\rangle
+\kappa_A f^{abc} |A^{\mu b}(t)\rangle |F_{\mu\nu}^c(t)\rangle
=\kappa_A \langle\Psi_\Gamma(t)|\gamma^{8x8}_\nu \tau_{8x8}^a|\Psi_\Gamma (t)\rangle,
\end{equation}
wobei der Dirac"=Zustand im Tensorraum der Ur"=Alternativen auf der rechten Seite natürlich
von dem Zustand im Tensorraum der Ur"=Alternativen zu unterscheiden ist, aus dem der abstrakte
Eichvektorbosonenzustand gemäß ($\ref{Zustand_Eichvektorboson}$) gebildet ist. Über diese
dynamische Gleichung ($\ref{Dynamik_Eichvektorbosonenzustand}$) wird vielmehr eine Verschränkung
des Eichvektorbosonenzustandes mit dem anderen Dirac"=Zustand induziert. Gemäß der Definition
($\ref{Analogon_Feldstaerketensor}$) kann ($\ref{Dynamik_Eichvektorbosonenzustand}$)
umgeschrieben werden zu folgender dynamischer Gleichung:

\begin{align}
\label{Dynamik_Eichvektorbosonzustand}
&i\left(P_{ABCD}\right)^\mu |F_{\mu\nu}^a(t)\rangle
+\kappa_A f^{abc} |A^{\mu b}(t)\rangle |F_{\mu\nu}^c(t)\rangle
=\kappa_A \langle\Psi_\Gamma(t)|\gamma^{8x8}_\nu \tau_{8x8}^a|\Psi_\Gamma (t)\rangle
\nonumber\\&\Leftrightarrow\
i\left(P_{ABCD}\right)^\mu \left[i\left(P_{ABCD}\right)_\mu |A_\nu^{a}(t)\rangle
-i\left(P_{ABCD}\right)_\nu |A_\mu^{a}(t)\rangle
+\kappa_A f^{abc}|A_\mu^{b}(t)\rangle |A_\nu^{c}(t)\rangle \right]
\nonumber\\&\quad\quad
+\kappa_A f^{abc} |A^{\mu b}(t)\rangle \left[i\left(P_{ABCD}\right)_\mu |A_\nu^{c}(t)\rangle
-i\left(P_{ABCD}\right)_\nu |A_\mu^{c}(t)\rangle
+\kappa_A f^{cde}|A_\mu^{d}(t)\rangle |A_\nu^{e}(t)\rangle \right]
\nonumber\\&\quad\quad
=\kappa_A \langle\Psi_\Gamma(t)|\gamma^{8x8}_\nu \tau_{8x8}^a|\Psi_\Gamma (t)\rangle.
\end{align}
Man kann die unter Eichtransformationen der in ($\ref{Eichtransformationen}$) gegebenen
Form invariante dynamische Gleichung ($\ref{Dynamik_Eichvektorbosonzustand}$) nun mit
Hilfe von ($\ref{Energieoperator_Darstellung}$), ($\ref{Ort_Impuls_Darstellung}$) und
($\ref{Zustand_Darstellung_Raum-Zeit}$) in die Raum"=Zeit abbilden:

\begin{align}
&\partial^\mu \left(F_{\mu\nu}^a\right)_{N2}\left(\mathbf{x},t\right)
+\kappa_A f^{abc} \left(A^{\mu b}\right)_{N2}\left(\mathbf{x},t\right)
\left(F_{\mu\nu}^c\right)_{N2}\left(\mathbf{x},t\right)
=\kappa_A \left(\bar \Psi_\Gamma\right)_{N1}\left(\mathbf{x},t\right)
\gamma^{8x8}_\nu \tau_{8x8}^a\left(\Psi_\Gamma\right)_{N1}\left(\mathbf{x},t\right)
\nonumber\\&\Leftrightarrow\
\partial^\mu \left[\partial_\mu \left(A_\nu^{a}\right)_{N2}\left(\mathbf{x},t\right)
-\partial_\nu \left(A_\mu^{a}\right)_{N2}(\mathbf{x},t)
+\kappa_A f^{abc}\left(A_\mu^{b}\right)_{N2}\left(\mathbf{x},t\right)
\left(A_\nu^{c}\right)_{N2}\left(\mathbf{x},t\right)\right]
\nonumber\\&
+\kappa_A f^{abc} \left(A^{\mu b}\right)_{N2}\left(\mathbf{x},t\right)
\left[\partial_\mu \left(A_\nu^{c}\right)_{N2}\left(\mathbf{x},t\right)
-\partial_\nu\left(A_\mu^{c}\right)_{N2}\left(\mathbf{x},t\right)
+\kappa_A f^{cde}\left(A_\mu^{d}\right)_{N2}\left(\mathbf{x},t\right)
\left(A_\nu^{e}\right)_{N2}\left(\mathbf{x},t\right)\right]
\nonumber\\&
=\kappa_A \left(\bar \Psi_\Gamma\right)_{N1}\left(\mathbf{x},t\right)\gamma^{8x8}_\nu
\tau_{8x8}^a\left(\Psi_\Gamma\right)_{N1}\left(\mathbf{x},t\right),
\label{Dynamik_Eichvektorbosonzustand_Raum-Zeit-Darstellung}
\end{align}
wobei $\left(\bar \Psi_\Gamma\right)_{N1}\left(\mathbf{x},t\right)$ die zu
$\left(\Psi_\Gamma\right)_{N1}\left(\mathbf{x},t\right)$ adjungierte Größe
darstellt. Die Indizes mit den Gesamtbesetzungszahlen $N1$ und $N2$ sind deshalb
voneinander unterschieden, weil der Dirac"=Zustand ($\ref{Dirac-Zustand_chi8}$)
und der Eichvektorbosonenzustand ($\ref{Zustand_Eichvektorboson}$) voneinander
unterschieden und durch zwei unterschiedliche Besetzungszahlen und damit
auch durch zwei unterschiedliche Gesamtinformationsmengen charakterisiert sind.
Genaugenommen müssen, wenn man die Selbstwechselwirkung beziehungsweise
Verschränkung verschiedener Eichvektorbosonen untereinander betrachtet,
bei einem Produkt mit sich selbst die verschiedenen Einzelzustände in
($\ref{Dynamik_Eichvektorbosonzustand_Raum-Zeit-Darstellung}$) auch mit
unterschiedlichen Besetzungszahlen gekennzeichnet werden. Entscheidend
ist weiter, dass der Zustand eines Eichvektorbosons ($\ref{Zustand_Eichvektorboson}$)
wie die Zustände der Fermionen auch aus gewöhnlichen Dirac"=Zuständen im Tensorraum der
Ur"=Alternativen gebildet ist ($\ref{Dirac-Zustand_chi8}$), die einen gegenüber
Permutationen symmetrischen und antisymmetrischen Anteil sowie eine Komponente
zu positiver und negativer Energie aufweisen. Wenn man in der abstrakten
Dirac"=Gleichung im Tensorraum der Ur"=Alternativen mit Wechselwirkungsterm
($\ref{Dirac-Gleichung_Gamma_Wechselwirkung}$) die Definitionen
($\ref{Zustand_Eichvektorboson}$) und ($\ref{Zustand_Eichvektorboson_Summierung}$)
verwendet und die in ($\ref{Zustand_Eichvektorboson}$) auftauchenden Dirac"=Zustände
mit einem Index von dem bereits in ($\ref{Dirac-Gleichung_Gamma_Wechselwirkung}$)
auftauchenden Dirac"=Zustand unterscheidet, so ergibt sich der folgende Ausdruck
mit einem Selbstwechselwirkungsterm:

\begin{align}
&\gamma_{8x8}^\mu \left[\left(P_{ABCD}\right)_\mu
+\kappa_A \langle \Psi_{\Gamma}\left(t\right)|_2 \gamma^{8x8}_0 \gamma^{8x8}_\mu
\tau_{8x8}^a|\Psi_{\Gamma}\left(t\right)\rangle_2 \tau_{8x8}^a\right]|\Psi_\Gamma(t)\rangle_1=0,
\label{Dirac-Gleichung_Gamma_Wechselwirkung_nur_Dirac-Zustaende}\\
&\Leftrightarrow\quad \gamma_{8x8}^\mu \left(P_{ABCD}\right)_\mu |\Psi_\Gamma(t)\rangle_1
+\kappa_A \langle \Psi_{\Gamma}\left(t\right)|_2 \gamma^{8x8}_0 \gamma^{8x8}_\mu
\tau_{8x8}^a|\Psi_{\Gamma}\left(t\right)\rangle_2 \gamma_{8x8}^\mu \tau_{8x8}^a |\Psi_\Gamma(t)\rangle_1=0,
\nonumber
\end{align}
wobei die beiden auftauchenden Dirac"=Zustände der Form ($\ref{Dirac-Zustand_chi8}$)
mit Zahlen nummeriert sind, um sie voneinander zu unterscheiden. Mit Hilfe von
($\ref{Energieoperator_Darstellung}$), ($\ref{Ort_Impuls_Darstellung}$) und
($\ref{Zustand_Darstellung_Raum-Zeit}$) kann die dynamische Gleichung
($\ref{Dirac-Gleichung_Gamma_Wechselwirkung_nur_Dirac-Zustaende}$)
dann in der Raum"=Zeit dargestellt werden:

\begin{equation}
\gamma_{8x8}^\mu \partial_\mu \left(\Psi_{\Gamma}\right)_{N1}\left(\mathbf{x},t\right)
+i\kappa_A\left[\left(\bar \Psi_{\Gamma}\right)_{N2}\left(\mathbf{x},t\right)
\left[\gamma^{8x8}_0 \gamma^{8x8}_\mu \tau_{8x8}^a\right]\left(\Psi_{\Gamma}\right)_{N2}\left(\mathbf{x},t\right)\right]
\gamma_{8x8}^\mu \tau_{8x8}^a \left(\Psi_{\Gamma}\right)_{N1}\left(\mathbf{x},t\right)=0.
\label{Dirac-Gleichung_Gamma_Wechselwirkung_nur_Dirac-Zustaende_raum-zeitliche_Darstellung}
\end{equation}
Diese dynamische Gleichung weist bezüglich ihrer grundlegenden Gestalt beinahe eine zur
Grundgleichung der Heisenbergschen nichtlinearen Spinorfeldtheorie \cite{Heisenberg:1967}
analoge Gestalt auf, deren Urfeld $\psi\left(\mathbf{x},t\right)$ einen Spin- und einen
Isospinfreiheitsgrad enthält, und aus dem Heisenberg die Existenz aller Objekte
und deren Eigenschaften herzuleiten trachtete:

\begin{equation}
\gamma^\mu \partial_\mu \psi\left(\mathbf{x},t\right) \pm l^2 \gamma^\mu \gamma^5
\psi\left(\mathbf{x},t\right)\left[\bar \psi\left(\mathbf{x},t\right)
\gamma_\mu \gamma_5 \psi\left(\mathbf{x},t\right)\right]=0,
\label{Heisenberg_einheitliche_Spinorquantenfeldtheorie}
\end{equation}
wobei die $\gamma^\mu,\ \mu=0,1,2,3$, hier die gewöhnlichen Dirac"=Matrizen
beschreiben, $l$ eine fundamentale Längeneinheit beschreibt und desweiteren
gilt: $\gamma^5=i\gamma^0 \gamma^1 \gamma^2 \gamma^3$. Aber natürlich
muss in aller Deutlichkeit daran erinnert werden, dass die Gleichung
($\ref{Dirac-Gleichung_Gamma_Wechselwirkung_nur_Dirac-Zustaende}$)
sich auf reine Quantenlogik bezieht, während die Heisenbergsche Gleichung
($\ref{Heisenberg_einheitliche_Spinorquantenfeldtheorie}$) eine nachträglich
einer Quantisierung unterworfene Feldgleichung darstellt. Denn man muss bei all
diesen Betrachtungen immer wieder beachten, dass die raum"=zeitliche Darstellung in
($\ref{Dirac-Gleichung_Gamma_Wechselwirkung_nur_Dirac-Zustaende_raum-zeitliche_Darstellung}$)
in keiner Weise eine Rückkehr ins feldtheoretische Denken bedeutet, weil die
entsprechenden darin enthaltenen abstrakten Zustände durch Ur"=Alternativen und
damit durch reine Quantenlogik definiert bleiben und lediglich raum"=zeitlich
dargestellt werden.

\subsection{Zeitentwicklung und Verschränkung der Zustände}

Um die Zeitentwicklung des Zustandes $|\Psi_\Gamma(t)\rangle$ eines Quantenobjektes
explizit betrachten zu können, muss man von der abstrakten Dirac"=Gleichung mit
Wechselwirkungsterm ausgehen ($\ref{Dirac-Gleichung_Gamma_Wechselwirkung}$)
und diese in die Gestalt der allgemeinen Schrödingergleichung bringen:

\begin{align}
&\gamma_{8x8}^\mu \left[\left(P_{ABCD}\right)_\mu+\kappa_A|A_\mu^\tau(t)\rangle\right]|\Psi_\Gamma(t)\rangle=0\nonumber\\
&\Leftrightarrow\quad \left[\gamma_{8x8}^0 \left(P_{ABCD}\right)_0+\gamma_{8x8}^i \left(P_{ABCD}\right)_i
+\kappa_A \gamma^\mu |A_\mu^\tau(t)\rangle\right]|\Psi_\Gamma(t)\rangle=0\nonumber\\
&\Leftrightarrow\quad i\partial_t|\Psi_\Gamma(t)\rangle=\left[\gamma_{8x8}^0\gamma_{8x8}^i \left(P_{ABCD}\right)_i
+\kappa_A \gamma_{8x8}^0\gamma_{8x8}^\mu |A_\mu^\tau(t)\rangle\right]|\Psi_\Gamma(t)\rangle.
\label{Dirac-Schroedinger-Gleichung_Wechselwirkung}
\end{align}
Wenn man die in ($\ref{Dirac-Schroedinger-Gleichung_Wechselwirkung}$) gegebene dynamische
Gleichung mit ($\ref{Energieoperator_Darstellung}$), ($\ref{Ort_Impuls_Darstellung}$) und ($\ref{Zustand_Darstellung_Raum-Zeit}$) in die Raum"=Zeit abbildet, so ergibt sich die
folgende Gleichung:

\begin{align}
&\gamma_{8x8}^\mu\left[\partial_\mu+i\kappa_A\left(A_\mu^\tau\right)_N\left(\mathbf{x},t\right)\right]
\left(\Psi_\Gamma\right)_N\left(\mathbf{x},t\right)=0\nonumber\\
&\Leftrightarrow\quad \left[\gamma_{8x8}^0\partial_t+\gamma_{8x8}^i \partial_i
+i\kappa_A\gamma_{8x8}^\mu\left(A_\mu^\tau\right)_N\left(\mathbf{x},t\right)\right]
\left(\Psi_\Gamma\right)_N\left(\mathbf{x},t\right)=0\nonumber\\
&\Leftrightarrow\quad i\partial_t \left(\Psi_\Gamma\right)_N\left(\mathbf{x},t\right)
=-i\left[\gamma_{8x8}^0 \gamma_{8x8}^i \partial_i+i\kappa_A \gamma_{8x8}^0 \gamma_{8x8}^\mu
\left(A_\mu^\tau\right)_N\left(\mathbf{x},t\right)\right]\left(\Psi_\Gamma\right)_N\left(\mathbf{x},t\right).
\label{Dirac-Schroedinger-Gleichung_Wechselwirkung_Raum-Zeit-Darstellung}
\end{align}
Der abstrakten Dirac"=Gleichung im Tensorraum der Ur"=Alternativen mit Wechselwirkungsterm
($\ref{Dirac-Gleichung_Gamma_Wechselwirkung}$) beziehungsweise der ihr entsprechenden
abstrakten Schrödingergleichung im Tensorraum der Ur"=Alternativen
($\ref{Dirac-Schroedinger-Gleichung_Wechselwirkung}$) entspricht
die folgende Zeitentwicklung:

\begin{equation}
|\Psi_\Gamma(t)\rangle=\mathcal{O}_t\exp\left[-i\int_{0}^{t}\left(\gamma_{8x8}^0\gamma_{8x8}^i \left(P_{ABCD}\right)_i
+\kappa_A \gamma_{8x8}^0\gamma_{8x8}^\mu |A_\mu^\tau(t)\rangle\right)dt\right]|\Psi_\Gamma(0)\rangle,
\end{equation}
wobei die Zeitentwicklung des darin enthaltenen Zustandes $|A_\mu^\tau(t)\rangle$
($\ref{Zustand_Eichvektorboson}$) ihrerseits durch ($\ref{Jacobi_Identitaet_Dynamik}$)
und ($\ref{Dynamik_Eichvektorbosonzustand}$) determiniert ist. Die Verschränkung
der abstrakten Zustände $|\Psi_\Gamma(t)\rangle$ ($\ref{Dirac-Zustand_chi8}$)
und $|A_\mu^\tau(t)\rangle$ ($\ref{Zustand_Eichvektorboson}$), die sich durch
die Zeitentwicklung ergibt, ist also durch ein ziemlich kompliziertes
Gleichungssystem bestimmt. Natürlich ist die Verschränkung grundsätzlich
von der in ($\ref{Zustand_N_Objekte_A}$), ($\ref{Zustand_N_Objekte_B}$) beziehungsweise
($\ref{Zeitentwicklung_N_Objekte}$) gegebenen Form. Die in den dynamischen Gleichungen
($\ref{Jacobi_Identitaet_Dynamik}$), ($\ref{Dynamik_Eichvektorbosonzustand}$) und
($\ref{Dirac-Schroedinger-Gleichung_Wechselwirkung}$) enthaltenen Hamiltonoperatoren,
welche eine Verschränkung der Zustände induzieren, was das quantentheoretische Analogon der
Wechselwirkung darstellt, sind von der in ($\ref{Hamiltonoperator_Wechselwirkung_Produkt}$)
gegebenen grundsätzlichen Form. Wenn man also die freie Dynamik der durch
$|\Psi_\Gamma(t)\rangle$  und $|A_\mu^\tau(t)\rangle$ beschriebenen Quantenobjekte
gemäß ($\ref{Zeitentwicklung_frei}$) separat betrachtet, dann ist das Gleichungssystem,
das die Zeitentwicklung der abstrakten Zustände beschreibt, eine spezielle Manifestation
desjenigen aus ($\ref{Gleichungssystem_Wechselwirkung}$). Desweiteren kann man diese
Dynamik im Prinzip auch in der in ($\ref{Feldoperatoren_Tensorraum_Ur-Alternativen}$),
($\ref{Feldoperatoren_Tensorraum_Ur-Alternativen_Dynamik}$), ($\ref{Propagator_Tensorraum_Ur-Alternativen_frei}$)
und ($\ref{Propagator_Tensorraum_Ur-Alternativen_Wechselwirkung}$) angedeuteten Weise
störungstheoretisch behandeln, wobei diese abstrakte quantenlogische Manifestation
der Störungstheorie zunächst noch näher ausgearbeitet werden muss.

\section{Die Wechselwirkung der Gravitation}

\subsection{Der klassische Grenzfall der Gravitation als Eichtheorie der Translationen}

Das Korrespondenzprinzip Niels Bohrs besagt, dass die quantentheoretische Formulierung
einer Theorie in einer solchen Weise geschehen muss, dass sich als Grenzfall die
klassische Formulierung ergibt. Da die fundamentalen Wechselwirkungen des Standardmodells
der Elementarteilchenphysik als lokale Eichtheorien formuliert sind, muss in einer
einheitlichen Beschreibung auch die Gravitation als Eichtheorie beschrieben werden.
Tatsächlich kann die Gravitation, wie sie im Rahmen der klassischen allgemeinen
Relativitätstheorie beschrieben wird, als Eichtheorie der lokalen Translationen
formuliert werden. Dass die klassische Gravitation im Rahmen der allgemeinen
Relativitätstheorie mit der Eichtheorie der lokalen Translationen identisch
ist und nicht mit der Eichtheorie der lokalen $SO(3,1)$"=Symmetriegruppe,
also der lokalen Lorentzgruppe, kann gemäß Holger Lyre \cite{Lyre:2004}
durch drei zentrale Argumente begründet werden:\\
\\
\textbf{Argument A:} Die Eichgruppe, die einer Eichtheorie zugrunde liegt,
entspricht immer der grundlegenden Symmetriegruppe dieser Theorie. Die fundamentale
Symmetriegruppe der allgemeinen Relativitätstheorie ist die Diffeomorphismengruppe
$Diff\left(\mathcal{M}\right)$, also die Gruppe beliebiger kontinuierlicher
Koordinatentransformationen in Bezug auf die Raum"=Zeit"=Mannigfaltigkeit
$\mathcal{M}$. Die Diffeomorphismenruppe $Diff\left(\mathcal{M}\right)$ aber
ist identisch mit der Gruppe lokaler Translationen in Bezug auf die Raum"=Zeit.
Die $SO(3,1)$"=Symmetriegruppe ist deshalb in der lokalen Translationsgruppe
implizit enthalten.\\
\\
\textbf{Argument B:} Die zu einer Symmetriegruppe gemäß dem Noetherschen Theorem
\cite{Noether:1918} gehörige Erhaltungsgröße ist zugleich diejenige Größe, an welche
das entsprechende Eichfeld gekoppelt ist. Gemäß der Einsteinschen Feldgleichung
bestimmt der Energie"=Impuls"=Tensor die Gestalt des metrischen Feldes,
welches das Gravitationsfeld darstellt. Der Energie"=Impuls"=Tensor ist
aber genau die Noethersche Erhaltungsgröße, welche zur Symmetriegruppe
der Raum"=Zeit"=Translationen gehört und damit in der Eichtheorie
lokaler Translationen an das Gravitationsfeld koppelt. Damit besteht
diesbezüglich eine vollkommene Übereinstimmung zwischen der gewöhnlichen
Formulierung der allgemeinen Relativitätstheorie und der Eichtheorie
lokaler Translationen. Die Noethersche Erhaltungsgröße der
$SO(3,1)$"=Symmetriegruppe ist hingegen ein Tensor dritter
Stufe und dies stimmt ganz eindeutig nicht mit der allgemeinen
Relativitätstheorie überein.\\
\\
\textbf{Argument C:} Das Eichfeld der Eichtheorie der lokalen Translationen
ist ein Tetradenfeld. Die dazugehörige Feldstärke ist die Torsion, während die
Riemann"=Krümmung in der Translationseichtheorie verschwindet. Aber die aus der
Torsion gebildete Wirkung ist nicht nur quadratisch in den Feldstärken, was mit
den Yang"=Mills"=Eichtheorien der Elementarteilchenphysik übereinstimmt, sondern
die sich daraus ergebende Feldgleichung ist auch isomorph zu der Einsteinschen
Feldgleichung. Dies bedeutet, dass trotz der unterschiedlichen Art der Darstellung
die Eichtheorie lokaler Translationen mit der gewöhnlichen Formulierung der
allgemeinen Relativitätstheorie mathematisch identisch ist.\\
\\
Die Gravitation kann also als die Eichtheorie der lokalen Translationen angesehen werden,
die formuliert ist in \cite{Cho:1975}, \cite{Hehl:1976}. Die Gruppe lokaler Translationen
ist faktisch isomorph zur Diffeomorphismengruppe $Diff\left(\mathcal{M}\right)$ und enthält
auch die konforme Gruppe. In der Quantentheorie der Ur"=Alternativen gibt es auf der
fundamentalen Ebene allerdings keine Hintergrund"=Raum"=Zeit, nicht nur keine
Hintergrundmetrik, sondern auch keine differenzierbare Mannigfaltigkeit, auch kein
abstraktes Netzwerk, sondern überhaupt gar keine Hintergrundstruktur. Denn der physikalische
Raum ist hier lediglich ein Darstellungsmedium der diskreten und abstrakten quantenlogischen
Zustände im Tensorraum der Ur"=Alternativen und die Zeit entspricht mathematisch gesehen
einem Automorphismus des Tensorraumes der Ur"=Alternativen. Dennoch sind in Bezug auf
den Tensorraum der Ur"=Alternativen Operatoren aus den Erzeugungs- und Vernichtungsoperatoren ($\ref{Operatoren_symmetrisch}$),($\ref{ErzeugungsVernichtungsOperatorenXYZ}$) definiert, die sich
algebraisch wie Impulsoperatoren verhalten ($\ref{Ort_Impuls_Operatoren}$),($\ref{Heisenbergsche_Algebra}$).
Diese sind gemeinsam mit dem Energieoperator ($\ref{Energie-Impuls-Relation}$) in der Größe
$P_{ABCD}$ definiert in ($\ref{Viererimpuls_Tensorraum}$) zusammengefasst. Den drei darin
enthaltenen unabhängigen Freiheitsgraden entsprechen die drei Freiheitsgrade des Tensorraumes
der Ur"=Alternativen, wobei der mit der Besetzungszahl $N_n$ verbundene Freiheitsgrad, der bei
gegebenen Besetzungszahlen $N_x$, $N_y$, $N_z$ direkt mit der Gesamtinformationsmenge $N$
($\ref{Gesamtzahl_Ur-Alternativen}$) korreliert ist ($\ref{Besetzungszahlen_Informationsmenge}$),
hier ausgenommen wird. Die Dirac"=Gleichung im Tensorraum der Ur"=Alternativen
($\ref{Dirac-Gleichung_chi8}$) weist eine Invarianz unter Transformationen auf, welche die
Operatoren $P_{ABCD}$ als Generatoren zugrunde legen ($\ref{Transformationsoperator_Poincare-Gruppe}$),
($\ref{Transformation_Poincare-Gruppe}$) und die durch diese abstrakten Transformationen sich
konstituierende Symmetrie, welche die Zustände im Tensorraum der Ur"=Alternativen aufweisen,
entspricht demgemäß einer Symmetrie unter der Translationsgruppe. Wenn man also im Sinne
des Korrespondenzprinzips die Beschreibung der Gravitation im begrifflichen rein
quantentheoretischen Rahmen der Quantentheorie der Ur"=Alternativen formulieren möchte,
so muss man in Analogie zur klassischen Translationseichtheorie Symmetrie unter
Transformationen fordern, bei der die Parameter der Translationsgeneratoren im
Tensorraum der Ur"=Alternativen, also den Komponenten des Energie"=Impuls"=Operators $P_{ABCD}$,
von den aus den Erzeugungs- und Vernichtungsoperatoren im Tensorraum der Ur"=Alternativen
($\ref{ErzeugungsVernichtungsOperatorenXYZ}$) gemäß ($\ref{Ort_Impuls_Operatoren}$)
und ($\ref{Ortsoperator_ABCD}$) konstruierten Viererortsoperatoren $X_{ABCD}$ abhängen.
Dies entspricht dann zugleich einer Symmetrie unter beliebigen Transformationen im
symmetrischen Anteil des Tensorraumes der Ur"=Alternativen, wobei $N_n$ als mit
der Gesamtinformationsmenge $N$ korrelierter Freiheitsgrad ausgenommen ist.
Diese Symmetrie kann damit als das quantenlogische Analogon zur Diffeomorphismeninvarianz
$Diff\left(\mathcal{M}\right)$ in der allgemeinen Relativitätstheorie als dem
entsprechenden klassischen feldtheoretischen Grenzfall der hier betrachteten rein
quantentheoretischen Gravitationstheorie angesehen werden. Diese rein quantentheoretische
Fassung der Gravitation ist in einem viel grundsätzlicheren Sinne hintergrundunabhängig
als die klassische allgemeine Relativitätstheorie. Denn sie setzt, wie erwähnt, nicht nur
keine Hintergrund"=Metrik voraus, sondern noch nicht einmal eine Raum"=Zeit"=Mannigfaltigkeit
und auch keine andere Hintergrundstruktur. Die sich daraus ergebenden dynamischen Größen der
Gravitation konstituieren sich innerhalb des Rahmens der Quantentheorie der Ur"=Alternativen,
wie alle anderen Größen auch, aus rein quantenlogischen Informationseinheiten. Was in diesem
Abschnitt also letztendlich vollzogen wird, das ist die Übertragung der Translationseichtheorie
der Gravitation, die in \cite{Cho:1975} ausformuliert wurde, wo auch deren Äquivalenz zur
allgemeinen Relativitätstheorie gezeigt wurde, in einen rein quantenlogischen Begriffsrahmen
ohne jeden Raum"=Zeit"=Hintergrund.

\subsection{Abstrakte Translationen dargestellt im Tensorraum der Ur-Alternativen}

Die Gruppe der Translationen wird durch den abstrakten quantenlogischen
Energie"=Impuls"=Operator im Tensorraum der Ur"=Alternativen $P_{ABCD}$
($\ref{Viererimpuls_Tensorraum}$) generiert. Ohne Abhängigkeit von dem
abstrakten quantenlogischen Viererortsoperator im Tensorraum der
Ur"=Alternativen $X_{ABCD}$ ($\ref{Ortsoperator_ABCD}$) lautet der
abstrakte Translationsoperator in seiner allgemeinen Form wie folgt:

\begin{equation}
U_P\left[\lambda\right]=\exp\left[i\lambda^m \left(P_{ABCD}\right)_m\right],\quad m=0,...,3,
\label{Transformation_Translation}
\end{equation}
wobei der Vektor $\lambda$ aus reellen Parametern besteht. Die abstrakte Dirac"=Gleichung
($\ref{Dirac-Gleichung_chi8}$) ist natürlich invariant unter quantenlogischen Translationen,
die nicht von den Komponenten des quantenlogischen Viererortsoperators $X_{ABCD}$ abhängen,
denn die quantenlogischen Translationsgeneratoren, also die Komponenten des quantenlogischen
Energie"=Impuls"=Operators $P_{ABCD}$, kommutieren untereinander. Durch Transformation
der abstrakten Dirac"=Gleichung im Tensorraum der Ur"=Alternativen ($\ref{Dirac-Gleichung_chi8}$)
über den Transformationsoperator ($\ref{Transformation_Translation}$) ergibt sich als Teil
der Transformation ($\ref{Transformation_Poincare-Gruppe_Dirac-Gleichung}$) Folgendes:

\begin{align}
\gamma_{8x8}^\mu (P_{ABCD})_\mu |\Psi_\Gamma (t)\rangle \quad\longrightarrow\quad&
\gamma_{8x8}^\mu (P_{ABCD})_\mu U_P\left[\lambda\right]|\Psi_\Gamma (t)\rangle
=U_P\left[\lambda\right] \gamma_{8x8}^\mu (P_{ABCD})_\mu |\Psi_\Gamma (t)\rangle=0.
\label{Transformation_Translation_Dirac-Gleichung_Gamma}
\end{align}
Wenn man die Komponenten des Parameters $\lambda$ nun in eine Abhängigkeit
von den abstrakten Viererortsoperatoren $X_{ABCD}$ stellt, dann erhält man
folgende allgemeine Gestalt des quantenlogischen Translationsoperators:

\begin{equation}
U_P\left[\lambda\right]\quad\longrightarrow\quad
U_P\left[\lambda\left(X_{ABCD}\right)\right]
=\exp\left[i\lambda^m \left(X_{ABCD}\right)\left(P_{ABCD}\right)_m\right],\quad m=0,...,3.
\label{Transformationsoperator_Translationen_Ortsoperator}
\end{equation}
Dieser allgemeine quantenlogische Translationsoperator ($\ref{Transformationsoperator_Translationen_Ortsoperator}$),
welcher von den Komponenten des abstrakten Viererortsoperators im Tensorraum der Ur"=Alternativen
$X_{ABCD}$ abhängig ist, enthält im Prinzip alle beliebigen Transformationen im gegenüber Permutation
der Ur"=Alternativen symmetrischen Teil des Tensorraumes, wobei wie erwähnt der mit der Besetzungszahl
$N_n$ verbundene und bei festen Besetzungszahlen $N_x$, $N_y$ und $N_z$ mit der Gesamtinformationsmenge
$N$ ($\ref{Gesamtzahl_Ur-Alternativen}$) korrelierte Freiheitsgrad hier ausgelassen wird. Denn wie oben
thematisiert entsprechen den drei Translationsfreiheitsgraden die drei Freiheitsgrade des Tensorraumes
der Ur"=Alternativen ohne die Gesamtinformationsmenge $N$ ($\ref{Gesamtzahl_Ur-Alternativen}$).
Die Zeittranslation stellt einen Automorphismus dar, der über die abstrakte Schrödingergleichung
($\ref{Schroedingergleichung_Tensorraum}$) beziehungsweise Dirac"=Gleichung ($\ref{Dirac-Gleichung_chi8}$)
im Tensorraum der Ur"=Alternativen definiert ist, und die deshalb wie oben bereits erörtert keinen
unabhängigen Freiheitsgrad darstellt. Dass die von $X_{ABCD}$ abhängigen quantenlogischen Translationen
($\ref{Transformationsoperator_Translationen_Ortsoperator}$) beliebige Transformationen im gegenüber
Permutation der einzelnen Ur"=Alternativen symmetrischen Anteil des Tensorraumes der Ur"=Alternativen
repräsentieren, welche die mit der Gesamtinformationsmenge $N$ ($\ref{Gesamtzahl_Ur-Alternativen}$)
korrelierte Besetzungszahl $N_n$ unberührt lassen, entspricht im klassischen Grenzfall
die Tatsache, dass die lokale Translationsgruppe identisch mit der Diffeomorphismengruppe
$Diff\left(\mathcal{M}\right)$ ist. Wenn man eine solche durch einen Transformationsoperator
der Form ($\ref{Transformationsoperator_Translationen_Ortsoperator}$) dargestellte Transformation
auf die abstrakte Dirac"=Gleichung im Tensorraum der Ur"=Alternativen ($\ref{Dirac-Gleichung_chi8}$)
anwendet, so ergibt sich in Analogie zu ($\ref{Eichtransformation_mit_Ortsoperator_Dirac-Gleichung}$)
das Folgende:

\begin{align}
\gamma_{8x8}^\mu \left(P_{ABCD}\right)_\mu |\Psi_\Gamma (t)\rangle\quad \longleftrightarrow \quad &
\gamma_{8x8}^\mu \left(P_{ABCD}\right)_\mu U_P\left[\lambda\left(X_{ABCD}\right)\right]|\Psi_\Gamma (t)\rangle\nonumber\\
&=\gamma_{8x8}^\mu \left[\left(P_{ABCD}\right)_\mu,
U_P\left[\lambda\left(X_{ABCD}\right)\right]\right]|\Psi_\Gamma (t)\rangle\nonumber\\
&\quad+U_P\left[\lambda\left(X_{ABCD}\right)\right]\gamma_{8x8}^\mu \left(P_{ABCD}\right)_\mu
|\Psi_\Gamma (t)\rangle\nonumber\\
&=\gamma_{8x8}^\mu \left[\left(P_{ABCD}\right)_\mu,U_P\left[\lambda\left(X_{ABCD}\right)\right]\right]
|\Psi_\Gamma (t)\rangle \neq 0.
\end{align}

\subsection{Abstrakter Tetradenzustand im Tensorraum der Ur-Alternativen}

In Analogie zur Eichtheorie in Bezug auf die $SU(N_T)$"=Symmetrien des antisymmetrischen Anteiles
des Tensorraumes der Ur"=Alternativen muss ein weiterer Zustand, gebildet aus Ur"=Alternativen,
in die abstrakte Dirac"=Gleichung im Tensorraum der Ur"=Alternativen ($\ref{Dirac-Gleichung_chi8}$)
eingefügt werden, um die Eichsymmetrie auch unter Transformationen zu gewährleisten, die von
den quantenlogischen Viererortsoperatoren im Tensorraum der Ur"=Alternativen $X_{ABCD}$ abhängen
($\ref{Transformationsoperator_Translationen_Ortsoperator}$). Dieser muss das abstrakte Analogon
zu einer Tetrade in der gewöhnlichen Translationseichtheorie der Gravitation sein. Zunächst muss
daher in Analogie zum Zustand eines quantenlogischen Eichvektorbosons ($\ref{Zustand_Eichvektorboson}$)
ein Zustand eines quantenlogischen gravitativen Potentials aus einem allgemeinen Dirac"=Zustand im
Tensorraum der Ur"=Alternativen ($\ref{Dirac-Zustand_chi8}$) in der folgenden Weise gebildet werden:

\begin{equation}
|H_\mu^m(t)\rangle=\langle \Psi_\Gamma(t)|\left[\gamma^{8x8}_0 \gamma^{8x8}_\mu
\left(P_{ABCD}\right)^m\right]|\Psi_\Gamma(t)\rangle,
\quad m=0,...,3,
\label{Definition_Gravitationspotentialzustand_Komponenten}
\end{equation}
wobei die $\gamma^{8x8}$"=Matrizen in ($\ref{Definition_Gravitationspotentialzustand_Komponenten}$)
die Dirac"=Matrizen aus ($\ref{Dirac-Matrizen_8x8}$) sind. In Analogie zu
($\ref{Potential_Komponenten}$) ausführlicher geschrieben können die einzelnen
Komponenten des in ($\ref{Definition_Gravitationspotentialzustand_Komponenten}$)
definierten abstrakten Gravitationspotentials im Tensorraum der Ur"=Alternativen
in der folgenden Weise ausgedrückt werden:

\begin{align}
|H_0^m(t)\rangle=&\left[\langle \Psi_{E}\left(t\right)|_S \otimes
\langle \left(\chi_{8}\right)_{E}\left(t\right)|\right]
\left[\left(P_{ABCD}\right)^m\right] \left[|\Psi_{E}\left(t\right)\rangle_S \otimes
|\left(\chi_{8}\right)_{E}\left(t\right)\rangle\right]\\
&+\left[\langle\Psi_{-E}\left(t\right)|_S \otimes \langle\left(\chi_{8}\right)_{-E}\left(t\right)|\right]
\left[\left(P_{ABCD}\right)^m\right] \left[|\Psi_{-E}\left(t\right)\rangle_S
\otimes |\left(\chi_{8}\right)_{-E}\left(t\right)\rangle\right],\nonumber\\
|H_1^m(t)\rangle=&\left[\langle \Psi_{E}\left(t\right)|_S \otimes \langle \left(\chi_{8}\right)_{E}\left(t\right)|\right]
\left[\sigma^1_{8x8} \left(P_{ABCD}\right)^m\right] \left[|\Psi_{E}\left(t\right)\rangle_S \otimes
|\left(\chi_{8}\right)_{E}\left(t\right)\rangle\right]\nonumber\\
&-\left[\langle\Psi_{-E}\left(t\right)|_S \otimes \langle\left(\chi_{8}\right)_{-E}\left(t\right)|\right]
\left[\sigma^1_{8x8} \left(P_{ABCD}\right)^m\right] \left[|\Psi_{-E}\left(t\right)\rangle_S \otimes |\left(\chi_{8}\right)_{-E}\left(t\right)\rangle\right],\nonumber\\
|H_2^m(t)\rangle=&\left[\langle \Psi_{E}\left(t\right)|_S \otimes \langle \left(\chi_{8}\right)_{E}\left(t\right)|\right]
\left[\sigma^2_{8x8} \left(P_{ABCD}\right)^m\right] \left[|\Psi_{E}\left(t\right)\rangle_S \otimes
|\left(\chi_{8}\right)_{E}\left(t\right)\rangle\right]\nonumber\\
&-\left[\langle\Psi_{-E}\left(t\right)|_S \otimes \langle\left(\chi_{8}\right)_{-E}\left(t\right)|\right]
\left[\sigma^2_{8x8} \left(P_{ABCD}\right)^m\right] \left[|\Psi_{-E}\left(t\right)\rangle_S \otimes |\left(\chi_{8}\right)_{-E}\left(t\right)\rangle\right],\nonumber\\
|H_3^m(t)\rangle=&\left[\langle \Psi_{E}\left(t\right)|_S \otimes \langle \left(\chi_{8}\right)_{E}\left(t\right)|\right]
\left[\sigma^3_{8x8} \left(P_{ABCD}\right)^m\right] \left[|\Psi_{E}\left(t\right)\rangle_S \otimes
|\left(\chi_{8}\right)_{E}\left(t\right)\rangle\right]\nonumber\\
&-\left[\langle\Psi_{-E}\left(t\right)|_S \otimes \langle\left(\chi_{8}\right)_{-E}\left(t\right)|\right]
\left[\sigma^3_{8x8} \left(P_{ABCD}\right)^m\right] \left[|\Psi_{-E}\left(t\right)\rangle_S \otimes |\left(\chi_{8}\right)_{-E}\left(t\right)\rangle\right].\nonumber
\end{align}
Man kann nun wie im Falle des Eichvektorbosonenzustandes ($\ref{Zustand_Eichvektorboson}$)
in Analogie zu ($\ref{Zustand_Eichvektorboson_Summierung}$) auch bezüglich eines quantenlogischen
Gravitationspotentialzustandes ($\ref{Definition_Gravitationspotentialzustand_Komponenten}$)
die Summe über alle Generatoren bilden, was zu einem Zustand der folgenden Form führt:

\begin{equation}
|H_\mu^P(t)\rangle=|H_\mu^m(t)\rangle \left(P_{ABCD}\right)_m.
\label{Definition_Gravitationspotentialzustand_Gesamt}
\end{equation}

\subsection{Abstrakte Dirac-Gleichung mit gravitativem Wechselwirkungsterm}

Mit Hilfe dieses Gravitationspotentialzustandes ($\ref{Definition_Gravitationspotentialzustand_Gesamt}$)
kann man die entsprechende abstrakte Dirac"=Gleichung im Tensorraum der Ur"=Alternativen in
Analogie zu ($\ref{Dirac-Gleichung_Gamma_Wechselwirkung}$) formulieren:

\begin{equation}
\gamma_{8x8}^\mu \left[\left(P_{ABCD}\right)_\mu+\kappa_H|H_\mu^P(t)\rangle\right]|\Psi_\Gamma(t)\rangle=0,
\label{Dirac-Gleichung_Gamma_Wechselwirkung_Gravitation}
\end{equation}
wobei $|H_\mu^P(t)\rangle |\Psi_\Gamma(t)\rangle=|H_\mu^P(t)\rangle \otimes |\Psi_\Gamma(t)\rangle$
gilt und $\kappa_H$ ein reeller Parameter ist, der die Stärke der Gravitation determiniert.
Ein Gravitationspotentialzustand ($\ref{Definition_Gravitationspotentialzustand_Gesamt}$)
im Tensorraum der Ur"=Alternativen muss sich desweiteren in der folgenden Weise transformieren:

\begin{align}
|H_\mu^P(t)\rangle \quad \longrightarrow \quad &U_P\left[\lambda\left(X_{ABCD}\right)\right]
|H_\mu^P(t)\rangle U^{\dagger}_P\left[\lambda\left(X_{ABCD}\right)\right]\nonumber\\
&-\frac{1}{\kappa_H}\left[\left(P_{ABCD}\right)_\mu U_P\left[\lambda\left(X_{ABCD}\right)\right]\right]
U^{\dagger}_P\left[\lambda\left(X_{ABCD}\right)\right].
\label{Transformation_Gravitationspotentialzustand}
\end{align}
Da es sich um eine Abelsche Eichtheorie handelt, und die Generatoren der Translationsgruppe im Tensorraum
der Ur"=Alternativen, also $P_{ABCD}$, miteinander kommutieren ($\ref{Kommutatoren_Energie-Impuls}$),
ist die Transformation ($\ref{Transformation_Gravitationspotentialzustand}$) faktisch identisch mit:

\begin{equation}
|H_\mu^P(t)\rangle \quad \longrightarrow \quad |H_\mu^P(t)\rangle
-\frac{1}{\kappa_H}\left[\left(P_{ABCD}\right)_\mu \lambda\left(X_{ABCD}\right)\right].
\end{equation}
Gemäß ($\ref{Transformation_Gravitationspotentialzustand}$) weist eine gesamte Eichtransformation
in Bezug auf vom Viererortsoperator im Tensorraum der Ur"=Alternativen $X_{ABCD}$ abhängige Translationen,
was dem Analogon der Diffeomorphismengruppe $Diff\left(\mathcal{M}\right)$ in Bezug auf abstrakte
diskrete quantenlogische Zustände entspricht, in Analogie zu ($\ref{Eichtransformationen}$)
die folgende allgemeine Gestalt auf:

\begin{align}
\label{Eichtransformation_Gravitation}
|\Psi_\Gamma(t)\rangle \quad\longrightarrow\quad &U_P\left[\lambda\left(X_{ABCD}\right)\right]|\Psi_\Gamma(t)\rangle
\nonumber\\
|H_\mu^P(t)\rangle \quad\longrightarrow\quad &U_P\left[\lambda\left(X_{ABCD}\right)\right]
|H_\mu^P(t)\rangle U^{\dagger}_P\left[\lambda\left(X_{ABCD}\right)\right]
\nonumber\\&
-\frac{1}{\kappa_H}\left[\left(P_{ABCD}\right)_\mu U_P\left[\lambda\left(X_{ABCD}\right)\right]\right]
U^{\dagger}_P\left[\lambda\left(X_{ABCD}\right)\right].
\end{align}
Bei einer Transformation der abstrakten Dirac"=Gleichung mit gravitativem Wechselwirkungsterm
($\ref{Dirac-Gleichung_Gamma_Wechselwirkung_Gravitation}$) unter der Eichtransformation
($\ref{Eichtransformation_Gravitation}$) ergibt sich in Analogie zu
($\ref{Eichtransformation_Dirac-Gleichung_Wechselwirkung}$) das Folgende:

\begin{align}
\label{Eichtransformation_Dirac-Gleichung_Wechselwirkung_Gravitation}
&\gamma_{8x8}^\mu \left[\left(P_{ABCD}\right)_\mu+\kappa_H|H_\mu^P(t)\rangle\right]|\Psi_\Gamma(t)\rangle=0\nonumber\\
&\longrightarrow \gamma_{8x8}^\mu \left\{\left(P_{ABCD}\right)_\mu+\kappa_H U_P\left[\lambda\left(X_{ABCD}\right)\right]
|H_\mu^P(t)\rangle U^{\dagger}_P\left[\lambda\left(X_{ABCD}\right)\right]\right.\nonumber\\
&\left.\quad\quad\quad\quad-\left[\left(P_{ABCD}\right)_\mu U_P\left[\lambda\left(X_{ABCD}\right)\right]\right]
U^{\dagger}_P\left[\lambda\left(X_{ABCD}\right)\right]\right\}
U_P\left[\lambda\left(X_{ABCD}\right)\right]|\Psi_\Gamma(t)\rangle
\nonumber\\&\quad\quad
=\gamma_{8x8}^\mu \left\{U_P\left[\lambda\left(X_{ABCD}\right)\right]\left(P_{ABCD}\right)_\mu
+\left[\left(P_{ABCD}\right)_\mu U_P\left[\lambda\left(X_{ABCD}\right)\right]\right]\right.\nonumber\\
&\left.\quad\quad\quad\quad\quad
+\kappa_H U_P\left[\lambda\left(X_{ABCD}\right)\right]|H_\mu^P(t) \rangle
U^{\dagger}_P\left[\lambda\left(X_{ABCD}\right)\right]
U_P\left[\lambda\left(X_{ABCD}\right)\right]\right.\nonumber\\
&\left.\quad\quad\quad\quad\quad
-\left[\left(P_{ABCD}\right)_\mu U_P\left[\lambda\left(X_{ABCD}\right)\right]\right]
U^{\dagger}_P\left[\lambda\left(X_{ABCD}\right)\right]
U_P\left[\lambda\left(X_{ABCD}\right)\right]\right\}|\Psi_\Gamma(t)\rangle\nonumber\\
&\quad\quad=\gamma_{8x8}^\mu \left\{U_P\left[\lambda\left(X_{ABCD}\right)\right]\left(P_{ABCD}\right)_\mu
+\left[\left(P_{ABCD}\right)_\mu U_P\left[\lambda\left(X_{ABCD}\right)\right]\right]
\right.\nonumber\\&\left.
\quad\quad\quad\quad\quad
+\kappa_H U_P\left[\lambda\left(X_{ABCD}\right)\right]|H_\mu^P(t)\rangle U^{\dagger}_P\left[\lambda\left(X_{ABCD}\right)\right]
-\left[\left(P_{ABCD}\right)_\mu U_P\left[\lambda\left(X_{ABCD}\right)\right]\right]\right\}
|\Psi_\Gamma(t)\rangle\nonumber\\
&\quad\quad=U_P\left[\lambda\left(X_{ABCD}\right)\right]
\gamma_{8x8}^\mu \left[\left(P_{ABCD}\right)_\mu+\kappa_H|H_\mu^P(t)\rangle\right]|\Psi_\Gamma(t)\rangle=0.
\end{align}
Dies bedeutet, dass man in Bezug auf die Gravitation einen kovarianten quantenlogischen
Energie"=Impuls"=Operator $\mathcal{D}_{ABCD}^P$ in der folgenden Weise definieren kann:

\begin{equation}
\left(\mathcal{D}_{ABCD}^P\right)_\mu=\left(P_{ABCD}\right)_\mu+\kappa_H|H_\mu^P(t)\rangle,
\label{Kovarianter_Impulsoperator_Gravitation}
\end{equation}
der sich aufgrund von ($\ref{Eichtransformation_Dirac-Gleichung_Wechselwirkung_Gravitation}$)
bei einer Eichtransformation der Form ($\ref{Eichtransformation_Gravitation}$) in der
folgenden Weise transformiert:

\begin{equation}
\left(\mathcal{D}_{ABCD}^P\right)_\mu \quad\longrightarrow \quad U_P\left[\lambda\left(X_{ABCD}\right)\right]\left(\mathcal{D}_{ABCD}^P\right)_\mu
U^{\dagger}_P\left[\lambda\left(X_{ABCD}\right)\right].
\label{Transformation_Kovarianter_Impulsoperator_Gravitation}
\end{equation}
Dies kann in Analogie zu ($\ref{Beweis_Kovarianter_Impulsoperator_Transformation}$)
wie folgt gezeigt werden:

\begin{align}
&\left(\mathcal{D}_{ABCD}^P\right)_\mu=\left(P_{ABCD}\right)_\mu \mathbf{1}+\kappa_H|H_\mu^{P}(t)\rangle
\nonumber\\&\longrightarrow
\left(P_{ABCD}\right)_\mu U_P\left[\lambda\left(X_{ABCD}\right)\right]
U_P^{\dagger}\left[\lambda\left(X_{ABCD}\right)\right]
+\kappa_H U_P\left[\lambda\left(X_{ABCD}\right)\right]|H_\mu^{P}(t)\rangle
U_P^{\dagger}\left[\lambda\left(X_{ABCD}\right)\right]
\nonumber\\&\quad\quad
-\left[\left(P_{ABCD}\right)_\mu U_P\left[\lambda\left(X_{ABCD}\right)\right]\right]
U_P^{\dagger}\left[\lambda\left(X_{ABCD}\right)\right]
\nonumber\\&\quad\quad
=\left[\left(P_{ABCD}\right)_\mu U_P\left[\lambda\left(X_{ABCD}\right)\right]\right] U_P^{\dagger}\left[\lambda\left(X_{ABCD}\right)\right]
+U_P\left[\lambda\left(X_{ABCD}\right)\right] 
\left(P_{ABCD}\right)_\mu U_P^{\dagger}\left[\lambda\left(X_{ABCD}\right)\right]
\nonumber\\&\quad\quad\quad
+\kappa_H U_P\left[\lambda\left(X_{ABCD}\right)\right]|H_\mu^{P}(t)\rangle
U_P^{\dagger}\left[\lambda\left(X_{ABCD}\right)\right]
-\left[\left(P_{ABCD}\right)_\mu U_P\left[\lambda\left(X_{ABCD}\right)\right]\right]
U_P^{\dagger}\left[\lambda\left(X_{ABCD}\right)\right]
\nonumber\\&\quad\quad
=U_P\left[\lambda\left(X_{ABCD}\right)\right] 
\left[\left(P_{ABCD}\right)_\mu U_P^{\dagger}\left[\lambda\left(X_{ABCD}\right)\right]\right]
+\kappa_H U_P\left[\lambda\left(X_{ABCD}\right)\right]|H_\mu^{P}(t)\rangle
U_P^{\dagger}\left[\lambda\left(X_{ABCD}\right)\right]
\nonumber\\&\quad\quad
=U_P\left[\lambda\left(X_{ABCD}\right)\right]\left[\left(P_{ABCD}\right)_\mu
+\kappa_H|H_\mu^{P}(t)\rangle\right] U_P^{\dagger}\left[\lambda\left(X_{ABCD}\right)\right]
\nonumber\\&\quad\quad
=U_P\left[\lambda\left(X_{ABCD}\right)\right]\left(\mathcal{D}_{ABCD}^P\right)_\mu
U_P^{\dagger}\left[\lambda\left(X_{ABCD}\right)\right].
\label{Transformation_kovarianter-Energie-Impuls-Operator_Gravitation}
\end{align}
Das Analogon zur gewöhnlichen Tetrade, also ein abstrakter Tetradenzustand im
Tensorraum der Ur"=Alternativen, ergibt sich dann in der folgenden Weise:

\begin{equation}
|E_\mu^m(t)\rangle=\delta_\mu^m+\kappa_H|H_\mu^m(t)\rangle.
\label{Tetradenzustand}
\end{equation}
Wenn man die zu den Generatoren der quantenlogischen Translationen $P_{ABCD}$
gehörigen Komponenten in Analogie zu ($\ref{Zustand_Eichvektorboson_Summierung}$)
und ($\ref{Definition_Gravitationspotentialzustand_Gesamt}$) aufsummiert,
dann ergibt sich der folgende Zustand, der identisch mit dem entsprechenden zur
Gravitation gehörigen kovarianten Energie"=Impuls"=Operator $\mathcal{D}_{ABCD}^P$
gegeben in ($\ref{Kovarianter_Impulsoperator_Gravitation}$) ist:

\begin{align}
|E_\mu^P(t)\rangle&=|E_\mu^m(t)\rangle\left(P_{ABCD}\right)_m=
\left(\delta_\mu^m+\kappa_H|H_\mu^m(t)\rangle\right)\left(P_{ABCD}\right)_m\nonumber\\&
=\left(P_{ABCD}\right)_\mu+\kappa_H|H_\mu^P(t)\rangle
=\left(\mathcal{D}_{ABCD}^P\right)_\mu.
\label{Tetradenzustand_Summierung}
\end{align}
Dies bedeutet also, dass der aufsummierte quantenlogische Tetradenzustand ($\ref{Tetradenzustand_Summierung}$)
identisch ist mit dem kovarianten Energie"=Impuls"=Operator $\mathcal{D}_{ABCD}^P$, zumindest
solange die Spinkonnektion im Tensorraum der Ur"=Alternativen noch nicht miteinbezogen ist,
was weiter unten noch geschehen soll. Demnach transformiert er sich unter einer
Eichtransformation ($\ref{Eichtransformation_Gravitation}$) auch in der gleichen Weise:

\begin{equation}
|E_\mu^P(t)\rangle\quad \longrightarrow\quad U_P\left[\lambda\left(X_{ABCD}\right)\right]|E_\mu^P(t)\rangle
U^{\dagger}_P\left[\lambda\left(X_{ABCD}\right)\right].
\label{Transformation_Tetradenzustand}
\end{equation}
Damit kann die abstrakte Dirac"=Gleichung im Tensorraum der Ur"=Alternativen mit gravitativem
Wechselwirkungsterm ($\ref{Dirac-Gleichung_Gamma_Wechselwirkung_Gravitation}$) in der
folgenden Weise umgeschrieben werden:

\begin{align}
&\gamma_{8x8}^\mu \left[\left(P_{ABCD}\right)_\mu+i\kappa_H|H_\mu^P(t)\rangle\right]|\Psi_\Gamma(t)\rangle=0,
\quad\Leftrightarrow\quad \gamma_{8x8}^\mu |E_\mu^m(t)\rangle \left(P_{ABCD}\right)_m |\Psi_\Gamma(t)\rangle=0,\nonumber\\
&\Leftrightarrow\quad \gamma_{8x8}^\mu \left(\mathcal{D}_{ABCD}^P\right)_\mu |\Psi_\Gamma(t)\rangle=0.
\label{Diracgleichung_Wechselwirkung_Gravitation}
\end{align}
Natürlich ist mit dem abstrakten quantenlogischen Tetradenzustand ($\ref{Tetradenzustand}$)
zugleich auch der Zustand einer abstrakten quantenlogischen Metrik definiert:

\begin{equation}
|G_{\mu\nu}(t)\rangle=\left[|E_\mu^m(t)\rangle \otimes |E_\nu^n(t)\rangle\right]\eta_{mn}
=|E_\mu^m(t)\rangle |E_\nu^n(t)\rangle \eta_{mn}.
\label{Definition_Zustand_Metrik}
\end{equation}

\subsection{Die Spinkonnektion im Tensorraum der Ur-Alternativen}

Genaugenommen muss der abstrakte kovariante Energie"=Impuls"=Operator der Gravitation
($\ref{Kovarianter_Impulsoperator_Gravitation}$) aber noch um eine Spinkonnektion
erweitert werden. Denn wie oben bereits erwähnt entspricht einer lokalen Translation
ein beliebiger Diffeomorphismus. Dies aber bedeutet in Bezug auf den Tensorraum der
Ur"=Alternativen, dass eine von den Viererortsoperatoren $X_{ABCD}$ abhängige Eichtransformation
bezüglich der Translationsgruppe im Tensorraum der Ur"=Alternativen mit einem dementsprechenden
Transformationsoperator der Form ($\ref{Transformationsoperator_Translationen_Ortsoperator}$)
dem nach einem Übergang gemäß ($\ref{Uebergang_Transformation}$) von den Viererortsoperatoren
$X_{ABCD}$ abhängigen Analogon zu einer Eichtransformation der Poincare"=Gruppe im
Tensorraum der Ur"=Alternativen mit einem Transformationsoperator der Form
($\ref{Transformationsoperator_Poincare-Gruppe}$) entspricht:

\begin{align}
&U_P\left[\lambda\left(X_{ABCD}\right)\right]
=\exp\left[i\lambda^\mu\left(X_{ABCD}\right)\left(P_{ABCD}\right)_\mu\right]\nonumber\\
&\widehat{=}\quad U_{PM}\left\{\bar \lambda\left[\lambda\left(X_{ABCD}\right)\right],
\bar \Sigma\left[\lambda\left(X_{ABCD}\right)\right]\right\}\nonumber\\
&\quad\quad=\exp\left\{i\bar \lambda^\mu\left[\lambda\left(X_{ABCD}\right)\right]\left(P_{ABCD}\right)_\mu
+i\bar \Sigma^{\mu\nu}\left[\lambda\left(X_{ABCD}\right)\right]\left(M_{ABCD}\right)_{\mu\nu}\right\}.
\end{align}
Aus diesem Grunde muss zusätzlich das quantenlogische Analogon zu einer Spinkonnektion eingeführt werden,
welche sich auf Spin"=Freiheitsgrad im Tensorraum der Ur"=Alternativen bezieht. Im Tensorraum der
Ur"=Alternativen entspricht dem Freiheitsgrad des Spin die relative Ausrichtung der Ur"=Alternativen
des gegenüber Permutation der einzelnen Ur"=Alternativen antisymmetrischen Teilraumes $\mathcal{H}_{TAS}$
in Bezug auf die Ur"=Alternativen des gegenüber Permutation der einzelnen Ur"=Alternativen symmetrischen
Teilraumes $\mathcal{H}_{TS}$ des Tensorraumes der Ur"=Alternativen. Wenn allerdings beliebige von den
quantenlogischen Viererortsoperatoren im Tensorraum $X_{ABCD}$ abhängige quantenlogische Translationen
auch quantenlogische Rotationen im Tensorraum implizieren können, dann bedeutet dies, dass dies auch
eine Auswirkung auf den Spin haben muss. Wenn also über die von den quantenlogischen Viererortsoperatoren
$X_{ABCD}$ abhängigen quantenlogischen Translationen auch eine quantenlogische Rotation im symmetrischen
Anteil des Tensorraumes induziert wird, der in den dreidimensionalen Ortsraum abgebildet werden kann,
so bewirkt dies eine relative Veränderung in Bezug auf die Ur"=Alternativen des antisymmetrischen
Anteiles, dem die Quantenzahlen entsprechen. Da es sich aber um eine Symmetrietransformation handelt,
müssen natürlich die relativen Beziehungen gleich bleiben, weshalb auch eine analoge Transformation
im antisymmetrischen Anteil des Tensorraumes der Ur"=Alternativen durchgeführt werden muss,
also in Bezug auf den Spin. Demnach muss eine quantenlogische Spinkonnektion in Abhängigkeit von
dem abstrakten Tetradenzustand im Tensorraum der Ur"=Alternativen ($\ref{Tetradenzustand}$) gebildet
und in den quantenlogischen kovarianten Energie"=Impuls"=Operator $\mathcal{D}^P_{ABCD}$ integriert werden.
Zu diesem Behuf müssen zunächst die Generatoren der Lorentz"=Gruppe im quantenlogischen Dirac"=Spinor"=Raum
gebildet werden, dessen Zustände in ($\ref{Dirac-Zustand_chi8}$) enthalten sind, was wie
üblich mit Hilfe der Dirac"=Matrizen ($\ref{Dirac-Matrizen_8x8}$) geschieht:

\begin{equation}
\left(M_\Gamma\right)^{ab}=-\frac{i}{4}\left[\gamma_{8x8}^a,\gamma_{8x8}^b\right],\quad a,b=0,...,3.
\end{equation}
Der Transformationsoperator ($\ref{Transformationsoperator_Translationen_Ortsoperator}$)
muss in der folgenden Weise erweitert werden:

\begin{align}
\label{Transformationsoperator_Spinkonnektion_Ortsoperator}
&U_P\left[\lambda\left(X_{ABCD}\right)\right]=\exp\left[i\lambda^m \left(X_{ABCD}\right)\left(P_{ABCD}\right)_m \right]\\
&\longrightarrow \quad \exp\left[i\lambda^m \left(X_{ABCD}\right)\left(P_{ABCD}\right)_m
+i\Sigma^{ab}\left[\lambda^P\left(X_{ABCD}\right)\right]\left(M_{\Gamma}\right)_{ab}\right]
=\left(U_{P}\right)_{\Gamma}\left[\lambda^P\left(X_{ABCD}\right)\right].\nonumber
\end{align}
Desweiteren muss auch der kovariante Energie"=Impuls"=Operator erweitert
werden und nimmt damit folgende Gestalt an:

\begin{align}
&\left(\mathcal{D}_{ABCD}^{P}\right)_\mu=\gamma_{8x8}^\mu |E_\mu^m(t)\rangle \left(P_{ABCD}\right)_m
=\gamma_{8x8}^m|E_m^\mu(t)\rangle \left(P_{ABCD}\right)_\mu\nonumber\\
&\longrightarrow\quad \gamma_{8x8}^m|E_m^\mu(t)\rangle \left[\left(P_{ABCD}\right)_\mu
+|\omega_\mu^{ab}\left(|E(t)\rangle\right)\rangle\left(M_\Gamma\right)_{ab}\right],
\label{Kovarianter_Impulsoperator_Gravitation_Erweiterung}
\end{align}
wobei die Komponenten des Spin"=Konnektion"=Zustandes $|\omega_\mu^{ab}\left(|E(t)\rangle\right)\rangle$,
die natürlich von den Komponenten des Tetradenzustandes $|E_\mu^m(t)\rangle$ abhängig sind,
in Analogie zum entsprechenden klassischen Zustand in der folgenden Weise definiert sind:

\begin{eqnarray}
|\omega_\mu^{ab}(t)\rangle&=&2|E^{\nu a}(t)\rangle \left(P_{ABCD}\right)_\mu |E_{\nu}^b(t)\rangle
-2|E^{\nu b}(t)\rangle \left(P_{ABCD}\right)_\mu |E_{\nu}^a(t)\rangle\nonumber\\
&&-2|E^{\nu a}(t)\rangle \left(P_{ABCD}\right)_\nu |E_{\mu}^b(t)\rangle
+2|E^{\nu b}(t)\rangle \left(P_{ABCD}\right)_\nu |E_{\mu}^a(t)\rangle\nonumber\\
&&+|E_{\mu k}(t)\rangle |E^{\nu a}(t)\rangle |E^{\rho b}(t)\rangle \left(P_{ABCD}\right)_\rho |E_\nu^k(t)\rangle\nonumber\\
&&-|E_{\mu k}(t)\rangle |E^{\nu a}(t)\rangle |E^{\rho b}(t)\rangle \left(P_{ABCD}\right)_\nu |E_\rho^k(t)\rangle.
\end{eqnarray}
Die entsprechende abstrakte Dirac"=Gleichung im Tensorraum der Ur"=Alternativen
weist demnach also die folgende Gestalt auf:

\begin{align}
&\gamma_{8x8}^\mu \left(\mathcal{D}_{ABCD}^{P}\right)_\mu |\Psi_\Gamma(t)\rangle=0
\quad\Leftrightarrow\quad \gamma_{8x8}^m|E_m^\mu(t)\rangle \left(D_{ABCD}^M\right)_\mu |\Psi_\Gamma(t)\rangle=0
\nonumber\\&\Leftrightarrow\quad
\gamma_{8x8}^m|E_m^\mu(t)\rangle \left[\left(P_{ABCD}\right)_\mu
+|\omega_\mu^{ab}(t)\rangle\left(M_\Gamma\right)_{ab}\right]|\Psi_\Gamma(t)\rangle=0,
\label{Diracgleichung_Wechselwirkung_Gravitation_Spinkonnektion}
\end{align}
wobei definiert wurde:

\begin{equation}
\left(D_{ABCD}^M\right)_\mu=\left(P_{ABCD}\right)_\mu+|\omega_\mu^{ab}(t)\rangle
\left(M_\Gamma\right)_{ab}.
\label{Impulsoperator_Spinkonnektion}
\end{equation}
Natürlich kann man den kovarianten Energie"=Impuls"=Operator auch auf einen quantenlogischen
Vektor- oder Tensorzustand höherer Ordnung anwenden. Wenn $|A^\mu(t)\rangle$ ein beliebiger
Vektorzustand im Tensorraum der Ur"=Alternativen ist, also ein Zustand, der eine zeitliche
und drei räumliche Komponenten aufweist, die ihrerseits jeweils Zustände im Tensorraum der
Ur"=Alternativen repräsentieren, dann ergibt sich durch Anwendung des kovarianten
Energie"=Impuls"=Operators $\mathcal{D}_{ABCD}^{P}$ der folgende Ausdruck:

\begin{equation}
\left(\mathcal{D}_{ABCD}^{P}\right)_\mu |A^\nu(t)\rangle=\left(P_{ABCD}\right)_\mu |A^\nu(t)\rangle
+|\Gamma_{\mu\rho}^{\nu}(t)\rangle |A^\rho(t)\rangle,
\end{equation}
wobei die Komponenten des entsprechenden Konnektionszustandes, der das rein quantentheoretische Analogon zur
Levy"=Civita"=Konnektion darstellt, die natürlich von den Komponenen des Tetradenzustandes $|E_\mu^m(t)\rangle$
abhängig sind, in der folgenden Weise definiert sind:

\begin{eqnarray}
|\Gamma_{\mu\nu}^{\rho}(t)\rangle&=&|E^{\sigma}_m(t)\rangle |E^{m\rho}(t)\rangle
\left[|E_{\sigma n}(t)\rangle \left(P_{ABCD}\right)_\mu |E^{n}_\nu(t)\rangle
+|E_{\sigma n}(t)\rangle \left(P_{ABCD}\right)_\nu |E^{n}_\mu(t)\rangle\right.\nonumber\\
&&\left.+|E_{\nu n}(t)\rangle \left(P_{ABCD}\right)_\mu |E^{n}_\sigma(t)\rangle
-|E_{\nu n}(t)\rangle \left(P_{ABCD}\right)_\sigma |E^{n}_\mu(t)\rangle\right.\nonumber\\
&&\left.+|E_{\mu n}(t)\rangle \left(P_{ABCD}\right)_\nu |E^{n}_\sigma(t)\rangle
-|E_{\mu n}(t)\rangle \left(P_{ABCD}\right)_\sigma |E^{n}_\nu(t)\rangle\right]\nonumber\\
&=&|G^{\sigma\rho}(t)\rangle \left[|E_{\sigma n}(t)\rangle \left(P_{ABCD}\right)_\mu |E^{n}_\nu(t)\rangle
+|E_{\sigma n}(t)\rangle \left(P_{ABCD}\right)_\nu |E^{n}_\mu(t)\rangle\right.\nonumber\\
&&\left.+|E_{\nu n}(t)\rangle \left(P_{ABCD}\right)_\mu |E^{n}_\sigma(t)\rangle
-|E_{\nu n}(t)\rangle \left(P_{ABCD}\right)_\sigma |E^{n}_\mu(t)\rangle\right.\nonumber\\
&&\left.+|E_{\mu n}(t)\rangle \left(P_{ABCD}\right)_\nu |E^{n}_\sigma(t)\rangle
-|E_{\mu n}(t)\rangle \left(P_{ABCD}\right)_\sigma |E^{n}_\nu(t)\rangle\right],
\label{Erweiterung_kovarianter_Impulsoperator_Gravitation}
\end{eqnarray}
wobei ($\ref{Definition_Zustand_Metrik}$) verwendet wurde. Der kovariante Energie"=Impuls"=Operator
$\mathcal{D}_{ABCD}^{P}$ mit dem rein quantentheoretischen Analogon zur Levy"=Civita"=Konnektion
muss die Bedingung erfüllen, dass er angewandt auf den Zustand des metrischen Tensors
$|G_{\mu\nu}(t)\rangle$ verschwindet:

\begin{equation}
\left(\mathcal{D}_{ABCD}^{P}\right)_\mu |G_{\nu\rho}(t)\rangle=\left(P_{ABCD}\right)_\mu |G_{\nu\rho}\rangle
+|\Gamma_{\mu\nu\sigma}(t)\rangle |G_{\rho\sigma}(t)\rangle+|\Gamma_{\mu\rho\sigma}(t)\rangle |G_{\nu\sigma}(t)\rangle=0.
\label{Dynamik_Gravitation}
\end{equation}
Nachdem ein Tetradenzustand im Tensorraum der Ur"=Alternativen als entscheidende Größe des
rein quantentheoretischen Analogons einer Eichtheorie der Translationen konstruiert wurde
und weiter gezeigt wurde, dass er sich wie eine kovariante Größe unter Translationen im Tensorraum
der Ur"=Alternativen verhält ($\ref{Transformation_kovarianter-Energie-Impuls-Operator_Gravitation}$),
($\ref{Tetradenzustand}$), ergibt sich die Aufgabe, darauf basierend die dynamische
Grundgleichung für den Tetradenzustand im Tensorraum der Ur"=Alternativen
($\ref{Tetradenzustand}$) zu formulieren. Dies ist identisch mit der Formulierung der
Dynamik der Gravitation im begrifflichen Rahmen der Quantentheorie der Ur"=Alternativen
und damit dem rein quantentheoretischen Analogon zur Einsteinschen Feldgleichung.

\subsection{Der Torsionszustand im Tensorraum der Ur-Alternativen}

Mit Hilfe des kovarianten Energie"=Impuls"=Operators $\mathcal{D}_{ABCD}^P$
($\ref{Kovarianter_Impulsoperator_Gravitation}$) beziehungsweise des dazu
äquivalenten aufsummierten abstrakten Tetradenzustandes ($\ref{Tetradenzustand_Summierung}$)
kann man das rein quantentheoretische Analogon zur gewöhnlichen Torsion bilden,
das als abstraktes quantenlogisches Analogon zu einem gravitativen Feldstärketensor
angesehen werden kann. Zu diesem Behuf muss der Kommutator der Komponenten
des kovarianten Energie"=Impuls"=Operators $\mathcal{D}_{ABCD}^P$
($\ref{Kovarianter_Impulsoperator_Gravitation}$) gebildet werden:

\begin{align}
&\left[\left(\mathcal{D}_{ABCD}^{P}\right)_\mu, \left(\mathcal{D}_{ABCD}^{P}\right)_\nu\right]\nonumber\\
&=\kappa_H\left[|E_\mu^m(t) \rangle \left(P_{ABCD}\right)_m |E_\nu^n(t) \rangle \left(P_{ABCD}\right)_n
-|E_\nu^m(t) \rangle \left(P_{ABCD}\right)_m |E_\mu^n(t) \rangle
\left(P_{ABCD}\right)_n\right]\nonumber\\
&=\kappa_H\left[|E_\mu^m(t) \rangle \left(P_{ABCD}\right)_m |E_\nu^n(t) \rangle
- |E_\nu^m(t) \rangle \left(P_{ABCD}\right)_m |E_\mu^n(t) \rangle \right]\left(P_{ABCD}\right)_n\nonumber\\
&=\kappa_H\left[|E_\mu^m(t) \rangle \left(P_{ABCD}\right)_m |E_\nu^n(t) \rangle
-|E_\nu^m(t) \rangle \left(P_{ABCD}\right)_m |E_\mu^n(t) \rangle \right]
|E^\rho_n(t) \rangle \left(P_{ABCD}\right)_\rho\nonumber\\
&=\kappa_H |E^\rho_n(t)\rangle\left[\left(P_{ABCD}\right)_\mu |E_\nu^n(t) \rangle
-\left(P_{ABCD}\right)_\nu |E_\mu^n(t) \rangle \right]
\left(P_{ABCD}\right)_\rho\nonumber\\
&\equiv|T_{\ \ \mu\nu}^\rho(t)\rangle \left(P_{ABCD}\right)_\rho,
\label{Kommutator_Komponenten_Kovarianter_Impulsoperator}
\end{align}
wobei im letzten Schritt eine Definition des quantenlogischen Torsionszustandes
im Tensorraum der Ur"=Alternativen vorgenommen wurde:

\begin{equation}
|T_{\ \ \mu\nu}^{\rho}(t)\rangle=\kappa_H |E^\rho_n(t)\rangle \left[\left(P_{ABCD}\right)_\mu |E_\nu^n(t)\rangle
-\left(P_{ABCD}\right)_\nu |E_\mu^n(t)\rangle\right].
\label{Torsionszustand}
\end{equation}
Da die Komponenten des Energie"=Impuls"=Operators $P_{ABCD}$ miteinander kommutieren
($\ref{Kommutatoren_Energie-Impuls}$) und es sich demnach um eine Abelsche Eichtheorie
handelt, enthält der abstrakte Torsionszustand ($\ref{Torsionszustand}$) im Gegensatz
zu dem quantenlogischen Analogon der Feldstärketensoren der Eichtheorien bezüglich
interner Symmetrien ($\ref{Analogon_Feldstaerketensor}$) keinerlei zusätzliche Terme,
in denen ein Kommutator der Generatoren der Lie"=Algebra der Symmetriegruppe auftaucht.
Allerdings enthält der abstrakte Torsionszustand ($\ref{Torsionszustand}$) und auch
die daraus später gebildete dynamische Grundgleichung durchaus Tensorprodukte des
abstrakten Tetradenzustandes ($\ref{Tetradenzustand}$) mit sich selbst. Dies muss
natürlich der Fall sein, da die Gravitation eine Selbstkopplung enthält. Um die
Transformationseigenschaften des Torsionszustandes ($\ref{Torsionszustand}$)
unter Eichtransformationen der Form ($\ref{Eichtransformation_Gravitation}$)
zu bestimmen, müssen die Transformationseigenschaften des Kommutators
des kovarianten Energie"=Impuls"=Operators $\mathcal{D}_{ABCD}^P$
($\ref{Kovarianter_Impulsoperator_Gravitation}$) betrachtet werden:

\begin{align}
&\left[\left(\mathcal{D}_{ABCD}^{P}\right)_\mu, \left(\mathcal{D}_{ABCD}^{P}\right)_\nu\right]
=|T_{\ \ \mu\nu}^\rho(t)\rangle \left(P_{ABCD}\right)_\rho
\nonumber\\
&\longrightarrow\quad \left[U_P\left[\lambda\left(X_{ABCD}\right)\right]\left(\mathcal{D}_{ABCD}^{P}\right)_\mu
U^{\dagger}_P\left[\lambda\left(X_{ABCD}\right)\right], U_P\left[\lambda\left(X_{ABCD}\right)\right]
\left(\mathcal{D}_{ABCD}^{P}\right)_\nu U^{\dagger}_P\left[\lambda\left(X_{ABCD}\right)\right]\right]
\nonumber\\
&\quad\quad\quad=U_P\left[\lambda\left(X_{ABCD}\right)\right]\left[\left(\mathcal{D}_{ABCD}^{P}\right)_\mu,
\left(\mathcal{D}_{ABCD}^{P}\right)_\nu\right] U^{\dagger}_P\left[\lambda\left(X_{ABCD}\right)\right].\nonumber\\
&\quad\quad\quad=U_P\left[\lambda\left(X_{ABCD}\right)\right]|T_{\ \ \mu\nu}^{\rho}(t)\rangle \left(P_{ABCD}\right)_\rho U^{\dagger}_P\left[\lambda\left(X_{ABCD}\right)\right],
\label{Kommutator_Gravitation_Transformation}
\end{align}
wobei hier ($\ref{Transformation_Kovarianter_Impulsoperator_Gravitation}$) verwendet wurde. Aufgrund
der erwähnten Tatsache, dass es sich um eine Abelsche Eichtheorie handelt und die Generatoren der
quantenlogischen Translationsgruppe im Tensorraum der Ur"=Alternativen, also die Komponenten des
quantenlogischen Energie"=Impuls"=Operators $P_{ABCD}$ ($\ref{Viererimpuls_Tensorraum}$), miteinander
kommutieren ($\ref{Kommutatoren_Energie-Impuls}$), bedeutet ($\ref{Kommutator_Gravitation_Transformation}$)
faktisch, dass der Torsionszustand unter einer Eichtransformation der Form
($\ref{Eichtransformation_Gravitation}$) konstant bleibt:

\begin{equation}
|T_{\ \ \mu\nu}^\rho(t)\rangle\quad \longrightarrow\quad |T_{\ \ \mu\nu}^{\rho}(t)\rangle.
\label{Eichtransformation_Torsion}
\end{equation}

\subsection{Dynamische Grundgleichung für den Tetradenzustand}

In der klassischen Translationseichtheorie der Gravitation \cite{Cho:1975}
wird aus der klassischen Torsionsgröße eine Wirkung gebildet, die zur
Einsteinschen Feldgleichung ausgedrückt durch das Tetradenfeld führt.
Dies ist insofern nicht wirklich überraschend, weil die Einsteinsche
Feldgleichung ja gerade durch ihre Eigenschaft der Diffeomorphismeninvarianz
ausgezeichnet sind. Und im letzten Kapitel wurde bereits darauf hingewiesen,
dass die Diffeomorphismengruppe identisch ist mit der Gruppe lokaler Translationen.
Demnach muss also in diesem Zusammenhang aufgrund des Korrepondenzprinzips und aufgrund
der Notwendigkeit der Eichinvarianz der dynamischen Grundgleichung der Gravitation
unter Translationstransformationen im Tensorraum der Ur"=Alternativen, die von den
abstrakten Viererortsoperatoren $X_{ABCD}$ abhängen, das rein quantentheoretische Analogon
zur Einsteinschen Feldgleichung basierend auf dem Tetradenzustand ($\ref{Tetradenzustand}$)
gebildet werden. Man kann ein rein quantenlogisches Analogon zur klassischen Wirkung
der Translationseichtheorie der Gravitation basierend auf dem rein quantenlogischen
Zustand der Torsion ($\ref{Torsionszustand}$) in folgender Weise bilden:

\begin{equation}
S_{QG}=\frac{1}{\kappa_H^2}\int dt \left[|E(t)\rangle
\left(\frac{1}{4} |T^{\mu\nu\rho}(t)\rangle |T_{\mu\nu\rho}(t)\rangle
+\frac{1}{2}|T^{\mu\nu\rho}(t)\rangle |T_{\mu\rho\nu}(t)\rangle
-|T^{\mu\nu\nu}(t)\rangle |T_{\mu\rho\rho}(t)\rangle\right)\right],
\label{Wirkung_Gravitation_quantenlogisch}
\end{equation}
wobei die Determinante des Tetradenzustandes $|E(t)\rangle$
über ($\ref{Tetradenzustand}$) wie folgt definiert ist:

\begin{equation}
|E(t)\rangle=\det|E^P(t)\rangle=\epsilon_{mnpq}|E_0^m(t)\rangle |E_1^n(t)\rangle |E_2^p(t)\rangle |E_3^q(t)\rangle,
\end{equation}
wobei $\epsilon_{mnpq}$ der total antisymmetrische Tensor vierter Stufe ist.
Diese Wirkung ($\ref{Wirkung_Gravitation_quantenlogisch}$) enthält natürlich kein
Integral über die Raumkoordinaten, sondern nur ein Integral über die Zeit, da der rein
quantenlogische Torsionszustand ($\ref{Torsionszustand}$) keinen Raum"=Zeit"=Hintergrund
voraussetzt, sondern nur die Zeitentwicklung als einen eindimensionalen Automorphismus
enthält. Demnach gibt es auch keine Langrangedichte, sondern eine zeitabhängige
Langrangefunktion $L_{QG}\left(t\right)$, die nur quantenlogische Zustände enthält:

\begin{equation}
L_{QG}\left(t\right)=\left[\frac{1}{\kappa_H^2}|E(t)\rangle
\left(\frac{1}{4} |T^{\mu\nu\rho}(t)\rangle |T_{\mu\nu\rho}(t)\rangle
+\frac{1}{2}|T^{\mu\nu\rho}(t)\rangle |T_{\mu\rho\nu}(t)\rangle
-|T^{\mu\nu\nu}(t)\rangle |T_{\mu\rho\rho}(t)\rangle\right)\right].
\label{Lagrangefunktion-Gravitation_quantenlogisch}
\end{equation}
Die Wirkung ($\ref{Wirkung_Gravitation_quantenlogisch}$) beziehungsweise die Langrangefunktion
($\ref{Lagrangefunktion-Gravitation_quantenlogisch}$) sind aufgrund von ($\ref{Eichtransformation_Torsion}$)
invariant unter Eichtransformationen der Form ($\ref{Eichtransformation_Gravitation}$). Um aus
($\ref{Wirkung_Gravitation_quantenlogisch}$) beziehungsweise ($\ref{Lagrangefunktion-Gravitation_quantenlogisch}$)
die entsprechende dynamische Grundgleichung für die Gravitation zu erhalten, muss
($\ref{Wirkung_Gravitation_quantenlogisch}$) nach dem quantenlogischen Tetradenzustand
($\ref{Tetradenzustand}$) variert werden. Dies muss in der quantenlogisch analogen Weise
zu jenem klassischen Grenzfall geschehen, der in \cite{Golovnev:2017} gegeben ist.
Wenn man dies auf den quantenlogischen Fall überträgt, so ergibt sich für die
Variation der Größen $|E_\mu^m(t)\rangle$, $|E(t)\rangle$ und
$|T^{\mu}_{\nu\rho}(t)\rangle$ das Folgende:

\begin{align}
\label{Variation_A}
&\delta_E |E^\mu_m(t)\rangle=-|E^\mu_n(t)\rangle |E^\nu_m(t)\rangle \delta |E_\nu^n(t)\rangle,
\quad \delta_E |E(t)\rangle=|E(t)\rangle |E^\mu_m(t)\rangle \delta |E_\mu^m(t)\rangle,\\
&\delta_E |T^{\mu}_{\ \ \nu\rho}(t)\rangle=
-|E^\mu_m(t)\rangle |T^{\sigma}_{\ \ \nu\rho}(t)\rangle \delta |E^m_\sigma(t)\rangle
+|E^\mu_m(t)\rangle \left[\left(D_{ABCD}^M\right)_\nu \delta |E_\rho^m(t)\rangle
-\left(D_{ABCD}^M\right)_\rho \delta |E_\nu^m(t)\rangle\right].\nonumber
\end{align}
Um die zu ($\ref{Wirkung_Gravitation_quantenlogisch}$) beziehungsweise
($\ref{Lagrangefunktion-Gravitation_quantenlogisch}$) gehörige dynamische
Grundgleichung für den quantenlogischen Tetradenzustand im Tensorraum der
Ur"=Alternativen ($\ref{Tetradenzustand}$) als rein quantentheoretisches
Analogon zur Einsteinschen Feldgleichung zu formulieren, ist es sinnvoll,
in Analogie zum klassischen Fall \cite{Golovnev:2017} folgende
quantenlogische Größen zu definieren:

\begin{eqnarray}
|T(t)\rangle&=&\frac{1}{4}|T^{\mu\nu\rho}(t)\rangle |T_{\mu\nu\rho}(t)\rangle
+\frac{1}{2}|T^{\mu\nu\rho}(t)\rangle |T_{\mu\rho\nu}(t)\rangle
-|T^{\mu\nu\nu}(t)\rangle |T_{\mu\rho\rho}(t)\rangle, \nonumber\\
|T^{\mu}(t)\rangle&=&|T^{\mu\nu\nu}(t)\rangle, \nonumber\\
|K^{\mu\nu\rho}(t)\rangle&=&\frac{1}{2}\left[|T^{\mu\nu\rho}(t)\rangle
+|T^{\rho\mu\nu}(t)\rangle+|T^{\nu\mu\rho}(t)\rangle\right], \nonumber\\
|S^{\mu\nu\rho}(t)\rangle&=&|K^{\nu\mu\rho}(t)\rangle+|G^{\mu\nu}(t)\rangle |T^{\rho}(t)\rangle
-|G^{\mu\rho}(t)\rangle |T^{\nu}(t)\rangle,
\label{Definitionen_Groessen_Gravitation}
\end{eqnarray}
wobei der quantenlogische Zustand $|G^{\mu\nu}(t)\rangle$ der in ($\ref{Definition_Zustand_Metrik}$)
definierte metrische Zustand in Abhängigkeit des quantenlogischen Tetradenzustandes ist
($\ref{Tetradenzustand}$). Zudem müssen die in ($\ref{Wirkung_Gravitation_quantenlogisch}$)
beziehungsweise ($\ref{Lagrangefunktion-Gravitation_quantenlogisch}$) enthaltenen Terme
variiert werden, um die dynamische Grundgleichung zu erhalten. Zu diesem Behuf müssen die
Variationen aus ($\ref{Variation_A}$) verwendet werden:

\begin{eqnarray}
\delta_E \left(|T_{\mu\nu\rho}(t)\rangle |T^{\mu\nu\rho}(t)\rangle\right)
&=&\delta_E \left(|T_{\mu}(t)\rangle |T^{\mu}(t)\rangle\right)
\nonumber\\
&=&-2\left(|T^{\rho}(t)\rangle |T^{\nu}_{\rho\mu}(t)\rangle
+|T^{\nu}(t)\rangle |T_{\mu}(t)\rangle\right)
|E^\mu_m(t)\rangle \delta |E_\nu^m(t)\rangle
\nonumber\\
&&+2\left(|T^{\nu}(t)\rangle |E^\mu_m\rangle
-|T^{\mu}(t)\rangle |E^\nu_m\rangle\right)
\left(D_{ABCD}\right)_\nu \delta |E_\mu^m(t)\rangle,
\nonumber\\
\delta_E \left(|T_{\rho\mu\nu}(t)\rangle |T^{\mu\rho\nu}(t)\rangle\right)
&=&-2\left(|T^{\rho\mu\nu}(t)\rangle-|T^{\nu\mu\rho}(t)\rangle\right)|T_{\mu\nu\sigma}(t)\rangle
|E_m^\sigma(t)\rangle \delta |E^m_\rho(t)\rangle\nonumber\\
&&+\left(|T^{\nu\ \ \rho}_{\ \ \mu}(t)\rangle-|T^{\rho\ \ \nu}_{\ \ \mu}(t)\rangle\right)
|E^\mu_m(t)\rangle \left(D_{ABCD}\right)_\nu \delta |E_\rho^m(t)\rangle,
\nonumber\\
\delta_E \left(|T_{\sigma\mu\nu}(t)\rangle |T^{\sigma\rho\nu}(t)\rangle\right)
&=&4 |T_{\rho}^{\ \ \mu\nu}(t)\rangle |E^\rho_m(t)\rangle \left(D_{ABCD}\right)_\mu \delta |E^m_\nu(t)\rangle
\nonumber\\
&&-4 |T^{\rho\mu\nu}(t)\rangle |T_{\rho\mu\sigma}(t)\rangle |E^\sigma_m(t)\rangle \delta |E^m_\nu(t)\rangle.
\label{Variation_B}
\end{eqnarray}
Mit Hilfe der Definitionen ($\ref{Definitionen_Groessen_Gravitation}$) kann die dynamische
Grundgleichung für den Tetradenzustand ($\ref{Tetradenzustand}$), der den gravitativen
Freiheitsgrad darstellt, als rein quantentheoretisches Analogon zur Einsteinschen
Feldgleichung, die in der klassischen durch Torsionsgrößen ausgedrückten Gestalt
zum Beispiel in \cite{Golovnev:2017} behandelt wird, in einer rein quantentheoretischen
durch Ur"=Alternativen ausgedrückten Gestalt in der folgenden Weise formuliert werden:

\begin{equation}
\left(\mathcal{D}_{ABCD}^P\right)_\mu |S_\nu^{\ \ \rho\mu}(t)\rangle
-|S^{\mu\rho\sigma}(t)\rangle\left(|T_{\mu\nu\sigma}(t)\rangle
+|K_{\mu\sigma\nu}(t)\rangle\right)
+\frac{1}{2}|T(t)\rangle\delta^{\rho}_{\nu}=0.
\label{Dynamische_Gleichung_Gravitation}
\end{equation}
Natürlich muss die Dynamik der quantenlogischen Tetradenzustände ($\ref{Tetradenzustand}$)
auch in einer Beziehung zu den Zuständen ($\ref{Dirac-Zustand_chi8}$) und
($\ref{Zustand_Eichvektorboson}$) stehen. Dazu muss ein Zustand im Tensorraum
der Ur"=Alternativen betrachtet werden, der als das Analogon zum Energie"=Impuls"=Tensor
in gewöhnlichen Quantenfeldtheorien angesehen werden kann. Dieser Zustand kann
über die Wirkungen eines quantenlogischen Dirac"=Zustandes ($\ref{Dirac-Zustand_chi8}$)
und eines quantenlogischen Eichvektorbosonenzustandes ($\ref{Zustand_Eichvektorboson}$)
definiert werden, die wie folgt lauten:

\begin{eqnarray}
S_{DQG}&=&\int dt \left[|E(t)\rangle \langle \Psi_\Gamma(t)|\gamma_{8x8}^m |E_m^\mu(t)\rangle
\left[\left(D_{ABCD}^M\right)_\mu+\kappa_A|A_\mu^\tau(t)\rangle\right]|\Psi_\Gamma(t)\rangle\right],
\label{Wirkung_Dirac_QG}
\end{eqnarray}
beziehungsweise

\begin{eqnarray}
S_{EVBQG}&=&\int dt \left[|E(t)\rangle |G^{\mu\nu}(t)\rangle |G^{\rho\sigma}(t)\rangle
\left[|F_{\mu\rho}^a(t)\rangle |F_{\nu\sigma}^a(t)\rangle\right]\right].
\label{Wirkung_Eichvektorboson_QG}
\end{eqnarray}
Der Energie"=Impuls"=Tensor ist in Abhängigkeit der Wirkungen ($\ref{Wirkung_Dirac_QG}$)
und ($\ref{Wirkung_Eichvektorboson_QG}$) definiert:

\begin{equation}
|\mathcal{T}^\mu_m\left[|\Psi_\Gamma(t)\rangle, |A_\mu^\tau(t)\rangle \right]\rangle
=\frac{1}{|E(t)\rangle}\frac{\delta S_{DQG}}{\delta |E_\mu^m(t)\rangle}
+\frac{1}{|E(t)\rangle}\frac{\delta S_{EVBQG}}{\delta |E_\mu^m(t)\rangle}.
\label{Zustand_Energie-Impuls-Tensor}
\end{equation}
Wenn man die dynamische Grundgleichung der Gravitation ($\ref{Dynamische_Gleichung_Gravitation}$)
als rein quantentheoretisches Analogon zur Einsteinschen Feldgleichung auf der rechten Seite
durch den Zustand ($\ref{Zustand_Energie-Impuls-Tensor}$) als quantenlogisches Analogon des
Energie"=Impuls"=Tensors ergänzt, so erhält man:

\begin{align}
&\left(\mathcal{D}_{ABCD}^P\right)_\mu |S_\nu^{\ \ \rho\mu}(t)\rangle
-|S^{\mu\rho\sigma}(t)\rangle\left(|T_{\mu\nu\sigma}(t)\rangle
+|K_{\mu\sigma\nu}(t)\rangle\right)+\frac{1}{2}|T(t)\rangle\delta^{\rho}_{\nu}\nonumber\\
&=\frac{\kappa_H^2}{2}|\mathcal{T}^\rho_\nu\left[|\Psi_\Gamma(t)\rangle, |A_\mu^\tau(t)\rangle \right]\rangle.
\label{Dynamische_Gleichung_Gravitation_Energie-Impuls}
\end{align}
Vielleicht könnte man die Gleichung ($\ref{Dynamische_Gleichung_Gravitation_Energie-Impuls}$)
als Weizsäcker"=Einstein"=Gleichung bezeichnen. Natürlich ist auch im Falle der Gravitation
die sich aus der in den quantenlogischen Wirkungen ($\ref{Wirkung_Gravitation_quantenlogisch}$),
($\ref{Wirkung_Dirac_QG}$) und ($\ref{Wirkung_Eichvektorboson_QG}$) beziehungsweise
in ($\ref{Dynamische_Gleichung_Gravitation_Energie-Impuls}$) enthaltenen Dynamik
ergebende Verschränkung der einzelnen Quantenobjekte im Tensorraum der
Ur"=Alternativen ($\ref{Dirac-Zustand_chi8}$), die sich zu Zuständen der Form
($\ref{Zustand_Eichvektorboson}$) und ($\ref{Tetradenzustand}$) organisieren
können, grundsätzlich von der in ($\ref{Zustand_N_Objekte_A}$), ($\ref{Zustand_N_Objekte_B}$)
beziehungsweise ($\ref{Zeitentwicklung_N_Objekte}$) gegebenen grundlegenden Gestalt.
Wenn man ($\ref{Dynamische_Gleichung_Gravitation_Energie-Impuls}$) mit Hilfe von
($\ref{Energieoperator_Darstellung}$), ($\ref{Ort_Impuls_Darstellung}$) und
($\ref{Zustand_Darstellung_Raum-Zeit}$) in die Raum"=Zeit abbildet, dann ergibt sich:

\begin{align}
&\left(\mathcal{D}_\mu^P\right)_{N3}\left(\mathbf{x},t\right) \left(S_\nu^{\ \ \rho\mu}\right)_{N3}\left(\mathbf{x},t\right)
-\left(S^{\mu\rho\sigma}\right)_{N3}\left(\mathbf{x},t\right)\left(T_{\mu\nu\sigma}\right)_{N3}\left(\mathbf{x},t\right)
+\left(K_{\mu\sigma\nu}\right)_{N3}\left(\mathbf{x},t\right)
+\frac{1}{2}T_{N3}\left(\mathbf{x},t\right)\delta^{\rho}_{\nu}\nonumber\\
&=\mathcal{T}^\rho_\nu\left[\left(\Psi_{\Gamma}\right)_{N1}\left(\mathbf{x},t\right)\rangle,
\left(A_\mu^\tau\right)_{N2}\left(\mathbf{x},t\right)\right],
\label{Dynamische_Gleichung_Gravitation_Ortsraum}
\end{align}
wobei aufgrund von ($\ref{Energieoperator_Darstellung}$), ($\ref{Viererimpuls_Tensorraum}$),
($\ref{Ort_Impuls_Darstellung}$), ($\ref{Zustand_Darstellung_Raum-Zeit}$)
und ($\ref{Tetradenzustand}$) gilt:

\begin{equation}
\left(\mathcal{D}_\mu^P\right)_N\left(\mathbf{x},t\right)=\left(E_\mu^m\right)_N\left(\mathbf{x},t\right)\partial_m.
\end{equation}
Auch bezüglich ($\ref{Dynamische_Gleichung_Gravitation_Ortsraum}$) müssen analog zu
($\ref{Dynamik_Eichvektorbosonzustand_Raum-Zeit-Darstellung}$) bezüglich der Betrachtung
der Selbstwechselwirkung beziehungsweise Verschränkung verschiedener Tetradenobjekte
untereinander bei den Produkten mit sich selbst genaugenommen die verschiedenen
Einzelzustände mit unterschiedlichen Besetzungszahlen gekennzeichnet werden.

\section{Symmetriestruktur, Einheit der Wechselwirkungen und Kosmologie}

\subsection{Die Algebra der $E_8$-Symmetriegruppe}

In diesem Kapitel soll gezeigt werden, dass die Quantentheorie der Ur"=Alternativen neben
der von den abstrakten Viererortsoperatoren $X_{ABCD}$ ($\ref{Ortsoperator_ABCD}$) abhängigen
Translationsgruppe, welche das rein quantentheoretische Analogon zur lokalen Translationsgruppe
und damit zur Diffeomorphismengruppe $\mathcal{D}iff\left(\mathcal{M}\right)$ der klassischen
allgemeinen Relativitätstheorie darstellt, aus welcher die quantenlogische Translationseichtheorie
der Gravitation hervorgeht, und den von den abstrakten Viererortsoperatoren $X_{ABCD}$ abhängigen
$SU(N)$"=Symmetrien als Teilsymmetrien der Symmetrie $U(8)$ beziehungsweise $SO(16)$,
aus welchen das quantenlogische Analogon zu den Yang"=Mills"=Eichtheorien der Elementarteilchenphysik
hervorgeht, noch eine weitere fundamentale Symmetrie aufweist. Diese bezieht sich auf beliebige
Transformationen im gegenüber Permutation der einzelnen Ur"=Alternativen symmetrischen Teilraum
des Tensorraumes der Ur"=Alternativen, welche vom Zustand im antisymmetrischen Teilraum abhängen,
und auf beliebige Transformationen im antisymmetrischen Teilraum, welche vom Zustand im
symmetrischen Teilraum allerdings unabhängig sind. Es wird sich zeigen, dass diese
Symmetriegruppe einer bestimmten Art der Erweiterung der $E_8$ entspricht.
Die $E_8$"=Symmetriegruppe ist bekanntlich die größte unter den außergewöhnlichen Lie"=Gruppen
$G_2$, $F_4$, $E_6$, $E_7$ und $E_8$. Sie besteht einerseits aus einer $SO(16)$"=Teilgruppe mit $120$
Generatoren $J_{ij},\ i,j=1,...,16$, und andererseits aus einer zusätzlichen Teilgruppe mit $128$
Generatoren als Komponenten eines $128$"=dimensionalen Spinors $Q_\alpha,\ \alpha=1,...,128$,
der aus $16$ Majorana"=Weyl"=Spinoren besteht. Damit hat die $E_8$"=Symmetriegruppe die Dimension
$248=120 \oplus 128$. Um die Algebra zu formulieren, benötigt man das Analogon zu den gewöhnlichen
Dirac"=Matrizen gegeben in ($\ref{Dirac-Matrizen}$) in Bezug auf $16$ euklidische Dimensionen,
also Größen $\Gamma^i$, $i=1,...,16$, welche die Clifford"=Algebra in Bezug auf einen
$16$"=dimensionalen reellen euklidischen Raum erfüllen:

\begin{equation}
\left\{\Gamma^i,\Gamma^j\right\}=2\delta^{ij},\quad i,j=1,...,16.
\end{equation}
Basierend auf diesen Größen, $J_{ij},\ i,j=1,...,16$, beziehungsweise $Q_\alpha,\ \alpha=1,...,128$,
kann die Algebra der $E_8$"=Gruppe in der folgenden Weise formuliert werden \cite{Cacciatori:2012}:

\begin{eqnarray}
\left[J_{ij},J_{kl}\right]&=&\delta_{il}J_{jk}-\delta_{ik}J_{jl}-\delta_{jl}J_{ik}+\delta_{jk}J_{il},\nonumber\\
\left[J_{ij},Q_\alpha\right]&=&\frac{1}{4}\left[\Gamma_i, \Gamma_j\right]_{\alpha\beta} Q_\beta,\nonumber\\
\left[Q_\alpha,Q_\beta\right]&=&\frac{1}{8}\left[\Gamma^i, \Gamma^j\right]_{\alpha\beta}J_{ij}.
\label{Algebra_E8}
\end{eqnarray}
Die Größen $\Gamma^i$, $i=1,...,16$ können mit Hilfe der gewöhnlichen Dirac"=Matrizen
gegeben in ($\ref{Dirac-Matrizen}$), der Einheitsmatrix in $8$ Dimensionen $\mathbf{1}_8$
und $\gamma_5$ konstruiert werden, wobei $\gamma_5$ wie folgt geschrieben werden kann:

\begin{equation}
\gamma_5=\left(\begin{matrix}\sigma^0 & 0\\ 0 & -\sigma^0\end{matrix}\right).
\end{equation}
Die Größen $\Gamma^i$, $i=1,...,16$, sind also als $128 \times 128$ Matrizen
darstellbar, die in der folgenden Weise definiert sind:

\begin{align}
&\Gamma^1=\gamma^0 \otimes \gamma^0 \otimes \mathbf{1}_8,\nonumber\\
&\Gamma^2=\gamma^0 \otimes \gamma^5 \otimes \gamma^1 \otimes \mathbf{1}_8,\nonumber\\
&\Gamma^3=\gamma^0 \otimes \gamma^5 \otimes \gamma^2 \otimes \mathbf{1}_8,\nonumber\\
&\Gamma^4=\gamma^0 \otimes \gamma^5 \otimes \gamma^3 \otimes \mathbf{1}_8,\nonumber\\
&\Gamma^5=\gamma^5 \otimes \gamma^1 \otimes \gamma^0 \otimes \mathbf{1}_8,\nonumber\\
&\Gamma^6=\gamma^5 \otimes \gamma^1 \otimes \gamma^5 \otimes \gamma^1 \otimes \mathbf{1}_8,\nonumber\\
&\Gamma^7=\gamma^5 \otimes \gamma^1 \otimes \gamma^5 \otimes \gamma^2 \otimes \mathbf{1}_8,\nonumber\\
&\Gamma^8=\gamma^5 \otimes \gamma^1 \otimes \gamma^5 \otimes \gamma^3 \otimes \mathbf{1}_8,\nonumber\\
&\Gamma^9=\gamma^5 \otimes \gamma^2 \otimes \gamma^0 \otimes \mathbf{1}_8,\nonumber\\
&\Gamma^{10}=\gamma^5 \otimes \gamma^2 \otimes \gamma^5 \otimes \gamma^1 \otimes \mathbf{1}_8,\nonumber\\
&\Gamma^{11}=\gamma^5 \otimes \gamma^2 \otimes \gamma^5 \otimes \gamma^2 \otimes \mathbf{1}_8,\nonumber\\
&\Gamma^{12}=\gamma^5 \otimes \gamma^2 \otimes \gamma^5 \otimes \gamma^3 \otimes \mathbf{1}_8,\nonumber\\
&\Gamma^{13}=\gamma^5 \otimes \gamma^3 \otimes \gamma^0 \otimes \mathbf{1}_8,\nonumber\\
&\Gamma^{14}=\gamma^5 \otimes \gamma^3 \otimes \gamma^5 \otimes \gamma^1 \otimes \mathbf{1}_8,\nonumber\\
&\Gamma^{15}=\gamma^5 \otimes \gamma^3 \otimes \gamma^5 \otimes \gamma^2 \otimes \mathbf{1}_8,\nonumber\\
&\Gamma^{16}=\gamma^5 \otimes \gamma^3 \otimes \gamma^5 \otimes \gamma^3 \otimes \mathbf{1}_8.
\label{Gamma-Matrizen_erweitert}
\end{align}
Es wird sich zeigen, dass sich die Algebra der $E_8$ ($\ref{Algebra_E8}$) in einem erweiterten
Rahmen aus den im symmetrischen und antisymmetrischen Teilraum des Tensorraumes der Ur"=Alternativen
wirkenden Generatoren, aus denen sich alle Transformationen zusammensetzen, in natürlicher Weise
ergibt, wenn man zugrundelegt, dass die Transformationen im antisymmetrischen Teilraum unabhängig
sind von denen im symmetrischen Teilraum aber nicht umgekehrt.

\subsection{Konstituierung der $E_8$"=Symmetriegruppe im Tensorraum der Ur-Alternativen}

Ein Zustand im Tensorraum der Ur"=Alternativen repräsentiert ein Quantenobjekt und besteht
gemäß ($\ref{Aufspaltung_symmetrisch_antisymmetrisch}$) aus einem gegenüber Permutation
der einzelnen Ur"=Alternativen symmetrischen und einem gegenüber Permutation der einzelnen
Ur"=Alternativen antisymmetrischen Zustand. Dabei führt der symmetrische Anteil durch die Abbildung
($\ref{Zustand_Darstellung_Ortsraum}$) zu den räumlichen Freiheitsgraden eines Quantenobjektes
und der antisymmetrische Anteil zu den Quantenzahlen, wobei die internen Quantenzahlen eine
relative Ausrichtung der Ur"=Alternativen des antisymmetrischen Zustandes zueinander beschreiben
und der Spin die relative Ausrichtung des antisymmetrischen Zustandes zum symmetrischen Zustand
oder anders ausgedrückt eine relative $SU(2)$"=Transformation aller Ur"=Alternativen des
antisymmetrischen Zustandes relativ zu denen des symmetrischen Zustandes einer
Spin"=Transformation entspricht. Die Erzeugungs- und Vernichtungsoperatoren
des symmetrischen Teilraumes ($\ref{Operatoren_symmetrisch}$), die aus einer
iterierten Quantisierung einer bereits mit komplexen Wahrheitswerten belegten einzelnen
quantentheoretischen Ur"=Alternative ($\ref{Ur-Alternative}$) gemäß den Vertauschungsrelationen ($\ref{Vertauschungsrelationen_UrAlternativen}$) entstehen, können natürlich wie die quantentheoretische
Alternative selbst gemäß ($\ref{Operator_Ur-Alternative}$) auch zu einem Weyl"=Spinor"=Operator
zusammengefasst werden:

\begin{equation}
\hat \varphi_{S}=\left(\begin{matrix} A_S+B_Si\\ C_S+D_Si \end{matrix}\right).
\end{equation}
Aus einem solchen Weyl"=Spinor"=Operator kann in der folgenden Weise ein
Majorana"=Spinor"=Operator konstruiert werden:

\begin{eqnarray}
\hat \varphi_{SM}&=&\left(\begin{matrix} \hat \varphi_{S}\\ i\sigma^2 \hat \varphi_{S}^{*} \end{matrix}\right)
=\left(\begin{matrix} A_S+B_S i\\ C_S+D_S i \\ C_S-D_S i \\ -A_S+B_S i\end{matrix}\right)
=\left(\begin{matrix} A_{SM}\\ B_{SM}\\ C_{SM}\\ D_{SM}\end{matrix}\right).
\label{Majorana_Spinor_Operator}
\end{eqnarray}
Dieser weist die vier unabhängigen Komponenten $A_{SM}$, $B_{SM}$, $C_{SM}$, $D_{SM}$ auf.
Der hermitesch adjungierte Operator zu dem Operator ($\ref{Majorana_Spinor_Operator}$)
ergibt sich dementsprechend zu:

\begin{eqnarray}
\hat \varphi_{SM}^{\dagger}&=&\left(\begin{matrix} \hat \varphi_{S}^{\dagger},
-i\sigma^2 \left(\hat \varphi_{S}^{\dagger}\right)^{*}\end{matrix}\right)
=\left(\begin{matrix} A_S^{\dagger}-B_S^{\dagger}i, & C_S^{\dagger}-D_S^{\dagger}i,
& C_S^{\dagger}+iD_S^{\dagger}, & -A_S^{\dagger}-B_S^{\dagger}i \end{matrix}\right)\nonumber\\
&=&\left(\begin{matrix} A_{SM}^{\dagger}, & B_{SM}^{\dagger}, & C_{SM}^{\dagger}, & D_{SM}^{\dagger}\end{matrix}\right).
\label{Majorana_Spinor_Operator_adjungiert}
\end{eqnarray}
Dieser weist die vier unabhängigen Komponenten $A_{SM}^{\dagger}$, $B_{SM}^{\dagger}$, $C_{SM}^{\dagger}$,
$D_S^{\dagger}$ auf. Und die Vertauschungsrelationen zwischen dem in ($\ref{Majorana_Spinor_Operator}$)
definierten Majorana"=Spinor"=Operator und dem in ($\ref{Majorana_Spinor_Operator_adjungiert}$)
definierten adjungierten Majorana"=Spinor"=Operator lauten natürlich:

\begin{equation}
\left[\left(\varphi_{SM}\right)_s,\left(\varphi_{SM}^{\dagger}\right)_r\right]=\delta_{rs},\quad r,s=1,...,4.
\end{equation}
Da ein allgemeiner Dirac"=Zustand ($\ref{Dirac-Zustand_chi8}$) sowohl eine Komponente
zu positiver alsauch zu negativer Energie aufweist, kann der Majorana"=Spinor"=Operator
($\ref{Majorana_Spinor_Operator}$) verdoppelt werden, indem man eine Komponente zu positiver
und eine Komponente zu negativer Energie betrachtet, wobei die Operatoren der entsprechenden
Komponente nur im jeweiligen Teilraum wirken. Damit ergibt sich der folgende
achtdimensionale Spinor:

\begin{equation}
\left(\hat \varphi_{SM}\right)_{\pm E}
=\left(\begin{matrix}\left[\varphi_{SM}\right]_E\\ \left[\varphi_{SM}\right]_{-E}\end{matrix}\right),
\label{Majorana_Spinor_Operator_verdoppelt}
\end{equation}
beziehungsweise die dazu adjungierte Größe:

\begin{equation}
\left(\hat \varphi_{SM}^{\dagger}\right)_{\pm E}
=\left(\begin{matrix}\left[\varphi_{SM}^{\dagger}\right]_E\\
\left[\varphi_{SM}^{\dagger}\right]_{-E}\end{matrix}\right).
\label{Majorana_Spinor_Operator_verdoppelt_adjungiert}
\end{equation} 
Der gegenüber Permutation der Ur"=Alternativen antisymmetrische Teilraum des Tensorraumes der
Ur"=Alternativen enthält gemäß ($\ref{Linearkombination_antisymmetrische_Basiszustaende}$)
$16$ Basiszustände, entspricht also einem $16$"=dimensionalen reellen Vektorraum. Die Menge
der orthonormalen Drehungen in diesem Raum ist demnach die $SO(16)$ und die entsprechenden
Generatoren, die in Bezug auf den antisymmetrischen Teilraum des Tensorraumes der
Ur"=Alternativen wirken, seien mit $\left(J_{AS}\right)_{ij},\ i,j=1,...,16$ bezeichnet
und erfüllen gemäß ($\ref{Algebra_E8}$) die Vertauschungsrelationen:

\begin{equation}
\left[\left(J_{AS}\right)_{ij},\left(J_{AS}\right)_{kl}\right]
=\delta_{il}\left(J_{AS}\right)_{jk}-\delta_{ik}\left(J_{AS}\right)_{jl}
-\delta_{jl}\left(J_{AS}\right)_{ik}+\delta_{jk}\left(J_{AS}\right)_{il}.
\label{Algebra_JAS}
\end{equation}
Man kann nun den gesamten Tensorraum der Ur"=Alternativen in der Weise in $16$ Teilräume
aufspalten, dass man eine Linearkombination der Tensorprodukte des gesamten symmetrischen
Teilraumes des Tensorraumes der Ur"=Alternativen mit der jeweiligen Komponente in Bezug
auf die Basiszustände im antisymmetrischen Teilraum des Tensorraumes der Ur"=Alternativen
bildet, also ($\ref{Linearkombination_antisymmetrische_Basiszustaende}$) in ($\ref{Zustand_gesamt}$)
verwendet, wobei man gemäß der Lösung ($\ref{Dirac-Zustand_chi8}$) der freien abstrakten
Dirac"=Gleichung im Tensorraum der Ur"=Alternativen ($\ref{Dirac-Gleichung_chi8}$),
jeweils die Aufspaltung in eine Komponente zu positiver und zu negativer
Energie betrachten muss:

\begin{eqnarray}
&&|\Psi_{\pm E} \rangle=|\Psi_{\pm E} \rangle_{S}\otimes |\Psi_{\pm E} \rangle_{AS}\nonumber\\
&&=\sum_{N_{ABCD}=0}^{N}\left(\psi_{\pm E}\right)_S\left(N_{ABCD}\right)|N_{ABCD}\rangle_S \otimes
\left[\left(\alpha_{\pm E}\right)_1 |0,0,0,0\rangle_{AS}+\left(\alpha_{\pm E}\right)_2 |1,1,1,1\rangle_{AS}
\right.\nonumber\\&&\left.\quad\quad\quad\quad
+\left(\alpha_{\pm E}\right)_3 |1,0,0,0 \rangle_{AS}+\left(\alpha_{\pm E}\right)_4 |0,1,0,0 \rangle_{AS}
+\left(\alpha_{\pm E}\right)_5 |0,0,1,0 \rangle_{AS}
\right.\nonumber\\&&\left.\quad\quad\quad\quad
+\left(\alpha_{\pm E}\right)_6 |0,0,0,1 \rangle_{AS}+\left(\alpha_{\pm E}\right)_7 |1,1,1,0\rangle_{AS}
+\left(\alpha_{\pm E}\right)_8 |1,1,0,1\rangle_{AS}
\right.\nonumber\\&&\left.\quad\quad\quad\quad
+\left(\alpha_{\pm E}\right)_9 |1,0,1,1\rangle_{AS}+\left(\alpha_{\pm E}\right)_{10} |0,1,1,1\rangle_{AS}
+\left(\alpha_{\pm E}\right)_{11} |1,1,0,0\rangle_{AS}
\right.\nonumber\\&&\left.\quad\quad\quad\quad
+\left(\alpha_{\pm E}\right)_{12} |1,0,1,0\rangle_{AS}+\left(\alpha_{\pm E}\right)_{13} |1,0,0,1\rangle_{AS}
+\left(\alpha_{\pm E}\right)_{14} |0,1,1,0\rangle_{AS}
\right.\nonumber\\&&\left.\quad\quad\quad\quad
+\left(\alpha_{\pm E}\right)_{15} |0,1,0,1\rangle_{AS}+\left(\alpha_{\pm E}\right)_{16} |0,0,1,1\rangle_{AS}\right].
\label{Tensorprodukt_Aufspaltung_antisymmetrischer_Teilraum}
\end{eqnarray}
In ($\ref{Tensorprodukt_Aufspaltung_antisymmetrischer_Teilraum}$) wird gemäß
($\ref{Dirac-Zustand_chi8}$) die Komponente zu positiver beziehungsweise negativer
Energie des symmetrischen Anteiles des Tensorraumes der Ur"=Alternativen nur mit
der Komponente zu positiver beziehungsweise negativer Energie des antisymmetrischen
Anteiles des Tensorraumes der Ur"=Alternativen multipliziert und umgekehrt.
Basierend auf diesem Zustandsraum kann nun ein $128$"=dimensionalen Spinor
$Q_{\Xi}$ konstruiert werden, dessen Komponenten aus der Bildung des Tensorproduktes
der Komponenten des Majorana"=Spinor"=Operators mit Komponenten zu positiver
beziehungsweise negativer Energie ($\ref{Majorana_Spinor_Operator_verdoppelt}$)
mit den $16$ Basiszuständen des antisymmetrischen Teilraumes hervorgehen und
die $128$ Generatoren entsprechen, welche jeweils eine Transformation im gegenüber
Permutation symmetrischen Teilraum des Tensorraumes der Ur"=Alternativen induzieren,
die vom Zustand im gegenüber Permutation antisymmetrischen Teilraum des Tensorraumes
der Ur"=Alternativen abhängig ist. Dieser $128$"=dimensionale Spinoroperator
$Q_{\Xi}$ kann in der folgenden Weise geschrieben werden:

\begin{eqnarray}
Q_{\Xi}=\left(\begin{matrix}
\left[\hat\varphi_{SM} \left(|0,0,0,0\rangle_{AS}\right)\right]_{\pm E}\\
\left[\hat\varphi_{SM} \left(|1,0,0,0\rangle_{AS}\right)\right]_{\pm E}\\
\left[\hat\varphi_{SM} \left(|0,1,0,0\rangle_{AS}\right)\right]_{\pm E}\\
\left[\hat\varphi_{SM} \left(|0,0,1,0\rangle_{AS}\right)\right]_{\pm E}\\
\left[\hat\varphi_{SM} \left(|0,0,0,1\rangle_{AS}\right)\right]_{\pm E}\\
\left[\hat\varphi_{SM} \left(|1,1,0,0\rangle_{AS}\right)\right]_{\pm E}\\
\left[\hat\varphi_{SM} \left(|1,0,1,0\rangle_{AS}\right)\right]_{\pm E}\\
\left[\hat\varphi_{SM} \left(|1,0,0,1\rangle_{AS}\right)\right]_{\pm E}\\
\left[\hat\varphi_{SM} \left(|0,1,1,0\rangle_{AS}\right)\right]_{\pm E}\\
\left[\hat\varphi_{SM} \left(|0,1,0,1\rangle_{AS}\right)\right]_{\pm E}\\
\left[\hat\varphi_{SM} \left(|0,0,1,1\rangle_{AS}\right)\right]_{\pm E}\\
\left[\hat\varphi_{SM} \left(|1,1,1,0\rangle_{AS}\right)\right]_{\pm E}\\
\left[\hat\varphi_{SM} \left(|1,1,0,1\rangle_{AS}\right)\right]_{\pm E}\\
\left[\hat\varphi_{SM} \left(|1,0,1,1\rangle_{AS}\right)\right]_{\pm E}\\
\left[\hat\varphi_{SM} \left(|0,1,1,1\rangle_{AS}\right)\right]_{\pm E}\\
\left[\hat\varphi_{SM} \left(|1,1,1,1\rangle_{AS}\right)\right]_{\pm E}
\end{matrix}\right).
\label{Q_Operator}
\end{eqnarray}
Da die Vernichtungsoperatoren des symmetrischen Anteiles des Tensorraumes der
Ur"=Alternativen, welche in den einzelnen Komponenten des Majorana"=Spinor"=Operators
mit positiver und negativer Energie ($\ref{Majorana_Spinor_Operator_verdoppelt}$)
enthalten sind, miteinander kommutieren, bilden die $128$ Komponenten des
Spinoroperators $Q_{\Xi}$ ($\ref{Q_Operator}$) gemeinsam mit den Generatoren
der $SO(16)$, $\left(J_{AS}\right)_{ij},\ i,j=1,...,16$ aus ($\ref{Algebra_JAS}$)
in Bezug auf den antisymmetrischen Teilraum des Tensorraumes der Ur"=Alternativen
eine $E_8$"=Algebra:

\begin{eqnarray}
\left[\left(J_{AS}\right)_{ij},\left(J_{AS}\right)_{kl}\right]
&=&\delta_{il}\left(J_{AS}\right)_{jk}-\delta_{ik}\left(J_{AS}\right)_{jl}
-\delta_{jl}\left(J_{AS}\right)_{ik}+\delta_{jk}\left(J_{AS}\right)_{il}\nonumber\\
\left[\left(J_{AS}\right)_{ij},\left(Q_{\Xi}\right)_\alpha\right]
&=&\frac{1}{4}\left[\left(\Gamma_{\Xi}\right)_i,
\left(\Gamma_{\Xi}\right)_j\right]_{\alpha\beta} \left(Q_{\Xi}\right)_\beta\nonumber\\
\left[\left(Q_{\Xi}\right)_\alpha,\left(Q_{\Xi}\right)_\beta\right]
&=&\frac{1}{8}\left[\left(\Gamma_{\Xi}\right)^i,\left(\Gamma_{\Xi}\right)^j\right]_{\alpha\beta} \left(J_{AS}\right)_{ij},
\label{Algebra_E8_Q_JAS}
\end{eqnarray}
wobei die Größen $\left(\Gamma_{\Xi}\right)^i,\ i=1,...,16$ die $128 \times 128$"=Matrizen
aus ($\ref{Gamma-Matrizen_erweitert}$) in Bezug auf den durch den Spinoroperator
($\ref{Q_Operator}$) definierten $128$"=dimensionalen Raum sind. Die Größen
($\ref{Q_Operator}$), welche neben den Größen $\left(J_{AS}\right)_{ij},\ i,j=1,...,16$
die Algebra ($\ref{Algebra_E8_Q_JAS}$) konstituieren, sind insofern besonders, als ihre
Komponenten im Gegensatz zum gewöhnlichen Fall in ($\ref{Algebra_E8}$) selbst Operatoren
darstellen, nämlich die Vernichtungsoperatoren aus ($\ref{Majorana_Spinor_Operator_verdoppelt}$).
Da im Spinoroperator $Q_{\Xi}$ aus ($\ref{Q_Operator}$) nur Vernichtungsoperatoren
enthalten sind, induziert dieser keine vom Zustand des antisymmetrischen Teilraumes des
Tensorraumes der Ur"=Alternativen abhängige Transformationen im symmetrischen Teilraum
des Tensorraumes, welche die Zahl der darin enthaltenen Ur"=Alternativen erhöht.
Daher ist es notwendig, einen adjungierten Spinoroperator $Q^{\dagger}_{\Xi}$
einzuführen, welcher statt den Größen ($\ref{Majorana_Spinor_Operator_verdoppelt}$)
die Größen ($\ref{Majorana_Spinor_Operator_verdoppelt_adjungiert}$) enthält:

\begin{eqnarray}
Q^{\dagger}_{\Xi}=
\left(\begin{matrix}\left[\hat\varphi^{\dagger}_{SM} \left(|0,0,0,0\rangle_{AS}\right)\right]_{\pm E}\\
\left[\hat\varphi^{\dagger}_{SM} \left(|1,0,0,0\rangle_{AS}\right)\right]_{\pm E}\\
\left[\hat\varphi^{\dagger}_{SM} \left(|0,1,0,0\rangle_{AS}\right)\right]_{\pm E}\\
\left[\hat\varphi^{\dagger}_{SM} \left(|0,0,1,0\rangle_{AS}\right)\right]_{\pm E}\\
\left[\hat\varphi^{\dagger}_{SM} \left(|0,0,0,1\rangle_{AS}\right)\right]_{\pm E}\\
\left[\hat\varphi^{\dagger}_{SM} \left(|1,1,0,0\rangle_{AS}\right)\right]_{\pm E}\\
\left[\hat\varphi^{\dagger}_{SM} \left(|1,0,1,0\rangle_{AS}\right)\right]_{\pm E}\\
\left[\hat\varphi^{\dagger}_{SM} \left(|1,0,0,1\rangle_{AS}\right)\right]_{\pm E}\\
\left[\hat\varphi^{\dagger}_{SM} \left(|0,1,1,0\rangle_{AS}\right)\right]_{\pm E}\\
\left[\hat\varphi^{\dagger}_{SM} \left(|0,1,0,1\rangle_{AS}\right)\right]_{\pm E}\\
\left[\hat\varphi^{\dagger}_{SM} \left(|0,0,1,1\rangle_{AS}\right)\right]_{\pm E}\\
\left[\hat\varphi^{\dagger}_{SM} \left(|1,1,1,0\rangle_{AS}\right)\right]_{\pm E}\\
\left[\hat\varphi^{\dagger}_{SM} \left(|1,1,0,1\rangle_{AS}\right)\right]_{\pm E}\\
\left[\hat\varphi^{\dagger}_{SM} \left(|1,0,1,1\rangle_{AS}\right)\right]_{\pm E}\\
\left[\hat\varphi^{\dagger}_{SM} \left(|0,1,1,1\rangle_{AS}\right)\right]_{\pm E}\\
\left[\hat\varphi^{\dagger}_{SM} \left(|1,1,1,1\rangle_{AS}\right)\right]_{\pm E}
\end{matrix}\right).
\label{Q_Operator_adjungiert}
\end{eqnarray}
Da auch die Erzeugungsoperatoren im symmetrischen Teilraum des Tensorraumes
der Ur"=Alternativen miteinander kommutieren, bilden auch die $128$ Generatoren
des Spinoroperators $Q^{\dagger}_{\Xi}$ in ($\ref{Q_Operator_adjungiert}$)
gemeinsam mit den Generatoren der $SO(16)$ $\left(J_{AS}\right)_{ij},\quad i,j=1,...,16$
aus ($\ref{Algebra_JAS}$) in Bezug auf den antisymmetrischen Teilraum eine $E_8$"=Algebra:

\begin{eqnarray}
\left[\left(J_{AS}\right)_{ij},\left(J_{AS}\right)_{kl}\right]&=&
\delta_{il}\left(J_{AS}\right)_{jk}-\delta_{ik}\left(J_{AS}\right)_{jl}
-\delta_{jl}\left(J_{AS}\right)_{ik}+\delta_{jk}\left(J_{AS}\right)_{il}\nonumber\\
\left[\left(J_{AS}\right)_{ij},\left(Q^{\dagger}_{\Xi}\right)_\alpha\right]
&=&\frac{1}{4}\left[\left(\bar \Gamma_{\Xi}^{\dagger}\right)_i,\left(\bar \Gamma_{\Xi}^{\dagger}\right)_j\right]_{\alpha\beta}
\left(Q^{\dagger}_{\Xi}\right)_\beta\nonumber\\
\left[\left(Q^{\dagger}_{\Xi}\right)_\alpha,\left(Q^{\dagger}_{\Xi}\right)_\beta\right]
&=&\frac{1}{8}\left[\left(\bar \Gamma_{\Xi}^\dagger\right)^i,\left(\bar \Gamma_{\Xi}^\dagger\right)^j\right]_{\alpha\beta}
\left(J_{AS}\right)_{ij},
\label{Algebra_E8_Q_adjungiert_JAS}
\end{eqnarray}
wobei die Größen $\left(\Gamma^\dagger_{\Xi}\right)^i,\ i=1,...,16$ die
$128 \times 128$"=Matrizen aus ($\ref{Gamma-Matrizen_erweitert}$) in Bezug auf den
durch den Spinoroperator ($\ref{Q_Operator_adjungiert}$) definierten $128$"=dimensionalen
Raum sind, also die zu den Größen $\left(\Gamma_{\Xi}\right)^i,\ i=1,...,16$ adjungierten
Größen. Da die Vernichtungsoperatoren im symmetrischen Teilraum des Tensorraumes
der Ur"=Alternativen aber nicht mit den entsprechenden Erzeugungsoperatoren
kommutieren, erfüllt der Spinoroperator ($\ref{Q_Operator}$) mit den in ihm
enthaltenen $128$ Generatoren mit dem Spinoroperator ($\ref{Q_Operator_adjungiert}$)
und den darin enthaltenen $128$ Operatoren die folgende Vertauschungsrelation:

\begin{eqnarray}
\left[\left(Q_{\Xi}\right)_\alpha,\left(Q^{\dagger}_{\Xi}\right)_\beta\right]
&=&\frac{1}{8}\left[\left(\Gamma_{\Xi}\right)^i,\left(\bar \Gamma_{\Xi}^{\dagger}\right)^j\right]
_{\alpha\beta}\left(J_{AS}\right)_{ij}
+\delta_{\alpha\beta}.
\label{Vertauschungsrelation_Q-Operator_Q-Operator-adjungiert}
\end{eqnarray}
Die durch die Vertauschungsrelationen in ($\ref{Algebra_E8_Q_JAS}$), ($\ref{Algebra_E8_Q_adjungiert_JAS}$)
und ($\ref{Vertauschungsrelation_Q-Operator_Q-Operator-adjungiert}$) definierte erweiterte
$E_8$"=Symmetriegruppe ist in der abstrakten Dirac"=Gleichung im Tensorraum der Ur"=Alternativen
($\ref{Dirac-Gleichung_chi8}$) zunächst nicht gewährleistet. Sie kann aber wie im Falle
der weiter oben in dieser Arbeit entwickelten quantenlogischen Fassung der
Yang"=Mills"=Eichtheorien und der Translationseichtheorie der Gravitation
durch Einführung zusätzlicher Zustände nachträglich hergestellt werden.
Dass sich eine erweiterte $E_8$"=Symmetriegruppe aus der Quantentheorie
der Ur"=Alternativen ergibt, ist insofern interessant, als sie die größte
exceptionelle Lie"=Gruppe ist und auch unabhängig davon in Bezug auf die
Frage der Vereinheitlichung der theoretischen Physik zu Rate gezogen wurde,
beispielsweise in \cite{Lisi:2010}, \cite{Bars:1980}.

\subsection{Die Einheit aller Wechselwirkungen}

Es erscheint als wichtig, noch einmal deutlich zu formulieren, dass die hier
gegebene rein quantentheoretische Fassung der Beschreibung der $SU(N)$"=Eichtheorien
der Elementarteilchenphysik und der rein quantentheoretischen Fassung der Beschreibung
der Gravitation als abstraktes Analogon zur Translationseichtheorie im Tensorraum
der Ur"=Alternativen sich als Konsequenz einer möglichst hohen Symmetrieforderung
im Tensorraum der Ur"=Alternativen ergeben. Da der Tensorraum der Ur"=Alternativen
gemäß ($\ref{Aufspaltung_symmetrisch_antisymmetrisch}$) selbst aufgespalten ist
in den Teilraum der gegenüber Permutation der einzelnen Ur"=Alternativen
symmetrischen und den gegenüber Permutation der einzelnen Ur"=Alternativen
antisymmetrischen Anteil, was einer Aufspaltung in Bose"=Statistik und
Fermi"=Statistik entspricht, ist auch die möglichst hohe Symmetrie im Tensorraum
der Ur"=Alternativen aufgespalten in einen Teil, der sich auf den symmetrischen und
einen Teil der sich auf den antisymmetrischen Teilraum bezieht. Dieser Aufspaltung
entspricht im Einklang mit den beiden letzten Kapiteln die Aufspaltung der realen
Wechselwirkungen in die Gravitation, welche sich als Eichtheorie auf Raum"=Zeit"=Symmetrien
bezieht und die Wechselwirkungen der Elementarteilchenphysik, welche sich als Eichtheorien
auf interne Symmetrien beziehen. Im Falle der rein quantentheoretischen Formulierung
der Gravitation handelt es sich durch die Abhängigkeit der zu den Komponenten des
Energie"=Impuls"=Operators $P_{ABCD}$ ($\ref{Viererimpuls_Tensorraum}$), die als
Generatoren der Translationsgruppe im Tensorraum der Ur"=Alternativen wirken,
gehörigen Parameter von den Viererortsoperatoren $X_{ABCD}$ ($\ref{Ortsoperator_ABCD}$)
demnach um die größte mögliche Symmetriegruppe im Teilraum der gegenüber Permutation
symmetrischen Zustände im Tensorraum der Ur"=Alternativen. Dies gilt unter Voraussetzung,
dass der mit der Gesamtinformationsmenge $N$ ($\ref{Gesamtzahl_Ur-Alternativen}$) unmittelbar
korrelierte Freiheitsgrad $N_n$ ausgenommen wird. Im Falle der rein quantentheoretischen
Formulierung der $SU(N)$"=Wechselwirkungen der Elementarteilchenphysik handelt es sich
demnach um die größte mögliche Symmetriegruppe im Teilraum der gegenüber Permutation
antisymmetrischen Zustände im Tensorraum der Ur"=Alternativen, wobei im letzteren
Falle aufgrund der Abhängigkeit der zu den $SU(N)$"=Generatoren gehörigen Parametern von
den Viererortsoperatoren $X_{ABCD}$ ($\ref{Ortsoperator_ABCD}$) die vorhandene Symmetrie
sogar zu einer Symmetrie erweitert wird, die den gegenüber Permutation symmetrischen
mit dem gegenüber Permutation antisymmetrischen Teilraum verbindet. Die beiden
entsprechenden Eichtheorien wurden in jeweils einem der letzten beiden Kapitel
betrachtet. Wenn man nun umgekehrt beliebige Transformationen im symmetrischen
Teilraum vom Zustand im antisymmetrischen Teilraum abhängig macht, wobei hierbei
nun alle Freiheitsgrade des symmetrischen Teilraumes vollkommen miteinbezogen sind,
dann erhält man jene erweiterte $E_8$"=Algebra, die durch die in ($\ref{Algebra_E8_Q_JAS}$), ($\ref{Algebra_E8_Q_adjungiert_JAS}$) und ($\ref{Vertauschungsrelation_Q-Operator_Q-Operator-adjungiert}$)
gegebenen Vertauschungsrelationen definiert ist. Man kann nun eine beliebige Transformation
der fundamentalen Symmetriegruppe aller Wechselwirkungen gemäß der Quantentheorie
der Ur"=Alternativen, welche eine sehr allgemeine Symmetriegruppe im Tensorraum
der Ur"=Alternativen darstellt, durch Kombination der allgemeinen Transformationsoperatoren
($\ref{Transformationsoperator_SUN_Ortsoperator}$), ($\ref{Transformationsoperator_Translationen_Ortsoperator}$)
und ($\ref{Transformationsoperator_Spinkonnektion_Ortsoperator}$) erhalten:

\begin{align}
\label{Transformationsoperator_TP}
&U_{\tau P}\left[\alpha^\tau\left(X_{ABCD}\right),\lambda^P\left(X_{ABCD}\right)\right]\\
&=\exp\left[i\alpha^a\left(X_{ABCD}\right)\tau_{8x8}^a\right]\exp\left[i\lambda^m\left(X_{ABCD}\right)\left(P_{ABCD}\right)_m
+i\Sigma^{ab}\left[\lambda^P\left(X_{ABCD}\right)\right]\left(M_{\Gamma}\right)_{ab}\right].\nonumber
\end{align}
Ein allgemeiner allgemeiner Zustand im Tensorraum der Ur"=Alternativen $|\Psi\rangle$ ist durch
($\ref{Zustand_gesamt}$) beziehungsweise ($\ref{Zustand_gesamt_chi8}$) gegeben. Wenn man
Lösungen der abstrakten Dirac"=Gleichung im Tensorraum betrachtet, welche eine Aufspaltung
in positive und negative Energiekomponente beinhalten und zeitabhängig sind, dann ergibt sich
ein Zustand $|\Psi_\Gamma(t)\rangle$, der gemäß ($\ref{Dirac-Zustand_chi8}$) definiert ist.
Symmetrie unter Anwendung des Transformationsoperators ($\ref{Transformationsoperator_TP}$)
auf $|\Psi_\Gamma(t)\rangle$:

\begin{align}
&|\Psi_\Gamma(t)\rangle \longrightarrow U_{\tau P}|\Psi_\Gamma(t)\rangle
=U_{\tau P}\left[\alpha^\tau\left(X_{ABCD}\right),\lambda^P\left(X_{ABCD}\right)\right]|\Psi_\Gamma(t)\rangle\\
&=\exp\left[i\alpha^a\left(X_{ABCD}\right)\tau_{8x8}^a\right]
\exp\left[i\lambda^\mu\left(X_{ABCD}\right)\left(P_{ABCD}\right)_\mu
+i\Sigma^{\mu\nu}\left[\lambda^P\left(X_{ABCD}\right)\right]\left(M_{\Gamma}\right)_{\mu\nu}\right]|\Psi_\Gamma(t)\rangle,
\nonumber
\end{align}
kann nur gewährleistet werden durch Einführung zusätzlicher Zustände, sodass
sich die entsprechende abstrakte Dirac"=Gleichung mit den Wechselwirkungstermen
sich durch Kombination von ($\ref{Dirac-Gleichung_Gamma_Wechselwirkung}$) und
($\ref{Diracgleichung_Wechselwirkung_Gravitation_Spinkonnektion}$) ergibt:

\begin{align}
\gamma_{8x8}^m|E_m^\mu(t)\rangle \left[\left(D_{ABCD}^M\right)_\mu
+\kappa_A|A_\mu^\tau (t)\rangle\right]|\Psi_\Gamma(t)\rangle=0,
\end{align}
wobei $D_{ABCD}^M$ gemäß ($\ref{Impulsoperator_Spinkonnektion}$) definiert ist und die abstrakte
Spinkonnektion enthält. Die dynamischen Grundgleichungen für die Eichvektorbosonenzustände
($\ref{Zustand_Eichvektorboson}$) und für die Tetradenzustände ($\ref{Tetradenzustand}$) sind
definiert durch die quantenlogischen Wirkungen ($\ref{Wirkung_Gravitation_quantenlogisch}$)
($\ref{Wirkung_Dirac_QG}$) und ($\ref{Wirkung_Eichvektorboson_QG}$). Beide Gleichungen sind
invariant unter der gesamten Symmetriegruppe im Tensorraum der Ur"=Alternativen, die durch
Transformationsoperatoren der in ($\ref{Transformationsoperator_TP}$) gegebenen Gestalt
ausgedrückt wird. Die allgemeine Gestalt eines Transformationsoperators in Bezug auf
die durch ($\ref{Algebra_E8_Q_JAS}$), ($\ref{Algebra_E8_Q_adjungiert_JAS}$) und
($\ref{Vertauschungsrelation_Q-Operator_Q-Operator-adjungiert}$)
definierte erweiterte $E_8$"=Algebra ist gegeben durch:

\begin{equation}
U_{Q}\left[\chi^Q,\bar \chi^Q\right]
=\exp\left[i\chi^{\alpha}\left(Q_{\Xi}\right)_\alpha+i\bar \chi^{\alpha}\left(Q_{\Xi}^{\dagger}\right)_\alpha\right].
\label{Transformationsoperator_E8_erweitert}
\end{equation}
In ($\ref{Transformationsoperator_E8_erweitert}$) sind auch Modifikationen der Besetzungszahl $N_n$
und der damit korrelierten Gesamtinformationsmenge $N$ ($\ref{Gesamtzahl_Ur-Alternativen}$) enthalten.
Symmetrie unter Anwendung des Transformationsoperators ($\ref{Transformationsoperator_E8_erweitert}$)
auf $|\Psi_\Gamma(t)\rangle$:

\begin{equation}
|\Psi_\Gamma(t)\rangle \longrightarrow U_{Q}|\Psi_\Gamma(t)\rangle
=U_{Q}\left[\chi^Q,\bar \chi^Q\right]|\Psi_\Gamma(t)\rangle=\exp\left[i\chi^{\alpha}\left(Q_{\Xi}\right)_\alpha
+i\bar \chi^{\alpha}\left(Q_{\Xi}^{\dagger}\right)_\alpha\right]
|\Psi_\Gamma(t)\rangle,
\end{equation}
kann ebenfalls nur durch Einfügung zusätzlicher Zustände in die abstrakte Dirac"=Gleichung
im Tensorraum der Ur"=Alternativen ($\ref{Dirac-Gleichung_chi8}$) erreicht werden:

\begin{equation}
\gamma_{8x8}^\mu\left[\left(P_{ABCD}\right)_\mu+i\kappa_Q|\left(A_Q\right)_\mu^\alpha(t)\rangle \left(Q_{\Xi}\right)_\alpha
+i\kappa_{Q^{\dagger}}|\left(A_{Q^{\dagger}}\right)_\mu^\alpha(t)\rangle \left(Q_{\Xi}^{\dagger}\right)_\alpha
\right]|\Psi_\Gamma(t)\rangle=0,
\end{equation}
wobei die Dynamik der Zustände $|\left(A_Q\right)_\mu^\alpha(t)\rangle$ und
$|\left(A_{Q^{\dagger}}\right)_\mu^\alpha(t)\rangle$ hier nicht behandelt
werden soll und $\kappa_Q$ sowie $\kappa_{Q^{\dagger}}$ reelle Parameter sind.

\subsection{Die umfassende Symmetriestruktur der Quantentheorie der Ur-Alternativen}

Die Symmetriestruktur der Quantentheorie der Ur"=Alternativen kann ähnlich wie der
grundlegende Weg der Rekonstruktion in einem Baumdiagramm veranschaulicht werden:

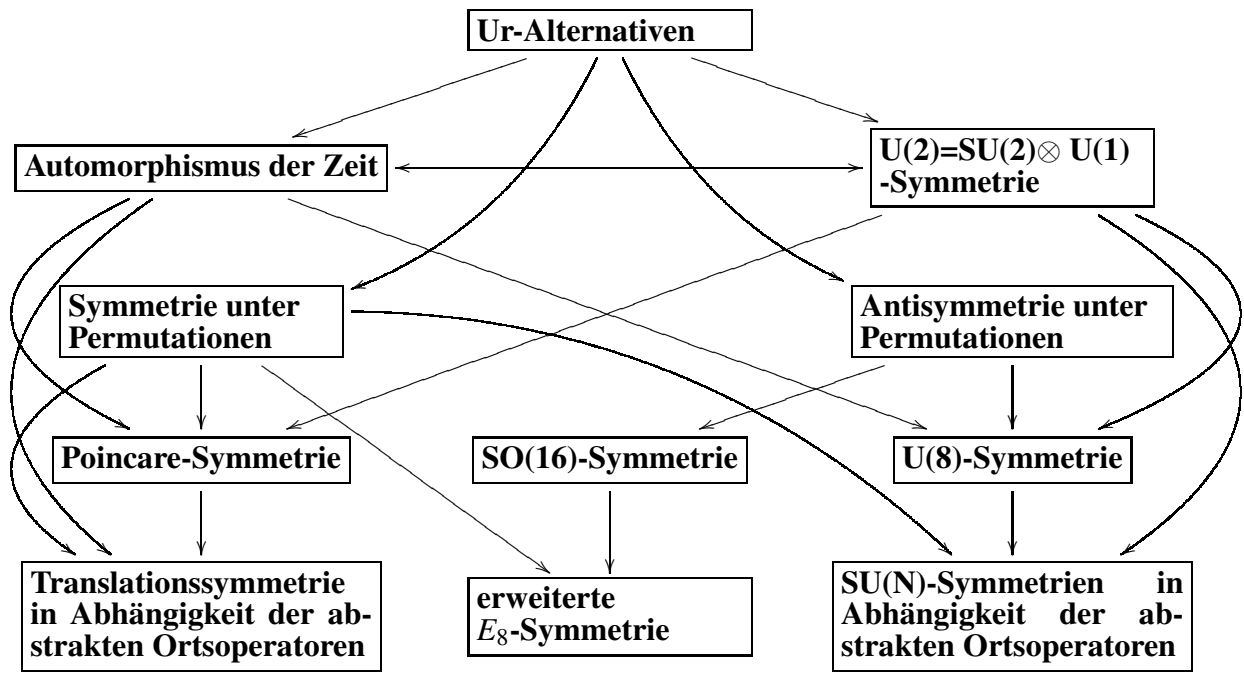
\fontsize{12}{12}\selectfont
\begin{center}
\begin{xy}
\xymatrix{&\fbox{\parbox[c]{35mm}{\textbf{Ur-Alternativen}}}\ar[dr]\ar[dl]
\ar@/_1cm/[ddr]\ar@/^1cm/[ddl]&\\
\fbox{\textbf{Automorphismus der Zeit}}\ar[rr]\ar[ddrr]\ar@/_2.5cm/[dd]\ar@/_2.5cm/[ddd]&&
\fbox{\parbox[c]{35mm}{\textbf{U(2)=SU(2)$\otimes$ U(1)\\-Symmetrie}}}
\ar[ll]\ar@/^3cm/[ddd]\ar@/^3cm/[dd]\ar[ddll]\\
\fbox{\parbox[c]{35mm}{\textbf{Symmetrie unter\\ Permutationen}}}
\ar@/^1.5cm/[rrdd]\ar[rdd]\ar[d]\ar@/_2.5cm/[dd]& &
\fbox{\parbox[c]{40mm}{\textbf{Antisymmetrie unter\\ Permutationen}}}\ar[d]\ar[ld]\\
\fbox{\textbf{Poincare-Symmetrie}}\ar[d]
&\fbox{\textbf{SO(16)-Symmetrie}}\ar[d] &\fbox{\textbf{U(8)-Symmetrie}}\ar[d]
\\\fbox{\parbox[c]{45mm}{\textbf{Translationssymmetrie in Abhängigkeit der abstrakten Ortsoperatoren}}}
&\fbox{\parbox[c]{35mm}{\textbf{erweiterte\\ $E_8$-Symmetrie}}}&
\fbox{\parbox[c]{45mm}{\textbf{SU(N)-Symmetrien in Abhängigkeit der abstrakten Ortsoperatoren}}}}
\end{xy}
\end{center}

\fontsize{12.8pt}{14.8pt}\selectfont

\subsection{Die Struktur des Kosmos und das Hierarchieproblem}

Es wurde in dieser Arbeit gezeigt, dass man die gegenüber Permutation symmetrischen
Zustände im Tensorraum der Ur"=Alternativen in einen dreidimensionalen reellen Ortsraum
abbilden kann, der daher mit dem physikalischen Ortsraum identifiziert wurde.
Quantenobjekte setzen sich aus Ur"=Alternativen zusammen und stellen sich indirekt
im dreidimensionalen Ortsraum dar. Die Frage der Struktur des Kosmos als einem
Ganzen wurde aber bisher noch nicht thematisiert. Carl Friedrich von Weizsäcker
interpretierte die $SU(2)$"=Gruppe, neben der $U(1)$"=Gruppe die fundamentale Symmetriegruppe
einer Ur"=Alternative, welche einer $\mathbb{S}^3$ entspricht, als die räumliche
Struktur des gesamten Kosmos. Demnach wiese der Kosmos eine kompakte Topologie auf.
Die quantentheoretischen Objekte, die wir in der Natur untersuchen, können aber
nur aus vielen Ur"=Alternativen gebildet werden. Demnach muss eine Beziehung des
Tensorraumes vieler Ur"=Alternativen zur räumlichen Struktur des Kosmos hergestellt werden.
Aus der Abbildung der abstrakten Zustände im Tensorraum der Ur-Alternativen in den
Ortsraum beziehungsweise in die Raum"=Zeit, wie sie in ($\ref{Zustand_Darstellung_Ortsraum}$)
beziehungsweise in ($\ref{Zustand_Darstellung_Raum-Zeit}$) vollzogen wird, ergibt sich
für die globale räumliche Struktur des Kosmos eine nicht"=kompakte Topologie, was einem
offenen Kosmos entspricht. Denn die Wellenfunktionen laufen in alle Richtungen in
die Unendlichkeit. Allerdings sind sie normiert und divergieren nicht, sodass der
Kosmos bei endlicher Informationsmenge trotz seiner offenen Struktur faktisch
dennoch von endlicher Größe ist. Dies deckt sich wohl mit den Schlüssen, die man
im Hinblick auf gegenwärtig zugängliche empirische Hinweise bezüglich der globalen
Struktur des Kosmos gezogen hat. Genaugenommen sind alle Objekte im Kosmos, die jeweils
aus einem symmetrischen und einem antisymmetrischen Teilzustand zusammengesetzt sind,
in eine Gesamtbeziehung aus Verschränkungen eingebunden und daher nicht als einzelne
Objekte voneinander trennbar. Die Trennbarkeit der Objekte ist nur eine Näherung,
in der sich das Phänomen der Wechselwirkung als Konsequenz der Verschränkungen ergibt.
Der Kosmos müsste demnach eigentlich durch einen kohärenten Quantenzustand beschrieben
werden, der aus miteinander verschränkten Quanteninformationszuständen als Darstellung
einer abstrakten quantenlogischen Beziehungsstruktur besteht.

Carl Friedrich von Weizsäcker unternahm desweiteren eine Abschätzung in Bezug auf die
Zahl der Ur"=Alternativen, die in einem gewöhnlichen Quantenobjekt enthalten sind und
in Bezug auf die Gesamtmenge der Ur"=Alternativen des gesamten Kosmos. Im Rahmen dieser
Abschätzung wird nun die Räumlichkeit des Kosmos in die Betrachtung genommen, um von dort
aus auf die Informationsmenge zu schließen. Dies erscheint insofern nicht ganz konsequent,
als die Information in Gestalt der Ur"=Alternativen ja gerade noch keinen physikalischen
Ortsraum voraussetzt, sondern diesen indirekt begründet. Aber wenn die Exsietnz des
physikalischen Ortsraumes als eines Darstellungsmediums ersteinmal hergeleitet ist,
dann kann man natürlich als eine Art Rückkopplung auch eine Beziehung zwischen der
räumlichen Struktur des Kosmos und der Informationsmenge herzustellen versuchen.
Dies kann nun in der folgenden Weise geschehen. Der Durchmesser des Kosmos, in der
Quantentheorie der Ur"=Alternativen wäre das der Bereich im dreidimensionalen Ortsraum,
in dem Wellenfunktionen als Darstellung der Tensorraumzustände einen von null
signifikant unterschiedenen Wert aufweisen, wird auf etwa $10^{40}$ Comptonwellenlängen
des Protons geschätzt, also die Ortsunbestimmtheit, mit der man ein Proton mit Hilfe
eines Photons lokalisieren kann. Nun kann man die räumliche Lokalisierung eines Protons
dadurch bestimmen, dass man in allen drei Raumrichtungen angibt, in welchem der $10^{40}$
Teilbereiche dieser Raumdimension sich ein Proton befindet. Das bedeutet, dass der Zustand
des Protons in Bezug auf den Raum dadurch definiert ist, dass ein Teilbereich besetzt ist
und alle anderen nicht. Das sind also in jeder Raumdimension $10^{40}$ Informationseinheiten,
also $3\times 10^{40}$, wobei der Faktor drei bei dieser groben Abschätzung vernachlässigt
werden kann. Nun kann man den gesamten Kosmos in dreidimensionale Zellen mit der Kantenlänge
der Comptonwellenlänge des Protons einteilen. Da es hier nicht mehr um den Zustand eines
einzelnen Protons geht, kann nicht nur eine Zelle besetzt sein, sondern alle können besetzt
oder unbesezt sein, womit jeder Zelle eine Informationseinheit entspricht, und dies entspricht
der Zahl $\left(10^{40}\right)^3=10^{120}$. Dem entspricht dann die Informationsmenge des
ganzen Kosmos. Wenn man nun die Informationsmenge des ganzen Kosmos durch die
Informationsmenge eines einzelnen Protons dividiert, also $10^{120}$ durch $3\times 10^{40}$,
so erhält man bezüglich der Größenordnung in etwa $10^{80}$, was in etwa der geschätzten
Zahl der Nukleonen im Kosmos entspricht.

Wenn man desweiteren bedenkt, dass der antisymmetrische Teilraum des Tensorraumes nur
maximal vier Ur"=Alternativen enthält und der symmetrische Teilraum des Tensorraumes
gemäß der obigen Abschätzung bei einem gewöhnlichen Quantenobjekt in etwa $10^{40}$
Ur"=Alternativen, dann kann man das Verhältnis bezüglich der abgeschätzten Zahl
der Ur"=Alternativen, die in dem räumlichen Freiheitsgrad eines Quantenobjektes in
etwa enthalten sind und der Zahl der Ur"=Alternativen, die in seinen inneren Freiheitsgraden
enthalten ist, auf etwa $10^{40}$ schätzen. Dies entspricht bezüglich der Größenordnung
erstaunlicherweise ziemlich gut dem Verhältnis der Stärke des Elektromagnetismus zur Stärke
der Gravitation. Dies könnte damit zu tun haben, dass die Gravitation sich als Eichtheorie
in Bezug auf die abstrakte von $X_{ABCD}$ abhängige Translationsgruppe formulieren lässt,
die sich auf den symmetrischen Teilraum des Tensorraumes bezieht, und der Elektromagnetismus
wie die anderen Wechselwirkungen der Elementarteilchenphysik als Eichtheorie in Bezug auf
die abstrakten von $X_{ABCD}$ abhängigen $SU(N)$"=Gruppen, die sich auf den antisymmetrischen
Teilraum des Tensorraumes beziehen. Wenn man diesen Zusammenhang noch genauer aufklären könnte,
dann hätte man eine Erklärung für das Hierarchieproblem gefunden.

\section{Zusammenfassung und Diskussion}

In dieser Arbeit wurde zunächst in einer zusammenfassenden Weise die begriffliche Basis der
Quantentheorie der Ur"=Alternativen des Carl Friedrich von Weizsäcker dargestellt. Demgemäß
besteht die einfachste, abstrakteste und fundamentalste Art und Weise, die Natur im menschlichen
Geist zu objektivieren, in einer Schematisierung der Erfahrungswirklichkeit durch abstrakte
quantentheoretische Alternativen, wobei den einzelnen Elementen dieser Alternativen
Wahrscheinlichkeiten zugeordnet werden. Mit einigen Postulaten führt dies zu der
abstraktesten und allgemeinsten Fassung der Quantentheorie als Theorie des Hilbertraumes,
die in diesem begrifflichen Rahmen als eine Darstellung logischer Beziehungsstrukturen in der
Zeit verstanden wird. Die elementarsten überhaupt denkbaren quantenlogischen Einheiten
als fundamentalste Einheiten der Objektivation der physikalischen Wirklichkeit im
menschlichen Geist, in welche sich alle quantenlogischen Beziehungsstrukturen auflösen lassen,
sind die Ur"=Alternativen. Basierend auf dem Begriff der Ur"=Alternative können Zustände
aus vielen Ur"=Alternativen gebildet werden, welche einen Tensorraum bilden.
Wenn man eine einzelne Alternative beziehungsweise einen dieser Alternative entsprechenden
Tensorraumzustand betrachtet, so muss dieser zunächst als näherungsweise trennbar vom Rest
des Kosmos angesehen werden, sodass die empirische Entscheidung dieser Alternative unabhängig
von der Entscheidung aller anderen Alternativen ist. In Wirklichkeit stehen die Alternativen aber
in einem Beziehungsgeflecht, sodass die entsprechenden Zustände miteinander verschränkt sind.
Jene Verschränkung hat das Phänomen zur Konsequenz, dass wir in einer Näherung als Wechselwirkung
voneinander trennbarer Objekte beschreiben. In Wirklichkeit sind alle Alternativen in einen
einzigen kosmischen Quantenzustand eingebunden, aber wenn einzelne Quantenobjekte als
näherungsweise voneinander trennbar beschrieben werden, so hat dies das Phänomen
der Wechselwirkung zur Konsequenz.

Bereits in \cite{Kober:2017} wurde gezeigt, dass man Zustände im Tensorraum der Ur"=Alternativen
als Wellenfunktionen in einem reellen dreidimensionalen Raum darstellen kann, was unter
Einbeziehung der Zeitentwicklung als eines einparametrigen Automorphismus des Tensorraumes
vieler Ur"=Alternativen zur Raum"=Zeit"=Struktur der Relativitätstheorie führt.
In der hier vorliegenden Arbeit wurde zunächst eine Aufspaltung in einen unter Permutation
der Ur"=Alternativen symmetrischen und antisymmetrischen Anteil vorgenommen.
Die Zustände des symmetrischen Teilraumes, welche beliebig viele Ur"=Alternativen
enthalten können, sind jene Zustände, welche gemäß \cite{Kober:2017} in einen
reellen dreidimensionalen Raum abgebildet werden können. Die antisymmetrischen Zustände,
welche aufgrund des Paulischen Ausschließungsprinzips nur maximal vier Ur"=Alternativen
enthalten können, nämlich eine in jedem der vier Basis"=Zustände einer einzelnen Ur"=Alternative,
wurden im Rahmen dieser Arbeit mit den diskreten Quantenzahlen der Elementarteilchenphysik
identifiziert, also des Spin und der inneren Symmetrien, auf welche sich die Eichtheorien
der Elementarteilchenphysik beziehen. Dies erklärt die Aufspaltung in äußere Symmetrien,
welche sich auf die Raum"=Zeit beziehen, und innere Symmetrien, welche sich auf die
Quantenzahlen beziehen, wobei der Spin die relative Ausrichtung der antisymmetrischen
gegenüber den symmetrischen Zuständen beschreibt. Damit ist gezeigt, dass auch hinter
der kontinuierlichen Raum"=Zeit mit ihren räumlichen Kausalitätsverhältnissen
letztendlich abstrakte Quantenzahlen stehen, also eine rein quantenlogische Realität.
Zudem wird dadurch sehr wahrscheinlich das Hierarchieproblem gelöst, da die
Differenz der Stärke der Gravitation zu den anderen Wechselwirkungen wohl mit der
viel höheren Informationsmenge im symmetrischen gegenüber dem antisymmetrischen
Teilraum der einzelnen Objekte zu tun hat.

Wenn man die zu diesen Symmetrien gehörigen Transformationen, also abstrakte quantenlogisch
dargestellte Translationen im Tensorraum der Ur"=Alternativen und die $SU(N)$"=Symmetrien
der Quantenzahlen der Elementarteilchenphysik, auf Symmetrietransformationen erweitert,
welche von den aus Erzeugungs- und Vernichtungsoperatoren im Tensorraum der Ur"=Alternativen
konstruierten Viererortsoperatoren $X_{ABCD}$ abhängen, so ergibt sich eine zur gewöhnlichen
Fassung der Gravitation gemäß der allgemeinen Relativitätstheorie und der Wechselwirkungen der
gewöhnlichen Elementarteilchenphysik analoge Struktur der dynamischen Grundgleichungen.
Der entscheidende Unterschied besteht allerdings darin, dass man es hier nicht mit
punktweisen Produkten von Feldern, sondern mit abstrakten Produkten zwischen Zuständen
im Tensorraum der Ur"=Alternativen zu tun hat, die zu unterschiedlichen Objekten gehören.
In diesen Sinne wurden in dieser Arbeit die fundamentalen Wechselwirkungen der Natur in
einem durch reine Quantenlogik charakterisierten begrifflichen Rahmen vereinheitlicht.
Das Phänomen der Wechselwirkung zwischen verschiedenen Objekten wird dabei jedoch durch
das Konzept der Verschränkung der Zustände ersetzt. Dabei verschmelzen in gewissem Sinne
zwei oder mehr Objekte zu einem einzigen Objekt, bezüglich dessen die Zustände, in denen
die Einzelobjekte als voneinander separierbare Objekte existieren, eine Menge des
Maßes null ist. Demnach bestimmt die Struktur der dynamischen Grundgleichungen,
welche durch die Anwendung des Eichprinzips im Tensorraum induziert wird, die Art
und Weise der Verschränkung der Zustände quantentheoretischer Objekte, welche damit
in eine einheitliche abstrakte quantenlogische Beziehungsstruktur eingebunden sind.

Das Entscheidende an all diesen Betrachtungen ist, dass sie sich alle auf die
abstrakten Informationsräume der Quantentheorie der Ur"=Alternativen beziehen und noch
keinerlei Raum"=Zeit, keinerlei Hintergrundstruktur überhaupt und damit keinerlei
feldtheoretische Begriffe vorausgesetzt werden. Die ganze Welt spannt sich wie bei
Hegel über der reinen Logik auf, genauer der reinen Quantenlogik. Es existiert neben
der Zeit zunächst wirklich nur die Logik, die im Nichts schwebt. Dies entspricht exakt
der Behauptung der christlichen Theologie, dass Gott die Welt aus dem Logos erschaffen habe.
Mit den Betrachtungen dieser Arbeit ist damit ein wesentlicher Beitrag dazu geleistet,
die in der Quantentheorie des Ur"=Alternativen seitens Carl Friedrich von Weizsäcker
vollzogene "`Kopernikanische Wende"' in der begrifflichen Basis der theoretischen Physik
auch in der mathematischen Ausformulierung zur Vollendung zu führen. Diese "`Kopernikanische
Wende"' besteht darin, dass sich in der Natur, so wie sie in unserem menschlichen Geist
objektiviert werden kann, nicht voneinander trennbare Objekte in einem vorgegebenen
Raum bewegen und miteinander wechselwirken, sondern dass umgekehrt durch reine Logik
charakterisierte Objekte in einer abstrakten quantentheoretischen Beziehungsstruktur
zueinander stehen, die nur indirekt räumlich dargestellt werden kann. Die Wirklichkeit
auf der fundamentalen Ebene ist demnach also nicht durch Geometrie, sondern durch
Logik charakterisiert, da die Geometrie sich als Ausfluss der Logik ergibt. 
Nicht der Raum existiert für sich, in dem sich die Objekte befinden, sondern
die Objekte existieren für sich, und der Raum ergibt sich als Konsequenz.
Diese Erkenntnis ist zwar sowohl in der allgemeinen Relativitätstheorie als
auch in der gewöhnlichen Quantentheorie implizit enthalten und wird auch seitens
Einstein und Heisenberg explizit benannt, wird aber erst bei von Weizsäcker in
diesem rein quantenlogischen Rahmen ohne jede Annahme räumlicher Kausalitätsstrukturen
begrifflich wirklich konsequent realisiert.

Natürlich müssen noch viele Probleme gelöst werden. Ich vermute, dass man zeigen kann,
dass sich bei störungstheoretischen Betrachtungen herausstellen wird, dass sich aufgrund
der Diskretheit der Basiszustände des Tensorraumes der Ur"=Alternativen keine unendlichen
Werte ergeben. Denn die Grundidee der Quantentheorie bestand ja in der seitens Planck
postulierten Diskretisierung der Energiezustände, um die Entstehung unendlicher Werte
zu vermeiden. Dies bedeutete, dass das Verfahren der Renormalisierung obsolet würde
und sich die Frage der Renormierbarkeit einer quantentheoretischen Formulierung der
Gravitation a priori gar nicht mehr stellen würde. Aber das ist bisher nur eine
Vermutung, die man noch konkret mathematisch beweisen muss. Zudem ist die Frage
der Entstehung der Massen durch das Higgsboson im Rahmen der Quantentheorie der
Ur"=Alternativen überhaupt noch nicht wirklich behandelt worden. Es könnte sein,
dass die Wechselwirkung beziehungsweise Verschränkung der Zustände des Higgsbosons
mit sich selbst und den anderen Quantenobjekten mit Symmetriebetrachtungen zusammenhängt,
die zwischen jenen Teilsymmetriegruppen vermitteln, die hier mit den verschiedenen
Sektoren der Elementarteilchenphysik identifiziert wurden. Auch dies ist bisher
nur eine Vermutung. Schließlich müssen aus der Quantentheorie der Ur"=Alternativen
irgendwann nach Möglichkeit konkrete Phänomene etwa in Zusammenhang mit der rein
quantentheoretischen Beschreibung der Gravitation vorhergesagt und beschrieben werden,
die bisher noch nicht bekannt sind. Aufgrund der Planck"=Skala ist dies wahrscheinlich
zunächst eine sehr schwierige Aufgabe. Dies ändert aber überhaupt nichts daran,
dass die Quantentheorie der Ur"=Alternativen aus vielen anderen Gründen, begrifflichen,
mathematischen und phänomenologischen, eigentlich als unumgänglich erscheint. Sie liefert
als einzige Theorie eine wirklich befriedigende Erklärung für das EPR"=Paradoxon und die
damit verbundene Überschreitung räumlich"=kausaler Beziehungen. Sie kann die Existenz
eines dreidimensionalen Ortsraumes herleiten. Sie basiert auf einem aus logischen Gründen
fundamentalen und unteilbaren Objekt. Sie ist in ihren Grundannahmen denkbar sparsam,
indem sie im Wesentlichen nur Zeit und Logik voraussetzt und ein paar Zusatzpostulate,
die sich direkt darauf beziehen. Und sie kann schließlich die Aufspaltung in die
raum"=zeitlichen und die inneren Symmetrien daraus mathematisch exakt begründen
und damit zugleich die Existenz der Gravitation und der $SU(N)$"=Wechselwirkungen
in einer rein quantentheoretischen Fassung. Damit sind diese Wechselwirkungen im Sinne
einer quantenlogischen Symmetriestruktur und Verschränkungsstruktur vereinheitlicht.

Deshalb scheint mir aufgrund der Überzeugungskraft der seitens Carl Friedrich von
Weizsäcker geschaffenen begrifflichen Basis der Quantentheorie der Ur"=Alternativen
und der in dieser Arbeit durchgeführten erweiterten mathematischen darauf
basierenden Betrachtungen gezeigt zu sein, dass die gewöhnliche Quantentheorie als
eine allgemeine Theorie der Dynamik beliebiger Objekte, die Relativitätstheorie als
Theorie von Raum und Zeit sowie der Gravitation und die konkrete quantentheoretische
Beschreibung der speziellen Objekte und Wechselwirkungen der Elementarteilchenphysik
im Prinzip in der Quantentheorie der Ur"=Alternativen vereinheitlicht werden.
Und dies bedeutet, dass sich die theoretische Physik in ihren grundlegenden
Strukturen aus reiner Quantenlogik in der Zeit ergibt.

\textbf{Danksagung:} Ich danke zunächst Bernd Henschenmacher für anregende und hilfreiche
Diskussionen über algebraische und begriffliche Grundfragen der Quantentheorie. Ich danke
desweiteren Thomas Görnitz, dem letzten treuen Mitarbeiter Carl Friedrich von Weizsäckers,
für Gedanken und Anregungen, die für die Entstehung dieser Arbeit von entscheidender
Bedeutung waren und über einige seiner zahlreichen wichtigen Beiträge zur Quantentheorie
der Ur"=Alternativen in diese Arbeit eingeflossen sind. Und ich danke schließlich meinen
Eltern Ute Kober und Karl"=Heinz Kober für ihre jahrelange finanzielle Unterstützung,
ohne welche diese Arbeit niemals hätte entstehen können und ohne welche die jahrelangen
vorbereitenden in Einsamkeit geführten Studien über die philosophischen Grundlagen
der Quantentheorie, die dazu unentbehrlich waren, niemals möglich gewesen wären.

\end{document}